\documentclass[12pt]{article}
\usepackage[utf8]{inputenc}
\usepackage{caption}
\usepackage{amsmath}
\usepackage{booktabs}
\usepackage{comment}
\usepackage{natbib}
\usepackage{subcaption}
\usepackage{setspace}
\usepackage[ruled,linesnumbered]{algorithm2e}
\usepackage{xcolor}
\usepackage{float}
\usepackage{graphicx,amsfonts,amsthm,amssymb}
\usepackage[colorlinks,citecolor=blue,urlcolor=blue]{hyperref}
\usepackage{amsmath}
\usepackage{mathrsfs}
\usepackage{bm}
\usepackage{wrapfig,lipsum,booktabs}
\usepackage{appendix}
\usepackage{enumitem}
\usepackage{multirow}
\usepackage{natbib}
\usepackage[font=small,skip=0pt]{caption}
\usepackage{algpseudocode} 
\newtheorem{definition}{Definition}
\newcommand{\blind}{1}

\def\spacingset#1{\renewcommand{\baselinestretch}%
{#1}\small\normalsize} \spacingset{1}

\newcommand{\pr}{\mathbb{P}}
\newcommand{\E}{\mathbb{E}}

\newcommand{\1}{{\rm 1}\kern-0.24em{\rm I}}
\newcommand{\R}{{\rm I}\kern-0.18em{\rm R}}

\newcommand{\cS}{\mathcal{S}}

\newcommand{\bX}{\mathbf{X}}
\newcommand{\bx}{\mathbf{x}}
\newcommand{\bY}{\mathbf{Y}}

\newcommand{\by}{\mathbf{y}}
\newcommand{\bI}{\mathbf{I}}

\newcommand{\diag}{\mathrm{diag}}

\newcommand*{\tran}{^{\mkern-1.5mu\mathsf{T}}}

\usepackage[bottom]{footmisc}
\newtheorem{theorem}{Theorem}
\newtheorem{example}{Example}
\newtheorem{assumption}{Assumption}

\newtheorem{lemma}{Lemma}

\newtheorem{simsetting}{Simulation Setting}

\def\bSigma{\boldsymbol\Sigma}
\def\bTheta{\boldsymbol\Theta}
\def\bbeta{\boldsymbol\beta}
\def\bnu{\boldsymbol\nu}

\def\bvarepsilon{\boldsymbol\varepsilon}

\renewcommand{\appendix}{
 \setcounter{section}{0}%
  \setcounter{subsection}{0}%
  \renewcommand\thesection{\Alph{section}}
  \setcounter{equation}{0}
  \renewcommand{\theequation}{S.\arabic{equation}}
  \setcounter{figure}{0}
  \renewcommand\thefigure{S\arabic{figure}}
  \setcounter{table}{0}
  \renewcommand\thetable{S\arabic{table}}  
  \renewcommand{\thelemma}{S\arabic{lemma}}
  }

\allowdisplaybreaks

\begin{document}

\if1\blind
{
  %\title{Nullstrap: An original-data-preservation  Framework for High-Power False Discovery Rate Control in High-dimensional Variable Selection}

  \title{
  Nullstrap: A Simple, High-Power, and Fast Framework for FDR Control in Variable Selection for Diverse High-Dimensional Models
  % Synthetic Null Parallelism (Nullstrap): A Simple, High-Power, and Fast Framework for FDR Control in Variable Selection for Diverse High-Dimensional Models
  }

  \author{
%  	Xinzhou Ge   \hspace{.2cm}\\
%    Dongyuan Song \hspace{.2cm}
%    \\
   Changhu Wang \hspace{.2cm}\\
   Department of Statistics and Data Science \\ University of California, Los Angeles \\
   Ziheng Zhang \hspace{.2cm}\\
   Department of Biostatistics 
    \\ University of California, Los Angeles \\
%%    Kexin Li \hspace{.2cm}
%    \\
   Jingyi Jessica Li\hspace{.1cm} \thanks{Correspondence should be addressed to Jingyi Jessica Li (jli@stat.ucla.edu, lijy03@fredhutch.org)}\\
 Department of Statistics and Data Science \\ University of California, Los Angeles \\
 Biostatistics
Program, Public Health Science
Division \\ Fred Hutchinson Cancer
Center
}
  \date{}
  \maketitle
} \fi

\if0\blind
{
 \title{
  Nullstrap: A Simple, High-Power, and Fast Framework for FDR Control in Variable Selection for Diverse High-Dimensional Models

  % Synthetic Null Parallelism (Nullstrap): A Simple, High-Power, and Fast Framework for FDR Control in Variable Selection for Diverse High-Dimensional Models
  }
    \date{}
  \maketitle
} \fi

\maketitle

\def\spacingset#1{\renewcommand{\baselinestretch}%
{#1}\small\normalsize} \spacingset{1}

%%%%%%%%%%%%%%%%%%%%%%%%%%%%%%%%%%%%%%%%%%%%%%%%%%%%%%%%%%%%%%%%%%%%%%%%%%%%%%

\begin{abstract}
Balancing false discovery rate (FDR) control with high statistical power remains a central challenge in high-dimensional variable selection. While several FDR-controlling methods have been proposed, many degrade the original data—by adding knockoff variables or splitting the data—which often leads to substantial power loss and hampers detection of true signals. We introduce Nullstrap, a novel framework that controls FDR without altering the original data. Nullstrap generates synthetic null data by fitting a null model under the global null hypothesis that no variables are important. It then applies the same estimation procedure in parallel to both the original and synthetic data. This parallel approach mirrors that of the classical likelihood ratio test, making Nullstrap its numerical analog. By adjusting the synthetic null coefficient estimates through a data-driven correction procedure, Nullstrap identifies important variables while controlling the FDR. We provide theoretical guarantees for asymptotic FDR control at any desired level and show that power converges to one in probability. Nullstrap is simple to implement and broadly applicable to high-dimensional linear models, generalized linear models, Cox models, and Gaussian graphical models. Simulations and real-data applications show that Nullstrap achieves robust FDR control and consistently outperforms leading methods in both power and efficiency.

\end{abstract}

% \newpage
% \tableofcontents

\spacingset{1.9}
\section{Introduction}
\label{sec:introduction}
Variable selection is a fundamental challenge in high-dimensional data analysis, aiming to identify a subset of relevant variables from a large pool of candidates. This task is crucial in various fields, such as bioinformatics, genetics, and neuroscience, where the number of variables often far exceeds the number of observations. %For example, genome-wide association studies seek to identify genetic markers linked to specific diseases or traits, with the goal of selecting a few variables (i.e., genetic markers) associated with the response variable (i.e., disease or trait) while disregarding irrelevant variables.
The variable selection problem is rigorously defined under high-dimensional linear models, and numerous methods have been proposed to address it, including LASSO \citep{tibshirani1996regression}, Elastic Net \citep{zou2005regularization}, SCAD \citep{fan2001variable}, and stability selection \citep{meinshausen2010stability}.
%In real-world applications, it is crucial to control false positives while maximizing true positives to ensure the reliability of the selected variables.
However, most of these methods concentrate on selecting relevant variables without explicitly considering the false discovery rate (FDR)—the expected proportion of false discoveries among the selected variables. 
%To address this, researchers often first select variables using these methods, followed by controlling the FDR through multiple testing correction procedures.
%However, when variable selection is data-dependent, applying classical inference methods to the selected variables may lead to double-dipping bias. This bias often leads to inflated FDR, compromising the validity of the results. To mitigate double-dipping bias, various approaches have been developed for FDR control in high-dimensional variable selection. 

% \subsection{Existing Methods and Challenges}
The Benjamini-Hochberg (BH) procedure \citep{benjamini1995controlling} was the first and remains the most widely used method for controlling the FDR in multiple testing, assuming valid and independent $p$-values. To address the common dependencies among $p$-values (e.g., in high-dimensional variable selection where variables are often correlated), the BHq \citep{benjamini2001control} and adaptive BH \citep{benjamini2006adaptive} procedures were developed. Both methods are designed to control the FDR under the assumption of positive dependence among variables while still requiring valid $p$-values.
In high-dimensional variable selection, however, obtaining valid $p$-values is challenging. When variable selection depends on the data, applying classical inference methods to the selected variables can introduce double-dipping bias, often resulting in invalid $p$-values. To tackle this challenge, various strategies have been proposed. For example, \cite{javanmard2019false} and \cite{ma2021global} utilized the debiased LASSO to compute asymptotically valid $p$-values for variables in high-dimensional linear and logistic regression models, followed by the application of the BHq procedure for FDR control. \cite{sur2019modern} demonstrated that in high-dimensional logistic models, the likelihood-ratio test statistic deviates from the classical asymptotic chi-square distribution. They proposed a framework to derive an accurate asymptotic distribution, enabling valid $p$-value computation.
Nonetheless, while these methods yield $p$-values that are asymptotically valid, their $p$-values often exhibit significant non-uniformity under the null hypothesis in finite samples. In parallel with these methods relying on asymptotic distributions, $p$-values can be computed through conditional randomization testing when the variables' joint distribution is assumed known \citep{candes2018panning}. However, this approach is computationally intensive and may become impractical in high-dimensional settings.

To address the challenges associated with \( p \)-value calculation, several approaches have been proposed. \citet{bogdan2015slope} introduced the Sorted \(\ell_1\) Penalized Estimation (SLOPE), which modifies the LASSO to achieve FDR control. However, its theoretical guarantees are limited to the setting where the design matrix is orthogonal. \cite{barber2015controlling} introduced the knockoff filter, a more general method that controls the FDR without relying on valid $p$-values in linear models under the fixed-X design, where the design matrix $\bX$ is treated as fixed. The Fixed-X knockoff filter constructs a set of ``knockoff variables" that mimic the correlation structure of the original variables. By comparing the original variables to their knockoff counterparts, it identifies relevant variables with FDR control. However, a limitation of the Fixed-X knockoff filter is that it requires the number of observations to be greater than the number of variables, limiting its applicability in high-dimensional settings.
To overcome this limitation, \cite{candes2018panning} proposed the Model-X knockoff filter, which extends the knockoff approach to high-dimensional settings by assuming knowledge of the joint distribution of the variables, $\bX$. However, if this distribution is unknown, studies in \cite{barber2020robust} and \cite{dai2023false} demonstrate that the Model-X knockoff filter can lead to inflated FDR and reduced power. Even when the joint distribution of $\bX$ is known, constructing the Model-X knockoff filter remains challenging and computationally intensive due to the stringent exchangeability condition, which requires that swapping any subset of variables with their knockoffs preserves the joint distribution of all variables and their knockoffs.
Recent advancements in generating high-quality knockoff variables include approaches using deep generative models \citep{romano2020deep, jordon2018knockoffgan}, sequential MCMC algorithms \citep{bates2021metropolized}, robust knockoff generation \citep{fan2023ark}, minimizing reconstructability \citep{spector2022powerful},  and derandomizing knockoffs \citep{ren2023derandomizing, ren2024derandomised}. Additionally, the knockoff filter has been adapted for various models, including Gaussian graphical models \citep{li2021ggm} and Cox regression \citep{li2023coxknockoff}.
In addition to the challenges of knowing the joint distribution of $\bX$ and satisfying the exchangeability condition, a significant issue with both Fixed-X and Model-X knockoff filters is that they double the size of the design matrix by concatenating the original variables with their knockoffs. This effectively degrades the original data and creates a linear model that differs from the one based solely on the original variables, potentially reducing statistical power \citep{xing2023controlling}.

Alongside the knockoff filters, the Gaussian Mirror (GM) approach \citep{xing2023controlling} represents an alternative line of research for controlling the FDR without relying on $p$-values. It computes a per-variable mirror statistic by fitting two linear models on two datasets that differ only in one perturbed variable—created by adding and subtracting Gaussian noise to form a pair of “mirror variables,” with each dataset containing one of the pair—while keeping all other variables unchanged. This results in smaller modifications to the original data compared to the knockoff filter. Since the GM method perturbs one variable at a time and requires $2p$ separate linear model fittings, the computational cost can become substantial as the number of variables $p$ increases. 
To address this computational issue, a subsequent data splitting (DS) method \citep{dai2023false} perturbs all variables simultaneously by randomly splitting the data into two halves, reducing computational demand to only two linear model fittings. However, the DS method inflates the variances of estimated regression coefficients, potentially leading to power loss. To mitigate this issue, the multiple data splitting (MDS) method \citep{dai2023false} aggregates variable selection results from independent replications of DS. Nonetheless, the computational cost of MDS can become substantial due to the need for multiple replications. Similar to the knockoff filters, the DS approach has been extended beyond linear models to logistic regression \citep{dai2023scale} and Cox regression \citep{ge2024false}.

Motivated by the limitations of existing methods for FDR control without $p$-values---including power reduction caused by degradation of the original data (through concatenation of knockoff variables or data splitting) and high computational cost---we propose a novel framework called \textbf{Nullstrap}. 
% (\textbf{Sy}nthetic \textbf{N}ull \textbf{Par}allelism). 
This framework offers three key advantages over existing methods: (1) it is easy to implement, (2) it achieves high-power FDR control by preserving the integrity of the original data, and (3) it is computationally efficient. Moreover, it is broadly applicable to various models, including linear, generalized linear, Cox regression, and Gaussian graphical models.

Nullstrap generates synthetic null data from a designated ``null model,'' and then applies the same estimation procedure in parallel to both the synthetic null data and original data to estimate the parameters of interest.
% \( F(\cdot \mid \bX; \, \bbeta_0, \bnu^*) \), where \(\bbeta_0\) represents the parameters of interest under the null hypothesis. 
% It then estimates the parameters of interest on both the synthetic and original data in parallel using the same estimation procedure. 
By comparing the parameters estimated from the synthetic null data to those obtained from the original data, Nullstrap effectively detects false positives, serving as a numerical analog of the likelihood ratio test. 
 %A detailed description of Nullstrap is provided in Section \ref{sec:method}. 
Notably, Nullstrap is computationally efficient, making it particularly suitable for high-dimensional data analysis. %We demonstrate the effectiveness of Nullstrap compared to existing methods, \textcolor{red}{including ...,} through extensive simulation studies and real data applications \textcolor{red}{to ...}. 

We compare Nullstrap with the knockoff filters, GM, and DS methods conceptually from two perspectives: their approach to creating contrasts from the original data and their strategy for fitting a model to the data. Table \ref{tab:comparison} summarizes the comparison. Both Nullstrap and the knockoff filters generate synthetic data where variable have no effect on the response. However, they differ in how the model is fitted: Nullstrap fits separate models to the original and synthetic null data in parallel, resulting in two fitted models, whereas the knockoff filter concatenates the original and knockoff variables into a single design matrix and fits one model to the concatenated data. In contrast, GM and DS do not generate synthetic data. Instead, they perturb the original data or split it into two datasets, fitting the model to these datasets in parallel.

 \begin{table}[ht]
     \centering
      \caption{Comparison of Nullstrap with the knockoff filters, GM, and DS methods.}
     \setlength{\tabcolsep}{20pt}
     \resizebox{\textwidth}{!}{
     \begin{tabular}{ccc}
         \toprule
         & Modeling Fitting to Parallel Data & Modeling Fitting to Concatenated Data\\ \midrule
     Data Synthesis  & Nullstrap & Knockoff Filters \\ \midrule
     Data Purturbation  & GM & -- \\ \midrule
     Data Splitting & DS & -- \\ \bottomrule
     \end{tabular}
     }
    
     \label{tab:comparison}
 \end{table}
\vspace{-3ex}

Our contributions are as follows: (1) We introduce Nullstrap, a conceptually novel and computationally efficient framework for FDR control in high-dimensional variable selection, %\textcolor{red}{under ?? models}, 
that achieves high power by preserving the integrity of the original data. (2) We evaluate Nullstrap through extensive simulations and real data applications, comparing it with existing methods, including the Fixed-X knockoff, Model-X knockoff, GM, DS, and MDS, demonstrating its superior performance in terms of FDR control and statistical power. (3) We provide theoretical guarantees showing that Nullstrap asymptotically controls the FDR at any desired level and achieves optimal power under mild conditions on the tail behavior of the estimation error distribution.

Section~\ref{sec:method} introduces the Nullstrap framework, detailing its model, methodology and theory.  
% Section~\ref{sec:theory} establishes FDR and power guarantees.  
Section~\ref{sec:LR} reports extensive simulations and a real linear-regression analysis.  
Section~\ref{sec:extension} extends Nullstrap to generalized linear, Cox, and Gaussian graphical models. 
% with additional simulations in Appendices~F–H. 

\vspace{-3ex}

\section{Nullstrap}
\label{sec:method}

In this section, we present Nullstrap, a general framework for variable selection with FDR control, applicable to a broad class of statistical models.  
The primary notations used in the Nullstrap framework are summarized in Table~\ref{tab:Nullstrap_notation}.
\begin{table}[h!]
    \footnotesize
\centering
\caption{Summary of notations in Nullstrap methodology.}
\label{tab:Nullstrap_notation}
\begin{tabular}{ll}
\toprule
\textbf{Notation} & \textbf{Description} \\
\midrule
$\mathbf{X} \in \mathbb{R}^{n \times p}$ & Design matrix (original data; $n$ observations; $p$ variables) \\
$\mathbf{y} \in \mathbb{R}^{n}$ & Response vector (original data) \\
\( F: \mathbb{R}^n \to [0,1] \) & Data-generating model: $\mathbf{y} \sim F(\cdot \mid \mathbf{X}; \bbeta, \bnu) $ \\
$\bbeta \in \mathbb{R}^{p}$ & True coefficient vector in the data-generating model \\
$\bnu$ & True nuisance parameter(s) or function(s) in the data-generating model \\
\( \mathcal{S}_0(F) \subset \{1,\dots, p\} \) & Null variable set: \(\mathcal{S}_0(F) := \{ j : \beta_j = 0 \}\) \\
%  & \\
% 
% \( F^*: \mathbb{R}^n \to [0,1] \) & True data-generating model: $F^* := F(\cdot \mid \mathbf{X}; \bbeta^*, \bnu^*) $ \\
 & \\
$\mathcal{E}(\cdot, \cdot): \mathbb{R}^{n} \times \mathbb{R}^{n \times p} \to \mathbb{R}^{p}$ & Estimation procedure mapping data to an estimated coefficient vector\\
$\hat{\bbeta} = \mathcal{E}(\mathbf{y}, \mathbf{X}) \in \mathbb{R}^{p}$ & Estimated coefficient vector from the original data \\
$\hat{\bnu}$ & Estimated nuisance parameter(s) or function(s) from the original data \\
\( \hat{F}: \mathbb{R}^n \to [0,1] \) & Fitted model: \( \hat{F} = F(\cdot \mid \mathbf{X}; \hat{\bbeta}, \hat{\bnu}) \) \\
 & \\
$\bbeta_0 \in \mathbb{R}^{p}$ &Coefficient vector under the null model; $\bbeta_0 = \mathbf{0}$ under the global null\\
\( \tilde{F}_0: \mathbb{R}^n \to [0,1] \) & Null model: \( \tilde{F}_0 := F(\cdot \mid \mathbf{X}; \bbeta_0, \hat{\bnu}) \) \\
$\tilde{\mathbf{y}} \in \mathbb{R}^{n}$ & Null response vector: $\tilde{\mathbf{y}} \sim \hat{F}_0$ (synthetic null data) \\
$\tilde{\bbeta} = \mathcal{E}(\tilde{\mathbf{y}}, \mathbf{X}) \in \mathbb{R}^{p}$ & Estimated null coefficient vector from the synthetic null data \\
$\gamma_{n,p} \in \mathbb{R}^{+}$ & Correction factor \\
$|\tilde{\beta}_j^\prime| \in \mathbb{R}^{+}$ & Corrected estimated null coefficient (absolute value) for variable $j$: \\ & $|\tilde{\beta}_j^\prime| = |\tilde{\beta}_j| + \gamma_{n,p}$, $j=1,\ldots,p$ \\
% $\widehat{\text{FDP}}(t)$ & FDP estimate: $\frac{\#\{ j : |\tilde{\beta}_j^\prime| \geq t \}}{\max( \#\{ j : |\hat{\beta}_j| \geq t \}, 1 )}$ \\
% $\widehat{\cS}(\tau_q)$ & Selected variables: $\{ j : |\hat{\beta}_j| > \tau_q \}$ \\
% $\tau_q$ & Data-driven threshold s.t. $\widehat{\text{FDP}}(\tau_q) \leq q$ \\
% $q$ & Target false discovery rate (FDR) level \\
% \( \mathcal{T}(F) \) & Expected number of null variables in \( \mathcal{S}_0(F) \) exceeding threshold \( t \):\\
% & \( \mathbb{E}\left[\#\left\{ j \in \mathcal{S}_0(F) : |\hat{\beta}_j(F)| \geq t \right\}\right] \) \\
\bottomrule
\end{tabular}
\end{table}
\vspace{-6ex}

\subsection{Modeling Framework}
We focus on high-dimensional variable selection under a general statistical model:
\vspace{-20pt}
\begin{equation}\label{eq:model}
F := F( \cdot \mid \bX; \bbeta, \bnu), \quad \by \sim F,
\vspace{-20pt}
\end{equation}
where \( \by = (y_1,\ldots,y_n)\tran \in \mathbb{R}^n \) represents the observable response, and \( \bX = [\bx_1, \ldots, \bx_n]\tran \in \mathbb{R}^{n \times p} \) is the design matrix, with each row corresponding to an observation and each column representing a variable. 
% We abbreviate the true data-generating distribution $F^* = F(\cdot \mid \bX; \bbeta^*, \bnu^*)$ as \( F^* \). 
Model~\eqref{eq:model} represents a fixed design, as it is about the randomness of $\by$ conditional on $\bX$. The coefficient vector \( \bbeta = (\beta_1,\ldots,\beta_p)\tran \in \mathbb{R}^p \) represents the parameters of interest and captures the effects of the variables on the response. The term \( \bnu \in \mathbb{R}^d \) contains the nuisance parameter(s) or function(s), which account for additional model structure or potential sources of variability that are not the main focus of inference. Here, \( d \) can either be finite, \( d < \infty \), indicating a parametric model, or infinite, \( d = \infty \), representing the inclusion of a non-parametric component. %In Section \ref{sec:method}, we show that the linear model, generalized linear model, Cox regression model, and Gaussian graphical model can all be formulated in the form of \eqref{eq:model}.
%The parameter \(\beta\) could either be a vector or a matrix.
% Below, we introduce four commonly used statistical models, all of which can be high-dimensional. 
% Below, we introduce linear regression, which is a special case of model~\eqref{eq:model}. 
% four commonly used statistical models, all of which can be high-dimensional. 
\begin{example}[Linear model]
A linear model can be written as:
$
\by = \bX\bbeta + \bvarepsilon,
$
where \( \bX \in \mathbb{R}^{n \times p} \) is the design matrix, \( \bbeta \in \mathbb{R}^p \) represents the coefficient vector, and \( \bvarepsilon \) is a random error term. When \( \bvarepsilon \sim \mathcal{N}(0, \sigma^2 {\bI}) \), model~\eqref{eq:model} becomes
$
\by \sim \mathcal{N}(\bX \bbeta, \sigma^{2} {\bI}),
$
where the nuisance parameter $\bnu$ in \eqref{eq:model} is $\sigma^{2}$.
\end{example}
\begin{example}[Generalized linear model]\label{ex:glm}
    A generalized linear model (GLM) extends a linear model by allowing the $i$-th response variable \(y_i\) to follow a one-dimensional exponential-family distribution $F_i$ with $g(\mathbb{E}[y_i]) = \bx_i\tran \bbeta$, where $g(\cdot)$ is the link function, \( \bx_i \in \mathbb{R}^{p} \) is the $i$-th row of $\bX$, and \( \bbeta \in \mathbb{R}^p \) represents the coefficient vector. The nuisance parameter(s) $\bnu$ in \eqref{eq:model} includes the additional parameters involved in $F_i$, $i=1,\ldots,n$.
    \end{example}
\begin{example}[Cox model]\label{ex:cox}
	The Cox proportional hazards model assumes that the response variable \( y_i \) follows a one-dimensional distribution with the hazard function given by
$
h(y_i \mid \bx_i) = h_0(y_i) \exp(\bx_i^\top \bbeta),
$
where \( h_0(y_i) \) is the baseline hazard function, \( \bx_i \in \mathbb{R}^{p} \) denotes the \( i \)-th row of \( \bX \), and \( \bbeta \in \mathbb{R}^p \) represents the coefficient vector. The nuisance function \( \bnu \) in Equation~\eqref{eq:model} corresponds the baseline hazard function \( h_0(\cdot) \).
    % is a widely used statistical method in survival analysis for studying the relationship between survival time and predictor variables. It 
    % is a semi-parametric model that does not assume a specific form for the baseline hazard function, providing flexibility in modeling time-to-event data. The model defines the hazard function \( h(y \mid \bx) \) as the product of a baseline hazard function \( h_0(y) \) and an exponential term involving the covariates \( \bx \), expressed as \( h(y \mid \bx) = h_0(y) \exp(\bx\tran \bbeta) \), where \( \bbeta \) denotes the regression coefficients. In this context, the baseline hazard is denoted as \(\bnu = h_0(y)\), and we write \( Y \sim \text{Cox}(h_0(y) \exp(\bx\tran \bbeta)) \).
\end{example}
% \begin{example}[Gaussian graphical model]
% % A GGM is a probabilistic model that represents the conditional dependence structure among random variables using an undirected graph, where each node corresponds to a variable, and edges signify conditional dependence.
% Let \( \by = (y_1, y_2, \dots, y_p)\tran \) follow a multivariate Gaussian distribution \( \by \sim \mathcal{N}(0, \bSigma) \), where \( \bSigma \) is the covariance matrix. In a GGM, the absence of an edge between \( y_i \) and \( y_j \) implies their conditional independence, captured by the precision matrix \( \bTheta = \bSigma^{-1} \), with \( \Theta_{ij} = 0 \) indicating conditional independence between \( y_i \) and \( y_j \).
% Furthermore, \( \by \) can be expressed as \( \by = \bTheta^{-1/2} \bvarepsilon \), where \( \bvarepsilon \sim \mathcal{N}(0, \bI) \). The vector \(\bbeta\) contains the off-diagonal elements of \( \bTheta \), and \( \bnu = \diag(\bTheta) \) represents the diagonal elements.
% \end{example}
In the context of variable selection, for a statistical model \( F(\cdot \mid \mathbf{X}; \bbeta, \bnu) \), we define the set of indices corresponding to the non-zero elements of \(\bbeta\) as the signal variable set, denoted by \(\cS(F)\), and we define the null variable set as \( \mathcal{S}_0(F) = \{ j : \beta_j = 0 \} \), which is the complement of \(\cS(F)\).
% Conversely, for a statistical model \( F(\cdot \mid \mathbf{X}; \bbeta, \bnu) \), we define the null variable set as \( \mathcal{S}_0(F) = \{ j : \beta_j = 0 \} \). 
% Specifically, we abbreviate \( \mathcal{S}_0(F^*) \) as \( \mathcal{S}_0 \) when the model is clear from the context.
Our objective is to provide a selected variable set \(\widehat{\cS} \subset \{1,\ldots,p\}\), an estimate of \(\cS(F)\), while controlling the FDR, defined as
$\text{FDR} = \mathbb{E}\left[{V}/{\max(R,1)} \right],
$
where 
\( V = \#\left( \widehat{\cS} \cap \mathcal{S}_0(F) \right) \) denotes the number of false positives, and  
\( R = \#\widehat{\cS} \) is the total number of selected variables. The quantity $\frac{V}{\max(R, 1)}$ is referred to as the false discovery proportion (FDP).
In Section~\ref{sec:LR}, we will show how Nullstrap controls the FDR asymptotically in a linear model. In Appendices F--H, we provide the detailed procedures and simulation results for the GLM, Cox model, and Gaussian graphical model (GGM), respectively.

\vspace{-3ex}
\subsection{Nullstrap methodology}
The core idea of Nullstrap involves generating synthetic null data and applying the same model fitting approach to both the original and synthetic null data in parallel to estimate the parameters of interest about variable importance. The parameter estimates from the synthetic null data serve as the negative control to those from the original data to identify important variables with FDR control.

\begin{definition}[Synthetic null data] \label{def:syn_null_data}
The synthetic null data used in Nullstrap retains the original design matrix $\mathbf{X}$ and incorporates a synthetic null response $\tilde{\mathbf{y}}$ generated from the fitted null model:
\vspace{-10pt}
\begin{equation}\label{eq:Nullstrap}
   \tilde{F}_0 = F(\cdot \mid \mathbf{X}; \, \boldsymbol{\beta}_0, \hat{\boldsymbol{\nu}}), \quad \tilde{\mathbf{y}} \sim \, \tilde{F}_0,
    \vspace{-15pt}
\end{equation}
where \(\hat{\boldsymbol{\nu}}\) is the nuisance parameter estimated jointly with the coefficient vector $\hat\bbeta$ from the original data \(\{\mathbf{y}, \mathbf{X}\}\). The vector \(\boldsymbol{\beta}_0\) represents the coefficient vector specified under the null hypothesis. For instance, \(\boldsymbol{\beta}_0 = (0, \dots, 0)^\top\) corresponds to the global null hypothesis, where no variables have an effect.
%  Alternatively, \(\boldsymbol{\beta}_0\) with its \(j\)-th element specified as zero represents the individual null hypothesis that the \(j\)-th variable has no effect\footnote{The other elements of \(\boldsymbol{\beta}_0\) need to be estimated from the original data $\{\by, \bX\}$.}.
\end{definition}

%A simplest way to obtain \(\tilde{{X}}\) is \(\tilde{{X}} \equiv {X}\).
%The difference between Nullstrap data and the original data is that the parameter of interest in the Nullstrap data is a zero vector. Therefore, as long as we can generate simulated data, we can use our method to control the FDR in model selection, or more generally, to estimate some unwanted bias introduced by double-dipping.  

 Let \( \mathcal{E}(\cdot, \cdot) \) denote an estimation procedure for $\bbeta$ such that
$\hat{\bbeta} = \mathcal{E}(\by, {\bX})$ estimates $\bbeta$, and $ \tilde{\bbeta} = \mathcal{E}(\tilde{\by}, {{\bX}})$ estimates $\bbeta_0$. 
Our goal is to use \( \tilde{\bbeta} = (\tilde\beta_1, \ldots, \tilde\beta_p)\tran \) as a negative control for \( \hat{\bbeta} = (\hat\beta_1, \ldots, \hat\beta_p)\tran \) to facilitate variable selection with FDR control. 

In this work, we define selected variables as those with large absolute coefficient estimates in \( \hat\bbeta \); specifically, we rank the \( p \) variables by \( \{|\hat\beta_j|\}_{j=1}^p \). We do not standardize the coefficient estimates by dividing them by their standard errors, as estimating standard errors reliably is itself a challenge in high-dimensional settings \citep{javanmard2019false}. Instead, we standardize the design matrix \( \bX \) by centering each variable at zero and scaling it to have unit variance, which ensures that the magnitude of \( \hat\beta_j \) is comparable across variables.
 
Ideally, for any null variable \( j \) with \( \beta_j = 0 \), we expect \( |\hat\beta_j| \) to be of similar or smaller magnitude than \( |\tilde\beta_j| \) with high probability. This allows us to decide if a non-zero \( |\hat\beta_j| \) is significant enough to reject the null hypothesis \( \beta_j = 0 \).  Formally, we require \( \pr(|\hat{\beta}_j| \geq t) \leq \pr(|\tilde{\beta}_j| \geq t) \) for all \( j \in \mathcal{S}_0(F) \), which implies
%         \vspace{-20pt}
% \begin{equation}\label{eq:gamma_intu}
$	\E\left[\#\{j: j \in \cS_0(F), |\hat{\beta}_j| \geq t\}\right] \leq \E\left[\#\{j: |\tilde{\beta}_j| \geq t\}\right].$ To ensure this inequality holds, we introduce a correction factor \( \gamma_{n,p} \) to modify $|\tilde{\beta}_j|$ as:
% %             \vspace{-20pt}
% % \end{equation}
% % With this inequality, the FDP can be approximated as $\frac{\#\{ j : |\tilde{\beta}_j| \geq t \}} {\max\left( \#\{ j : |\hat{\beta}_j| \geq t \}, 1 \right)}$. However, achieving this inequality is not straightforward due to the differences between \( \tilde{\by} \) and \( \by \), which result in the differences between \( \hat{\bbeta} \) and \( \tilde{\bbeta} \). 
% % \subsubsection*{Correction factor for Nullstrap}
% To ensure the false discovery rate (FDR) control of Nullstrap using 
% \( \hat{\boldsymbol{\beta}} = \mathcal{E}(\mathbf{y}, \mathbf{X}) \) 
% and 
% \( \tilde{\boldsymbol{\beta}} = \mathcal{E}(\tilde{\mathbf{y}}, \mathbf{X}) \), 
% where \( \tilde{\mathbf{y}} \) is generated under the global null hypothesis, 
% we introduce a correction factor term \( \gamma_{n,p} \) to model the underlying relationship 
% between \( \hat{\boldsymbol{\beta}} \) and \( \tilde{\boldsymbol{\beta}} \). 
% The correction factor ensures the validity of the inequality in \eqref{eq:gamma_intu}, which is critical for FDR control. With $\gamma_{n,p}$, to achieve the inequality in \eqref{eq:gamma_intu}, we modify $\tilde{\bbeta}$ as follows:
% \vspace{-20pt}
% \begin{equation}\label{eq:perturb}
$
	|\tilde{\beta}_j^\prime| = |\tilde{\beta}_j| + \gamma_{n,p}, \; j = 1, \ldots, p\,.
    $
    % \vspace{-20pt}
% \end{equation}
% This modification is sufficient to ensure that \eqref{eq:gamma_intu} holds. Specifically, 
% for \( j \in \mathcal{S}_0(F) \), we have \( \beta_j = \beta_{0j} = 0 \). Consequently, \( |\hat{\beta}_j| \) and \( |\tilde{\beta}_j| \) are both bounded by \( \gamma_{n,p} \) with high probability, as stated in Assumption~\ref{assum:gamma_estimator}. Therefore, it follows that \( |\hat{\beta}_j| \leq |\tilde{\beta}^\prime_j| \) with high probability. 
% Note that \( \beta_j = \beta_{0j} = 0 \) for all \( j \in \mathcal{S}_0(F) \). Under Assumption~\ref{assum:gamma_estimator}, which ensures that both \( |\hat{\beta}_j| \) and \( |\tilde{\beta}_j| \) are bounded by \( \gamma_{n,p} \) with high probability, it follows that \( |\hat{\beta}_j| \leq |\tilde{\beta}^\prime_j| \) with high probability.
% Then, each $|\hat\beta_j|$ would be compared against $|\tilde\beta_j^\prime|$ to decide if the $j$-th variable is a discovery. While this modification may make Nullstrap more conservative, our simulation results across all four examples (Examples 1-4) demonstrate that Nullstrap achieves the highest power compared to state-of-the-art approaches, including the knockoff and DS methods. Nevertheless, further research could focus on optimizing the  correction factor to enhance power.
% The correction factor \( \gamma_{n,p} \) plays a crucial role in ensuring the validity of the inequality in \eqref{eq:gamma_intu} and the FDR control of Nullstrap. 
In general, $\gamma_{n,p}$ should be chosen based on a well-specified statistical model and a reliable estimation procedure. An intuitive approach is to calibrate the correction factor using numerical simulations under the specified model and estimation procedure. A more principled strategy is to estimate $\gamma_{n,p}$ directly from the data. In this work, we develop a data-driven algorithm for selecting $\gamma_{n,p}$, detailed in Appendix B.1. 
Below, we provide a high-level overview of the algorithm. 

\subsubsection*{Data-driven selection of the correction factor $\gamma_{n,p}$}
We refer to the fitted model \( F(\cdot \mid \bX; \, \hat{\bbeta}, \hat{\bnu}) \) as \( \hat{F} \), where \( \hat{\bbeta} = \mathcal{E}(\by, \bX) \) denotes the estimated coefficients and \( \hat{\bnu} \) represents the estimated nuisance parameter(s) or function(s). 
To ensure valid FDR control, the correction factor \( \gamma_{n,p} \) should satisfy:
\vspace{-15pt}
\begin{equation}\label{eq:gamma_estimator}
\mathbb{E}\left[\#\left\{ j \in \mathcal{S}_0(F) : |\hat{\beta}_j| \geq t \right\}\right] 
\leq \mathbb{E}\left[\#\left\{ j : |\tilde{\beta}_j^\prime| \geq t \right\}\right],\; \text{with } |\tilde{\beta}_j^\prime| = |\tilde{\beta}_j| + \gamma_{n,p},
\vspace{-15pt}
\end{equation}
for all $j = 1, \ldots, p$.
% where .
% which is the corrected version of inequality~\eqref{eq:gamma_intu}.
In practice, the left-hand expectation in \eqref{eq:gamma_estimator} is unknown, since \( \mathcal{S}_0(F) \) depends on the true model. To address this challenge, we propose estimating \( \mathcal{S}_0(F) \) by \( \mathcal{S}_0(\hat{F}) \), the set of null variables under the fitted model. 
For any model \( F(\cdot \mid \bX; \bbeta, \bnu) \), define the statistical functional:
\vspace{-15pt}
\begin{equation}\label{eq:gamma_estimator_def}
\mathcal{T}[F] = \mathbb{E}_{\bY \sim F}\left[\#\left\{ j \in \mathcal{S}_0(F) : |\left[ \mathcal{E}(\bY, \bX) \right]_j| \geq t \right\}\right],
\vspace{-15pt}
\end{equation}
where 
% \( \hat{\bbeta} = \mathcal{E}(\by, \bX) \) with 
\( \bY \sim F(\cdot \mid \bX; \bbeta, \bnu) \), $\left[ \mathcal{E}(\bY, \bX) \right]_j$ represents the $j$-th element of the estimated coefficient vector $\mathcal{E}(\bY, \bX)$, and $t$ is a fixed threshold on the absolute coefficient estimates. Then the left-hand side of \eqref{eq:gamma_estimator} can be written as \( \mathcal{T}[F] \), which we can approximate using \( \mathcal{T}[\hat{F}] \). %\footnote{The estimator \( \hat{\bbeta} \) used for determining the correction factor does not need to match the procedure \( \mathcal{E}(\cdot, \cdot) \); any reasonable estimation method may be used.} 
To ensure this approximation is accurate, we require \( \hat{\bbeta}\) to be a consistent estimator of \( \bbeta \), a requirement that holds under a well-specified model and a reliable estimation procedure. Based on an estimate of \( \mathcal{T}[\hat{F}] \), we then compute the smallest value of \( \gamma_{n,p} \) that satisfies \eqref{eq:gamma_estimator}. To improve the stability of FDR control, the procedure can be repeated multiple times, with the 95th percentile of \( \gamma_{n,p} \) selected as the correction factor (Appendix B.1).

\subsubsection*{Threshold selection for Nullstrap}
Nullstrap selects variables whose \( |\hat{\beta}_j| \ge t \). The FDP is defined as:
\vspace{-10pt}
\[\text{FDP}(t) = \frac{\#\{j: j \in \cS_0(F), |\hat{\beta}_j| \geq t\}}{\max\left( \#\{ j : |\hat{\beta}_j| \geq t \}, 1 \right)}\,,
\vspace{-10pt}
\]
and is expected to be bounded from above with high probability by:
\vspace{-10pt}
\begin{equation}\label{eq:FDP}
\widehat{\text{FDP}}(t) = \frac{\#\{ j : |\tilde{\beta}_j^\prime| \geq t \}}{\max\left( \#\{ j : |\hat{\beta}_j| \geq t \}, 1 \right)}\,,
\vspace{-10pt}
\end{equation}
since the numerator of $\widehat{\text{FDP}}(t)$ \emph{overestimates} the unobservable numerator of $\text{FDP}(t)$. Based on this rationale, Nullstrap determines the threshold for \( |\hat{\beta}_j| \) as $
\tau_q = \min \{ t > 0 : \widehat{\text{FDP}}(t) \leq q \}$,
where \( q \) represents the target FDR level, and selects the variables in
$
\widehat{\cS} = \{ j : |\hat{\beta}_j| \geq \tau_q \}.
$ The Nullstrap procedure is summarized in Algorithm~\ref{alg:SNP}.
\begin{algorithm}
    \caption{Variable selection via Nullstrap}
    \label{alg:SNP}
        \textbf{Input:} original data $\{\by, \bX\}$; estimation procedure $\mathcal{E}(\cdot, \cdot)$; target FDR level $q \in (0,1)$; correction factor $\gamma_{n,p}$. {\bf{Note}}: For data-driven selection of $\gamma_{n,p}$, see Appendix B.1. 
        \\
    \textbf{Output:} The set of selected variables $\widehat{\cS}(\tau_q)$. \\
    
         Generate synthetic null data \( \{ \tilde{\mathbf{y}}, \mathbf{X} \} \) as in (\ref{eq:Nullstrap}). \\

         Compute parameter estimates \( \hat{\boldsymbol{\beta}} \) from the original data \( \{ \mathbf{y}, \mathbf{X} \} \) and the negative control \( \tilde{\boldsymbol{\beta}} \) from the synthetic null data \( \{ \tilde{\mathbf{y}}, \mathbf{X} \} \) using the same estimation procedure $\mathcal{E}(\cdot, \cdot)$. \\

         Add the correction factor \( \gamma_{n,p} \) to each element of \( |\tilde{{\beta}_j}| \), resulting in \( |\tilde{\beta}_j^\prime| \). 
         % {\bf{Note}}: The determination of \( \gamma_{n,p} \) depends on the specific statistical model.
        % \begin{itemize}
        %     \item 
        % \end{itemize} 

         Given a target FDR level \( q \in (0,1) \), calculate the threshold \( \tau_q \) as:
        \begin{equation}\label{eq:tauq}
        \tau_q = \min \left\{ t > 0 : \widehat{\text{FDP}}(t) = \frac{\#\{ j : |\tilde{\beta}_j^\prime| \geq t \}}{\max\left( \#\{ j : |\hat{\beta}_j| \geq t \}, 1 \right)} \leq q \right\}.
        \end{equation} \\

         Select the set of variables:
        \begin{equation}\label{eq:hat_cS}
        \widehat{\cS}(\tau_q) = \{ j : |\hat{\beta}_j| > \tau_q \}.
        \end{equation}
\end{algorithm}
\vspace{-3ex}

An alternative approach to estimate the FDP is based on \( W_j = |\hat{\beta}_j| - |\tilde{\beta}_j^\prime| \), defined as:
\vspace{-10pt}
\begin{equation}\label{eq:FDP2}
\widehat{\text{FDP}}(t) = \frac{1+\#\{ j : W_j \leq -t \}}{\max\left( \#\{ j : W_j \geq t \}, 1 \right)}\,,
\vspace{-10pt}
\end{equation}
which is widely used in the literature \citep{dai2023false, candes2018panning, ge2021clipper} and is applicable to Nullstrap. However, it is important to note that, compared to \( |\hat{\beta}_j| \), the difference \( W_j \) incorporates variability from \( |\tilde{\beta}_j^\prime| \) arising from synthetic null data generation, which may reduce the stability and reproducibility of the selected variables across replications. By replacing \(\widehat{\text{FDP}}(t)\) in \eqref{eq:tauq} of Algorithm~\ref{alg:SNP} with \eqref{eq:FDP2} and selecting variables in \(\widehat{\cS}(\tau_q) = \{ j : W_j > \tau_q \}\), we define this Nullstrap variant as Nullstrap-Diff, where ``Diff" refers to the difference \( W_j \). In our simulation studies (Appendix C.4), we compare the performance of Nullstrap with that of Nullstrap-Diff. The results show that the FDR control and power achieved by Nullstrap-Diff are slightly inferior to those achieved by Nullstrap, supporting the choice of using the FDP estimate in \eqref{eq:FDP} for Nullstrap.

\subsubsection*{An alternative approach to generate synthetic null data: Nullstrap (individual)}

Following Definition~\ref{def:syn_null_data}, we propose generating \( \tilde{\by} \) under the global null hypothesis. Alternatively, another approach is to generate synthetic null data for each variable, corresponding to the individual null hypothesis that the \( j \)-th variable has no effect. 
Specifically, \(\boldsymbol{\beta}_0\) with its \( j \)-th element set to zero represents the individual null hypothesis, indicating that the \( j \)-th variable has no effect.  
% \footnote{The other elements of \(\boldsymbol{\beta}_0\) need to be estimated from the original data $\{\by, \bX\}$.}.
% and the other based on individual null hypotheses, each corresponding to a single variable. 
We refer to the global null and individual null approaches as ``Nullstrap" and ``Nullstrap (individual)", respectively.
Nullstrap is computationally efficient, requiring only a single synthetic null dataset generated under the global null hypothesis \( H_0: \bbeta = 0 \) and a single model fitting for that dataset. In contrast, Nullstrap (individual) is computationally intensive because it generates \( p \) synthetic null datasets, each corresponding to one of the \( p \) individual null hypotheses \( H_{0j}: \beta_{0j} = 0 \) for \( j = 1, \ldots, p \), and performs \( p \) separate model fittings on these datasets. While Nullstrap (individual) is conceptually ideal, as it aligns with the individual null hypotheses that define the variable selection problem, its computational demands make it impractical. This parallels the distinction between GM and DS---GM perturbs one variable at a time, requiring \( p \) separate model fittings, whereas DS splits the data into two halves once, requiring just two model fittings.

For Nullstrap, we set \( \bbeta_0 = (0, \dots, 0)\tran \), with its detailed procedure described in Algorithm~1 and Section~\ref{sec:LR}. On the other hand, Nullstrap (individual) generates synthetic null data for the \( j \)-th variable by setting \( \bbeta_0 = \bbeta_0^j := \left(\hat{\bbeta}_{1:(j-1)}^{-j}, \, 0, \, \hat{\bbeta}_{j:(p-1)}^{-j}\right)\tran \), where \( \hat{\bbeta}^{-j} = \left(\hat{\bbeta}_{1:(j-1)}^{-j}, \, \hat{\bbeta}_{j:(p-1)}^{-j}\right)\tran \) is the estimated coefficient vector based on \( \by \) and the design matrix \( \bX^{-j} \), which excludes the \( j \)-th variable. Synthetic null data \( \tilde{\by}^j \) is then generated based on \( \bbeta_0^j \), and the \( j \)-th negative-control coefficient estimate \( \tilde{\beta}_j \), corresponding to \( \hat{\beta}_j \), is extracted as the \( j \)-th element of \( \mathcal{E}(\tilde{\by}^j, \bX) \). Repeating this procedure for \( j = 1, \ldots, p \), the data-driven threshold for Nullstrap (individual) is determined as the smallest \( t > 0 \) satisfying the inequality \(\frac{\#\{ j : |\tilde{\beta}_j| \geq t \}}{\max\left( \#\{ j : |\hat{\beta}_j| \geq t \}, 1 \right)} \leq q\), where \( q \) is the target FDR level. The detailed procedure for Nullstrap (individual) is provided in Appendix B.2. 
In Section \ref{sec:sim}, we numerically compare Nullstrap and Nullstrap (individual) 
% to evaluate whether Nullstrap achieves satisfactory performance in {FDR control} and {power} 
under the linear model  
$
\by = \bX\bbeta + \bvarepsilon
$, where \( \bvarepsilon \sim \mathcal{N}(0, \bI) \).

\vspace{-3ex}
\subsection{Nullstrap theory}\label{sec:theory}
 The correction factor \( \gamma_{n,p} \) plays a crucial role in ensuring the validity of the inequality in \eqref{eq:gamma_estimator} and the FDR control of Nullstrap. 
Assumption~\ref{assum:gamma_estimator} guarantees the existence of \(\gamma_{n,p}\).
% A straightforward approach is to define the correction factor based on 
% \( \|\hat{\boldsymbol{\beta}} - \boldsymbol{\beta} \|_{\infty} \) 
% and 
% \( \|\tilde{\boldsymbol{\beta}} - \boldsymbol{\beta}_0 \|_{\infty} \), 
% quantifying the maximum deviations of the two estimators from the respective true parameters. 
% This motivates the following assumption.

\begin{assumption}[High-probability upper bound on estimation error]\label{assum:gamma_estimator}
	If the nuisance parameter estimator \( \hat{\boldsymbol{\nu}} \) lies within a compact set with probability approaching one as \( n \) and \( p \) increase, assume that
            \vspace{-20pt}
	$$
 \pr \left( \left\|\hat{\bbeta} - \bbeta \right\|_{\infty} \geq \gamma_{n,p} \right) = \alpha_{n,p}  \mbox{ and } \ \pr \left( \left\|\tilde{\bbeta} - \bbeta_0 \right\|_{\infty} \geq \gamma_{n,p} \right) = \alpha_{n,p},
         \vspace{-20pt}
	$$
	where \( \gamma_{n,p} \) is the correction factor, and \( \alpha_{n,p} = o(1) \) as \( n, p \to \infty \).
\end{assumption}
Ensuring that the nuisance parameter estimator \( \hat{\boldsymbol{\nu}} \) lies within a compact set can be achieved by projecting \( \hat{\boldsymbol{\nu}} \) onto a pre-specified compact set. This condition ensures the synthetic null response \( \tilde{\by} \) is well-defined and avoids numerical singularities during its generation. Essentially, Assumption~\ref{assum:gamma_estimator} requires that the estimation errors \( \|\hat{\bbeta} - \bbeta\|_{\infty} \) and \( \|\tilde{\bbeta} - \bbeta_0\|_{\infty} \) are bounded above by \( \gamma_{n,p} \) with high probability. In many models—such as those in Examples 1–3—this type of bound can be justified using tools from high-dimensional statistics, such as concentration inequalities and empirical process theory.

In deriving the theoretical guarantees for FDR control and power of Nullstrap, we assume that the correction factor \( \gamma_{n,p} \), selected in a data-driven manner (see Appendix B.1), satisfies Assumption~\ref{assum:gamma_estimator}. A theoretical investigation of whether the data-driven selection procedure guarantees this assumption is left for future work.

% \begin{assumption}[Weak dependence among variables]\label{assu:var}
% Given a target FDR level \( q \in (0,1) \) and the threshold $\tau_q$ in \eqref{eq:tauq}, assume that 
% \[
% \var \left(\sum_{j=1}^{p} \mathbb{I}(|\hat{\beta}_j| \geq \tau_q)\right) \big/ p^2 \rightarrow 0 \mbox{ and }  \var \left(\sum_{j=1}^{p} \mathbb{I}( |\tilde{\beta}_j^\prime| \geq \tau_q )\right) \big/ p^2 \rightarrow 0 \quad \text{as } n, p \to \infty.
% \]
% \end{assumption}
%Next, we provide theoretical guarantees for the FDR control and power of Nullstrap.
\begin{theorem}\label{thm:main}
Under Assumption \ref{assum:gamma_estimator}, given a target FDR level \( q \in (0,1) \), the threshold \( \tau_q \) in \eqref{eq:tauq}, and the selected variable set \( \widehat{\cS}(\tau_q) \) in \eqref{eq:hat_cS}, as \( n, p \to \infty \), Nullstrap satisfies:
\vspace{-10pt}
\[
\mathrm{FDR}(\tau_q) = \mathbb{E} \left[ \frac{\#\left\{\widehat{\cS}(\tau_q) \cap \cS_0(F) \right\}}{\max(\#\widehat{\cS}(\tau_q), 1)} \right] \leq q + \alpha_{n,p} = q + o(1)\,,
\vspace{-10pt}
\]
where $\alpha_{n,p} = o(1)$ is the small probability defined in Assumption \ref{assum:gamma_estimator}.

Furthermore, if \( \min_{j \in \cS(F)} |{\beta}_{j}| > 3\gamma_{n,p} \), then
\vspace{-10pt}
\[
\mathrm{Power}(\tau_q) = \mathbb{E} \left[ \frac{\#\left\{ \widehat{\cS}(\tau_q) \cap \cS(F) \right\}}{\# \cS(F)} \right] \geq 1-2\alpha_{n,p} = 1 - o(1)\,.
\vspace{-10pt}
\]
\end{theorem}

Theorem \ref{thm:main} provides a theoretical guarantee for controlling the FDR in Nullstrap. Furthermore, it establishes that when the minimum signal strength satisfies \( \min_{j \in \mathcal{S}(F)} |{\beta}_{j}| > 3\gamma_{n,p} \), the power of Nullstrap approaches 1 as \( n \) and \( p \) tend to infinity. In other words, under Assumption~\ref{assum:gamma_estimator}, which ensures that the estimation procedure for \( \bbeta \) is reliable, 
% and Assumption~\ref{assu:var}, which guarantees that the \( p \) variables exhibit weak dependence,
 Nullstrap effectively controls the FDR in variable selection and achieves an asymptotic power of 1 when the minimum signal strength is sufficiently large.

\vspace{-3ex}

\section{Nullstrap for linear models}\label{sec:LR}
%\subsection{Problem statement and FDR control via Nullstrap}
In this section, we outline the specific steps for applying Nullstrap to perform variable selection in a high-dimensional linear model, 
\(\mathbf{y} = \mathbf{X}\boldsymbol{\beta} + \boldsymbol{\varepsilon}\). 
A crucial step in this process is estimating the distribution of \(\boldsymbol{\varepsilon}\) from the original data \(\{\mathbf{y}, \mathbf{X}\}\), which enables the generation of synthetic null data \(\{\tilde{\mathbf{y}}, \mathbf{X}\}\). For instance, the distribution of $\boldsymbol{\varepsilon}$ can either be specified parametrically (e.g., as Gaussian) or estimated nonparametrically.
\begin{definition}[Parametric synthetic null data for a Gaussian linear model]\label{def:syn_null_data_LM}
		For a Gaussian linear model $\mathbf{y} = \mathbf{X}\boldsymbol{\beta} + \boldsymbol{\varepsilon}$, where $\bvarepsilon \sim \mathcal{N}(0, \sigma^2\mathbf{I})$, Nullstrap defines \( \tilde{\by} \in \mathbb{R}^n \) as 
		$
		\tilde{\by} = {\bX} \bbeta_0 + \tilde{\bvarepsilon} = \tilde{\bvarepsilon},
		$
		where \( \bbeta_0 = (0, \dots, 0)\tran \in \mathbb{R}^p \) is the coefficient vector under the global null hypothesis, and  \( \tilde{\bvarepsilon} \sim \mathcal{N}(0,  \hat\sigma^2\mathbf{I}) \), where $(\hat\bbeta, \hat\sigma^2)$ are estimates of $(\bbeta, \sigma^2)$ from the original data $\{\by, \bX\}$.
\end{definition}
%
%\begin{definition}
%	Regression parallel knockoffs for the family of random variables $\left\{\mathbf{X} = (X_1, \dots, X_p), \bm{\varepsilon}\right\}$ are defined as a new family of random variables $\left\{\tilde{\mathbf{X}} = (\tilde{X}_1, \dots, \tilde{X}_p), \tilde{\bm{\varepsilon}}\right\}$ if
%	\[
%	\left\{\mathbf{X} = (X_1, \dots, X_p), \bm{\varepsilon}\right\} \overset{d}{=} \left\{\tilde{\mathbf{X}} = (\tilde{X}_1, \dots, \tilde{X}_p), \tilde{\bm{\varepsilon}}\right\},
%	\]
%	where \(\overset{d}{=}\) denotes equality in distribution.
%\end{definition}
%Using this new family $\left\{\tilde{{X}}, \tilde{\bm{\varepsilon}}\right\}$, we construct the null response as $\tilde{Y} = \tilde{{X}} {\beta}_0 + \tilde{{\varepsilon}} =  \tilde{{\varepsilon}}$, where ${\beta}_0 = (0, \dots, 0)\tran \in \mathbb{R}^p$ is the null parameter vector.
We consider using the LASSO as the estimation procedure for $\bbeta$:  
    \vspace{-15pt}
\begin{equation}\label{eq:lasso}
\hat{\bbeta}= \operatorname*{argmin}_{\bbeta \in \mathbb{R}^{p}} \frac{1}{2n} \| \by - \bX \bbeta \|_2^2 + \lambda_n \|\bbeta\|_1 \mbox{ and } \ \tilde{\bbeta} = \operatorname*{argmin}_{\bbeta \in \mathbb{R}^{p}} \frac{1}{2n} \| \tilde{\by} - {\bX} \bbeta \|_2^2 + \lambda_n \|\bbeta\|_1,
    \vspace{-15pt}
\end{equation}
where \( \lambda_n \) is the same regularization parameter applied to both the original data and the synthetic null data.
Other estimation procedures, such as the Elastic Net and SCAD, can also be used (Appendix C.2).  Here, we focus on the LASSO for demonstrative purposes. The nuisance parameter \(\hat{\bnu}\) is estimated from the scaled residuals, accounting for the degrees of freedom of the LASSO estimator \citep{reid2016study}.
\begin{lemma}\label{lem:lasso}
Under the conditions specified in Theorem 1 of \cite{lounici2008sup}, Assumption \ref{assum:gamma_estimator} holds for the LASSO estimator with 
$\gamma_{n,p} = \kappa \left( \lambda_n + \sqrt{\frac{\log p}{n}} \right)$, where \(\kappa\) is a constant.
\end{lemma}
Lemma \ref{lem:lasso}, derived from \cite{lounici2008sup}, suggests that the correction factor $\gamma_{n,p}$ for the LASSO estimator can be expressed as  
$
\gamma_{n,p} = \kappa \left( \lambda_n + \sqrt{\frac{\log p}{n}} \right).
$
We estimate the value of \( \kappa \) using the data-driven correction factor selection procedure described in Appendix B.1. 

Definition \ref{def:syn_null_data_LM} defines the synthetic null data for the linear model by generating the error term \( \tilde{\bvarepsilon} \) under a parametric Gaussian model. However, in practice, the true distribution of \( \bvarepsilon \) may be unknown, and the parametric assumption may not always hold.  
To address this issue, we introduce a {non-parametric version of Nullstrap}, where synthetic null data is generated by resampling the residuals of the LASSO estimator. This approach is analogous to bootstrap resampling, except that the scaled residuals of the LASSO estimator are used in place of the ordinary least squares residuals. Define the residuals as \( \hat{\bvarepsilon} = \by - \bX \hat{\bbeta} \), and scale them according to the degrees of freedom of the LASSO estimator \citep{reid2016study}.
\begin{definition}[Non-parametric synthetic null data for a linear model]\label{def:syn_null_data_LM_nonparam}
        For a linear model $\mathbf{y} = \mathbf{X}\boldsymbol{\beta} + \boldsymbol{\varepsilon}$, where $\bvarepsilon$ follows an unknown distribution, Nullstrap defines \( \tilde{\by} \in \mathbb{R}^n \) as 
        $
        \tilde{\by} = {\bX} \bbeta_0 + \tilde{\bvarepsilon} = \tilde{\bvarepsilon},
        $
        where \( \bbeta_0 = (0, \dots, 0)\tran \in \mathbb{R}^p \) is the coefficient vector under the global null hypothesis, and  \( \tilde{\bvarepsilon} \) is generated by resampling the scaled residuals obtained from fitting a linear model to the original data $\{\by, \bX\}$ using the LASSO.
\end{definition}

% The detailed procedure for {non-parametric Nullstrap} is provided in Appendix B.3. 
We refer to the parametric and nonparametric versions of Nullstrap for the linear model---based on the synthetic null data defined in Definitions~\ref{def:syn_null_data_LM} and~\ref{def:syn_null_data_LM_nonparam}---as Nullstrap (param) and Nullstrap (non-param), respectively. 
\vspace{-3ex}

\subsection{Comprehensive method comparison in small-scale simulation}\label{sec:sim}
In this subsection, we comprehensively evaluate the performance of Nullstrap and 10 other approaches in terms of FDR control, power, and AUPR (area under the precision-recall curve) under the following simulation setting. While FDR control and power reflect both the quality of variable ranking and the effectiveness of thresholding at a target FDR level, AUPR specifically reflects the quality of variable ranking.

\begin{simsetting}\label{sim:setting1}
We set \( n = 300 \) and \( p = 200 \). The design matrix \( \bX \) consists of i.i.d. rows and AR(1) columns, generated from \( \mathcal{N}(\mathbf{0}, \bSigma) \), where \( \bSigma \) is a Toeplitz correlation matrix with an autocorrelation parameter \( \rho \in (0,1) \), representing the correlation between two adjacent variables in \( \bX \). The first \( 30 \) elements of the coefficient vector \( \bbeta \) are assigned values with amplitude \( A = 0.3 \) and random signs, while the remaining $170$ elements are set to zero. The response vector $\by$ follows $\mathbf{y} = \mathbf{X}\boldsymbol{\beta} + \boldsymbol{\varepsilon}$, where $\bvarepsilon \sim \mathcal{N}(0, \mathbf{I})$. 
\end{simsetting}

We first numerically compare Nullstrap and Nullstrap (individual) to evaluate whether Nullstrap achieves satisfactory performance in {FDR control} and {power}.  
The estimation procedure \( \mathcal{E}(\cdot, \cdot) \) is the LASSO. Prior to applying the LASSO, we center and scale the columns of \( \bX \) and center the response \( \by \).
 The regularization parameter $\lambda_n$ in \eqref{eq:lasso} is selected via 10-fold cross-validation. The correction factor $\gamma_{n,p}$ for Nullstrap is selected using the data-driven procedure described in Appendix B.1. In contrast, for Nullstrap (individual), the correction factor is set to 0 because the other coefficient estimates are retained from the original data. Therefore, a global adjustment to the coefficient estimate from the synthetic null data is unlikely to be necessary for Nullstrap (individual). 
The synthetic null data for Nullstrap and Nullstrap (individual) are generated in a parametric way, according to Definition \ref{def:syn_null_data_LM}.  
% Prior to applying the LASSO, we center and scale the columns of \( \bX \) and center the response \( \by \). 
% Details of the simulation setup are provided in Appendix \textcolor{red}{X}.  
We compare the power and FDR of Nullstrap and Nullstrap (individual) at various autocorrelation \( \rho \) values under a target FDR of $q = 0.1$. 
Each setting is evaluated using 100 replications. The results, summarized in Table~\ref{tab:IN_Nullstrap}, show that both approaches perform similarly in terms of power and FDR, but Nullstrap is computationally more efficient. Excluding the cross-validation time for determining the regularization parameter, Nullstrap requires 0.078 seconds, compared to 1.48 seconds for Nullstrap (individual). This computational advantage becomes more significant as \( p \) increases.
Interestingly, as \( \rho \) increases from $0.1$ to $0.9$, Nullstrap (individual) initially outperforms Nullstrap in power but later underperforms. Identifying the crossover point between the two approaches with respect to \( \rho \) could be a valuable theoretical topic for future research. Given its computational efficiency and strong performance, Nullstrap under the global null hypothesis is used in the following sections. 
% When no ambiguity arises, we refer to Nullstrapsimply as Nullstrap.

\begin{table}[h]
\centering
\caption{Comparison of power and FDR at various autocorrelation $\rho$ values, with a target FDR of $q=0.1$ under Simulation Setting \ref{sim:setting1}. ``Ind" and ``Gbl" represent Nullstrap (individual) and Nullstrap, respectively. The synthetic null data for both methods are generated according to Definition \ref{def:syn_null_data_LM}. Higher power values are indicated by underlining.}
\footnotesize
\begin{tabular}{ccccccccccc}
\toprule
$\rho$ & 0.0 & 0.1 & 0.2 & 0.3 & 0.4 & 0.5 & 0.6 & 0.7 & 0.8 & 0.9 \\
\midrule
\multicolumn{11}{c}{{Power}} \\
\midrule
{ Ind }   & \underline{0.964} & \underline{0.964} & \underline{0.949} & 0.906 & 0.835 & 0.723 & 0.597 & 0.482 & 0.317 & 0.186 \\
{ Gbl }  & 0.952 & 0.964 & 0.944 & \underline{0.908} & \underline{0.850} & \underline{0.771} & \underline{0.617} & \underline{0.492} & \underline{0.359} & \underline{0.216} \\
{ Gbl$-$Ind }   & -0.012 & -0.000 & -0.005 & 0.002 & 0.015 & 0.048 & 0.020 & 0.011 & 0.043 & 0.031 \\
\midrule
\multicolumn{11}{c}{{FDR}} \\
\midrule
{ Ind }     & 0.074 & 0.085 & 0.083 & 0.068 & 0.058 & 0.049 & 0.051 & 0.036 & 0.032 & 0.021 \\
{ Gbl }    & 0.088 & 0.089 & 0.082 & 0.085 & 0.082 & 0.069 & 0.068 & 0.054 & 0.053 & 0.042 \\
\bottomrule
\end{tabular}
\label{tab:IN_Nullstrap}
\end{table}
% \vspace{-3ex}

% \subsection{Simulation results on a toy dataset}\label{sec:sim}
We then compare Nullstrap against nine alternative approaches. These include five $p$-value-free approaches---Fixed-X knockoff (Fixed-X), Model-X knockoff (Model-X), Gaussian Mirror (GM), Data Splitting (DS), and Multiple Data Splitting (MDS)---as well as two $p$-value-based procedures, Benjamini--Hochberg (BH) and its adaptive variant BHq. In addition, we consider the permutation approach, which constructs synthetic null data by permuting the response vector \( \mathbf{y} \), and SLOPE.

There are two versions of Nullstrap: Nullstrap (param) and Nullstrap (non-param). Nullstrap (param) generates parametric synthetic null data according to Definition \ref{def:syn_null_data_LM}, while Nullstrap (non-param) generates non-parametric synthetic null data according to Definition \ref{def:syn_null_data_LM_nonparam}.
Under Simulation Setting~\ref{sim:setting1}, \( n > p \), so the \( p \)-values for BH and BHq are computed using \( t \)-tests based on OLS.  
Nullstrap is compared with other methods in terms of FDR and power across varying autocorrelation \( \rho \) values under a target FDR of \( q = 0.1 \). The results, summarized in Figure~S1 (Appendix~D), show that Nullstrap achieves the highest power ($0.25$--$1$) while effectively controlling the FDR, especially in high-correlation settings. The knockoff filters (Fixed-X and Model-X) exhibit conservative behavior, controlling the FDR but with reduced power ($0$--$0.15$). The DS, MDS, and $p$-value-based BH and BHq methods attain slightly higher power than the knockoff filters but remain approximately $0.15$ less powerful than Nullstrap. The GM method shows a slight violation of FDR control and reaches power levels about $0.1$ lower than those of Nullstrap. 
% As reported in Section~\ref{sec:method}, Nullstrap-Diff performs slightly worse than Nullstrap, especially in high-correlation scenarios, while Nullstrap-Adjust has better power and FDR control results than Nullstrap overall. 
The two versions of Nullstrap, Nullstrap (param) and Nullstrap (non-param), demonstrate comparable performance in FDR control and power. The permutation approach exhibits low power (approximately $0.1$--$0.2$), much lower than Nullstrap, as expected, since it does not leverage information about the extent to which the design matrix \( \bX \) explains the variance in \( \by \) (Figure~S2 in Appendix~D). The SLOPE method, whose assumption of an orthogonal design is violated in this setting, exhibits relatively high power but shows a substantial violation of FDR control, with inflation ranging from $0.05$ to $0.2$. 

Note that the two versions of Nullstrap and the permutation approach use the same statistic—the absolute coefficient estimates from the original data—to rank variables. Consequently, they achieve the same AUPR, which is higher than that of all other approaches (Figure~\ref{fig:AUPR}(a)), highlighting the superior effectiveness of this statistic for variable ranking.

% \begin{figure}[h]
% 	\centering
% 	\includegraphics[scale=0.4]{simulation/result_in_power_lm.png}
% 	\caption{Empirical FDR and power vs. autocorrelation ($\rho$) under Simulation Setting~\ref{sim:setting1}.}
% 	\centering
% 	\label{fig:in_rho}
% \end{figure}

% \begin{table}[h!]
% 	\centering
%     \footnotesize
%     \setlength{\tabcolsep}{3pt}
% 	\begin{tabular}{cccccc}
% 		\toprule
% 		Nullstrap (param) &Nullstrap (non-param)    & Permutation & Model-X & Fixed-X & GM\\
% 		\midrule
% 		0.42 & 0.50 &0.25 &  15.82 & 10.85& 42.10 \\
%         \midrule
%         DS & MDS & BH & BHq & SLOPE\\
%         		\midrule
%         0.87 & 25.01 &0.42 & 0.39 & 0.09\\
% 		\bottomrule
% 	\end{tabular}
% 	\caption{Comparison of runtimes (s)  under Simulation Setting~\ref{sim:setting1}.}
% 	\label{ind_time}
% 	\end{table}

Table~S1 in Appendix~C.1 compares the runtimes of Nullstrap (including cross-validation for selecting the LASSO regularization parameter) with those of other methods. Among them, SLOPE is the fastest ($0.09$ seconds), while GM is the slowest ($42.1$ seconds). Nullstrap (param), Nullstrap (non-param), BH, and BHq exhibit comparable computational efficiency, with runtimes between $0.39$ and $0.5$ seconds. Due to its long runtime, the GM method is excluded from further comparisons in the following sections.
\vspace{-3ex}
\subsection{Method comparison in comprehensive simulations}\label{sec:sim2}
In this subsection, we conduct a large-scale simulation study comparing Nullstrap with six competing methods—Fixed-X knockoff, Model-X knockoff, DS, MDS, BH, and BHq—that demonstrated good FDR control and reasonable runtimes in the previous subsection. As expected, we also show in Appendix C.3 that using LASSO alone for variable selection fails to control the FDR.
\begin{simsetting}\label{sim:setting2}
	We set \( n = 2000 \). The design matrix \( \bX \), the coefficient vector \( \bbeta \), and the response vector $\by$ are generated as in Simulation Setting \ref{sim:setting1}. We consider four simulation parameters for adjustment:  
	(a) the autocorrelation parameter \( \rho \in [0, 0.9] \),  
	(b) the signal amplitude \( A \in [0.15, 0.35] \),  
	(c) the target FDR level \( q \in [0.05, 0.4] \), and  
	(d) the number of variables \( p \in \{500, 1000, \dots, 3500\} \).  
	For each scenario where one simulation parameter varies, the remaining parameters are held constant as:  
	\vspace{-20pt}  
	\begin{equation}\label{eq:para}  
	\rho = 0.8, \, A = 0.25, \, q = 0.1, \, \text{and } \, p = 1000.  
	\vspace{-20pt}  
	\end{equation}  
	% The first \( 30 \) elements of the coefficient vector \( \bbeta \) are randomly assigned values with amplitude \( A \) and random signs, while the remaining \( p - 30 \) elements are set to zero. 
    Appendix~E addes two additional data-generation schemes: (i) random assignment of non-zero coefficients in $\bbeta$ and (ii) inclusion of interaction effects.
\end{simsetting}
% with a sample size of  n = 2000.
% {\color{red} \bf Change the figure’s label to something contextually relevant instead of using numeric labels like 1, 2, 3, such as "fig:LM\_amp"}
% We consider a scenario with a sample size of \( n = 2000 \) and the number of non-zero regression coefficients \( s = 30 \). 
% The design matrix \( \bX \) has i.i.d. rows and auto-regressive AR(1) columns, generated by drawing from a multivariate normal distribution \( \mathcal{N}(0, \bSigma) \), where \( \bSigma \) is constructed using the Toeplitz structure with a correlation parameter \( \rho \in (0,1) \). The first \( s \) elements of the coefficient vector \( \bbeta^* \) are randomly assigned values with amplitude \( A \) and random signs, while the remaining entries are set to zero. The noise term \( \bvarepsilon \) is sampled from a normal distribution \( \mathcal{N}(0, \sigma^2 \bI) \), where \( \sigma^2 = 1 \). Nullstrap estimates \( \sigma^2 \) using the residuals from LASSO and subsequently uses this estimate to generate the null data. The regularization parameter \( \lambda_n \) is selected using cross-validation.
% In our study, we vary four parameters across different scenarios:
% \begin{itemize}
%     \item[(a)] $\rho $
%     \item[(b)] $A $
%     \item[(c)] $q \in $
%     \item[(d)] $p \in $
% \end{itemize}
% The parameter $p$ is set to $1000$ to satisfy the exact finite sample FDR control requirement $(n \geq 2p)$ for the Fixed-X setting. 

For each scenario under Simulation Setting~\ref{sim:setting2}, we compare the FDR, power, and AUPR of Nullstrap and six competing methods based on 100 simulation replications.
% In Appendix C, we also evaluate the Area Under the Precision-Recall Curve (AUPR) for each method.
% We compare {Nullstrap} with four state-of-the-art methods: Fixed-X  knockoff (Fixed-X) \citep{barber2015controlling}, Model-X knockoff (Model-X) \citep{candes2018panning}, DS \citep{dai2023false}, and MDS \citep{dai2023false}. Additionally, we include a variant of Nullstrap, referred to as {Nullstrap\(-\)}, which employs an alternative FDP estimation approach defined in \eqref{eq:FDP2}. 
For scenarios where $p$  is large, we exclude Fixed-X knockoff from the comparison as it requires  $n \geq 2p$ . When  $n > p$ , the  $p$-values for BH and BHq are computed using the debiased LASSO.
% due to the high computational cost of using the debiased LASSO for $p$-value calculation.

\begin{figure}[h]
\centering
\includegraphics[scale=0.35]{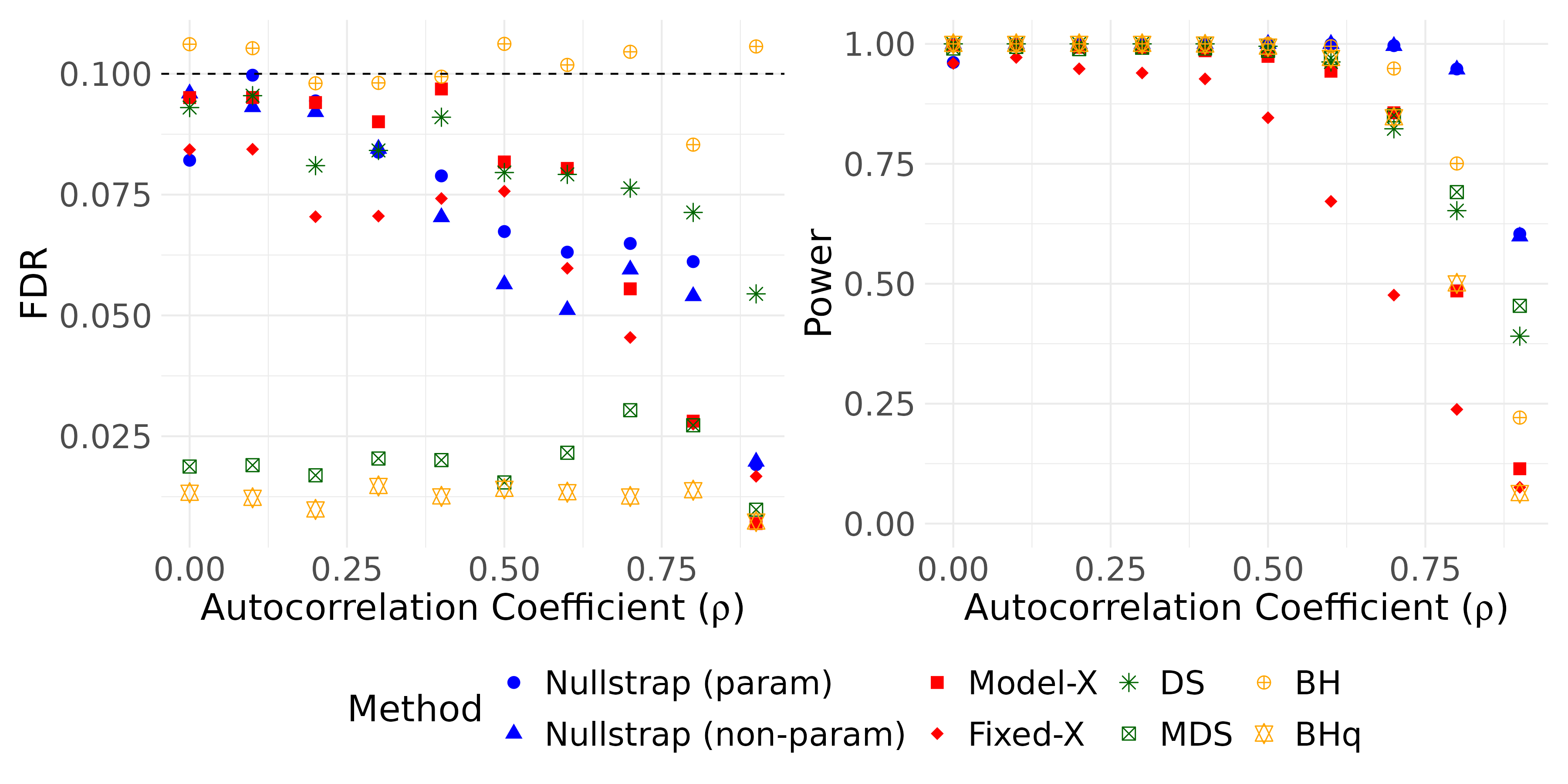}
\caption{Empirical FDR and power vs. autocorrelation ($\rho$) under Simulation Setting~\ref{sim:setting2}.}
\centering
\label{fig:LM_rho}
\end{figure}

\begin{figure}[h]
\centering
\includegraphics[scale=0.35]{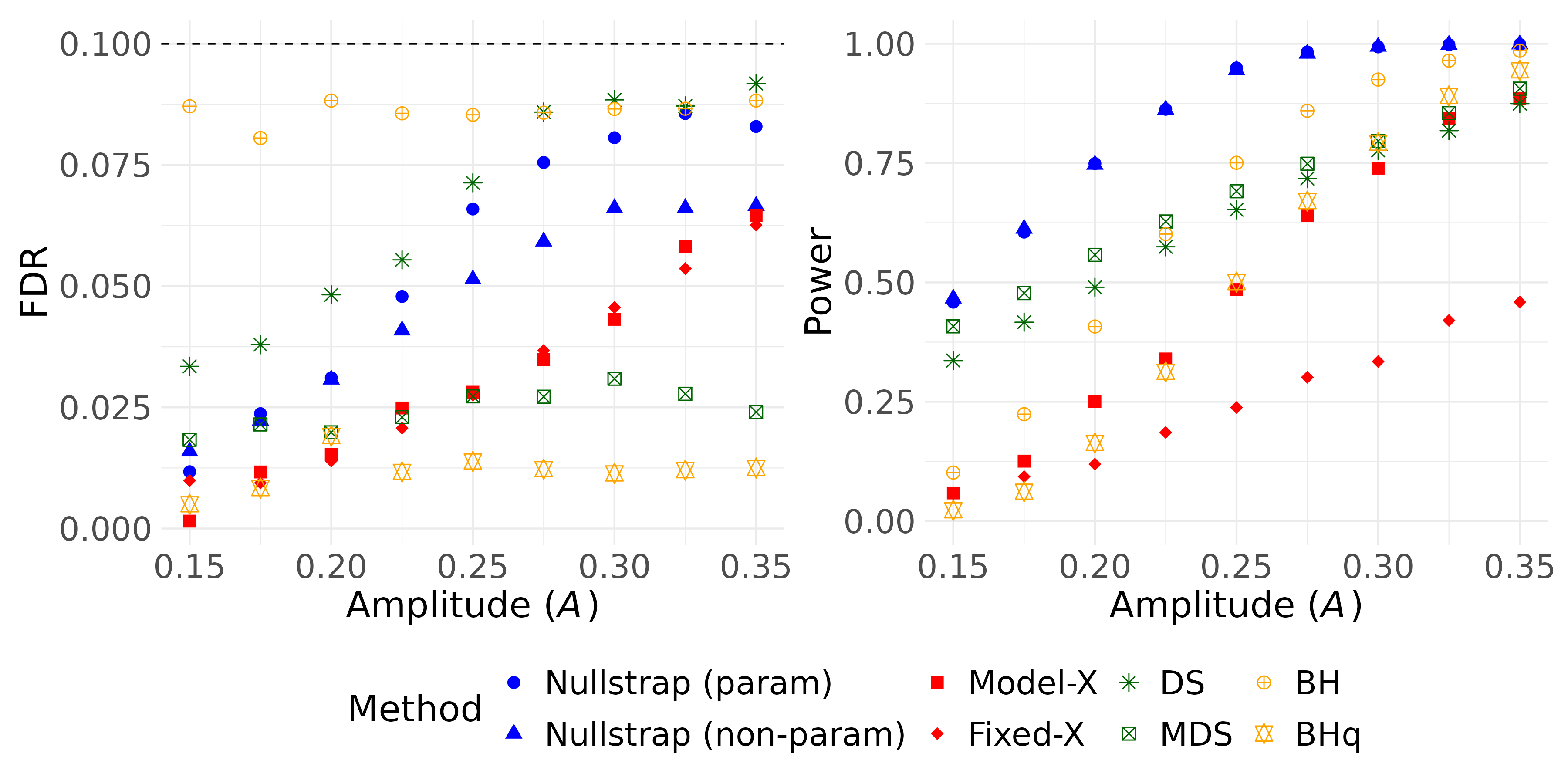}
\caption{Empirical FDR and power vs. signal amplitude ($A$) under Simulation Setting~\ref{sim:setting2}.}
\centering
\label{fig:LM_amp}
\end{figure}

\begin{figure}[h]
\centering
\includegraphics[scale=0.35]{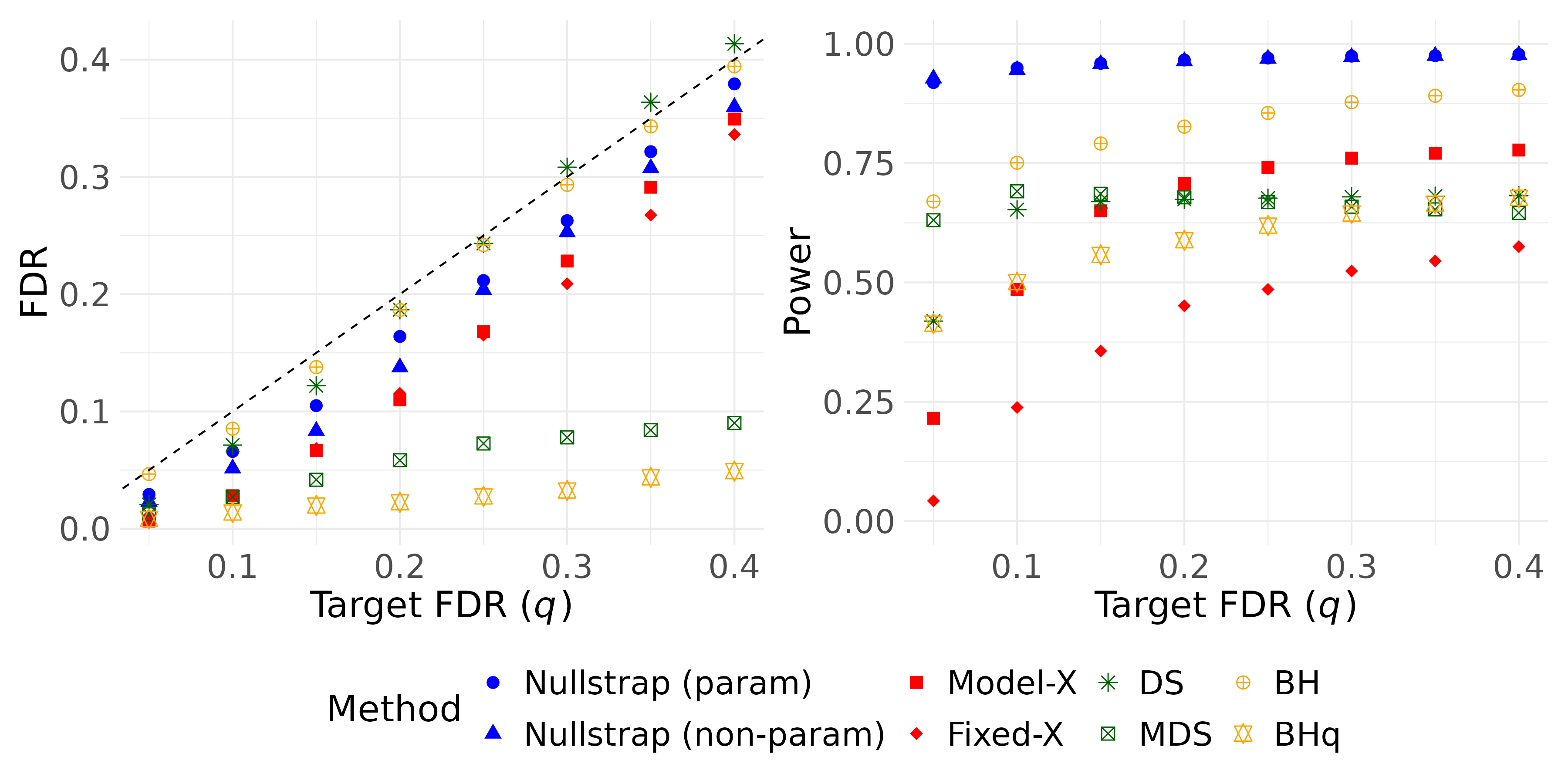}
\caption{Empirical FDR and power vs. target FDR level ($q$) under Simulation Setting~\ref{sim:setting2}.}
\centering
\label{fig:LM_fdr}
\end{figure}

\begin{figure}[h]
\centering
\includegraphics[scale=0.35]{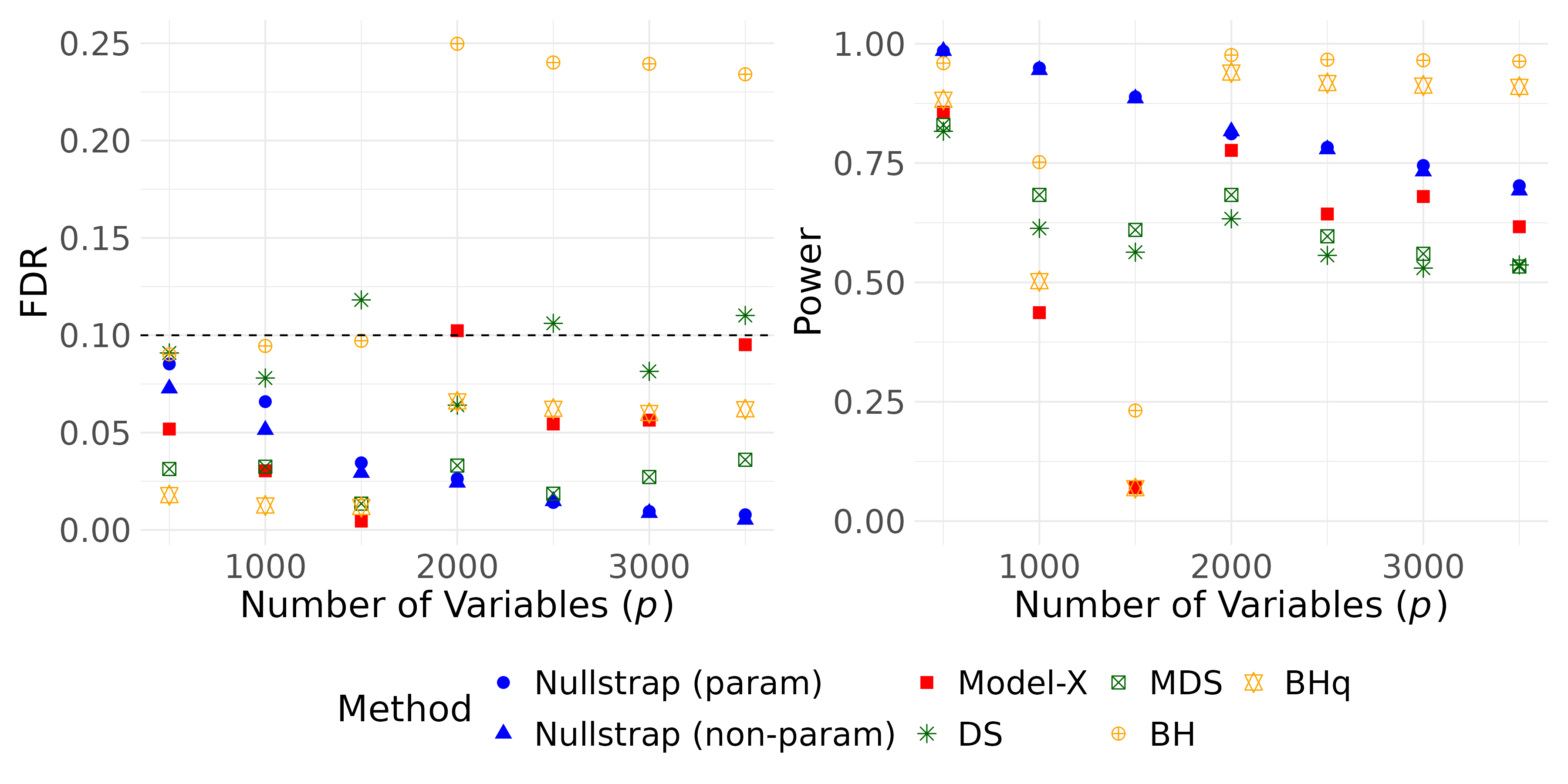}
\caption{Empirical FDR and power vs. number of variables ($p$) under Simulation Setting~\ref{sim:setting2}. BH and BHq are computed with OLS when \(p < n = 2000\), and with the debiased LASSO when \(p \ge n = 2000\).}
\centering
\label{fig:LM_var}
\end{figure}

\begin{figure}[h]
        \centering
        \begin{subfigure}{0.40\textwidth}
            \centering
            \includegraphics[width=\linewidth]{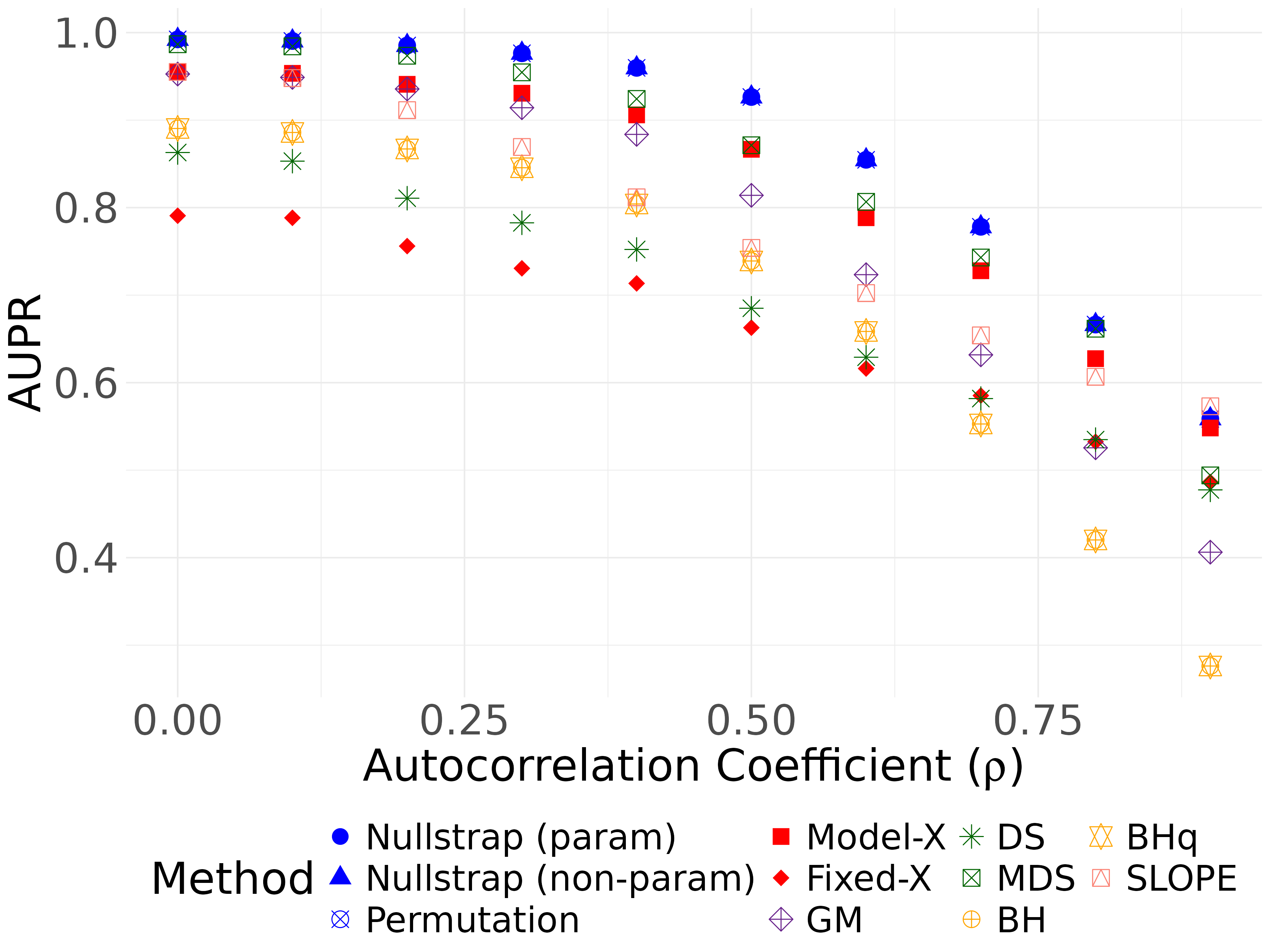} 
            \caption{Empirical AUPR vs. autocorrelation ($\rho$) under Simulation Setting~\ref{sim:setting1}.}
            \label{fig:sub1}
        \end{subfigure}
        \hfill
        \begin{subfigure}{0.40\textwidth}
            \centering
            \includegraphics[width=\linewidth]{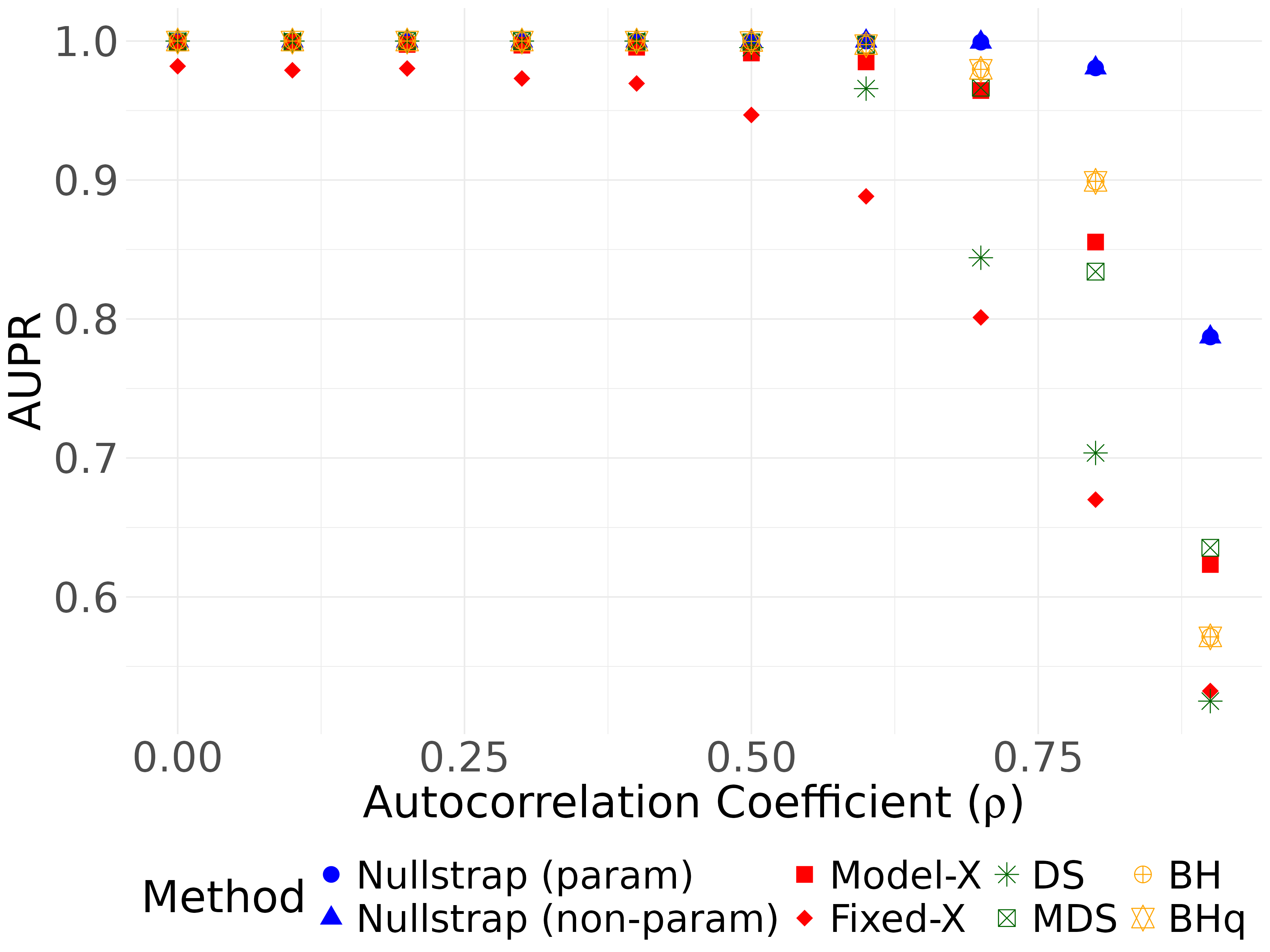} 
            \caption{Empirical AUPR vs. autocorrelation ($\rho$) under Simulation Setting~\ref{sim:setting2}.}
            \label{fig:sub2}
        \end{subfigure}

        \vspace{0.5cm}
        \begin{subfigure}{0.40\textwidth}
            \centering
            \includegraphics[width=\linewidth]{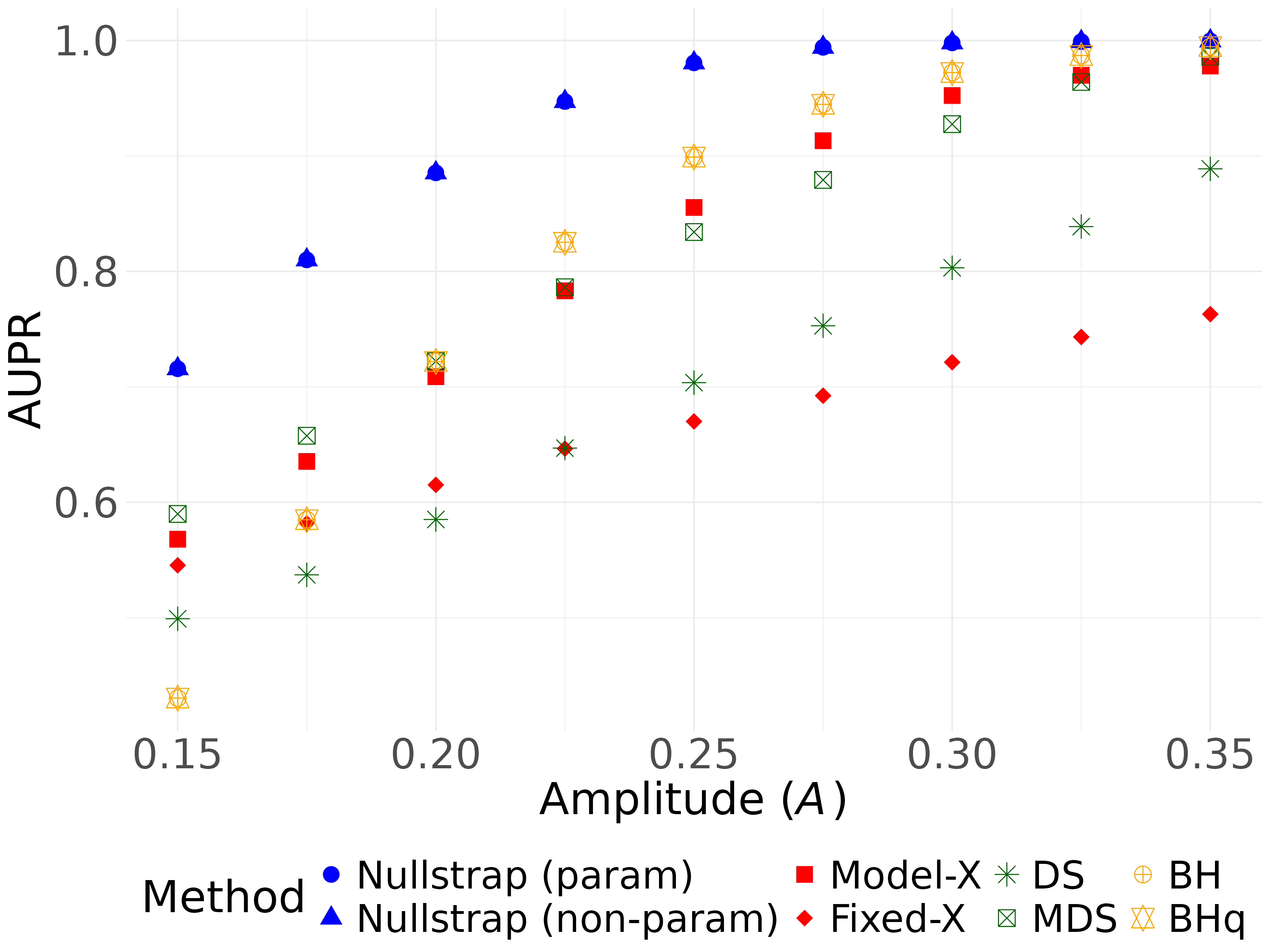} 
            \caption{Empirical AUPR vs. signal amplitude ($A$) under Simulation Setting~\ref{sim:setting2}.}
            \label{fig:sub3}
        \end{subfigure}
        \hfill
        \begin{subfigure}{0.40\textwidth}
            \centering
            \includegraphics[width=\linewidth]{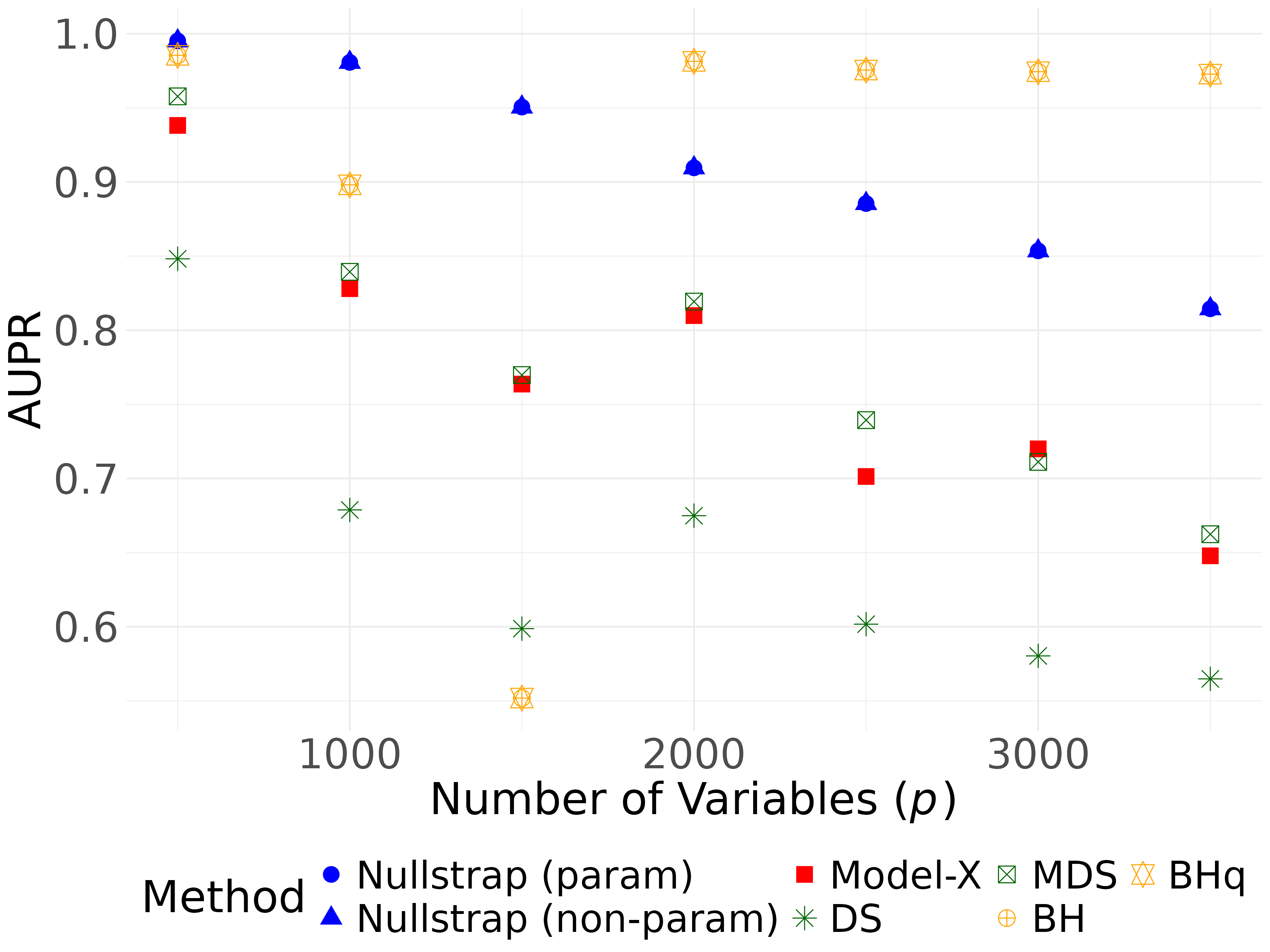} 
            \caption{Empirical AUPR vs. number of variables ($p$) under Simulation Setting~\ref{sim:setting2}.}
            \label{fig:sub4}
        \end{subfigure}
        
        \caption{Empirical AUPR under Simulation Settings~\ref{sim:setting1}--\ref{sim:setting2}}
        \label{fig:AUPR}
    \end{figure}

The empirical FDR and power of the above methods are presented in Figures \ref{fig:LM_rho}--\ref{fig:LM_var}. The AUPR results are provided in Figure~\ref{fig:AUPR}(b)--(d).
Overall, the FDR of most methods remain controlled across all scenarios, except for DS and BH, which sometimes slightly lose control. In all scenarios, Nullstrap consistently demonstrates reliable FDR control and, more importantly, achieves higher power and AUPR than all other methods except BHq with the debiased LASSO, which requires a long runtime. 
The two versions of Nullstrap---Nullstrap (param) and Nullstrap (non-param)---exhibit similar performance. Although the data are generated under the Gaussian linear model assumed by Nullstrap (param), Nullstrap (non-param) achieves only slightly lower power, demonstrating the robustness of Nullstrap (non-param) even without assuming Gaussian error distributions.
% Compared to Nullstrap, Nullstrap-Diff exhibits slightly lower power, particularly in more challenging scenarios such as those with high autocorrelations and low signal amplitudes. 
% This observation aligns with our analysis in Section \ref{sec:method}, which highlights the effectiveness of the metric in \eqref{eq:FDP} for FDP estimation in our method.

% \vspace{-3ex}

%\vspace{-3ex}

% \vspace{-3ex}

% \vspace{-3ex}

% \begin{figure}[h]
% 	\centering
% 	\includegraphics[scale=0.4]{simulation/result_rho_aupr_lm.png}
% 	\caption{Empirical AUPRs for the linear regression model (Correlation).}
% 	\centering
% 	\label{fig:LM_rho(aupr)}
% 	\end{figure}

% \begin{figure}[h]
% 	\centering
% 	\includegraphics[scale=0.4]{simulation/result_amp_aupr_lm.png}
% 	\caption{Empirical AUPRs for the linear regression model (Amplitude).}
% 	\centering
% 	\label{fig:LM_amp(aupr)}
% 	\end{figure}

% 	\begin{figure}[h]
% 	\centering
% 	\includegraphics[scale=0.4]{simulation/result_num_aupr_lm.png}
% 	\caption{Empirical AUPRs for the linear regression model (Number of variables).}
% 	\centering
% 	\label{fig:LM_num(aupr)}
% 	\end{figure}

Specifically, in Figure~\ref{fig:LM_rho}, where the autocorrelation \( \rho \) between variables increases, Nullstrap's power declines more slowly than that of other methods, demonstrating its greater robustness to high correlations among variables. Similarly, Figure~\ref{fig:AUPR}(b) shows that Nullstrap exhibits a slower decrease in AUPR as \( \rho \) increases.
In Figure~\ref{fig:LM_amp} and Figure~\ref{fig:AUPR}(c), where the amplitude \( A \) is varied, we observe that once \( A \) reaches 0.3, both the power and AUPR of Nullstrap attain \( 1 \) and remain constant thereafter.
% This can be attributed to the fact that beyond this amplitude, the signal strength is sufficient for Nullstrap to detect all true positives without error.  
In Figure \ref{fig:LM_fdr}, when varying the target FDR level \( q \), Nullstrap consistently achieves the highest power across all FDR levels compared to the other methods.
% , and Nullstrap-Diff also maintains relatively high power except at \( q = 0.05 \). 
%Furthermore, we observe that as $q$ increases, DS tends to slightly lose control over its actual FDR.
When varying the number of variables \( p \), Nullstrap consistently achieves the highest power among all methods that control the FDR (Figure~\ref{fig:LM_var}) and the highest AUPR (Figure~\ref{fig:AUPR}(d)), except for BHq when \( p \geq n = 2000 \). Notably, BH fails to control the FDR in this regime, even though BH and BHq share the same variable ranking, both relying on \( p \)-values from the debiased LASSO. However, as shown in Table~\ref{lm_time}, BHq incurs substantially higher computational cost---on average, two orders of magnitude greater than Nullstrap across values of \( p \)---particularly in high-dimensional settings. Moreover, the debiased LASSO is a model-specific method that may not generalize beyond linear models or LASSO-type estimators. In contrast, Nullstrap provides significantly faster computation while maintaining flexibility across a broad class of models and estimators.

For the two additional data-generation schemes, Figures~S4-S8 (Appendix E.1) present the results under random assignment of nonzero coefficients in \( \bbeta \), while Figures~S9-S11 (Appendix E.2) show the results for the setting with interaction effects included.

% Table \ref{lm_time} summarizes the runtimes under the specific setting described in \eqref{eq:para}.

%This can potentially be explained by the equality between the number of observations and that of variables $(n = p)$. 

% \begin{table}[h!]
% \centering
% \begin{tabular}{ccccc}
%     \toprule
%     Nullstrap &  Model-X & Fixed-X  & DS & MDS  \\
%     \midrule
%         5.72 &  40.92 &  62.92  & 2.53 & 86.56\\
%     \bottomrule
% \end{tabular}
% \caption{Comparison of runtimes (s) under \eqref{eq:para} in Simulation Setting~\ref{sim:setting2}.}
% \label{lm_time}
% \end{table}

\begin{table}[h!]
\centering
\caption{Comparison of total runtimes (in seconds) under varying $p$ in Simulation Setting~\ref{sim:setting2}.}
\footnotesize
\begin{tabular}{ccccccc}
    \toprule
    Nullstrap (param) & Nullstrap (non-param) & Model-X & DS & MDS & BH & BHq \\
    \midrule
        1319.53 & 1312.69 & 28{,}049.31 &  3172.26 &  36{,}011.37 & 108{,}543.96 & 108{,}997.01\\
    \bottomrule
\end{tabular}

\label{lm_time}
\end{table}
\vspace{-4ex}

\subsection{Robustness of Nullstrap to the error distribution}\label{sec:sim3}
In this subsection, we evaluate the robustness of Nullstrap to the distribution of the error term \( \bvarepsilon \) in the linear model. We consider the following simulation setting:
\begin{simsetting}\label{sim:setting_diff}
    The simulation setting is identical to Simulation Setting~\ref{sim:setting2}(b), except that the signal amplitude \( A \) is drawn from the interval \([0.3, 0.5]\), and the error term \( \boldsymbol{\varepsilon} \) follows a \( t \)-distribution with 3 degrees of freedom. Appendix~E considers alternative error distributions, including the \( t \)-distribution with 10 degrees of freedom, the Laplace distribution, and the centered, asymmetric Gamma distribution.
    % We set \( n = 2000 \), \( p = 1000 \) \( \rho = 0.8 \), and \( q = 0.1 \).  We vary one simulation parameter: the signal amplitude \( A \in [0.3, 0.5] \). The design matrix \( \bX \) and the true coefficient vector \( \bbeta \) are generated as in Simulation Setting~\ref{sim:setting2} 
    % % The design matrix \( \bX \) is generated as in Simulation Setting \ref{sim:setting1} with \( \rho = 0.8 \).
    % % The first 30 elements of the coefficient vector \( \bbeta \) are randomly assigned values with amplitude \( A \) and random signs, while the remaining \( p - 30 \) elements are set to zero.  
    % We consider the following error distribution: the \( t \)-distribution with 3 degrees of freedom, denoted as \( t_{3} \).  
    \end{simsetting}

 We compare the performance of the two versions of Nullstrap—Nullstrap (param) and Nullstrap (non-param)—with four competing methods (Fixed-X knockoffs, Model-X knockoffs, DS, and MDS) based on 100 simulation replications.
 Under Simulation Setting~\ref{sim:setting_diff}, Nullstrap (param) is subject to model misspecification. 
\begin{figure}[h]
\centering
\includegraphics[scale=0.35]{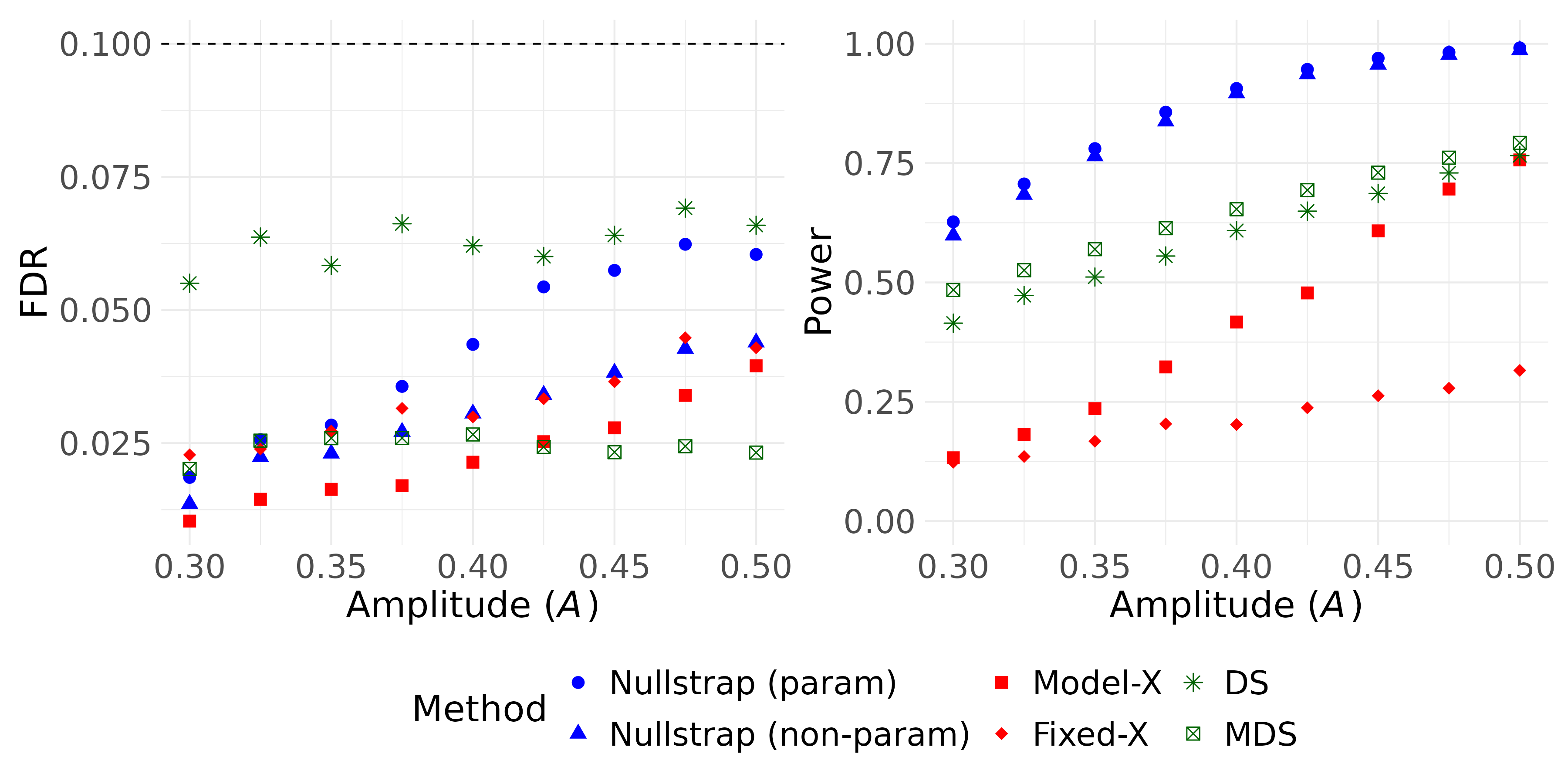}
\caption{Empirical FDR and power vs. signal amplitude ($A$) under Simulation Setting~\ref{sim:setting_diff}.}
\centering
\label{fig:amp_power_t3}
\end{figure}
% \vspace{-3ex}

The empirical FDR and power results are presented in Figure \ref{fig:amp_power_t3}. The AUPR results are provided in Figure S3 (Appendix~D). For alternative error distributions, the results are in Figures~S12--S20 (Appendix E.3). Nullstrap (param) and Nullstrap (non-param) exhibit similar performance, with Nullstrap (non-param) slightly more conservative in FDR control. Both versions of Nullstrap outperform the other methods in terms of power. Remarkably, Nullstrap (param) maintains FDR control despite model misspecification, highlighting its robustness to deviations in the error distribution, likely enabled by the data-driven correction factor. 
\vspace{-3ex}
\subsection{Comparison of method stability}\label{sec:stability}

In this subsection, we analyze the stability of Nullstrap in comparison with three randomized competing methods—Model-X knockoffs, DS, and MDS—under the linear model. Fixed-X knockoffs, BH, and BHq are excluded from this analysis as they do not involve any source of algorithmic randomness.
For each method, we perform 100 independent replications: synthetic null data generation for Nullstrap, knockoff variable generation for Model-X, and data splitting for DS and MDS. This results in 100 sets of selected variables per method.
The simulation data are generated under Simulation Setting~\ref{sim:setting2}, with parameters specified in~\eqref{eq:para}. To assess stability, we compute the Jaccard index, defined as the ratio of the intersection to the union of the 100 selected sets. This index quantifies the degree of overlap among selected variables across random initializations for each method.

\begin{table}[h!]
\centering
\caption{Comparison of Jaccard indices under the default parameter setting~\eqref{eq:para} in Simulation Setting~\ref{sim:setting2}.}
\footnotesize
\begin{tabular}{ccccc}
    \toprule
    Nullstrap (param) & Nullstrap (non-param) &  Model-X &  DS & MDS  \\
    \midrule
        0.980 & 0.993 & 0.000 & 0.416 & 0.864\\
    \bottomrule
\end{tabular}
\label{lm_jac}
\end{table}
% In Table \ref{lm_jac}, we plan to compare the Jaccard index to evaluate the overlap between the set of variables selected by a method. Since all methods involve some degree of randomization, 
Table~\ref{lm_jac} illustrates the stability of the two Nullstrap variants. Nullstrap (non-param) achieves the highest Jaccard index (0.993), followed by Nullstrap (param) at 0.980, demonstrating strong stability under randomization.
In contrast, DS and MDS yield lower Jaccard indices of 0.416 and 0.864, respectively. Model-X knockoffs exhibits a Jaccard index of 0.000 due to its low power---often selecting no variables under certain random initializations---which results in poor consistency across replications. These results highlight the superior stability of Nullstrap compared to existing randomized methods.

\vspace{-3ex}
\subsection{Real data analysis}\label{sec:real}
% \subsection{Data description}

In this section, we apply Nullstrap to a longitudinal time-to-labor dataset collected from pregnant women receiving antepartum and postpartum care at Stanford’s Lucile Packard Children’s Hospital \citep{stelzer2021integrated}. The dataset includes 63 participants in their second or third trimester of an uncomplicated pregnancy with a single fetus, each contributing 1 to 3 samples. Each sample comprises 6348 variables, including 3529 metabolites, 1317 plasma proteins, and 1502 single-cell immune variables derived from blood mass cytometry.

This dataset was previously analyzed using Stabl \citep{hedou2024discovery}, a method that integrates knockoff filters with stability selection. In that study, the dataset was split into training and validation datasets using a patient-wise shuffle-split approach: the training set includes 150 samples from 53 participants, and the validation set includes 27 samples from 10 participants. Because the validation dataset was not made available, our analysis focuses exclusively on the training dataset. For preprocessing, we removed variables that were zero across all observations, and the final dataset contains $n=150$ observations and $p=6331$ variables. As in the Stabl paper—which used linear models with LASSO, Elastic Net, and adaptive LASSO without accounting for the dataset's longitudinal structure—we also apply linear models here for method comparison, deferring more careful longitudinal modeling to future work.

%In this section, we apply Nullstrap to a longitudinal {time-to-labor} dataset, collected from pregnant women receiving antepartum and postpartum care at Stanford's Lucile Packard Children's Hospital \citep{stelzer2021integrated}. This dataset was previously analyzed with Stabl \citep{hedou2024discovery}, a method that combines knockoff filtering and stability selection. We do not include Stabl in our benchmarks because its computational demands are prohibitive.
% \textcolor{red}{EXPLAIN WHY WE DIDN'T COMPARE Nullstrap WITH STABL} 
%The dataset includes 63 participants in their second or third trimester of an uncomplicated pregnancy with a single fetus, each contributing 1 to 3 samples. Each sample comprises 6348 variables, including 3529 metabolites, 1317 plasma proteins, and 1502 single-cell immune variables derived from blood mass cytometry.
%The dataset was divided into training and validation sets via a patient-wise shuffle-split approach. The training set contains data from 53 women, totaling 150 samples, while the validation set includes 10 participants with 27 samples. Due to unavailability of the validation set in \cite{hedou2024discovery}, our analysis focuses exclusively on the training data.

The performance of Nullstrap, Model-X knockoffs, and MDS was evaluated using three metrics: model parsimony, prediction accuracy, and computational efficiency. Model parsimony reflects the preference for simpler models that use fewer variables, assuming similar predictive accuracy. Prediction accuracy is measured by the adjusted \( R^2 \) value, which captures how well the model explains variability in the response variable while penalizing for model complexity. Computational efficiency is assessed by runtime. Each metric was averaged over 70 replications of each method. We do not include Stabl due to its high computational cost; as shown in Table~\ref{lm_time}, even Model-X knockoffs—only one component of Stabl—require substantial runtime. We also exclude the Fixed-X knockoffs, whose applicability is restricted to \(n \ge 2p\), and DS, which MDS consistently supersedes in accuracy.

% \subsection{Analysis results}

\begin{figure}[h]
\centering
\footnotesize
\begin{subfigure}[t]{0.5\textwidth}
    \centering
    \includegraphics[width=0.8\textwidth]{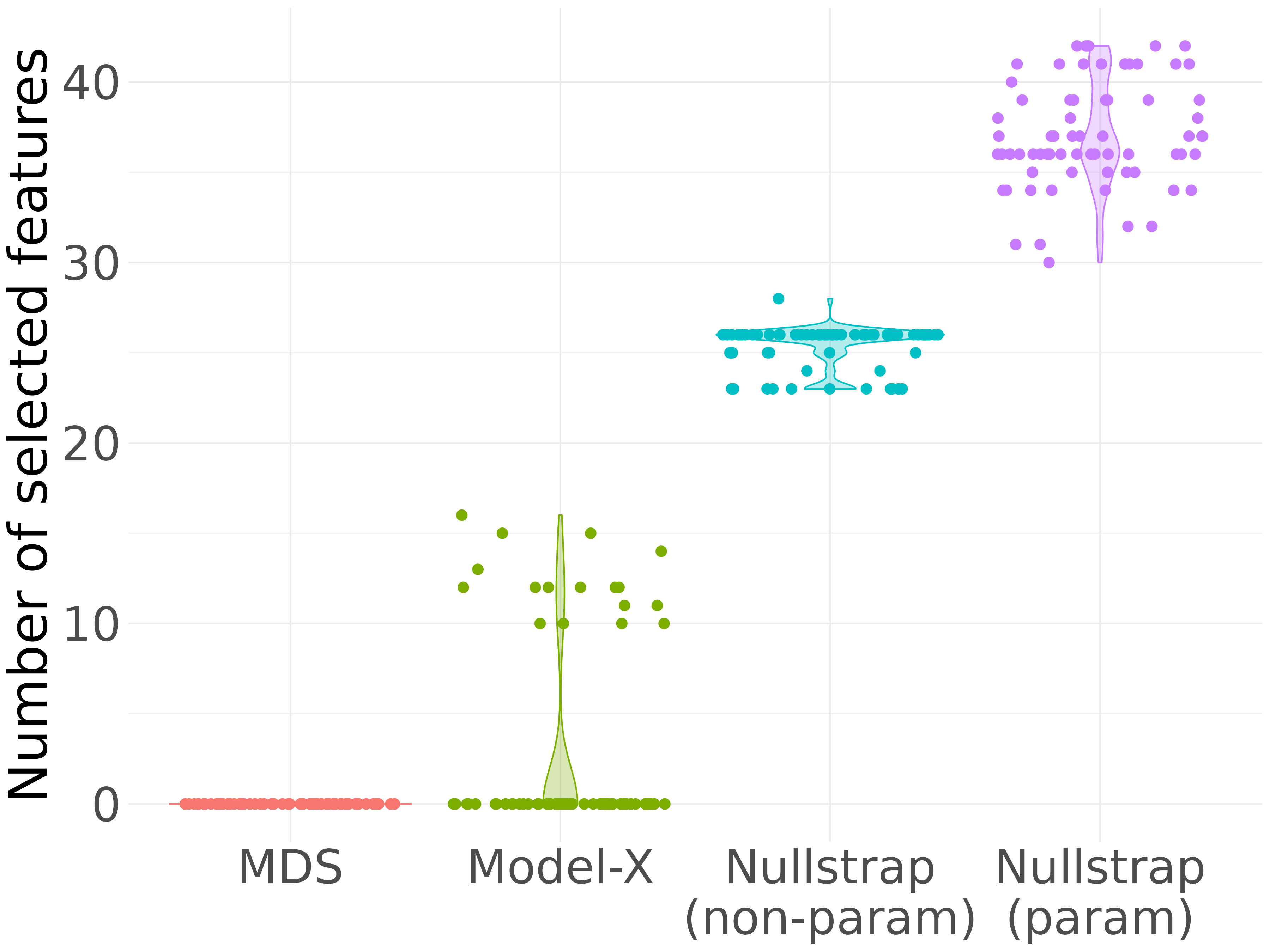}
    \caption{}
    \label{figure13:1}
\end{subfigure}%
\hfill
\begin{subfigure}[t]{0.5\textwidth}
    \centering
    \includegraphics[width=0.8\textwidth]{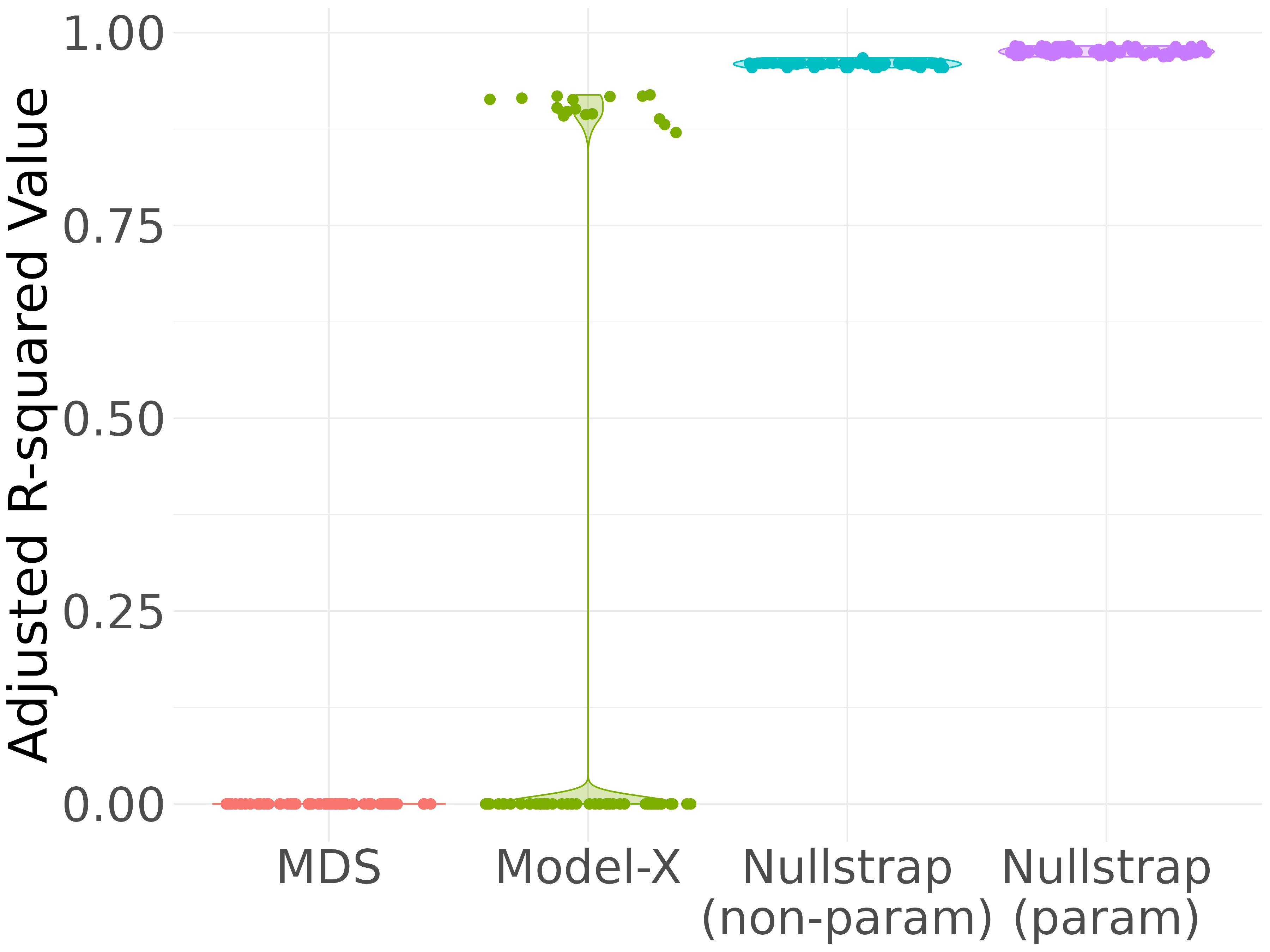}
    \caption{}
    \label{figure13:2}
\end{subfigure}%

\caption{Method performance on the time-to-labor dataset. (a) Number of selected variables (model parsimony). (b) Adjusted \( R^2 \) (prediction accuracy).}
\label{fig:real}
\end{figure}

The LASSO regularization parameter \( \lambda_n \) is selected using 10-fold cross-validation. %consistent with that in Simulation Setting~\ref{sim:setting3}.
The FDR level is \( q = 0.1 \) for all methods. Figure~\ref{fig:real} summarizes the performance of the methods. MDS selects no variables across all 70 replications, likely due to the small sample size (\(n = 150\)), which reduces power under data splitting, as the model is fit on only half of the data. Model-X knockoffs select variables in only 17 out of 70 replications, likely due to the high dimensionality (\(p = 6331\)), which poses challenges for the knockoff framework, as it doubles the number of variables in the linear model fitting.
In contrast, Nullstrap selects variables in every replication, consistent with its high power in variable selection observed in our simulation studies.
% MENTION THAT SUCH A SMALL SAMPLE SIZE DOES NOT AFFORD MDS

% \vspace{-3ex}

% \begin{table}[h!]
% \centering
% \setlength{\tabcolsep}{5mm}
% \begin{tabular}{cccccc}
%     \toprule
% 	Nullstrap &  Model-X knockoffs & Multiple Data Splitting  \\
%     \midrule
%      17.81 & 11432.09 & 421.47\\
%     \bottomrule
% \end{tabular}
% \caption{runtime (in seconds)}
% \label{onset_time}
% \end{table}

\begin{table}[h!]
	\centering
     \caption{Comparison of runtimes (s) on the time-to-labor dataset.}
	% \captionsetup{labelformat=empty, position=above}
	% \caption{runtime (in seconds)}
	\setlength{\tabcolsep}{5mm}
	\begin{tabular}{cccc}
		\toprule
		Nullstrap (param) & Nullstrap (non-param) & Model-X & MDS  \\
		\midrule
		 13.62 & 13.79 & 11432.09 & 421.47\\
		\bottomrule
	\end{tabular}
   
	\label{onset_time}
	\end{table}
% \vspace{-3ex}

Specifically, Nullstrap achieves higher adjusted \( R^2 \) values than Model-X knockoffs, reflecting its superior statistical power, even though Model-X attains slightly better model parsimony. In addition, Table~\ref{onset_time} highlights a key advantage of Nullstrap: computational efficiency, with a runtime approximately $1/800$ that of Model-X knockoffs and $1/30$ that of MDS. Overall, Nullstrap consistently outperforms the other two methods.

Next, we extract the variables selected by Nullstrap with a selection frequency exceeding 50\% across 70 replications. 
% Several variables consistently appeared in all 70 replications in both Nullstrap (param) and Nullstrap (non-param), highlighting their significance in relation to labor.
%Nullstrap identifies X metabolomic, X proteomic, and X immune cell variable, indicating a broader selection of key variables.
% From a biological perspective, the approaching labor is associated with a decreased responsiveness to inflammatory stimuli, such as the pSTAT1 signaling response to IFN$\alpha$ in natural killer (NK) cells \citep{shah2017changes, kraus2012characterizing}. 
Nullstrap identifies placental-derived proteins (e.g., Siglec-6) and immune-regulatory plasma proteins (e.g., IL-1R4 and SLPI), consistent with those reported by \citet{hedou2024discovery}. 
% Nullstrap selects placental-derived proteins (e.g. Siglec-6 \citep{brinkman2007human}), immune regulatory plasma proteins (e.g., IL-1R4 \citep{huang2017interleukin} and SLPI \citep{li2009alteration}). These findings are consistent with those reported with Stabl in \citet{hedou2024discovery}. 
Additionally, Nullstrap reveals increased Activin A and decreased hCG levels, consistent with previous findings \citep{petraglia1995abnormal,edelstam2007human}, neither of which were identified by Stabl \citep{hedou2024discovery}. Table~S10 summarizes the key variables identified by Nullstrap that may be predictive of labor timing.

\vspace{-3ex}

\section{Nullstrap for GLM, Cox model, and GGM}\label{sec:extension}
In this section, we apply Nullstrap for variable selection in the GLM, Cox proportional hazards model, and GGM. Detailed settings are provided in Appendices F–H. Below, we present representative simulation results for each model.
% , while the complete results can be found in the supplementary material.  

For the GLM, we use logistic regression as an example, where the response variable \( Y \) follows a Bernoulli distribution. As in the linear model setting, we compare Nullstrap with Fixed-X and Model-X knockoffs, DS, and MDS. Following the simulation setup in \citet{dai2023scale}, our results show that Nullstrap consistently achieves higher power than competing methods while maintaining FDR control (Figure~\ref{fig:GLM_amp} and Figures~S21-S28 in Appendix~F).
 
% The simulation setting considered in this section is as follows:  
% \begin{simsetting}\label{sim:settingglm}
%     We consider a logistic regression model with a sample size of \( n = 3000 \) and \( p = 500 \) variables. The design matrix \( \bX \) is generated as described in Simulation Setting \ref{sim:setting1}, with autocorrelation \( \rho = 0.6 \). Subsequently, \( \bX \) is centered and scaled by dividing each element by \( \sqrt{n} \). The coefficient vector \( \bbeta \) is defined in the same manner as in Simulation Setting \ref{sim:setting1}.  
%     We vary one simulation parameter: the signal amplitude \( A \in [6, 12] \). The target FDR level is set to \( q = 0.1 \).  
%     The first 30 elements of the coefficient vector \( \bbeta \) are randomly assigned values with amplitude \( A \) and random signs, while the remaining \( p - 30 \) elements are set to zero.  
%     \end{simsetting}
%     We perform 100 replications for each setting.  
      
    \begin{figure}[h]
        \centering
        \includegraphics[scale=0.35]{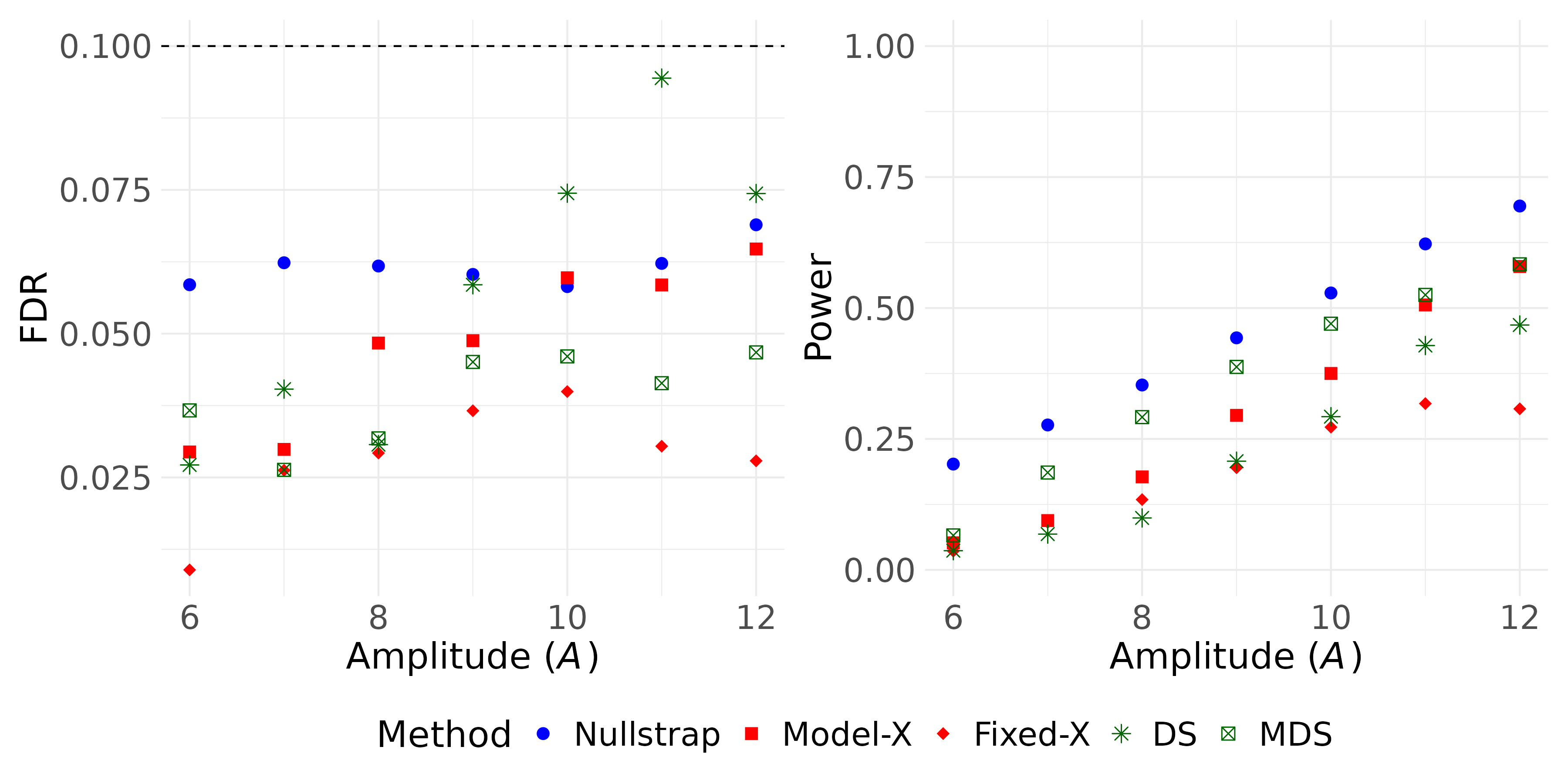}
        \caption{Empirical FDR and power vs. signal amplitude ($A$) under the GLM.}
        \centering
        \label{fig:GLM_amp}
        \end{figure}

For the Cox proportional hazards model, we compare the performance of Nullstrap with Fixed-X and Model-X knockoffs. DS and MDS are not included due to the lack of available implementations for the Cox model. As shown in Figure~\ref{fig:Cox_amp} and Figures~S29-S35 in Appendix~G, Nullstrap consistently outperforms the knockoff filters. In particular, when the signal amplitude \( A = 7 \), both the power and AUPR of Nullstrap reach 1 and remain stable. Moreover, for \( A < 7 \), Nullstrap’s power increases more rapidly than that of the knockoff methods.
        
%         The simulation setting is as follows:  
% \begin{simsetting}\label{setting:cox}
%     We consider a Cox regression model with a sample size of \( n = 400 \) and \( p = 200 \) variables. The design matrix \( \bX \) is generated as described in Simulation Setting \ref{sim:setting1}, with autocorrelation \( \rho = 0.4 \). Subsequently, \( \bX \) is centered and scaled by dividing each element by \( \sqrt{n} \).  
%     The baseline hazard function \( h_0(t) \) follows a Weibull distribution with shape parameter 1 and scale parameter 1. The coefficient vector \( \bbeta \) is defined in the same manner as in Simulation Setting \ref{sim:setting1}.  
%     We vary one simulation parameter: the signal amplitude \( A \in [2, 9] \). The target FDR level is set to \( q = 0.1 \).  
%     The first 30 elements of the coefficient vector \( \bbeta \) are randomly assigned values with amplitude \( A \) and random signs, while the remaining \( p - 30 \) elements are set to zero.   
%     \end{simsetting}
      
    \begin{figure}[h]
        \centering
        \includegraphics[scale=0.35]{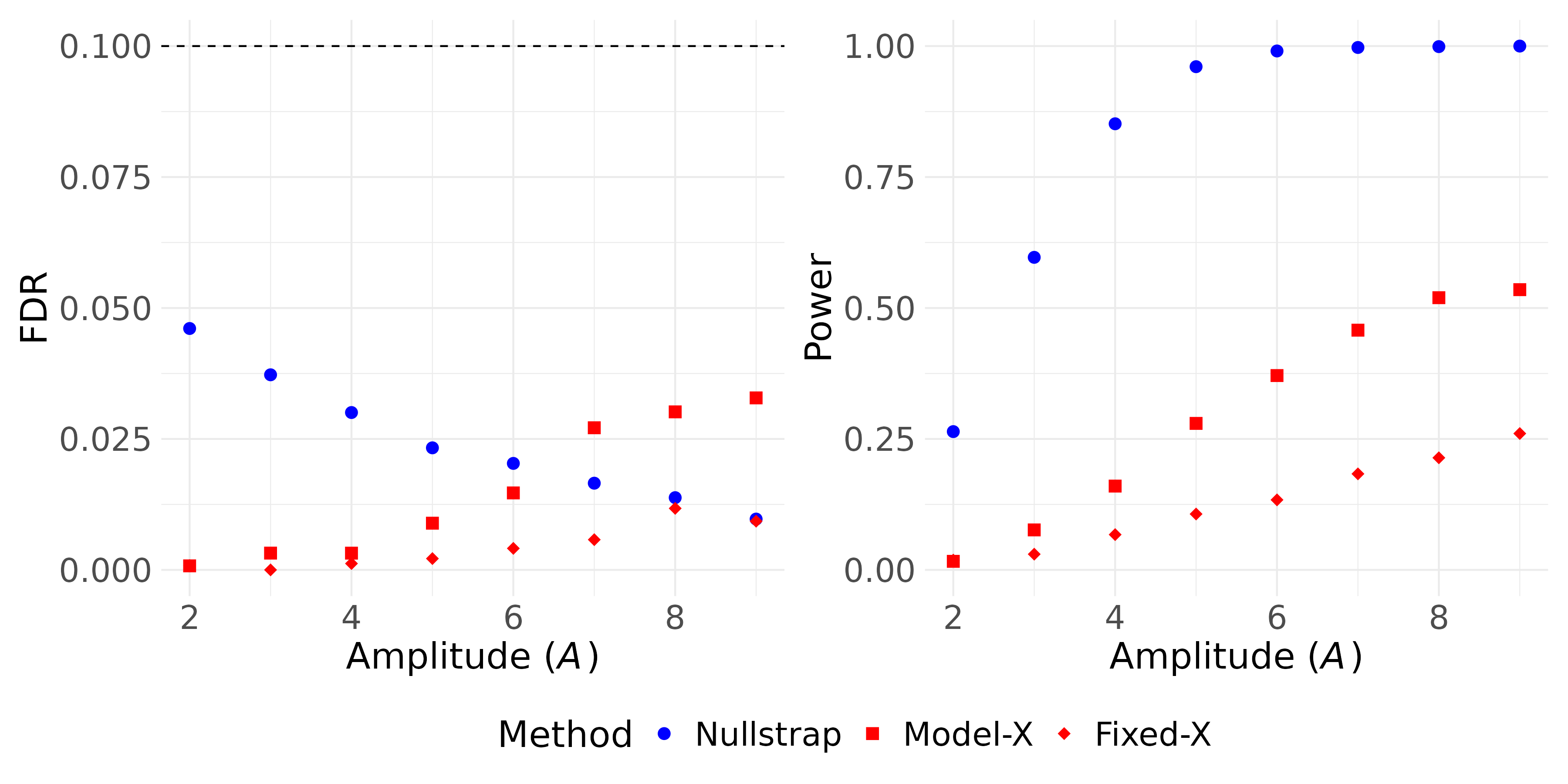}
        \caption{Empirical FDR and power vs. signal amplitude ($A$) under the Cox model.}
        \centering
        \label{fig:Cox_amp}
        \end{figure}

For the GGM, we compare Nullstrap with DS and three methods specifically designed for GGM variable selection: GFC-L \citep{698516a0-42f0-3d43-8108-2614216a2921}, GFC-SL \citep{698516a0-42f0-3d43-8108-2614216a2921}, and KO2 \citep{yu2021false}, which incorporates an in-house knockoff implementation. MDS is excluded from this comparison due to its prohibitive runtime: like DS, it requires fitting \( p \) node-wise linear regressions, each with \( p - 1 \) predictors, which becomes computationally infeasible in high-dimensional settings. Following the simulation setup in \cite{li2021ggm}—which also employs node-wise linear regressions and knockoffs but takes approximately 300 times longer to run than Nullstrap (see Appendix~H)—we find that Nullstrap outperforms all competing methods, achieving the highest power while maintaining FDR control across all sample sizes in three out of four graph-generating mechanisms (Figure~\ref{fig:GGM_n_block} and Figures~S36-S44 in Appendix~H).

%         The simulation setting is as follows:  
% \begin{simsetting}\label{setting:ggm}
%     We set the dimension of the precision matrix \( \bTheta \) to \( p = 200 \). We generate \( n \) independent samples from a multivariate normal distribution \( \mathcal{N}(0, \bTheta^{-1}) \), where  
% $
%     \bTheta := \bTheta^0 + (|\lambda_{\min}(\bTheta^0)| + 0.5) I,
% $
%     with \( \lambda_{\min}(\bTheta^0) \) being the minimum eigenvalue of \( \bTheta^0 \), ensuring that \( \bTheta \) is positive definite.  
%     The precision matrix \( \bTheta^0 \) follows a block graph structure. Specifically, \( \bTheta^0 \) is constructed by dividing the matrix into 10 blocks, each containing 20 consecutive nodes. Within each block, all diagonal elements are set to 1, and all off-diagonal elements are set to \( b = -0.8 \).  
%     We vary one parameter: the sample size \( n \in [1500, 4000] \). The target FDR level is set to \( q = 0.2 \).  
% \end{simsetting}
% We replicate 100 times for each setting. 

\begin{figure}[h]
    \centering
    \includegraphics[scale=0.35]{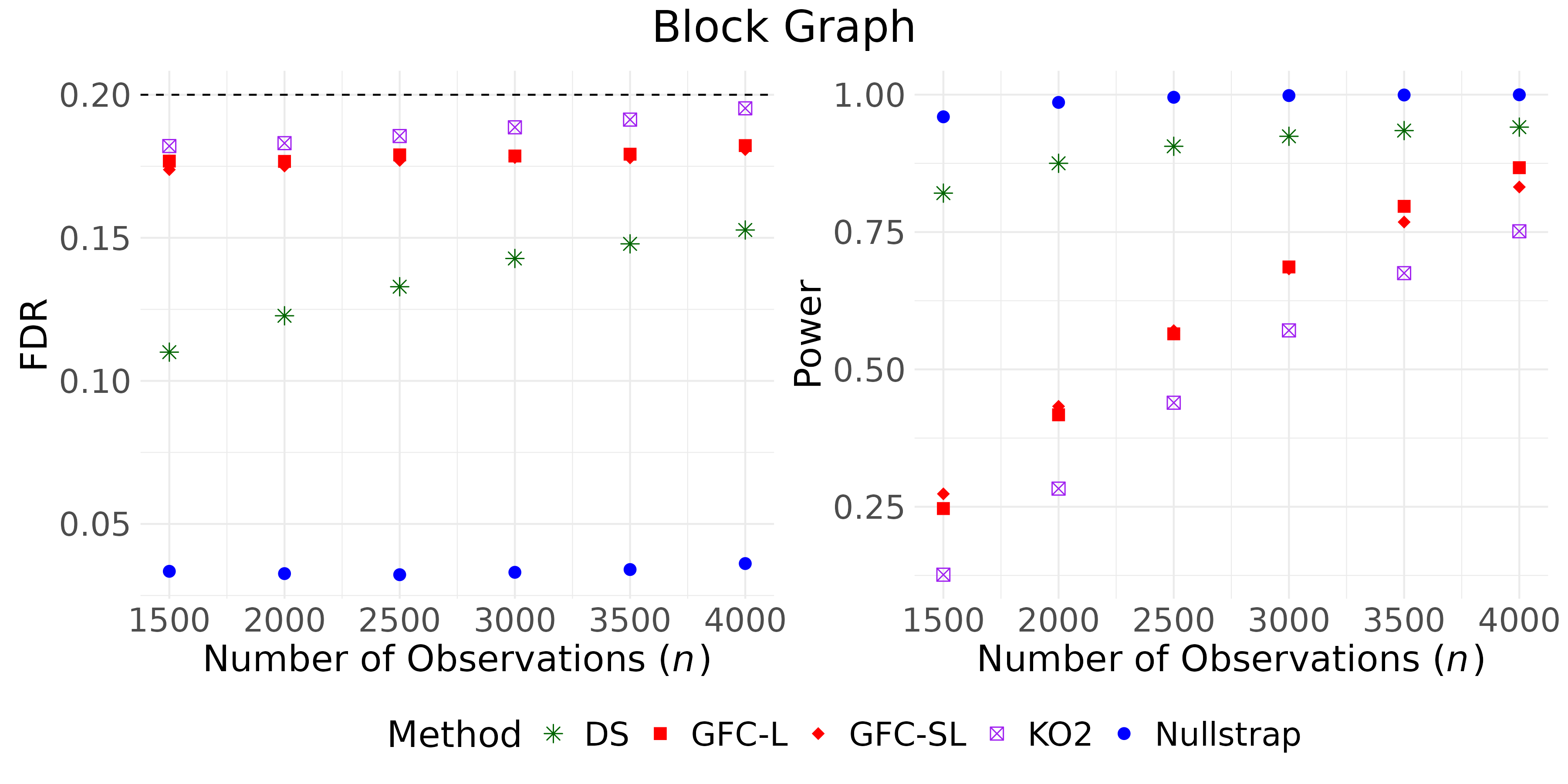}
    \caption{Empirical FDR and power vs. the number of observations ($n$) under the GGM with a block graph.}
    \centering
    \label{fig:GGM_n_block}
\end{figure}

Moreover, in Appendices~F and G, we extend our evaluation to settings with interaction effects in the GLM and Cox models. In these scenarios, Nullstrap consistently outperforms competing methods in terms of power—for example, achieving a power of 0.85 when knockoff filters attain only 0.05—while maintaining FDR control. These results highlight the versatility and robustness of Nullstrap for variable selection across diverse models.

% The complete simulation results for the GLM, Cox model, and GGM are provided in the supplementary material.

\vspace{-3ex}
\section{Discussion}
In this paper, we propose a statistical framework, Nullstrap, for controlling the FDR in high-dimensional variable selection. Unlike knockoff filters and data splitting methods, Nullstrap preserves the original data, resulting in higher statistical power. It also offers improved computational efficiency by enabling fast generation of synthetic null data—avoiding the costly knockoff construction and the need for repeated data splitting.  

Nullstrap relies on two key components: the generation of synthetic null data and an estimation procedure for variable coefficients. While its data generation strategy is closely related to the parametric bootstrap, the crucial distinction lies in the mechanism: the parametric bootstrap simulates data from the fitted model, whereas Nullstrap modifies the fitted model to generate synthetic data under the null hypothesis. With the synthetic null data, Nullstrap identifies false positives by comparing parameter estimates from the original and null datasets---serving as a numerical analog of a likelihood ratio test. We view Nullstrap as a special case of a broader simulation-based inference framework. Nullstrap illustrates the promise of this framework as a flexible alternative to conventional, theory-driven derivations in statistical method development.

First, Nullstrap is a versatile framework that can be extended to a broad class of statistical models, including quantile regression, linear and generalized linear mixed-effects models, and generalized additive models. Future research will explore the application of Nullstrap to these models, as well as its potential for emerging topics such as post-selection inference and conformal prediction.
Second, a key theoretical direction involves developing a principled selection of the data-driven correction factor used in Nullstrap. This includes investigating various selection strategies and conducting sensitivity analyses to understand their impact on inference.
\vspace{-3ex}
\section{Data Availability}
The R package \textit{Nullstrap}, along with code for simulations and data analyses, is available at the anonymous GitHub repository: \url{https://github.com/anonstats123/Nullstrap}, and on Zenodo: \url{https://doi.org/10.5281/zenodo.15881296}.

\vspace{-5ex}

\appendix
\section{Lemmas and proof of Theorem \ref{thm:main}}

\begin{lemma}\label{lem:tail}
	Under Assumption \ref{assum:gamma_estimator},  
	for any \(j \in \mathcal{S}_0(F)\),  there exists an event $\mathcal{G}$ that
%		\[
%		\mathbb{P}\left( |\hat{\beta}_j| \geq t \right) \leq \mathbb{P}\left( |\tilde{\beta}_j^\prime| \geq t \right) + o(p^{-1}).
%		\]
		\[
 |\hat{\beta}_j|  \leq  |\tilde{\beta}_j^\prime|,
\]
and the probability that $\mathcal{G}$ fails to hold satisfies \(\mathbb{P}(\mathcal{G}^{c}) = \alpha_{n,p}\),  
where \(\alpha_{n,p} \to 0\) as \(n, p \to \infty\).  

In other words, with high probability (i.e., on event \(\mathcal{G}\)), the estimated coefficient \(|\hat{\beta}_j|\) is upper-bounded by the synthetic-null estimate \(|\tilde{\beta}_j^\prime|\). The probability \(\alpha_{n,p}\) quantifies the chance that this upper bound fails and is asymptotically negligible.
\end{lemma}

\begin{proof}[Proof of Lemma \ref{lem:tail}]
By Assumption \ref{assum:gamma_estimator}, it follows that 
\begin{equation}\label{eq:tail1}
	\mathbb{P} \left( \left\|\hat{\bbeta} - \bbeta \right\|_{\infty} \geq \gamma_{n,p} \right) = \alpha_{n,p}.
\end{equation}
Define the event \( \mathcal{G} = \left\{ \left\| \hat{\bbeta} - \bbeta \right\|_{\infty} < \gamma_{n,p} \right\} \). For any \( j \in \mathcal{S}_0(F) \), we have \( \beta_j = 0 \), so on the event \( \mathcal{G} \), it follows that \( |\hat{\beta}_j| < \gamma_{n,p} \). By the definition \( |\tilde{\beta}_j^\prime| = |\tilde{\beta}_j| + \gamma_{n,p} \), we have
\[
|\tilde{\beta}_j^\prime| \geq \gamma_{n,p}.
\]
Therefore, on the event \( \mathcal{G} \), it holds that
\[
|\hat{\beta}_j| < \gamma_{n,p} \leq |\tilde{\beta}_j^\prime|,
\]
which completes the proof.

	%	for any \( |\hat{\beta}_j|  < \gamma_{n,p} \), it holds that 
	%	For any \(  \geq \gamma_{n,p} \), on , we have 
	%	\[
	%	\mathbb{P}\left( |\hat{\beta}_j| \geq t \right) \leq o(p^{-1}) \leq \mathbb{P}\left( |\tilde{\beta}_j^\prime| \geq t \right) + o(p^{-1}).
	%	\]
\end{proof}

\begin{lemma}\label{lem:FDR}
    Under Assumption~\ref{assum:gamma_estimator}, Nullstrap asymptotically controls the FDR at the target level \( q \in (0,1) \):
    \[
    \mathrm{FDR}(\tau_q) = \mathbb{E} \left[ \frac{\#\left\{\widehat{\mathcal{S}}(\tau_q) \cap \mathcal{S}_0(F) \right\}}{\max\left(\#\widehat{\mathcal{S}}(\tau_q), 1\right)} \right] \leq q + \alpha_{n,p},
    \]
    where \( \alpha_{n,p} \to 0 \) as \( n, p \to \infty \).
\end{lemma}

\begin{proof}[Proof of Lemma \ref{lem:FDR}]
    Let \( [p] := \{1, 2, \ldots, p\} \). By the definition \( \widehat{\mathcal{S}}(\tau_q) = \left\{j \in [p] : |\hat{\beta}_j| \geq \tau_q \right\} \), we have
    \[
    \begin{aligned}
        \mathrm{FDR}(\tau_q) & = \mathbb{E} \left[ \frac{\#\left\{j \in \mathcal{S}_0(F) : |\hat{\beta}_j| \geq \tau_q \right\}}{\max\left(\#\left\{j \in [p] : |\hat{\beta}_j| \geq \tau_q \right\}, 1\right)} \right] \\
        & = \mathbb{E} \left[ \frac{\#\left\{j \in \mathcal{S}_0(F) : |\hat{\beta}_j| \geq \tau_q \right\}}{\max\left(\#\left\{j \in [p] : |\hat{\beta}_j| \geq \tau_q \right\}, 1\right)} \mathbb{I}(\mathcal{G}) \right] \\
        & \quad + \mathbb{E} \left[ \frac{\#\left\{j \in \mathcal{S}_0(F) : |\hat{\beta}_j| \geq \tau_q \right\}}{\max\left(\#\left\{j \in [p] : |\hat{\beta}_j| \geq \tau_q \right\}, 1\right)} \mathbb{I}(\mathcal{G}^c) \right] \\
        & \leq \mathbb{E} \left[ \frac{\#\left\{j \in \mathcal{S}_0(F) : |\tilde{\beta}_j'| \geq \tau_q \right\}}{\max\left(\#\left\{j \in [p] : |\hat{\beta}_j| \geq \tau_q \right\}, 1\right)} \right] + \alpha_{n,p} \\
        & \leq \mathbb{E} \left[ \frac{\#\left\{j \in [p] : |\tilde{\beta}_j'| \geq \tau_q \right\}}{\max\left(\#\left\{j \in [p] : |\hat{\beta}_j| \geq \tau_q \right\}, 1\right)} \right] + \alpha_{n,p},
    \end{aligned}
    \]
    where the first inequality follows from Lemma~\ref{lem:tail} and and the fact that 
    $$\frac{\#\left\{j : j \in \cS_0(F)  \text{ and } |\hat{\beta}_j| \geq \tau_q \right\}}{\max\left(\#\left\{j : |\hat{\beta}_j| \geq \tau_q \right\}, 1\right) } \leq 1,$$ and the second inequality follows because
    \[
    \#\left\{j \in \mathcal{S}_0(F) : |\tilde{\beta}_j'| \geq \tau_q \right\} \leq \#\left\{j \in [p] : |\tilde{\beta}_j'| \geq \tau_q \right\}.
    \]
    By the definition of \( \tau_q \), we have
    \[
    \frac{\#\{ j \in [p] : |\tilde{\beta}_j'| \geq \tau_q \}}{\max\left( \#\{ j \in [p] : |\hat{\beta}_j| \geq \tau_q \}, 1 \right)} \leq q.
    \]
    Taking expectations on both sides yields
    \[
    \mathbb{E} \left[ \frac{\#\left\{j \in [p] : |\tilde{\beta}_j'| \geq \tau_q \right\}}{\max\left(\#\left\{j \in [p] : |\hat{\beta}_j| \geq \tau_q \right\}, 1\right)} \right] \leq q,
    \]
    which completes the proof.
\end{proof}

\begin{lemma}\label{lem:power}
	Under Assumption~\ref{assum:gamma_estimator} and the condition \( \min_{j \in \cS(F)} |{\beta}_{j}| > 3\gamma_{n,p} \), \textit{Nullstrap} achieves asymptotic power consistency:
	\[
	\text{Power}(\tau_q) := \mathbb{E} \left[ \frac{\#\left\{ \widehat{\cS}(\tau_q) \cap \cS(F) \right\}}{\#\cS(F)} \right] \geq 1 - 2\alpha_{n,p},
	\]
	where \( \alpha_{n,p} \to 0 \) as \( n, p \to \infty \).
\end{lemma}

\begin{proof}[Proof of Lemma \ref{lem:power}]
	By Assumption~\ref{assum:gamma_estimator} and the condition \( \min_{j \in \cS(F)} |{\beta}_{j}| > 3\gamma_{n,p} \), we have
	\[
	\mathbb{P} \left( \min_{j \in \cS(F)} |\hat{\beta}_{j}| \leq 2\gamma_{n,p} \right) \leq \alpha_{n,p}.
	\]
	Under the global null \( \bbeta_0 = \mathbf{0} \), Assumption~\ref{assum:gamma_estimator} further implies
	\[
	\mathbb{P} \left( \|\tilde{\bbeta}\|_{\infty} \geq \gamma_{n,p} \right) \leq \alpha_{n,p},
	\quad \text{so} \quad
	\mathbb{P} \left( \|\tilde{\bbeta}^\prime\|_{\infty} \geq 2\gamma_{n,p} \right) \leq \alpha_{n,p}.
	\]
	
	Define the event
	\[
	\mathcal{G}_2 := \left\{ \|\tilde{\bbeta}^\prime\|_{\infty} < 2\gamma_{n,p} \right\} \cap \left\{ \min_{j \in \cS(F)} |\hat{\beta}_{j}| > 2\gamma_{n,p} \right\}.
	\]
	Then \( \mathbb{P}(\mathcal{G}_2^c) \leq 2\alpha_{n,p} \). On the event \( \mathcal{G}_2 \), the estimated FDP at threshold \( t^* := 2\gamma_{n,p} \) satisfies \( \widehat{\text{FDP}}(t^*) = 0 \), so Nullstrap selects a threshold \( \tau_q \leq t^* \). By construction, \( \cS(F) \subseteq
 \widehat{\cS}(t^*) \subseteq
 \widehat{\cS}(\tau_q) \), implying
	\[
	\frac{\#\left\{ \widehat{\cS}(\tau_q) \cap \cS(F) \right\}}{\#\cS(F)} = 1 \quad \text{on } \mathcal{G}_2.
	\]
	Taking expectations,
	\[
	\mathbb{E} \left[ \frac{\#\left\{ \widehat{\cS}(\tau_q) \cap \cS(F) \right\}}{\#\cS(F)} \right]
	\geq \mathbb{E} \left[ \mathbb{I}(\mathcal{G}_2) \right]
	= 1 - \mathbb{P}(\mathcal{G}_2^c) \geq 1 - 2\alpha_{n,p},
	\]
	which completes the proof.
\end{proof}

\begin{proof}[Proof of Theorem \ref{thm:main}]
The result follows directly by combining Lemmas \ref{lem:FDR} and \ref{lem:power}, which establish the asymptotic FDR control and power consistency of \textit{Nullstrap}, respectively.
\end{proof}

% \subsection{Proof of Lemma \ref{lem:LASSO} }
% \section{Proofs of Lemmas}

\section{Additional algorithms for Nullstrap} 

	\subsection{Algorithm for data-driven selection of the correction factor}

We provide the detailed procedure for selecting the correction factor \( \gamma_{n,p} \) in a data-driven way, summarized in Algorithm~\ref{alg:gamma}.

	\begin{algorithm}[h]
		\caption{Data-driven selection of the correction factor $\gamma_{n,p}$}
		\label{alg:gamma}
		\textbf{Input:} original data $\{\by, \bX\}$; estimation procedure $\mathcal{E}(\cdot, \cdot)$; number of repetitions $B$ (default $B=5$); target FDR level $q \in (0,1)$; estimated coefficient vector $\hat{\bbeta}$ from applying $\mathcal{E}$ to the original data; estimated nuisance parameter $\Hat{\bnu}$ from the original data.\\
		\textbf{Output:} The correction factor $\gamma_{n,p}$. \\
		% Let $\hat{\bbeta}, \Hat{\bnu}$ be the sparsed estimated coefficients and nuisance parameters\;
            Compute the estimated null variable set \( \mathcal{S}_0(\hat{F}) = \{ j : |\hat{\beta}_j| = 0 \} \) based on the fitted model \( \hat{F} \), which includes the estimated parameters \( \hat{\bbeta} \) and \( \hat{\bnu} \)\;
		\For{\( b = 1, \dots, B \)}{
			Generate the $b$-th synthetic dataset \( \by^b \) from the fitted model \( \hat{F} = F(\cdot \mid \bX; \hat{\bbeta}, \hat{\bnu}) \)\;
			Compute \( \hat{\bbeta}^b= \mathcal{E}(\by^b, \bX) \), the estimated coefficient vector from the $b$-th synthetic dataset\;
			Generate the $b$-th synthetic null dataset
            \( \tilde{\by}^b \) from the null model \( F(\cdot \mid \bX; {\bbeta}_0, \hat{\bnu}) \), with $\bbeta_0 = \mathbf{0}$ under the global null\;
                Compute \( \tilde{\bbeta}^b= \mathcal{E}(\tilde{\by}^b, \bX) \), the estimated coefficient vector from the $b$-th synthetic null dataset\;
			Given a candidate correction factor \( \gamma > 0 \), assume that
\[
\mathbb{E}\left[\#\left\{ j \in \mathcal{S}_0(\hat{F}) : |\hat{\beta}_j^b| \geq t \right\}\right] 
\leq \mathbb{E}\left[\#\left\{ j : |\tilde{\beta}_j^b| + \gamma \geq t \right\}\right].
\]
Compute the threshold \( \tau_q^b(\gamma) \) for \( |\hat{\beta}_j^b| \) as follows:
\begin{equation}\label{eq:gamma1}
\tau_q^b(\gamma) = \min \left\{ t > 0 : \frac{\#\left\{ j : |\tilde{\beta}^b_j| + \gamma \geq t \right\}}{\max\left( \#\left\{ j : |\hat{\beta}^b_j| \geq t \right\}, 1 \right)} \leq q \right\},
\end{equation}
where \( q \in (0,1) \) is the target FDR level.

			Compute the selected variable set \( \widehat{\mathcal{S}}^b(\gamma) = \{ j : |\hat{\beta}_j^b| > \tau_q^b(\gamma) \} \) based on the \( b \)-th synthetic dataset and candidate correction factor \( \gamma \)\;
			Determine the $b$-th correction factor as the smallest value of \( \gamma \) such that the FDP of \( \widehat{\mathcal{S}}^b(\gamma) \), based on the fitted model, is controlled under the target level \( q \):
		\begin{equation}\label{eq:gamma2}
			\gamma_b = \min \left\{\gamma>0, \frac{\#\{ \widehat{\mathcal{S}}^b(\gamma) \cap {\mathcal{S}}_0(\hat{F})  \}}{\max\left( \#\{ \widehat{\mathcal{S}}^b(\gamma) \}, 1 \right)} \leq q \right\}.
		\end{equation}	
		}
		Select the correction factor as:
		$
		\gamma_{n,p} = \text{quantile}_{0.95} \left( \{ \gamma_b \}_{b=1}^B \right).
		$
	\end{algorithm}

\subsection{Algorithm of Nullstrap (individual)}
The detailed procedure of Nullstrap (individual), which generates synthetic null data for each variable under the individual null hypothesis that the \( j \)-th variable has no effect, is presented in Algorithm~\ref{alg:Nullstrap_individual}.
\begin{algorithm}[h]
	\caption{Variable Selection via Nullstrap (individual)}
	\label{alg:Nullstrap_individual}
    \textbf{Input:} original data $\{\by, \bX\}$; estimation procedure $\mathcal{E}(\cdot, \cdot)$; target FDR level $q \in (0,1)$.\\
    \textbf{Output:} The set of selected variables $\widehat{\cS}(\tau_q)$. \\
    Compute the estimated coefficient vector \( \hat{\boldsymbol{\beta}} \) and the estimated nuisance parameter $\hat{\bnu}$ from the original data $\{\by, \bX\}$\;
	\For{\( j = 1, \dots, p \)}{
    Estimate the coefficient vector for a reduced model $F(\cdot \mid \bX^{-j}; \bbeta^{-j}, \bnu)$, where \( \bX^{-j} \) and \( \bbeta^{-j} \) denote the design matrix and coefficients with the $j$-th variable removed:
    \( \hat{\bbeta}^{-j} = \left(\hat{\bbeta}_{1:(j-1)}^{-j}, \, \hat{\bbeta}_{j:(p-1)}^{-j}\right)\tran = \mathcal{E}(\bX^{-j}, \by) \)\;
    % Let \( \hat{\bbeta}^{-j} = \left(\hat{\bbeta}_{1:(j-1)}^{-j}, \, \hat{\bbeta}_{j:(p-1)}^{-j}\right)\tran \) be the estimated coefficients\;
    Set \( \bbeta_0^j = \left(\hat{\bbeta}_{1:(j-1)}^{-j}, \, 0, \, \hat{\bbeta}_{j:(p-1)}^{-j}\right)\tran \)\;
    Generate synthetic null data \( \tilde{\by}^j \) from the individual null model \( F(\cdot \mid \bX; \bbeta_0^j, \hat{\bnu}) \)\;
    Extract \( \tilde{\beta}_j \) as the $j$-th element from the estimated null coefficient vector \( \mathcal{E}(\tilde{\by}^j, \bX) \)\;
}

	 Given the target FDR level \( q \in (0,1) \), calculate the threshold \( \tau_q \) for $|\hat\beta_j|$ as:  
	\[
	\tau_q = \min \left\{ t > 0 : \widehat{\text{FDP}}(t) = \frac{\#\{ j : |\tilde{\beta}_j| \geq t \}}{\max\left( \#\{ j : |\hat{\beta}_j| \geq t \}, 1 \right)} \leq q \right\}.
	\]
	
	 Select the set of variables:  
	\[
	\widehat{\cS}(\tau_q) = \{ j : |\hat{\beta}_j| > \tau_q \}.
	\]
	
	\end{algorithm}

\clearpage
\section{Supplementary tables related to Nullstrap for linear models}

\subsection{Comparison of runtimes}
\begin{table}[htbp]
	\centering
    	\caption{Comparison of runtimes (in seconds)  under Simulation Setting~1.}
    \setlength{\tabcolsep}{3pt}
	\begin{tabular}{cccccc}
		\toprule
		Nullstrap (param) &Nullstrap (non-param)    & Permutation & Model-X & Fixed-X & GM\\
		\midrule
		0.42 & 0.50 &0.25 &  15.82 & 10.85& 42.10 \\
        \midrule
        DS & MDS & BH & BHq & SLOPE\\
        		\midrule
        0.87 & 25.01 &0.42 & 0.39 & 0.09\\
		\bottomrule
	\end{tabular}
	\label{ind_time}
	\end{table}
    
\subsection{Comparison of Nullstrap performance across regularized estimation procedures for high-dimensional linear models}
In this subsection, we compare the performance of Nullstrap across three regularized estimation procedures for high-dimensional linear models—LASSO, Elastic Net, and Smoothly Clipped Absolute Deviation (SCAD)—to evaluate its robustness under different regularization schemes. 

This simulation setting follows Simulation Setting 2 in the main text. All three estimation procedures are used to generate parametric synthetic null data as defined in Definition 2, corresponding to the parametric version of Nullstrap.  
The correction factor for LASSO, Elastic Net, and SCAD is selected in a data-driven manner for each estimation procedure using Algorithm~\ref{alg:gamma}. 
% is defined as \( \gamma_{n,p} = 0.2 \Hat{\sigma} \left(\lambda_n + \sqrt{\frac{\log p}{n}} \right) \). For Elastic Net and SCAD, it is given by \( \gamma_{n,p} = 0.2 \Hat{\sigma} 
% \sqrt{\frac{\log p}{n}} \). 
% {\color{red} \bf I suggest removing the Ridge regression from the table.}
% For Ridge regression, the correction factor is set to  
% \[
% \gamma_{n,p} = 0.8 \Hat{\sigma} \left(\frac{1}{\lambda_n + \frac{1}{p} \sum_{i=1}^{p} \left(\mathbf{X}\tran \mathbf{X}\right)_{ii}} + \sqrt{\frac{\log p}{n}} \right).
% \]
The regularization parameters for three procedures are selected using 10-fold cross-validation.
As shown in Tables~\ref{tab:ridge_Nullstrap_rho}--\ref{tab:ridge_Nullstrap_num}, Nullstrap achieves similar FDR control performance across LASSO, Elastic Net, and SCAD, with LASSO showing better power and AUPR.
% This discrepancy arises because {Ridge regression does not perform variable selection}, making it less effective in detecting true signals. Furthermore, the results from {Ridge regression} highlight the importance of the {data-driven correction factor}, as the actual {False Discovery Rate (FDR)} varies with the amplitude of the signal.

\begin{table}[htbp]
\centering
\caption{Comparison of FDR and power (under a target FDR level of $q = 0.1$), as well as AUPR, across different autocorrelation values $\rho$ under Simulation Setting~2, with $A = 0.25$, $p = 1000$, and $n = 2000$. All three regularized estimation procedures are used to generate parametric synthetic null data according to Definition~2 in the main text, corresponding to the parametric version of Nullstrap.}
\begin{tabular}{ccccccccccc}
\toprule
$\rho$ & 0.0 & 0.1 & 0.2 & 0.3 & 0.4 & 0.5 & 0.6 & 0.7 & 0.8 & 0.9 \\
\midrule
\multicolumn{11}{c}{{FDR ($q = 0.1$)}} \\
\midrule
{ SCAD }  & 0.056 & 0.064 & 0.069 & 0.059 & 0.059 & 0.057 & 0.048 & 0.039 & 0.059 & 0.029\\
{ Elastic Net }  & 0.102 & 0.111 & 0.115 & 0.104 & 0.097 & 0.051 & 0.034 & 0.023 & 0.014 & 0.006\\
{ LASSO }  & 0.086 & 0.102 & 0.098 & 0.088 & 0.081 & 0.071 & 0.068 & 0.067 & 0.066 & 0.022\\
\midrule
\multicolumn{11}{c}{{Power ($q = 0.1$)}} \\
\midrule
{ SCAD }  & 0.952 & 1.000 & 1.000 & 1.000 & 1.000 & 1.000 & 0.998 & 0.994 & 0.970 & 0.549\\
{ Elastic Net }  & 0.961 & 1.000 & 1.000 & 1.000 & 1.000 & 1.000 & 0.999 & 0.974 & 0.831 & 0.527\\
{ LASSO }  & 0.971 & 1.000 & 1.000 & 1.000 & 1.000 & 1.000 & 0.999 & 0.996 & 0.949 & 0.614\\
\midrule
\multicolumn{11}{c}{{AUPR}} \\
\midrule
{ SCAD }  & 1.000 & 1.000 & 1.000 & 1.000 & 1.000 & 1.000 & 0.999 & 0.997 & 0.978 & 0.588\\
{ Elastic Net }  & 1.000 & 1.000 & 1.000 & 1.000 & 1.000 & 1.000 & 0.999 & 0.993 & 0.907 & 0.690\\
{ LASSO }  & 1.000 & 1.000 & 1.000 & 1.000 & 1.000 & 1.000 & 1.000 & 0.999 & 0.981 & 0.787\\
\bottomrule
\end{tabular}
\label{tab:ridge_Nullstrap_rho}
\end{table}

\begin{table}[htbp]
\centering
\caption{Comparison of FDR and power (under a target FDR level of $q = 0.1$), as well as AUPR, across different signal amplitude values $A$ under Simulation Setting~2, with $\rho = 0.8$, $p = 1000$, and $n = 2000$. All three regularized estimation procedures are used to generate parametric synthetic null data according to Definition~2 in the main text, corresponding to the parametric version of Nullstrap.}
% \caption{
% Comparison of power, FDR and AUPR at different signal amplitude values $A$, with a target FDR of $q=0.1$. The simulation setting is $\rho = 0.8,\ p = 1000,\text{ and }n =2000$. All methods generate parametric synthetic null data according to Definition 2 in the main text, which corresponds to the parametric version of Nullstrap.}
\begin{tabular}{cccccccccc}
\toprule
$A$ & 0.150 & 0.175 & 0.200 & 0.225 & 0.250 & 0.275 & 0.300 & 0.325 & 0.350 \\
\midrule
\multicolumn{10}{c}{{FDR} ($q = 0.1$)} \\
\midrule
{ SCAD }  & 0.022 & 0.033 & 0.054 & 0.073 & 0.059 & 0.053 & 0.043 & 0.036 & 0.024\\
{ Elastic Net }  & 0.008 & 0.006 & 0.009 & 0.011 & 0.014 & 0.016 & 0.017 & 0.016 & 0.017\\
{ LASSO }    & 0.012 & 0.024 & 0.031 & 0.048 & 0.066 & 0.076 & 0.081 & 0.086 & 0.083\\
\midrule
\multicolumn{10}{c}{{Power} ($q = 0.1$)} \\
\midrule
{ SCAD }  & 0.384 & 0.514 & 0.719 & 0.917 & 0.970 & 0.987 & 0.993 & 0.998 & 0.998\\
{ Elastic Net }  & 0.448 & 0.532 & 0.632 & 0.727 & 0.831 & 0.908 & 0.953 & 0.980 & 0.992\\
{ LASSO }  & 0.459 & 0.605 & 0.749 & 0.863 & 0.949 & 0.983 & 0.993 & 0.998 & 0.999\\
% { Gbl$-$Ind }   & -0.009 & -0.011 & -0.009 & -0.008 & -0.006 & 0.023 & 0.037 & 0.052 & 0.089 & 0.092 \\
\midrule
\multicolumn{10}{c}{{AUPR}} \\
\midrule
{ SCAD }  & 0.464 & 0.581 & 0.763 & 0.933 & 0.978 & 0.990 & 0.995 & 0.999 & 0.999\\
{ Elastic Net }  & 0.623 & 0.693 & 0.773 & 0.835 & 0.907 & 0.958 & 0.984 & 0.995 & 0.998\\
{ LASSO }    & 0.716 & 0.810 & 0.885 & 0.947 & 0.981 & 0.994 & 0.998 & 0.999 & 1.000\\
\bottomrule
\end{tabular}
\label{tab:ridge_Nullstrap_amp}
\end{table}

\begin{table}[htbp]
\centering
\caption{Comparison of FDR and power (evaluated at various target FDR levels $q$), as well as AUPR, under Simulation Setting~2, with $A = 0.25$, $\rho = 0.8$, $p = 1000$, and $n = 2000$. All three regularized estimation procedures are used to generate parametric synthetic null data according to Definition~2 in the main text, corresponding to the parametric version of Nullstrap.}
\begin{tabular}{ccccccccc}
\toprule
$q$ & 0.05 & 0.10 & 0.15 & 0.20 & 0.25 & 0.30 & 0.35 & 0.40 \\
\midrule
\multicolumn{9}{c}{{FDR}} \\
\midrule
{ SCAD }  & 0.033 & 0.059 & 0.089 & 0.123 & 0.149 & 0.178 & 0.205 & 0.242\\
{ Elastic Net }  & 0.005 & 0.014 & 0.035 & 0.077 & 0.131 & 0.196 & 0.292 & 0.405\\
{ LASSO }    & 0.029 & 0.066 & 0.105 & 0.164 & 0.212 & 0.263 & 0.322 & 0.379\\
\midrule
\multicolumn{9}{c}{{Power}} \\
\midrule
{ SCAD }  & 0.970 & 0.970 & 0.970 & 0.970 & 0.970 & 0.970 & 0.970 & 0.970\\
{ Elastic Net }  & 0.798 & 0.831 & 0.853 & 0.865 & 0.877 & 0.886 & 0.893 & 0.899 \\
{ LASSO }  & 0.919 & 0.949 & 0.959 & 0.966 & 0.970 & 0.974 & 0.975 & 0.978\\
% { Gbl$-$Ind }   & -0.009 & -0.011 & -0.009 & -0.008 & -0.006 & 0.023 & 0.037 & 0.052 & 0.089 & 0.092 \\
\midrule
\multicolumn{9}{c}{{AUPR}} \\
\midrule
{ SCAD }  & \multicolumn{8}{c}{{0.978}}\\
{ Elastic Net }  & \multicolumn{8}{c}{{0.907}}\\
{ LASSO }    & \multicolumn{8}{c}{{0.981}}\\
\bottomrule
\end{tabular}
\label{tab:ridge_Nullstrap_fdr}
\end{table}

\begin{table}[htbp]
\centering
\caption{Comparison of FDR and power (under a target FDR level of $q = 0.1$), as well as AUPR, across different numbers of variables $p$ under Simulation Setting~2, with $A = 0.25$, $\rho = 0.8$, and $n = 2000$. All three regularized estimation procedures are used to generate parametric synthetic null data according to Definition~2 in the main text, corresponding to the parametric version of Nullstrap.}
\begin{tabular}{cccccccc}
\toprule
$p$ & 500 & 1000 & 1500 & 2000 & 2500 & 3000 & 3500 \\
\midrule
\multicolumn{8}{c}{{FDR} ($q = 0.1$)} \\
\midrule
{ SCAD }    & 0.063 & 0.059 & 0.066 & 0.023 & 0.015 & 0.023 & 0.020\\
{ Elastic Net }  & 0.030 & 0.014 & 0.014 & 0.010 & 0.009 & 0.008 & 0.017\\
{ LASSO }    & 0.085 & 0.066 & 0.034 & 0.026 & 0.014 & 0.009 & 0.008\\
\midrule
\multicolumn{8}{c}{{Power} ($q = 0.1$)} \\
\midrule
{ SCAD }    & 0.977 & 0.970 & 0.921 & 0.616 & 0.583 & 0.565 & 0.545\\
{ Elastic Net }  & 0.939 & 0.831 & 0.745 & 0.696 & 0.673 & 0.642 & 0.608\\
{ LASSO }  & 0.985 & 0.949 & 0.889 & 0.811 & 0.783 & 0.745 & 0.703\\
% { Gbl$-$Ind }   & -0.009 & -0.011 & -0.009 & -0.008 & -0.006 & 0.023 & 0.037 & 0.052 & 0.089 & 0.092 \\
\midrule
\multicolumn{8}{c}{{AUPR}} \\
\midrule
{ SCAD }    & 0.984 & 0.978 & 0.936 & 0.675 & 0.636 & 0.614 & 0.596\\
{ Elastic Net }  & 0.983 & 0.907 & 0.832 & 0.783 & 0.770 & 0.736 & 0.706\\
{ LASSO }    & 0.995 & 0.981 & 0.951 & 0.910 & 0.885 & 0.854 & 0.814\\
\bottomrule
\end{tabular}

\label{tab:ridge_Nullstrap_num}
\end{table}

\subsection{False discovery rate of the LASSO-only method}
In this subsection, Tables \ref{tab:non_Nullstrap_rho}--\ref{tab:non_Nullstrap_num} illustrate that the FDR of the LASSO-only method, where the selected variables are defined as 
\[
\hat{\mathcal{S}}_{\text{LASSO}} = \{ j : |\hat{\beta}_j| > 0 \},
\]
with \( \hat{\beta}_j \) being the estimated LASSO coefficient, is not controlled at the target level.
These results highlight the necessity of the proposed Nullstrap method, which effectively controls the FDR while maintaining high statistical power. In these results, Nullstrap generates synthetic null data according to Definition 2 in the main text, which corresponds to the parametric version.  

\begin{table}[htbp]
\centering
\caption{Comparison of FDR, the number of selected variables, and power (under a target FDR level of $q = 0.1$), as well as AUPR, across different autocorrelation values $\rho$ under Simulation Setting~2, with $s = 30$ (the number of true signal variables), $A = 0.25$, $p = 1000$, and $n = 2000$. The number of selected variables is rounded to the nearest integer. Nullstrap generates parametric synthetic null data according to Definition~2 in the main text, corresponding to the parametric version.}
\begin{tabular}{ccccccccccc}
\toprule
$\rho$ & 0.0 & 0.1 & 0.2 & 0.3 & 0.4 & 0.5 & 0.6 & 0.7 & 0.8 & 0.9 \\
\midrule
\multicolumn{11}{c}{{FDR} ($q = 0.1$)} \\
\midrule
{ LASSO-only }  & 0.920 & 0.920 & 0.920 & 0.921 & 0.922 & 0.922 & 0.921 & 0.923 & 0.913 & 0.882\\
{ Nullstrap (param) }  & 0.086 & 0.102 & 0.098 & 0.088 & 0.081 & 0.071 & 0.068 & 0.067 & 0.066 & 0.022\\
\midrule
\multicolumn{11}{c}{{Number of Selected Variables} ($q = 0.1$)} \\
\midrule
{ LASSO-only }  & 375 & 378 & 378 & 382 & 385 & 387 & 384 & 391 & 357 & 231\\
{ Nullstrap (param) }  & 32 & 34 & 34 & 33 & 33 & 32 & 32 & 32 & 31 & 19\\
\midrule
\multicolumn{11}{c}{{Power} ($q = 0.1$)} \\
\midrule
{ LASSO-only }  & 1.000 & 1.000 & 1.000 & 1.000 & 1.000 & 1.000 & 1.000 & 1.000 & 0.995 & 0.856\\
{ Nullstrap (param) }  & 0.971 & 1.000 & 1.000 & 1.000 & 1.000 & 1.000 & 0.999 & 0.996 & 0.949 & 0.614\\
% { Gbl$-$Ind }   & -0.009 & -0.011 & -0.009 & -0.008 & -0.006 & 0.023 & 0.037 & 0.052 & 0.089 & 0.092 \\
\midrule
\multicolumn{11}{c}{{AUPR}} \\
\midrule
{ LASSO-only }  & 1.000 & 1.000 & 1.000 & 1.000 & 1.000 & 1.000 & 1.000 & 0.999 & 0.981 & 0.787\\
{ Nullstrap (param) }  & 1.000 & 1.000 & 1.000 & 1.000 & 1.000 & 1.000 & 1.000 & 0.999 & 0.981 & 0.787\\
\bottomrule
\end{tabular}
\label{tab:non_Nullstrap_rho}
\end{table}

\begin{table}[htbp]
\centering
\caption{Comparison of FDR, the number of selected variables, and power (under a target FDR level of $q = 0.1$), as well as AUPR, across different signal amplitude values $A$ under Simulation Setting~2, with $s = 30$ (the number of true signal variables), $\rho = 0.8$, $p = 1000$, and $n = 2000$. The number of selected variables is rounded to the nearest integer. Nullstrap generates parametric synthetic null data according to Definition~2 in the main text, corresponding to the parametric version.}
\begin{tabular}{cccccccccc}
\toprule
$A$ & 0.150 & 0.175 & 0.200 & 0.225 & 0.250 & 0.275 & 0.300 & 0.325 & 0.350 \\
\midrule
\multicolumn{10}{c}{{FDR} ($q = 0.1$)} \\
\midrule
{ LASSO-only }    & 0.881 & 0.884 & 0.890 & 0.901 & 0.913 & 0.920 & 0.923 & 0.924 & 0.925\\
{ Nullstrap (param) }    & 0.012 & 0.024 & 0.031 & 0.048 & 0.066 & 0.076 & 0.081 & 0.086 & 0.083\\
\midrule
\multicolumn{10}{c}{{Number of Selected Variables} ($q = 0.1$)} \\
\midrule
{ LASSO-only }    & 213 & 243 & 271 & 314 & 357 & 382 & 394 & 399 & 402\\
{ Nullstrap (param) }    & 14 & 19 & 24 & 28 & 31 & 32 & 33 & 33 & 33\\
\midrule
\multicolumn{10}{c}{{Power} ($q = 0.1$)} \\
\midrule
{ LASSO-only }  & 0.819 & 0.892 & 0.940 & 0.979 & 0.995 & 0.999 & 1.000 & 1.000 & 1.000 \\
{ Nullstrap (param) }  & 0.459 & 0.605 & 0.749 & 0.863 & 0.949 & 0.983 & 0.993 & 0.998 & 0.999\\
% { Gbl$-$Ind }   & -0.009 & -0.011 & -0.009 & -0.008 & -0.006 & 0.023 & 0.037 & 0.052 & 0.089 & 0.092 \\

\midrule
\multicolumn{10}{c}{{AUPR}} \\
\midrule
{ LASSO-only }    & 0.716 & 0.810 & 0.885 & 0.947 & 0.981 & 0.994 & 0.998 & 0.999 & 1.000\\
{ Nullstrap (param) }    & 0.716 & 0.810 & 0.885 & 0.947 & 0.981 & 0.994 & 0.998 & 0.999 & 1.000\\
\bottomrule
\end{tabular}
\label{tab:non_Nullstrap_amp}
\end{table}

% \begin{table}[htbp]
% \centering
% \begin{tabular}{crrrrrrrr}
% \toprule
% $q$ & 0.05 & 0.10 & 0.15 & 0.20 & 0.25 & 0.30 & 0.35 & 0.40 \\
% \midrule
% \multicolumn{9}{c}{{Power}} \\
% \midrule
% { LASSO-only  }  & \multicolumn{8}{c}{{0.995}}\\
% { Nullstrap }  & 0.940 & 0.951 & 0.957 & 0.961 & 0.963 & 0.965 & 0.966 & 0.968\\
% % { Gbl$-$Ind }   & -0.009 & -0.011 & -0.009 & -0.008 & -0.006 & 0.023 & 0.037 & 0.052 & 0.089 & 0.092 \\
% \midrule
% \multicolumn{9}{c}{{FDR}} \\
% \midrule
% { LASSO-only  }    & \multicolumn{8}{c}{{0.913}}\\
% { Nullstrap }    & 0.033 & 0.054 & 0.077 & 0.098 & 0.117 & 0.137 & 0.156 & 0.178\\
% \midrule
% \multicolumn{9}{c}{{AUPR}} \\
% \midrule
% { LASSO-only  }    & \multicolumn{8}{c}{{0.981}}\\
% { Nullstrap }    & \multicolumn{8}{c}{{0.981}}\\
% \bottomrule
% \end{tabular}
% \caption{Comparison of power, FDR and AUPR at different target FDR values. The simulation setting is $\rho = 0.8, A = 0.25, p = 1000,\text{ and }n =2000$.}
% \label{tab:non_Nullstrap_fdr}
% \end{table}

\begin{table}[htbp]
\centering
\caption{Comparison of FDR, the number of selected variables, and power (under a target FDR level of $q = 0.1$), as well as AUPR, across different numbers of variables $p$ under Simulation Setting~2, with $s = 30$ (the number of true signal variables), $\rho = 0.8$, $A = 0.25$, and $n = 2000$. The number of selected variables is rounded to the nearest integer. Nullstrap generates parametric synthetic null data according to Definition~2 in the main text, corresponding to the parametric version.}
\begin{tabular}{cccccccc}
\toprule
$n$ & 500 & 1000 & 1500 & 2000 & 2500 & 3000 & 3500 \\
\midrule
\multicolumn{8}{c}{{FDR} ($q = 0.1$)} \\
\midrule
{ LASSO-only }    & 0.882 & 0.913 & 0.922 & 0.925 & 0.934 & 0.938 & 0.940\\
{ Nullstrap (param) }    & 0.085 & 0.066 & 0.034 & 0.026 & 0.014 & 0.009 & 0.008\\
\midrule
\multicolumn{8}{c}{{Number of Selected Variables} ($q = 0.1$)} \\
\midrule
{ LASSO-only }    & 256 & 357 & 397 & 402 & 441 & 458 & 453\\
{ Nullstrap (param) }    & 33 & 31 & 28 & 25 & 24 & 23 & 21\\
\midrule
\multicolumn{8}{c}{{Power} ($q = 0.1$)} \\
\midrule
{ LASSO-only }  & 0.999 & 0.995 & 0.981 & 0.955 & 0.940 & 0.920 & 0.879\\
{ Nullstrap (param) }  & 0.985 & 0.949 & 0.889 & 0.811 & 0.783 & 0.745 & 0.703\\
% { Gbl$-$Ind }   & -0.009 & -0.011 & -0.009 & -0.008 & -0.006 & 0.023 & 0.037 & 0.052 & 0.089 & 0.092 \\
\midrule
\multicolumn{8}{c}{{AUPR}} \\
\midrule
{ LASSO-only }    &  0.995 & 0.981 & 0.951 & 0.910 & 0.885 & 0.854 & 0.814\\
{ Nullstrap (param) }    & 0.995 & 0.981 & 0.951 & 0.910 & 0.885 & 0.854 & 0.814\\
\bottomrule
\end{tabular}
\label{tab:non_Nullstrap_num}
\end{table}

\subsection{Comparison of Nullstrap and Nullstrap-Diff}
In this subsection, we compare the performance of Nullstrap with that of Nullstrap-Diff, which estimates the FDP as fllows:
\begin{equation}\label{eq:FDP2}
\widehat{\text{FDP}}(t) = \frac{1+\#\{ j : W_j \leq -t \}}{\max\left( \#\{ j : W_j \geq t \}, 1 \right)}\,,
\end{equation}
where \( W_j = |\hat{\beta}_j| - |\tilde{\beta}_j^\prime| \).
Table~\ref{tab:Nullstrap_Diff_amp} presents the comparison between Nullstrap and Nullstrap-Diff across different signal amplitude values (\( A \)). The results show that Nullstrap-Diff yields lower power and AUPR than Nullstrap, particularly when the signal amplitude is small.

\begin{table}[htbp]
\centering
\caption{Comparison of FDR and power (under a target FDR level of $q = 0.1$), as well as AUPR, across different signal amplitude values $A$ under Simulation Setting~2, with $\rho = 0.8$, $p = 1000$, and $n = 2000$. Nullstrap-Diff represents estimating FDP using Equation~\eqref{eq:FDP2}.}
\begin{tabular}{cccccccccc}
\toprule
$A$ & 0.150 & 0.175 & 0.200 & 0.225 & 0.250 & 0.275 & 0.300 & 0.325 & 0.350 \\
\midrule
\multicolumn{10}{c}{{FDR} ($q = 0.1$)}  \\
\midrule
{ Nullstrap-Diff }    & 0.006 & 0.021 & 0.026 & 0.047 & 0.070 & 0.085 & 0.095 & 0.100 & 0.102\\
{ Nullstrap (param) }    & 0.012 & 0.024 & 0.031 & 0.048 & 0.066 & 0.076 & 0.081 & 0.086 & 0.083\\
\midrule
\multicolumn{10}{c}{{Power} ($q = 0.1$)} \\
\midrule
{ Nullstrap-Diff }  & 0.184 & 0.315 & 0.542 & 0.762 & 0.932 & 0.980 & 0.993 & 0.998 & 0.999 \\
{ Nullstrap (param) }  & 0.459 & 0.605 & 0.749 & 0.863 & 0.949 & 0.983 & 0.993 & 0.998 & 0.999\\
% { Gbl$-$Ind }   & -0.009 & -0.011 & -0.009 & -0.008 & -0.006 & 0.023 & 0.037 & 0.052 & 0.089 & 0.092 \\
\midrule
\multicolumn{10}{c}{{AUPR}} \\
\midrule
{ Nullstrap-Diff }    & 0.709 & 0.802 & 0.879 & 0.943 & 0.979 & 0.993 & 0.998 & 0.999 & 1.000\\
{ Nullstrap (param) }    & 0.716 & 0.810 & 0.885 & 0.947 & 0.981 & 0.994 & 0.998 & 0.999 & 1.000\\
\bottomrule
\end{tabular}
\label{tab:Nullstrap_Diff_amp}
\end{table}

\begin{table}[htbp]
\centering
\caption{Coefficients of key variables identified by Nullstrap in the time-to-labor dataset.}
\footnotesize
\begin{tabular}{ccccccc}
\toprule
\multicolumn{7}{c}{{Nullstrap (param)}}\\
\midrule
Variable & NK (STAT1, IFN-$\alpha$) &  Siglec-6 & IL-1R4 & SLPI & Activin A & hCG \\
Coefficient & 4.074 & 3.171 & 3.419 & 1.034 & 0.927 & -3.556 \\
\midrule
\multicolumn{7}{c}{{Nullstrap (non-param)}}\\
\midrule
Variable & NK (STAT1, IFN-$\alpha$) &  Siglec-6 & IL-1R4 & SLPI & hCG \\
Coefficient & 4.225 & 2.719 & 3.813 & 2.666 & -3.933 \\
\bottomrule
\end{tabular}

\label{fea_cof}
\end{table}

\clearpage
\section{Supplementary figures related to Nullstrap for linear models}
This section provides supplementary figures for the Nullstrap simulation results and its comparison with other variable selection methods for linear models.

\begin{figure}[H]
	\centering
	\includegraphics[scale=0.5]{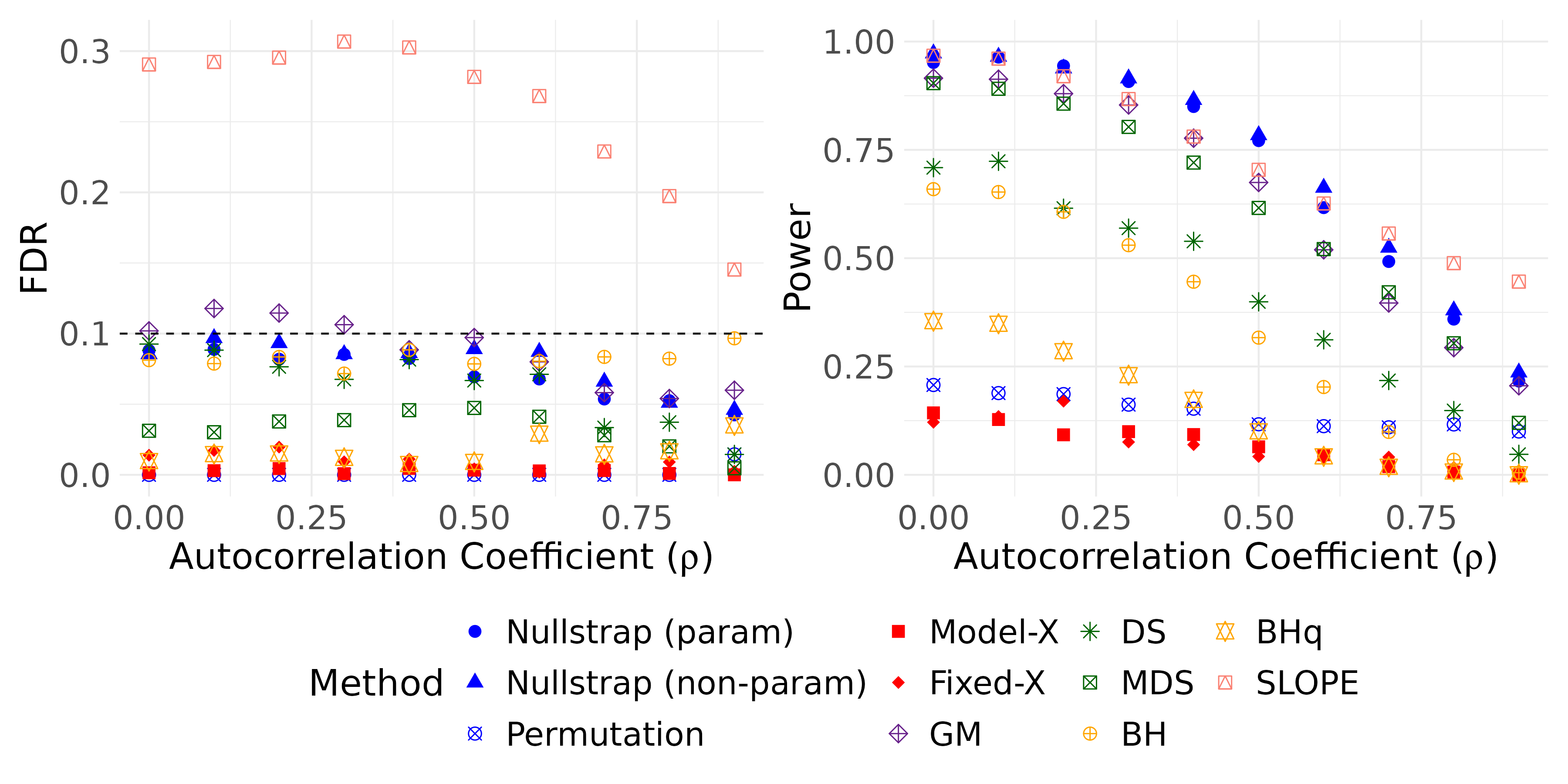}
	\caption{Empirical FDR and power vs. autocorrelation ($\rho$) under Simulation Setting~1.}
	\centering
	\label{fig:in_rho}
\end{figure}

\begin{figure}[H]
	\centering
	\includegraphics[scale=0.4]{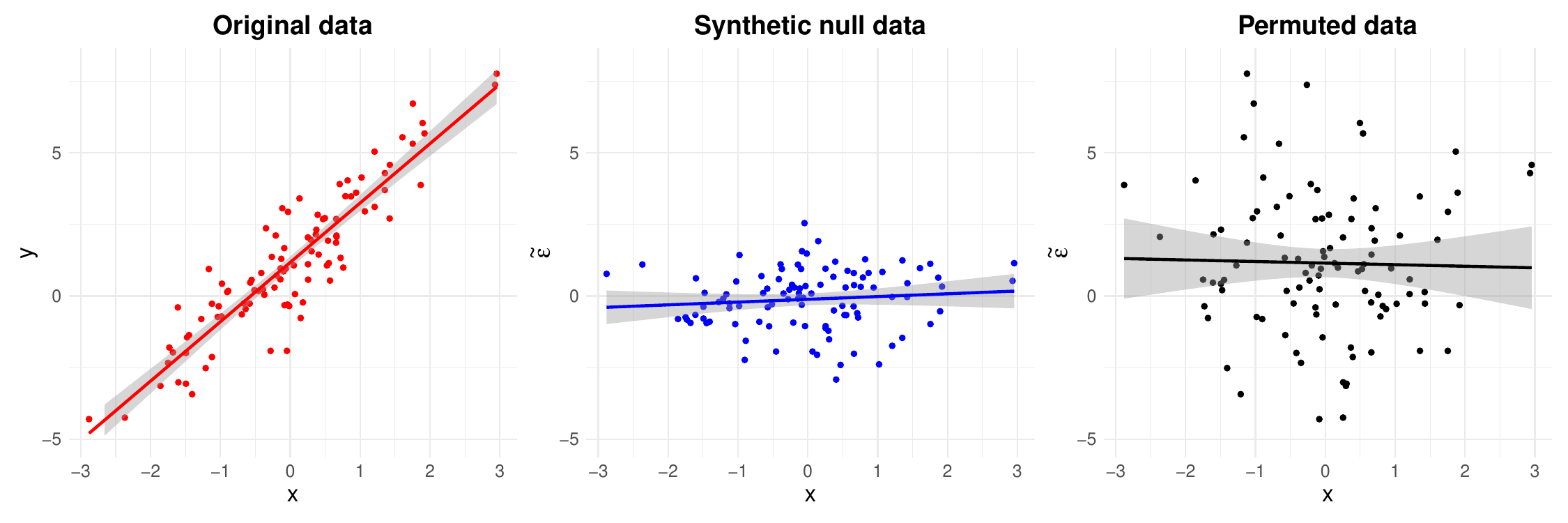}
	\caption{A graphical illustration showing why the permutation approach exhibits low power.}
	\centering
	\label{fig:in_rho}
\end{figure}

 % \begin{figure}[h]
 %        \centering
 %        \begin{subfigure}{0.45\textwidth}
 %            \centering
 %            \includegraphics[width=\linewidth]{simulation/result_in_aupr_lm.png} 
 %            \caption{Empirical AUPR vs. autocorrelation ($\rho$) under Simulation Setting~1.}
 %            \label{fig:sub1}
 %        \end{subfigure}
 %        \hfill
 %        \begin{subfigure}{0.45\textwidth}
 %            \centering
 %            \includegraphics[width=\linewidth]{simulation/result_rho_aupr_lm.png} 
 %            \caption{Empirical AUPR vs. autocorrelation ($\rho$) under Simulation Setting~2.}
 %            \label{fig:sub2}
 %        \end{subfigure}

 %        \vspace{0.5cm}
 %        \begin{subfigure}{0.45\textwidth}
 %            \centering
 %            \includegraphics[width=\linewidth]{simulation/result_amp_aupr_lm.png} 
 %            \caption{Empirical AUPR vs. signal amplitude ($A$) under Simulation Setting~2.}
 %            \label{fig:sub3}
 %        \end{subfigure}
 %        \hfill
 %        \begin{subfigure}{0.45\textwidth}
 %            \centering
 %            \includegraphics[width=\linewidth]{simulation/result_num_aupr_lm.png} 
 %            \caption{Empirical AUPR vs. number of variables ($p$) under Simulation Setting~2.}
 %            \label{fig:sub4}
 %        \end{subfigure}
        
 %        \caption{Empirical AUPR for the linear regression model under Simulation Setting~1 and 2.}
 %        \label{fig:AUPR}
 %    \end{figure}

   \begin{figure}[H]
            \centering
            \includegraphics[width=\linewidth]{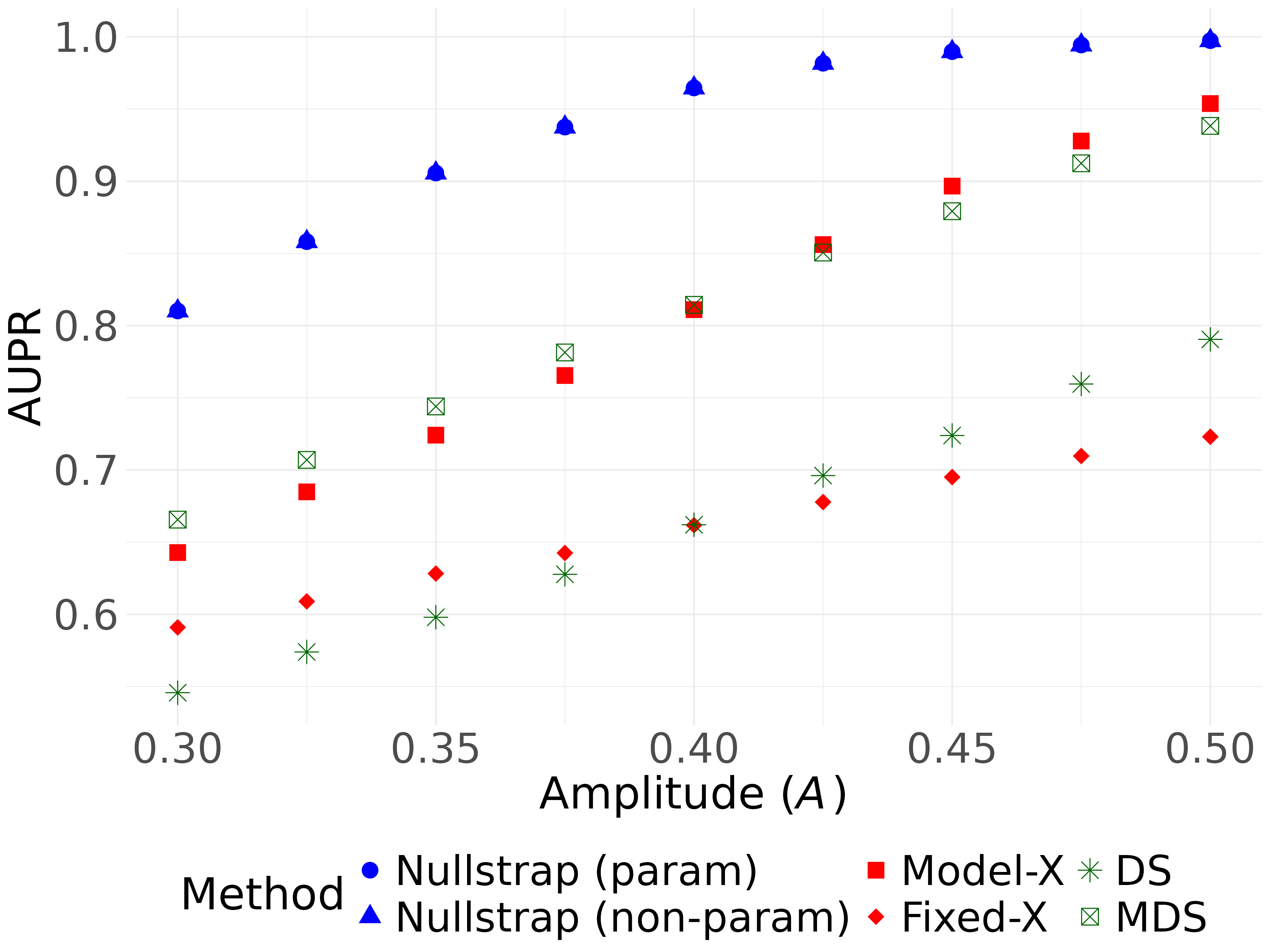} 
            \caption{Empirical AUPR vs. signal amplitude ($A$) under Simulation Setting~3.}
            \label{fig:amp_aupr_t3}
    \end{figure}

\clearpage
\section{Additional simulation settings for linear models}

In this section, we present three additional simulation settings along with the corresponding results for Nullstrap and competing methods.

\subsection{Non-consecutive signal variables: random index selection}
In the simulation settings presented in the main text, the first \( s = 30 \) elements of the coefficient vector \( \bbeta \) are set to be nonzero. In this subsection, we consider an alternative setting where the nonzero indices are selected randomly, as this alters the effect of autocorrelation between adjacent variables.

% First, we consider a random nonzero indices selection.
\begin{simsetting}\label{sim:setting_random}
    The coefficient vector \( \bbeta \) has \( 30 \) randomly selected elements assigned values with amplitude \( A \) and random signs, while the remaining \( p - 30 \) elements are set to zero. The autocorrelation parameter \( \rho \) ranges from \( 0 \) to \( 0.8 \). All other settings remain the same as in Simulation Setting~2 in the main text.
\end{simsetting}
% Following Algorithm~\ref{alg:gamma}, we set the correction factor \( \gamma_{n,p} \) to  
% $
% 0.3 \Hat{\sigma} \left(\lambda_n + \sqrt{\frac{\log p}{n}} \right)
% $
% in this simulation setting.  
By varying each parameter under Simulation Setting~\ref{sim:setting_random}, we compare the FDR, power, and AUPR of different methods using 100 replications. In scenarios with large \( p \) (number of variables), we exclude Fixed-X from the comparison, as it requires \( n \geq 2p \).
% We also omit BH and BHq due to the high computational cost associated with using the debiased LASSO for \( p \)-value calculation.  

\begin{figure}[H]
\centering
\includegraphics[scale=0.5]{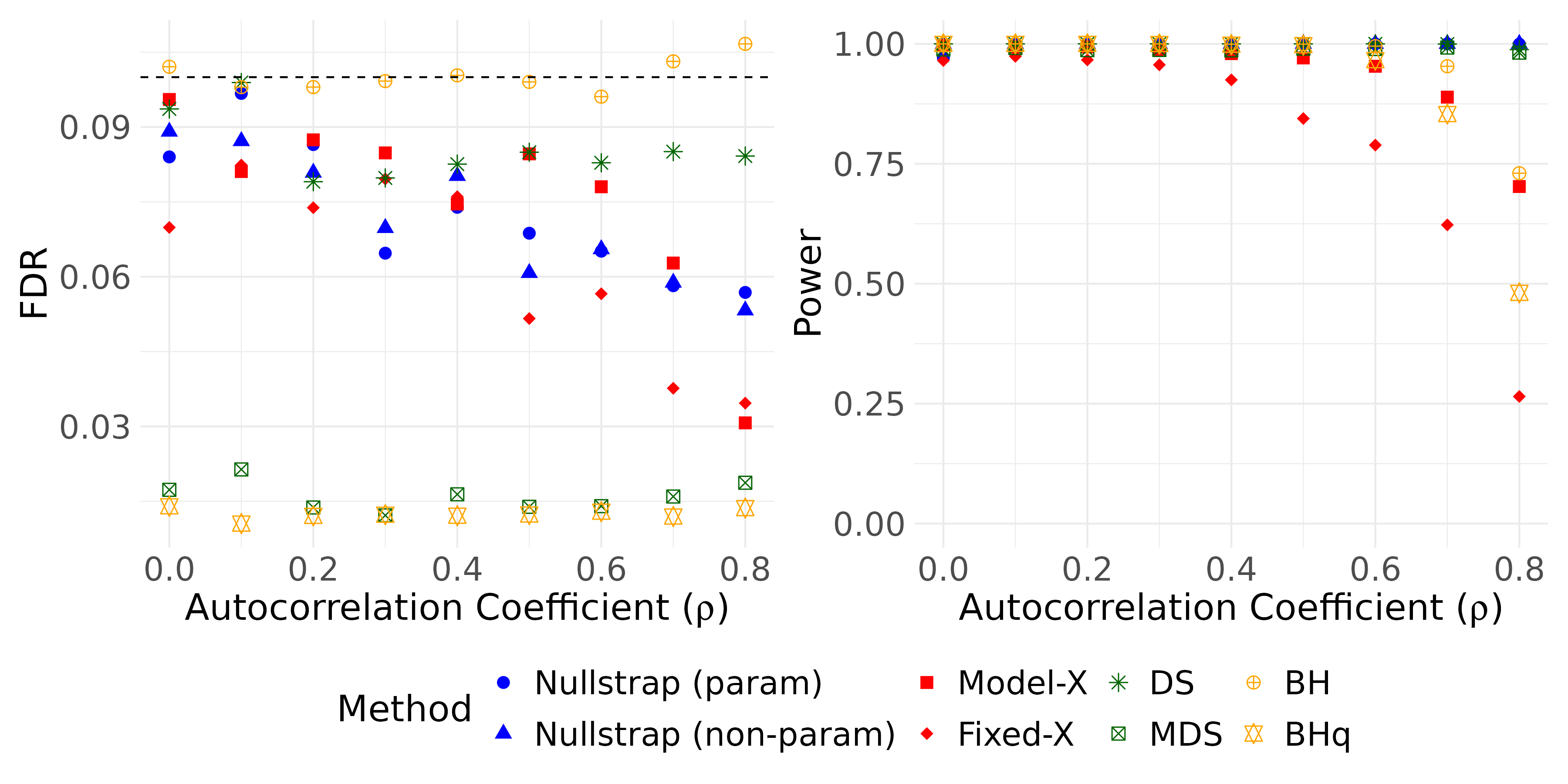}
\caption{Empirical FDR and power vs. autocorrelation ($\rho$) under Simulation Setting~\ref{sim:setting_random}.}
\centering
\label{fig:LM_rho_random}
\end{figure}

\begin{figure}[H]
\centering
\includegraphics[scale=0.5]{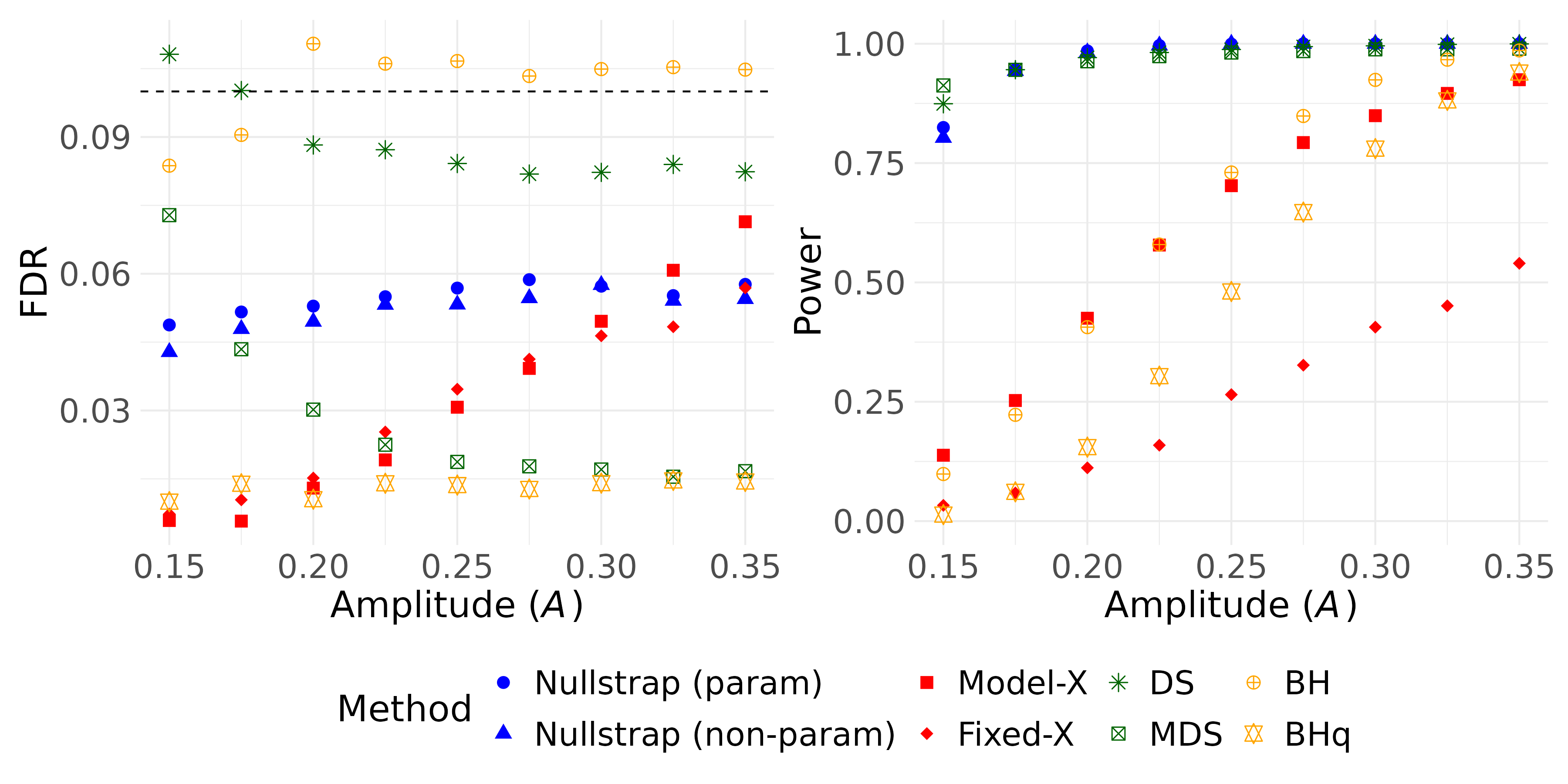}
\caption{Empirical FDR and power vs. signal amplitude ($A$) under Simulation Setting~\ref{sim:setting_random}.}
\centering
\label{fig:LM_amp_random}
\end{figure}

\begin{figure}[H]
\centering
\includegraphics[scale=0.5]{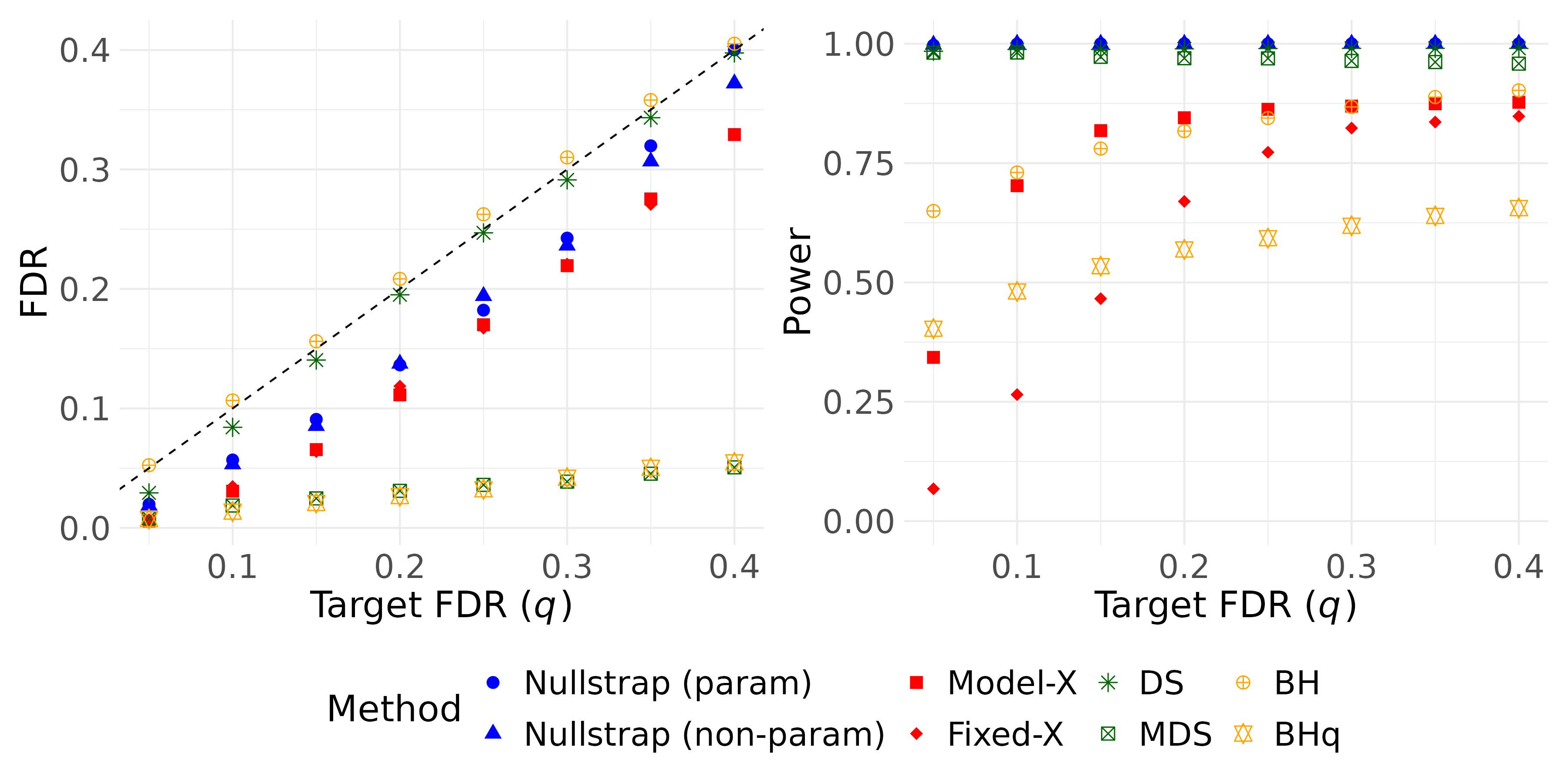}
\caption{Empirical FDR and power vs. target FDR level ($q$) under Simulation Setting~\ref{sim:setting_random}.}
\centering
\label{fig:LM_fdr_random}
\end{figure}

\begin{figure}[H]
\centering
\includegraphics[scale=0.5]{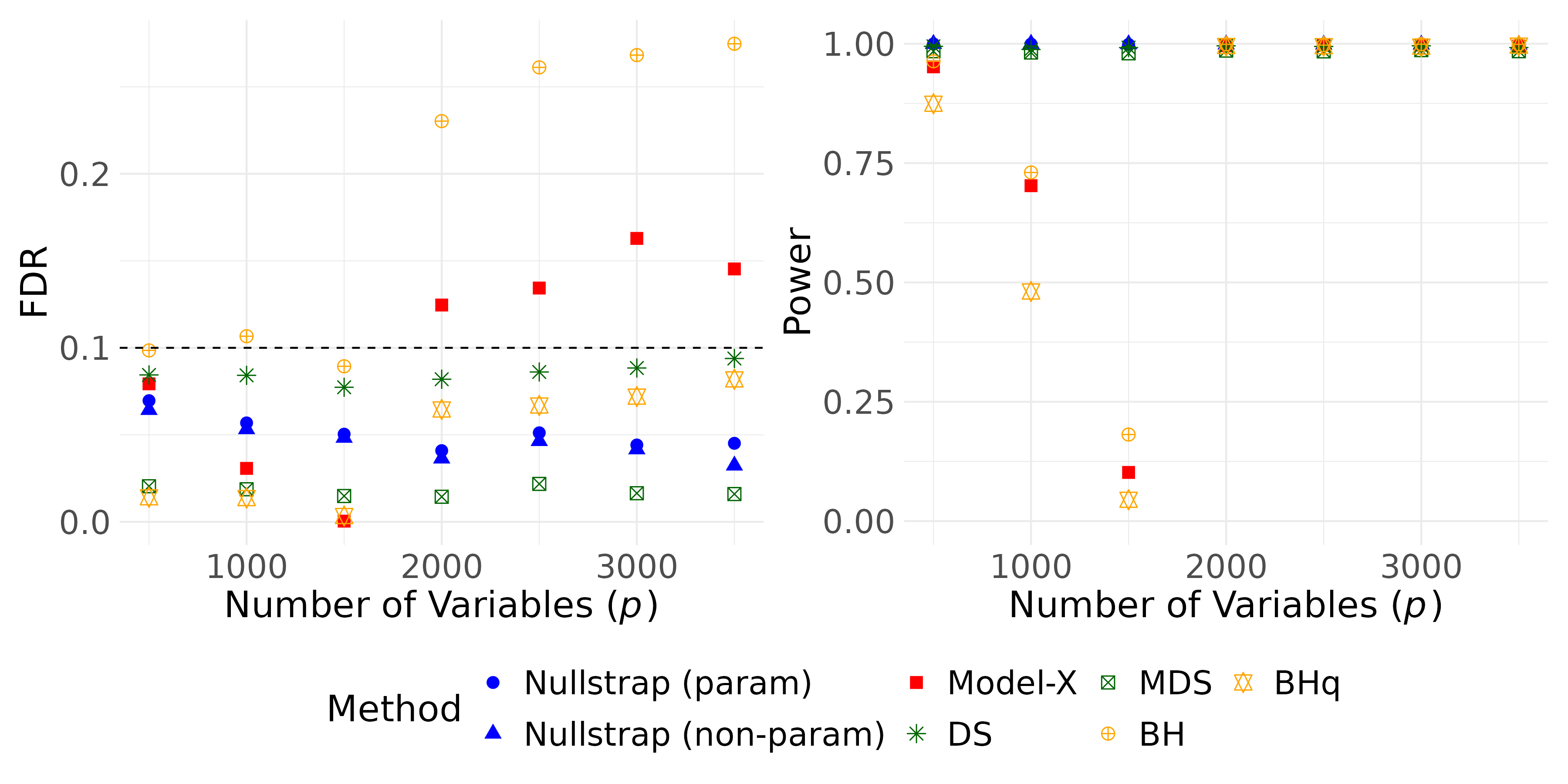}
\caption{Empirical FDR and power vs. number of variables ($p$) under Simulation Setting~\ref{sim:setting_random}.}
\centering
\label{fig:LM_var_random}
\end{figure}

\begin{figure}[H]
        \centering
        \begin{subfigure}{0.45\textwidth}
            \centering
            \includegraphics[width=\linewidth]{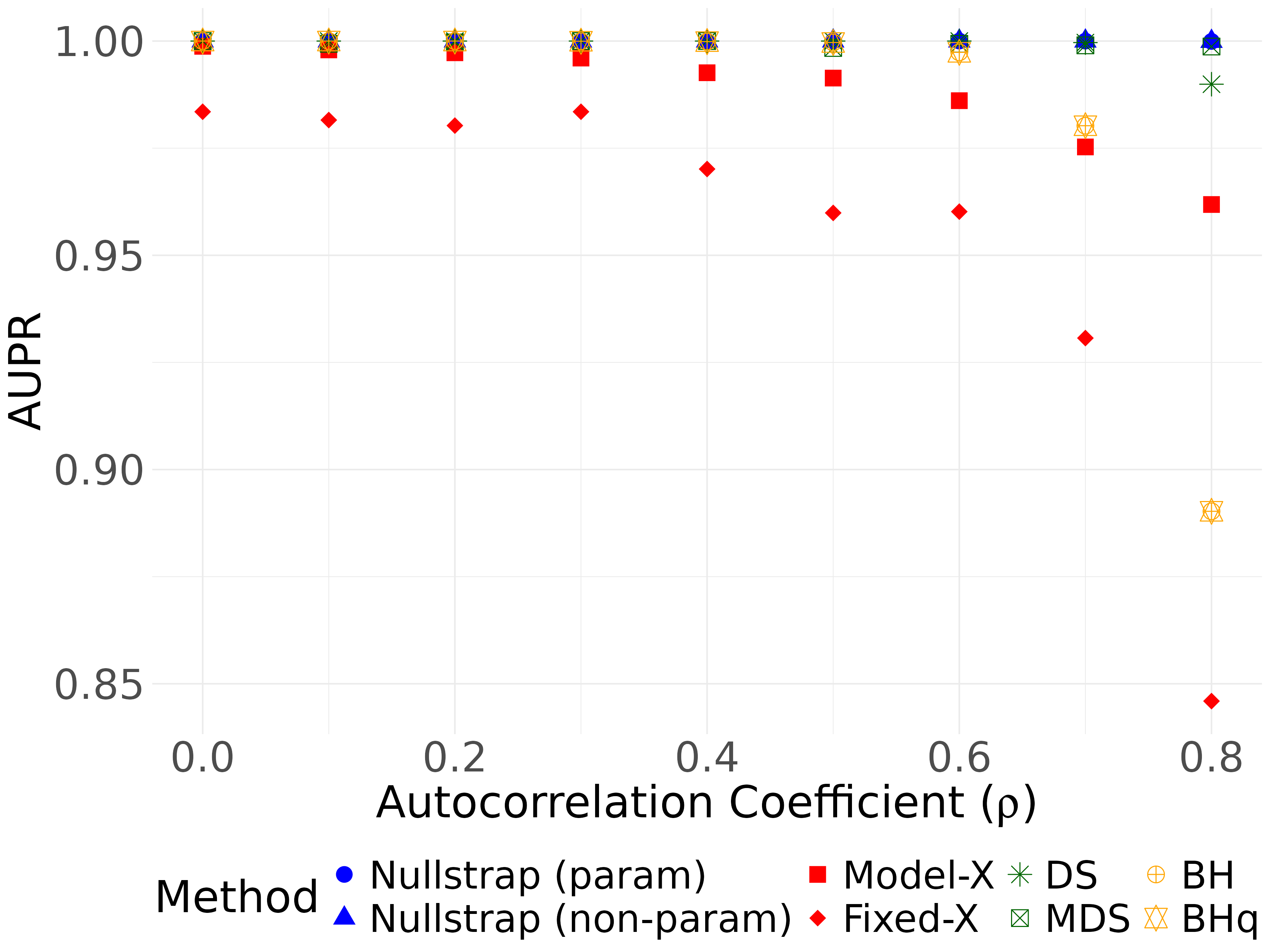} 
            \caption{Empirical AUPR vs. autocorrelation ($\rho$) under Simulation Setting~\ref{sim:setting_random}.}
            \label{fig:sub1}
        \end{subfigure}
        \hfill
        \begin{subfigure}{0.45\textwidth}
            \centering
            \includegraphics[width=\linewidth]{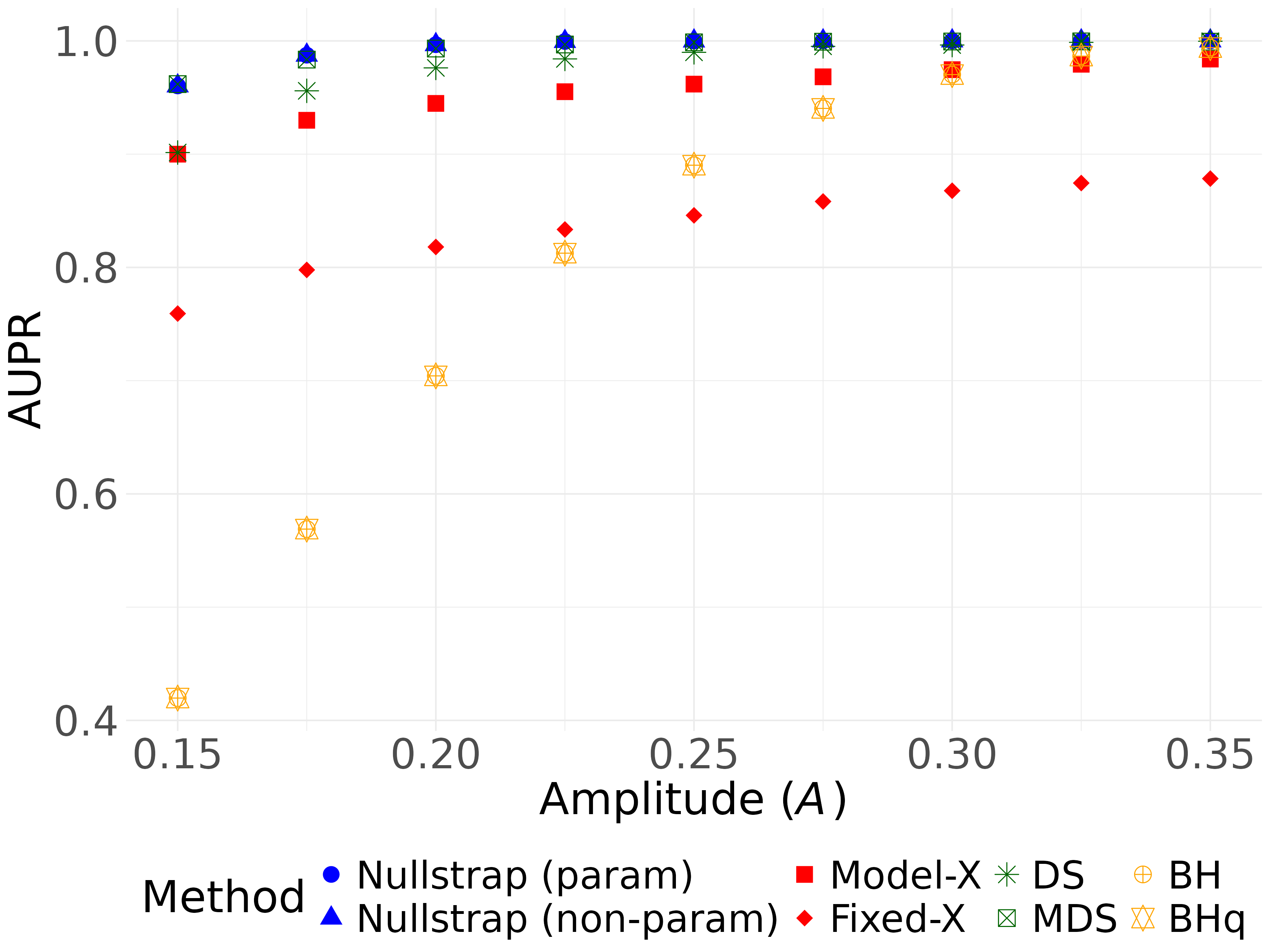} 
            \caption{Empirical AUPR vs. signal amplitude ($A$) under Simulation Setting~\ref{sim:setting_random}.}
            \label{fig:sub2}
        \end{subfigure}
        \vspace{0.5cm}
        \begin{subfigure}{0.45\textwidth}
            \centering
            \includegraphics[width=\linewidth]{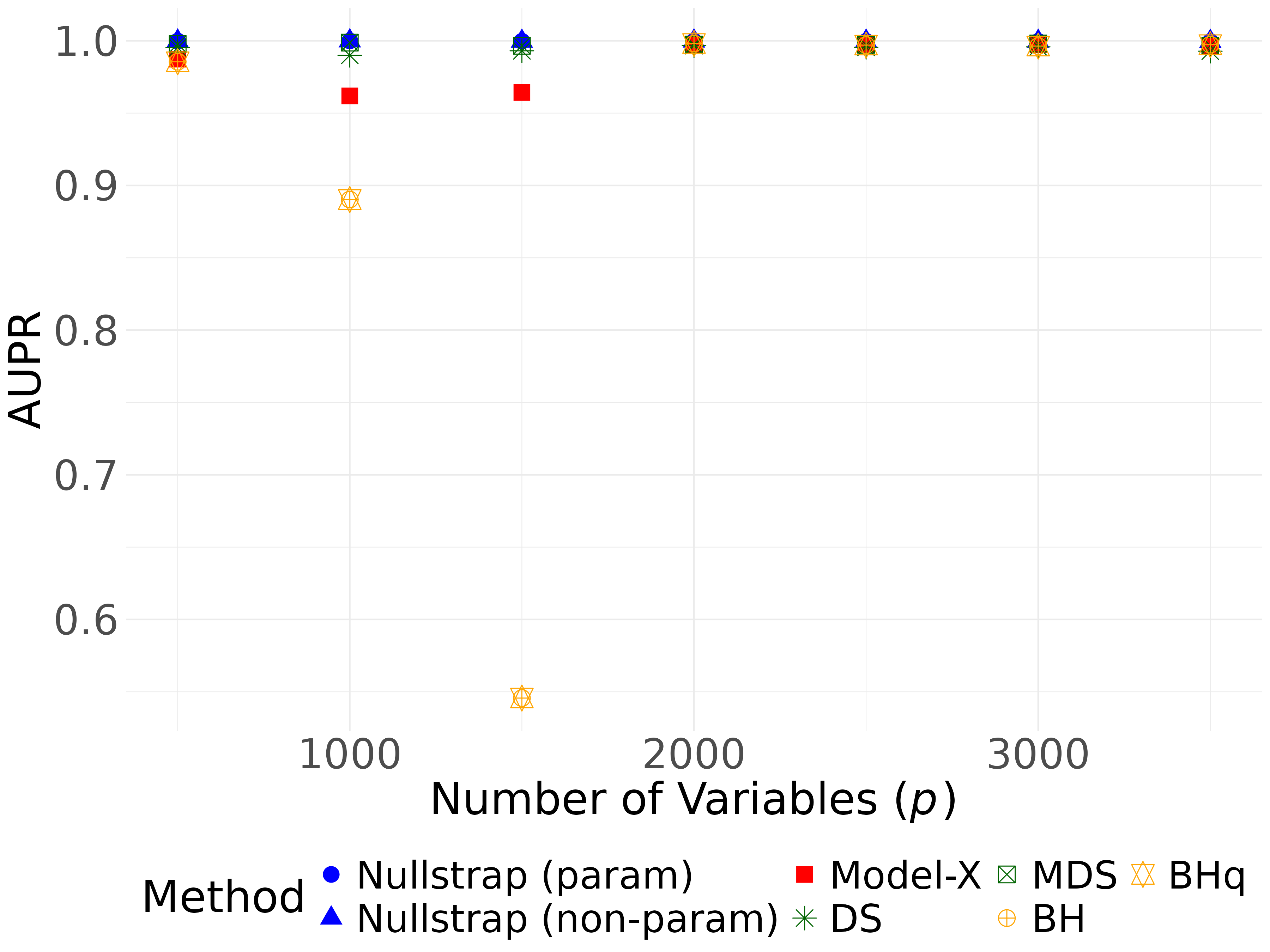} 
            \caption{Empirical AUPR vs. number of variables ($p$) under Simulation Setting~\ref{sim:setting_random}.}
            \label{fig:sub3}
        \end{subfigure}
        \hfill
        \begin{subfigure}{0.45\textwidth}
        \centering

        \end{subfigure}
        \caption{Empirical AUPR for the linear regression model with randomly selected nonzero indices.}
        \label{fig:AUPR_random}
    \end{figure}

The empirical FDR and power of the different methods are presented in Figures~\ref{fig:LM_rho_random}--\ref{fig:LM_var_random}, while the AUPR results are provided in Figure~\ref{fig:AUPR_random}.  
Overall, the FDR of most methods remain controlled across all scenarios, except for Model-X, DS, and BH, which occasionally exhibit slight violations. In all scenarios, Nullstrap consistently demonstrates reliable FDR control and, more importantly, achieves higher power and AUPR than other methods.  

% Compared to Nullstrap, Nullstrap-Diff exhibits slightly lower power, particularly in more challenging scenarios—such as those with high autocorrelations and low signal amplitudes—suggesting that Nullstrap is the preferred choice over Nullstrap-Diff.  

\subsection{Interactions between signal variables}
We next consider a simulation setting in which interactions between signal variables are incorporated into the design matrix, resulting in explicit correlations among its columns.

\begin{simsetting}\label{sim:setting_inter}
	We set \( n = 1000 \), \( p_{\rm{base}} = 40 \), and \( p = p_{\rm{base}} + \frac{p_{\rm{base}}(p_{\rm{base}}-1)}{2} \). 
The base design matrix \(\bX_{\text{base}}\) is drawn from \(\mathcal{N}(\mathbf{0},\boldsymbol{\Sigma}_{\text{base}})\), where \(\boldsymbol{\Sigma}_{\text{base}}\) is a Toeplitz correlation matrix with autocorrelation parameter \(\rho\in(0,1)\).
	We then construct interaction terms by computing pairwise products of the first \( p_{\rm{base}} \) variables, forming an interaction matrix \( \bX_{\rm{interact}} \). The first \( 5 \) elements of the coefficient vector \( \bbeta \) are randomly assigned values with amplitude \( A \) and random signs. Additionally, if both variables involved in an interaction term are among the first \( 5 \) variables, their corresponding coefficient is also randomly assigned values with amplitude \( A \) and random signs.  
	Finally, the full design matrix \( \bX \) is constructed by concatenating \( \bX_{\mathrm{base}} \) and \( \bX_{\mathrm{interact}} \) column-wise. We consider two simulation parameters for adjustment:  
	\begin{itemize}
		\item (a) the autocorrelation parameter \( \rho \in [0, 0.8] \),  
		\item (b) the signal amplitude \( A \in [0.25, 0.45] \).  
	\end{itemize}  
	For each scenario where one parameter varies, the remaining parameters are held constant as:  
	\begin{equation}\label{eq:para_inter}  
	\rho = 0.8, \quad A = 0.3.  
	\end{equation}  
        The response vector $\by$ are generated as in Simulation Setting~1.
	\end{simsetting}

% 	We still set the correction factor \( \gamma_{n,p} \) to
% $
% 0.2 \Hat{\sigma} \left(\lambda_n + \sqrt{\frac{\log p}{n}} \right)
% $
% in this simulation setting.

\begin{figure}[H]
\centering
\includegraphics[scale=0.5]{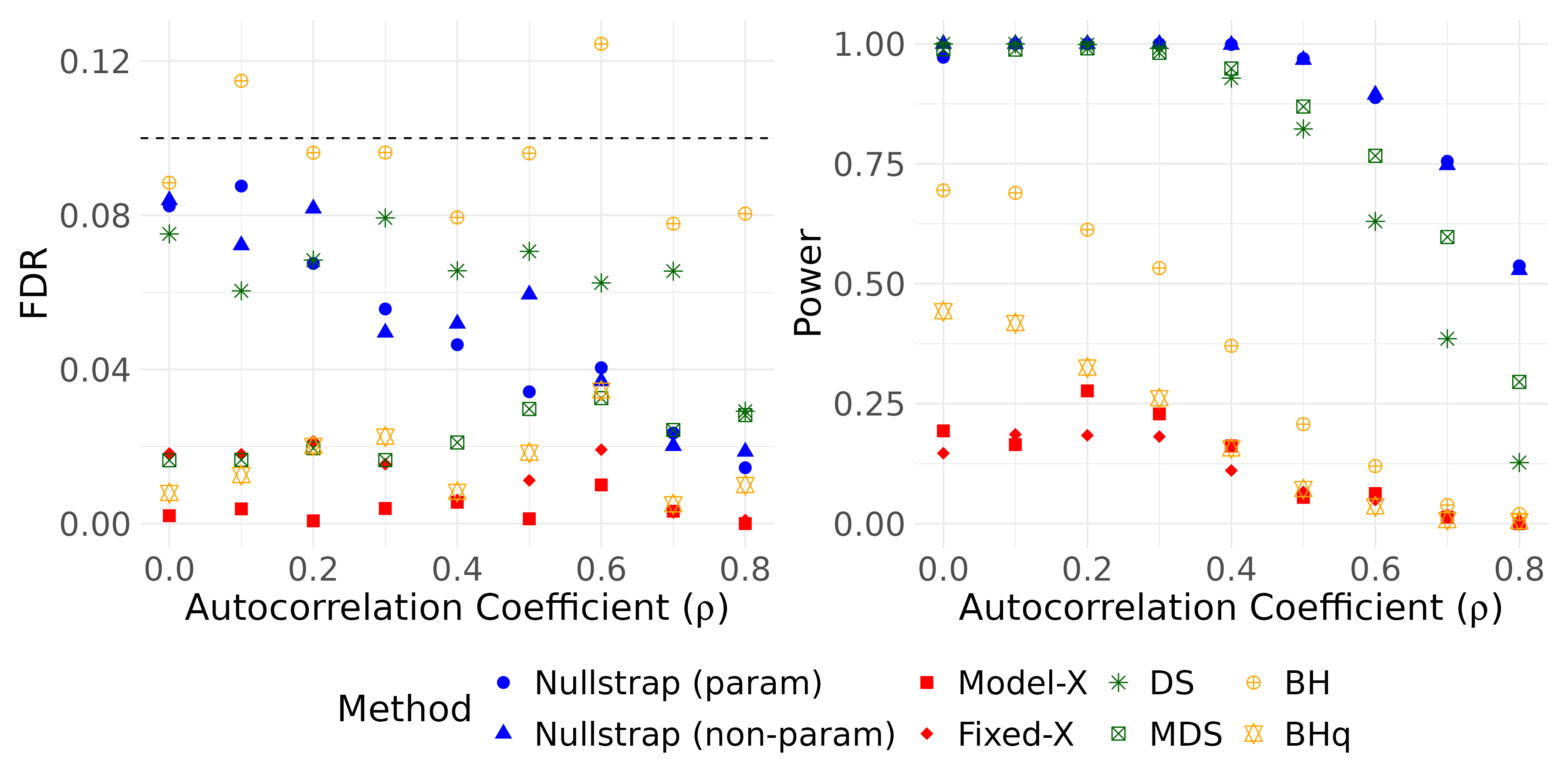}
\caption{Empirical FDR and power vs. autocorrelation ($\rho$) under Simulation Setting~\ref{sim:setting_inter}.}
\centering
\label{fig:LM_rho_inter}
\end{figure}

\begin{figure}[H]
\centering
\includegraphics[scale=0.5]{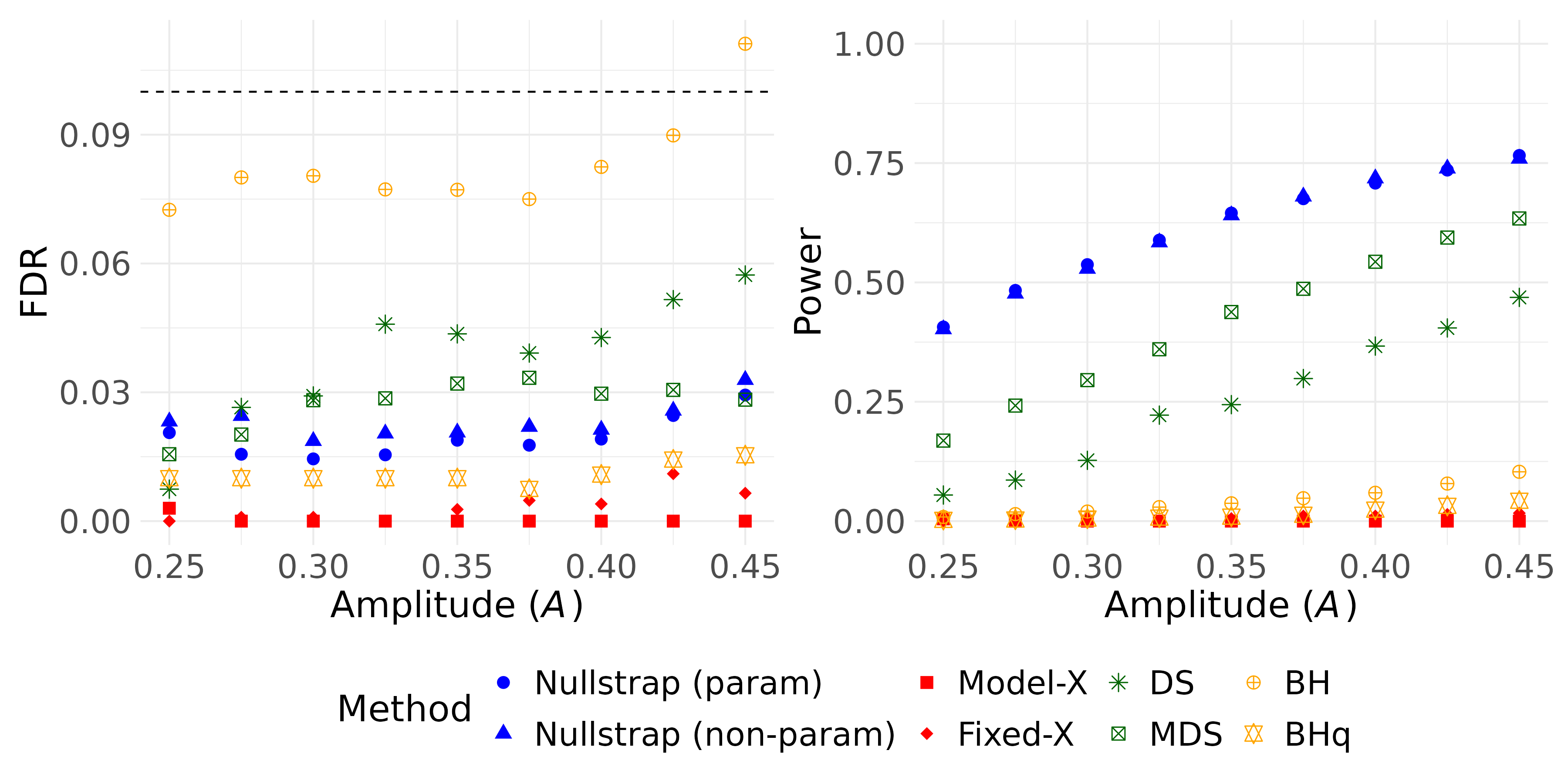}
\caption{Empirical FDR and power vs. signal amplitude ($A$) under Simulation Setting~\ref{sim:setting_inter}.}
\centering
\label{fig:LM_amp_inter}
\end{figure}

\begin{figure}[H]
        \centering
        \begin{subfigure}{0.45\textwidth}
            \centering
            \includegraphics[width=\linewidth]{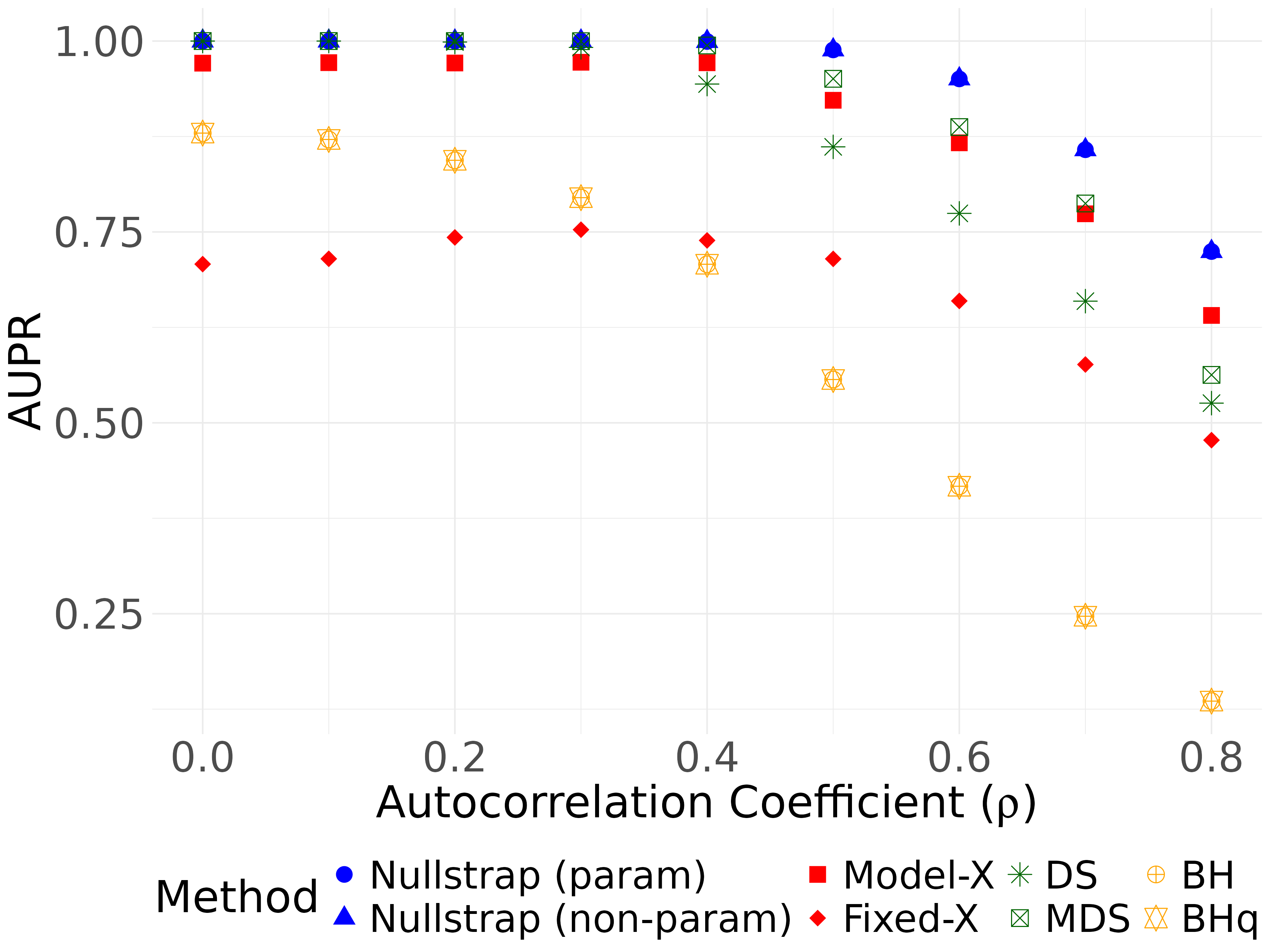} 
            \caption{Empirical AUPR vs. autocorrelation ($\rho$) under Simulation Setting~\ref{sim:setting_inter}.}
            \label{fig:sub1_inter}
        \end{subfigure}
        \hfill
        \begin{subfigure}{0.45\textwidth}
            \centering
            \includegraphics[width=\linewidth]{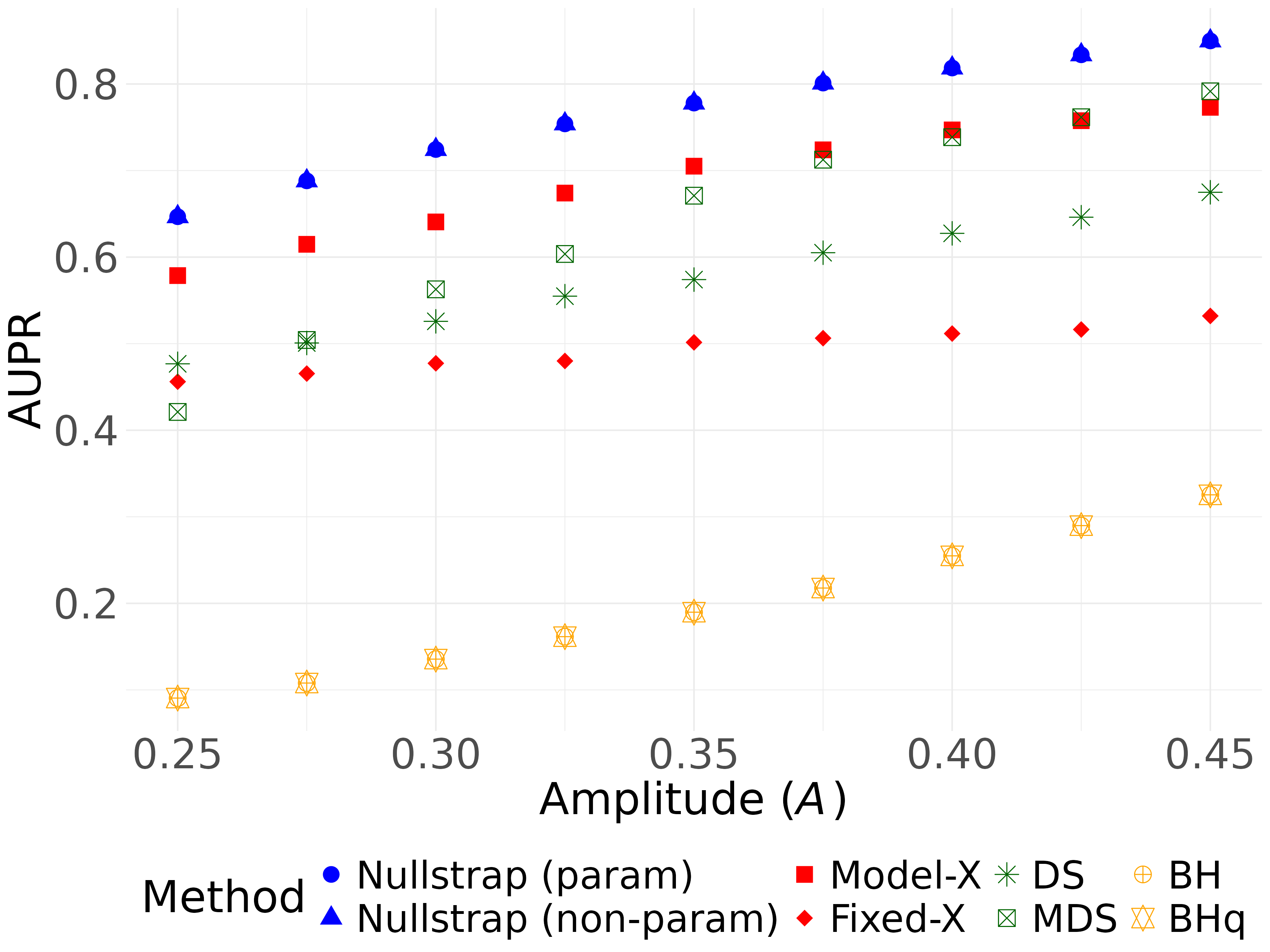} 
            \caption{Empirical AUPR vs. signal amplitude ($A$) under Simulation Setting~\ref{sim:setting_inter}.}
            \label{fig:sub2_inter}
        \end{subfigure}
        \caption{Empirical AUPR for the linear regression model with interaction terms.}
        \label{fig:AUPR_inter}
    \end{figure}

For each scenario under Simulation Setting~\ref{sim:setting_inter}, we compare the FDR, power, and AUPR of the different methods, using 100 replications. 
The empirical FDR and power of the different methods are presented in Figures \ref{fig:LM_rho_inter}--\ref{fig:LM_amp_inter}. The AUPR results are provided in Figure \ref{fig:AUPR_inter}.
Overall, the FDR of most methods remain controlled across all scenarios, except for BH, which sometimes slightly lose control. In all scenarios, Nullstrap once again consistently demonstrates reliable FDR control and, more importantly, achieves higher power and AUPR than other methods. 

% Compared to Nullstrap, Nullstrap-Diff exhibits lower power than DS and MDS, particularly in more challenging scenarios such as those with high autocorrelations and high signal amplitudes. 

\clearpage

\subsection{Alternative noise distributions}

\begin{simsetting}\label{sim:setting_diff}

We set \( n = 2000 \) and \( p = 1000 \). The design matrix \( \bX \) is generated as described in Simulation Setting~1 from the main text. We consider two simulation parameters for adjustment:  
	\begin{itemize}
		\item (a) the autocorrelation parameter \( \rho \in [0, 0.8] \),  
		\item (b) the signal amplitude \( A \in [0.3, 0.5] \).  
	\end{itemize}  
	For each scenario where one parameter varies, the remaining parameters are held constant as:  
	\begin{equation}\label{eq:para}  
	\rho = 0.8 \, \text{ and } \, A = 0.4.  
	\end{equation}  
    
 %    We consider two simulation parameters for adjustment: (1) the autocorrelation parameter \( \rho \in [0, 0.8] \) and (2) the signal amplitude \( A \in [0.3, 0.5] \).  
	% For each scenario where one parameter varies, the remaining parameters are held constant as:
	% \vspace{-20pt}  
	% \begin{equation}\label{eq:para}  
	% \rho = 0.8 \, \text{ and } \, A = 0.4.  
	% \vspace{-20pt}  
	% \end{equation}  
	The first \( 30 \) elements of the coefficient vector \( \bbeta \) are randomly assigned values with amplitude \( A \) and random signs, while the remaining \( p - 30 \) elements are set to zero. 
 We consider three noise distributions:
\begin{enumerate}[label=(\Roman*)]
    \item Laplace distribution, \(\mathrm{Laplace}(0,1)\);
    \item Student’s \(t\)-distribution with \(10\) degrees of freedom, \(t_{10}\);
    \item Student’s \(t\)-distribution with \(3\) degrees of freedom, \(t_{3}\).
\end{enumerate}
	% (a) Laplace distribution \(\text{Laplace}(0,1)\),  
	% (b) \(t\) distribution with \(10\) degrees of freedom \(t_{10}\), and  
	% (c) \(t\) distribution with \(3\) degrees of freedom \(t_{3}\).  
        The response vector $\by$ is generated as in Simulation Setting~2.
\end{simsetting}

\begin{figure}[H]
\centering
\includegraphics[scale=0.5]{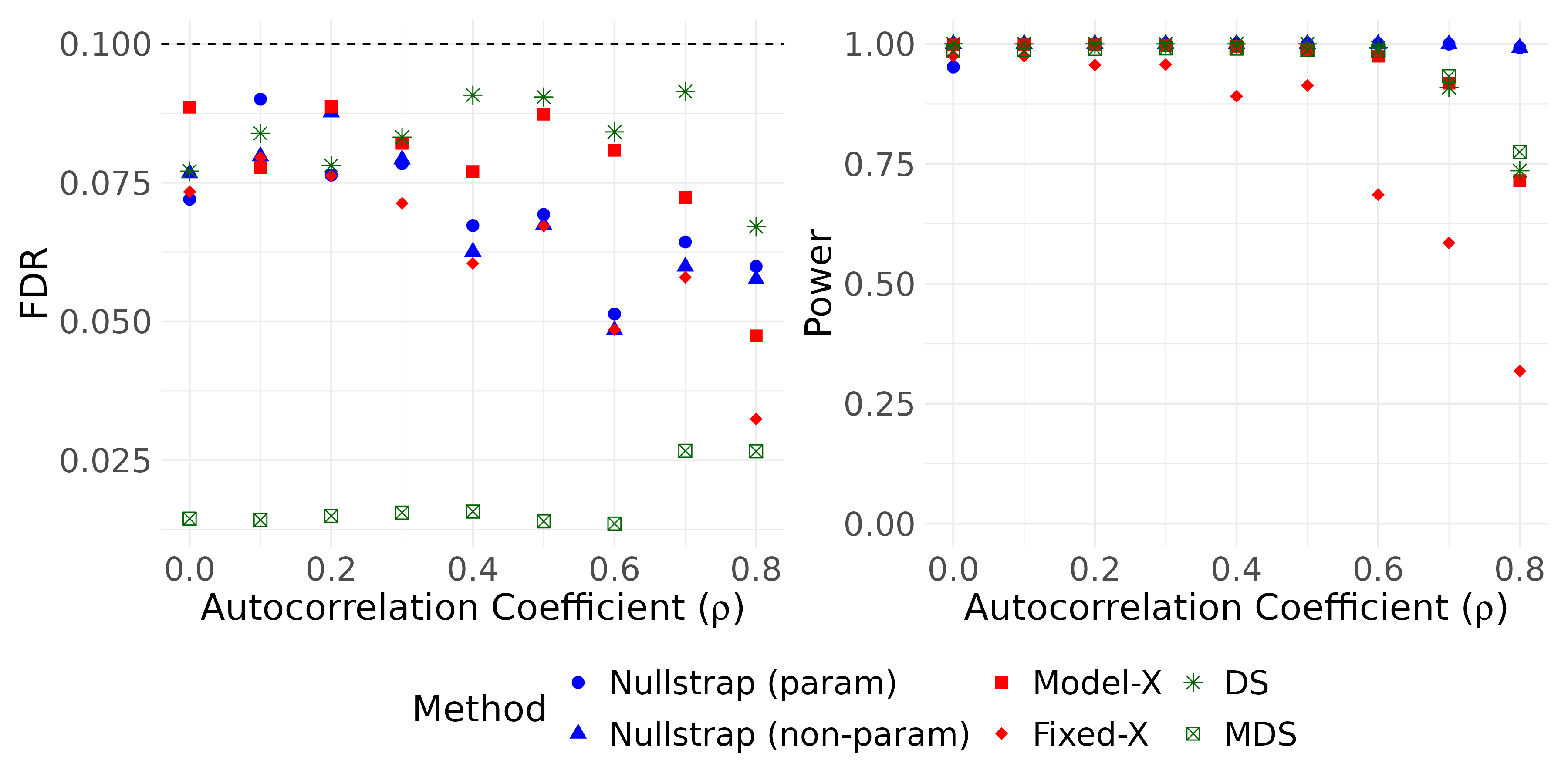}
\caption{Empirical FDR and power vs. autocorrelation ($\rho$) under Simulation Setting~\ref{sim:setting_diff} (I).}
\centering
\label{fig:rho_power_lap}
\end{figure}

\begin{figure}[H]
        \centering
         \begin{subfigure}{0.45\textwidth}
            \centering
            \includegraphics[width=\linewidth]{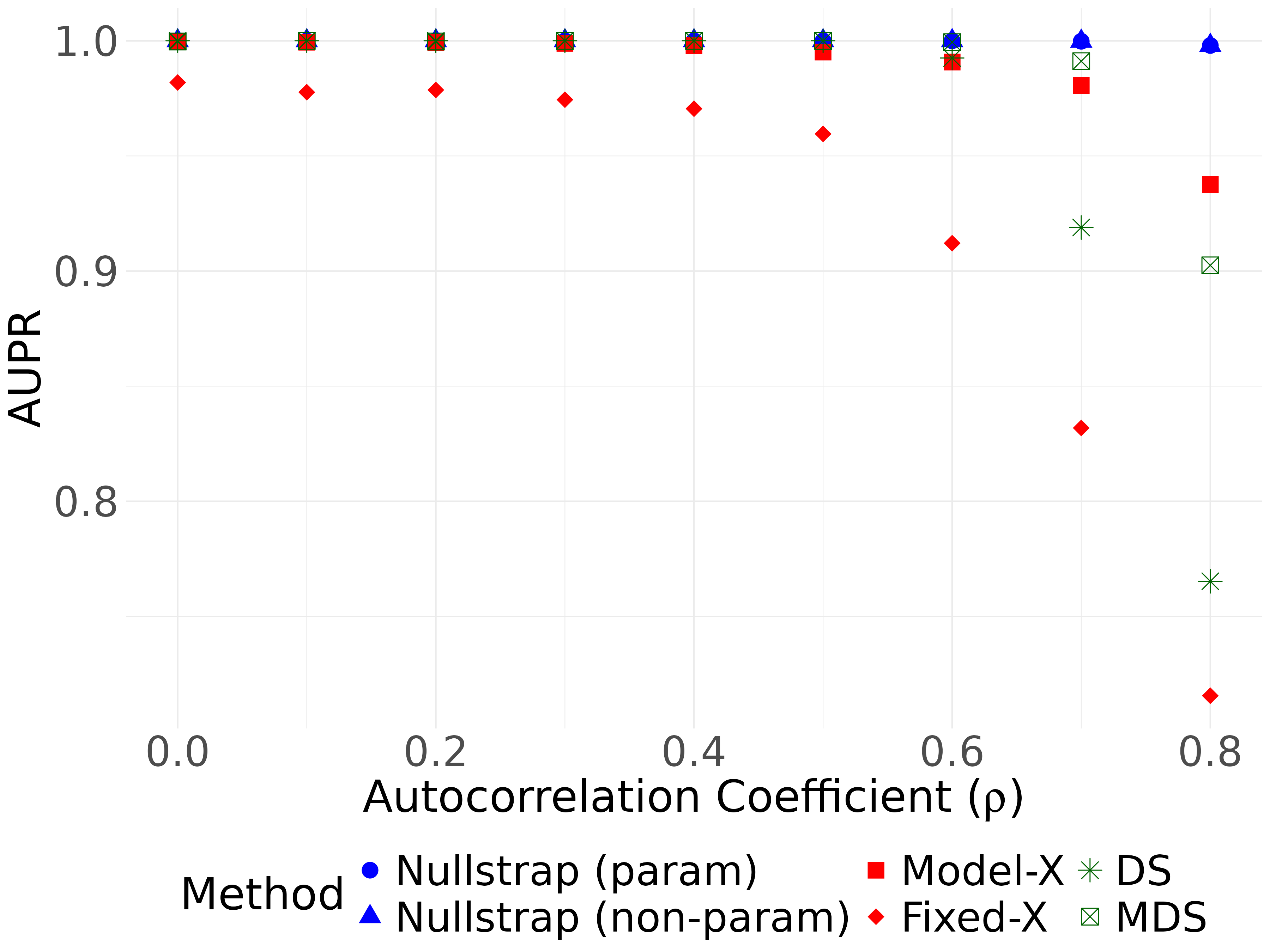} 
            \caption{Empirical AUPR vs. autocorrelation ($\rho$) under Simulation Setting~\ref{sim:setting_diff} (I).}
            \label{fig:rho_aupr_lap}
        \end{subfigure}
        \hfill
        \begin{subfigure}{0.45\textwidth}
            \centering
            \includegraphics[width=\linewidth]{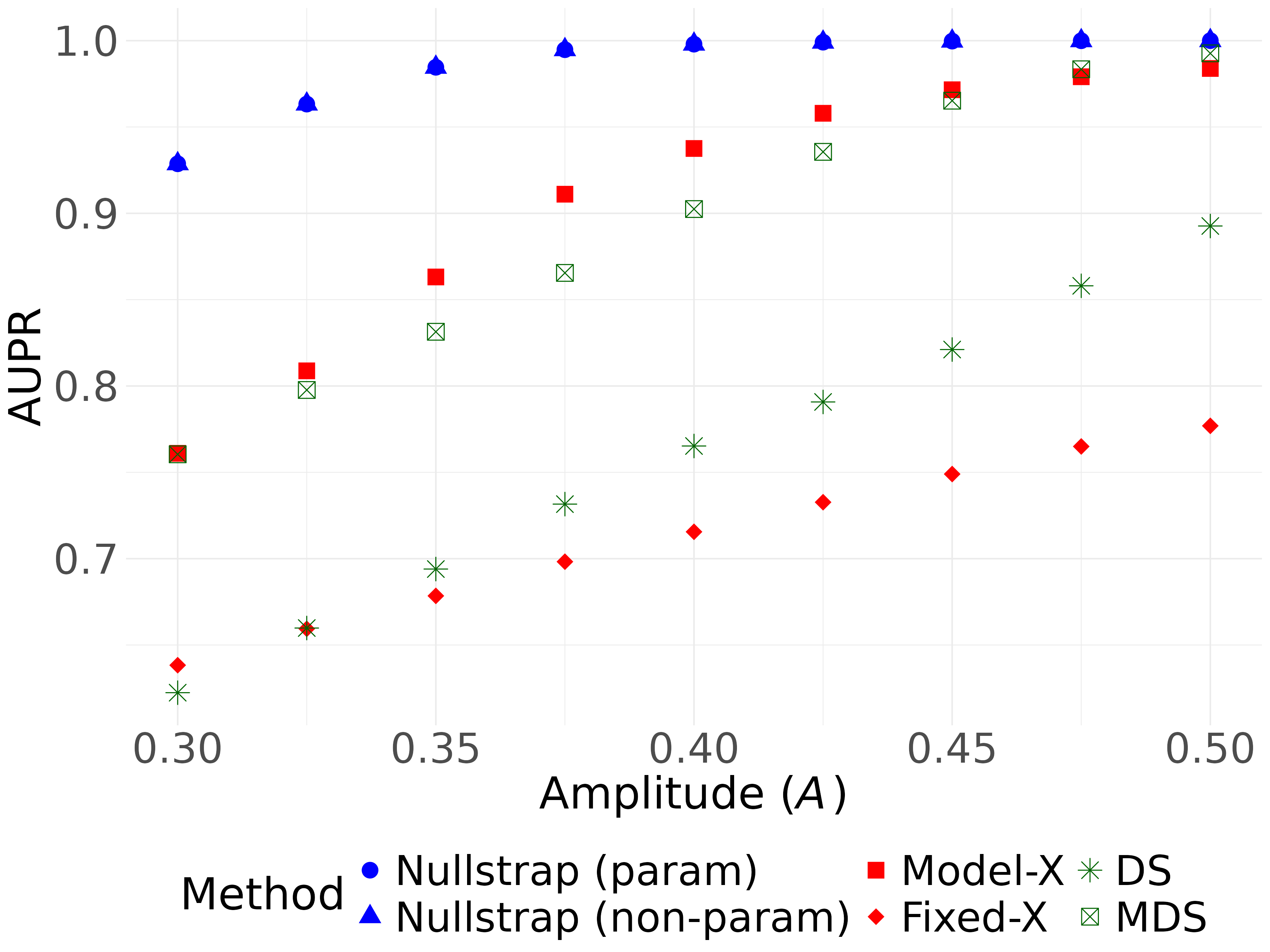} 
            \caption{Empirical AUPR vs. signal amplitude ($A$) under Simulation Setting~\ref{sim:setting_diff} (I).}
            \label{fig:amp_aupr_lap}
        \end{subfigure}
        
        \vspace{0.5cm}
        
       \begin{subfigure}{0.45\textwidth}
            \centering
            \includegraphics[width=\linewidth]{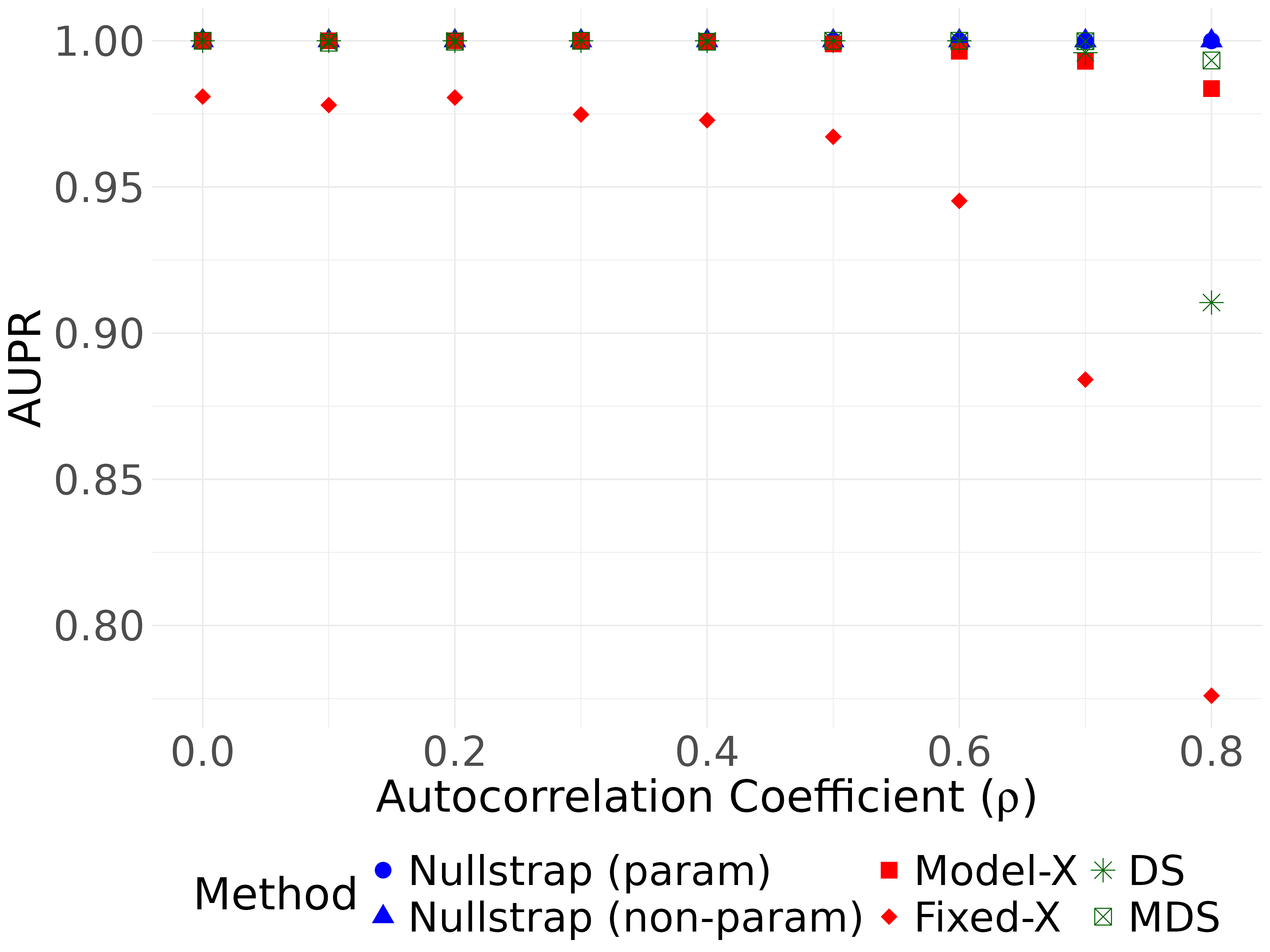} 
            \caption{Empirical AUPR vs. autocorrelation ($\rho$) under Simulation Setting~\ref{sim:setting_diff} (II).}
            \label{fig:rho_aupr_t10}
        \end{subfigure}
        \hfill
        \begin{subfigure}{0.45\textwidth}
            \centering
            \includegraphics[width=\linewidth]{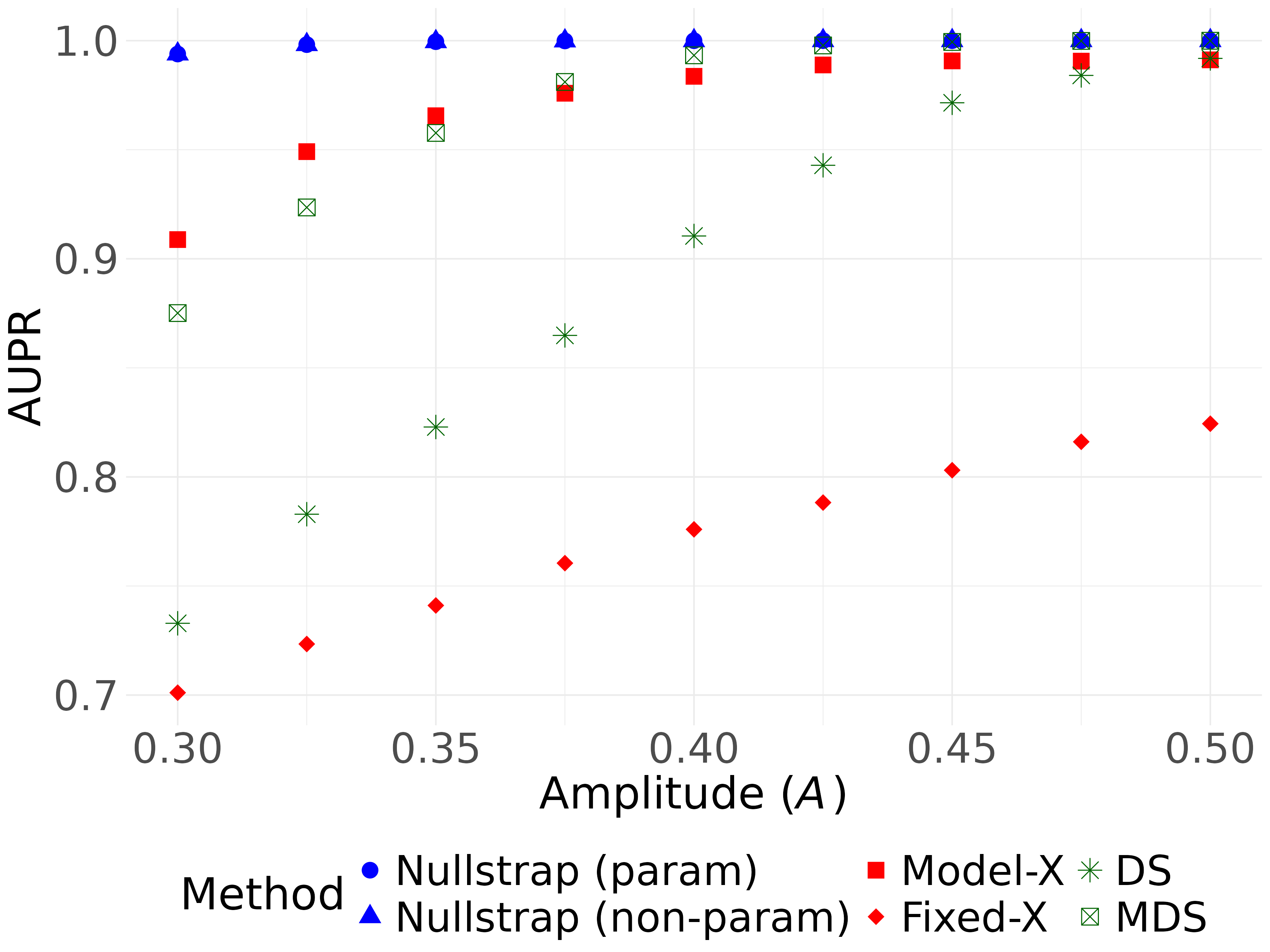} 
            \caption{Empirical AUPR vs. signal amplitude ($A$) under Simulation Setting~\ref{sim:setting_diff} (II).}
            \label{fig:amp_aupr_t10}
        \end{subfigure}
        
        \vspace{0.5cm}
        
        \begin{subfigure}{0.45\textwidth}
            \centering
            \includegraphics[width=\linewidth]{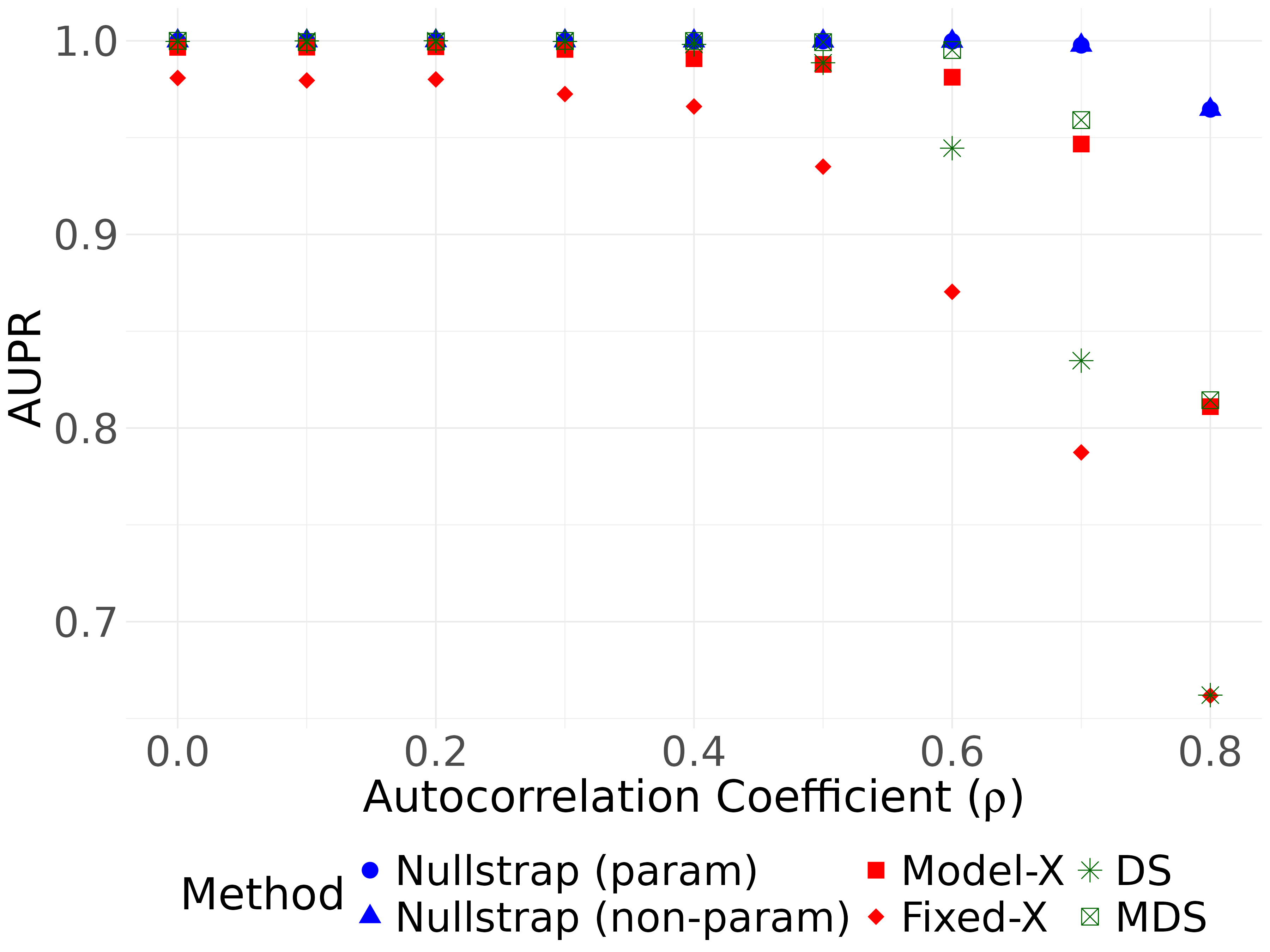} 
            \caption{Empirical AUPR vs. autocorrelation ($\rho$) under Simulation Setting~\ref{sim:setting_diff} (III).}
            \label{fig:rho_aupr_t3}
        \end{subfigure}
        \hfill
        \begin{subfigure}{0.45\textwidth}
            \centering
            \includegraphics[width=\linewidth]{simulation/result_amp_aupr_lm_t3.png} 
            \caption{Empirical AUPR vs. signal amplitude ($A$) under Simulation Setting~\ref{sim:setting_diff} (III).}
            \label{fig:amp_aupr_t3}
        \end{subfigure}

        \caption{Empirical AUPR for linear models with alternative error distributions.}
        \label{fig:aupr_diff}
    \end{figure}

For each scenario under Simulation Setting~\ref{sim:setting_diff}, we compare the FDR and power at the target FDR level \( q = 0.1 \), as well as the AUPR, across different methods using 100 replications.
The empirical FDR and power of the different methods are presented in Figures \ref{fig:rho_power_lap}--\ref{fig:amp_power_t3}. The AUPR results are provided in Figure \ref{fig:aupr_diff}.
Overall, all methods remain controlled for the FDR across all scenarios. In all scenarios, Nullstrap (param) and Nullstrap (non-param) once again consistently demonstrate reliable FDR control and, more importantly, achieves higher power and AUPR than other methods, especially in some challenging scenarios, such as high correlations among variables and low signal amplitude. 

Notably, under the more challenging conditions of \(t_3\) and Laplace distributions, where the noise term deviates significantly from normality, our methods exhibit even greater advantages. In these scenarios, Nullstrap (param) and Nullstrap (non-param) not only continue to control FDR effectively but also demonstrate a more substantial improvement in power and AUPR compared to competing methods. This robustness across different distributional settings highlights the adaptability and reliability of our approach, particularly in cases where the normality assumption is violated.

\begin{figure}[H]
\centering
\includegraphics[scale=0.5]{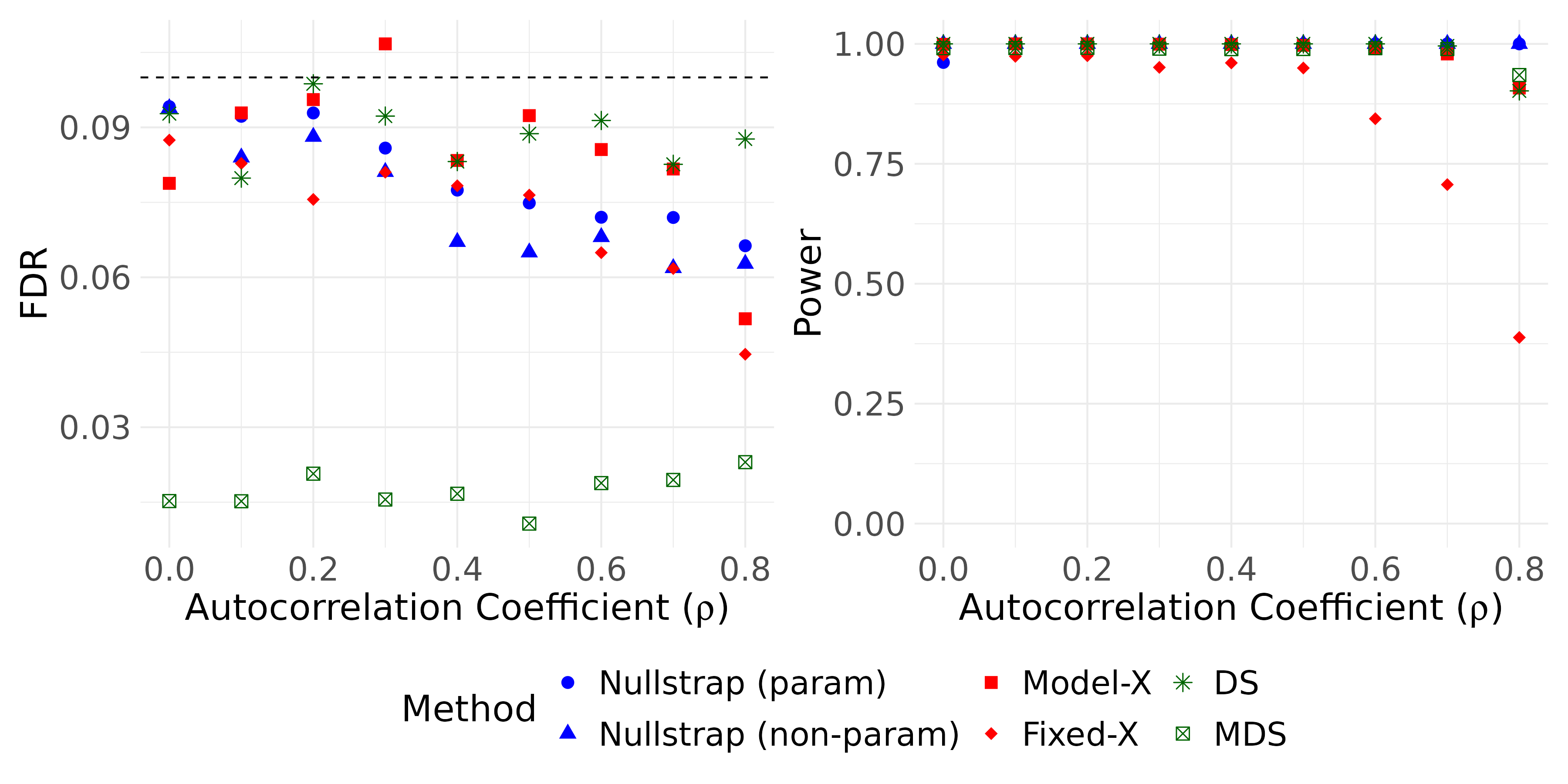}
\caption{Empirical FDR and power vs. autocorrelation ($\rho$) under Simulation Setting~\ref{sim:setting_diff} (II).}
\centering
\label{fig:rho_power_t10}
\end{figure}

\begin{figure}[H]
\centering
\includegraphics[scale=0.5]{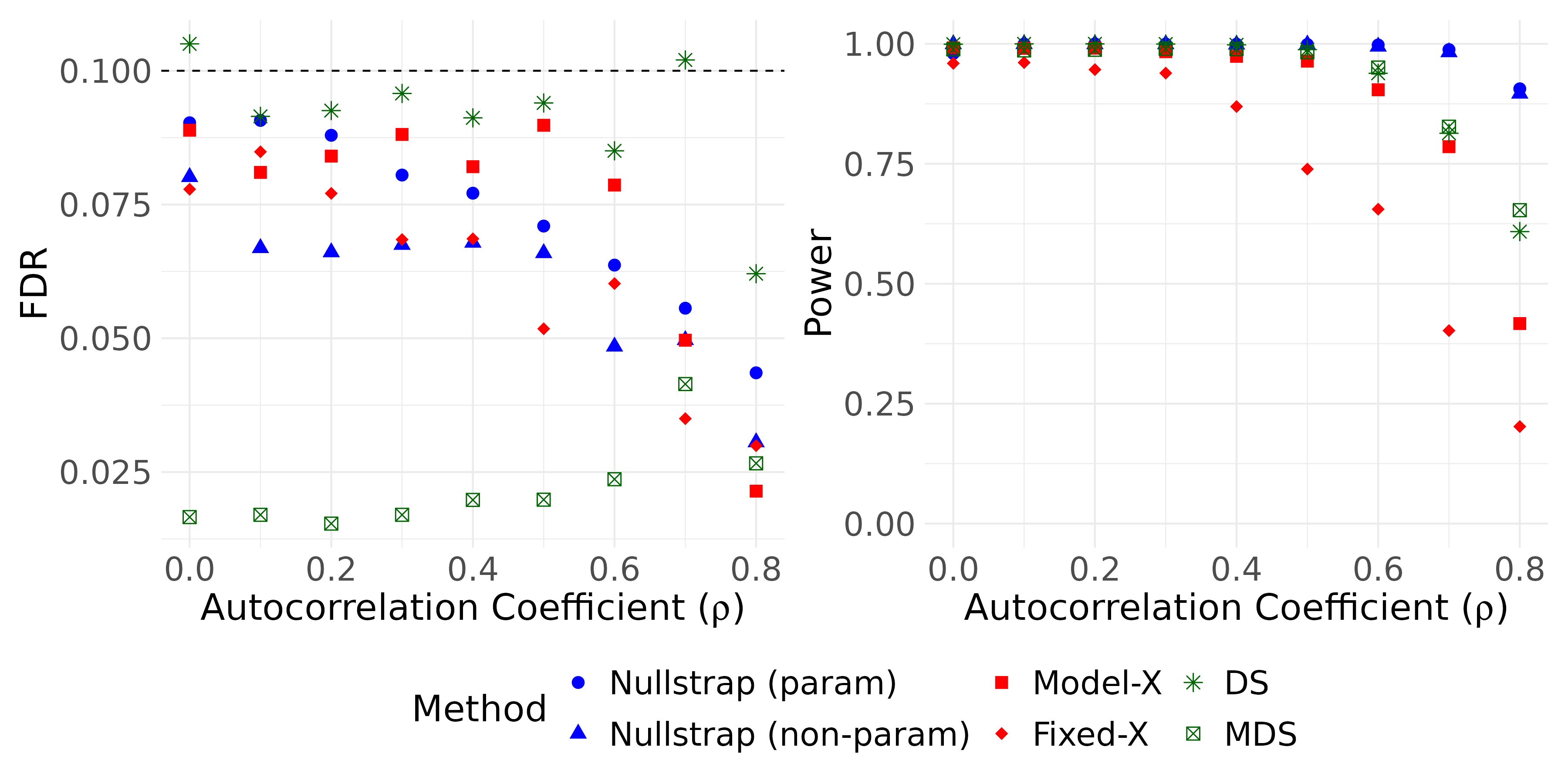}
\caption{Empirical FDR and power vs. autocorrelation ($\rho$) under Simulation Setting~\ref{sim:setting_diff} (III).}
\centering
\label{fig:rho_power_t3}
\end{figure}

\begin{figure}[H]
\centering
\includegraphics[scale=0.5]{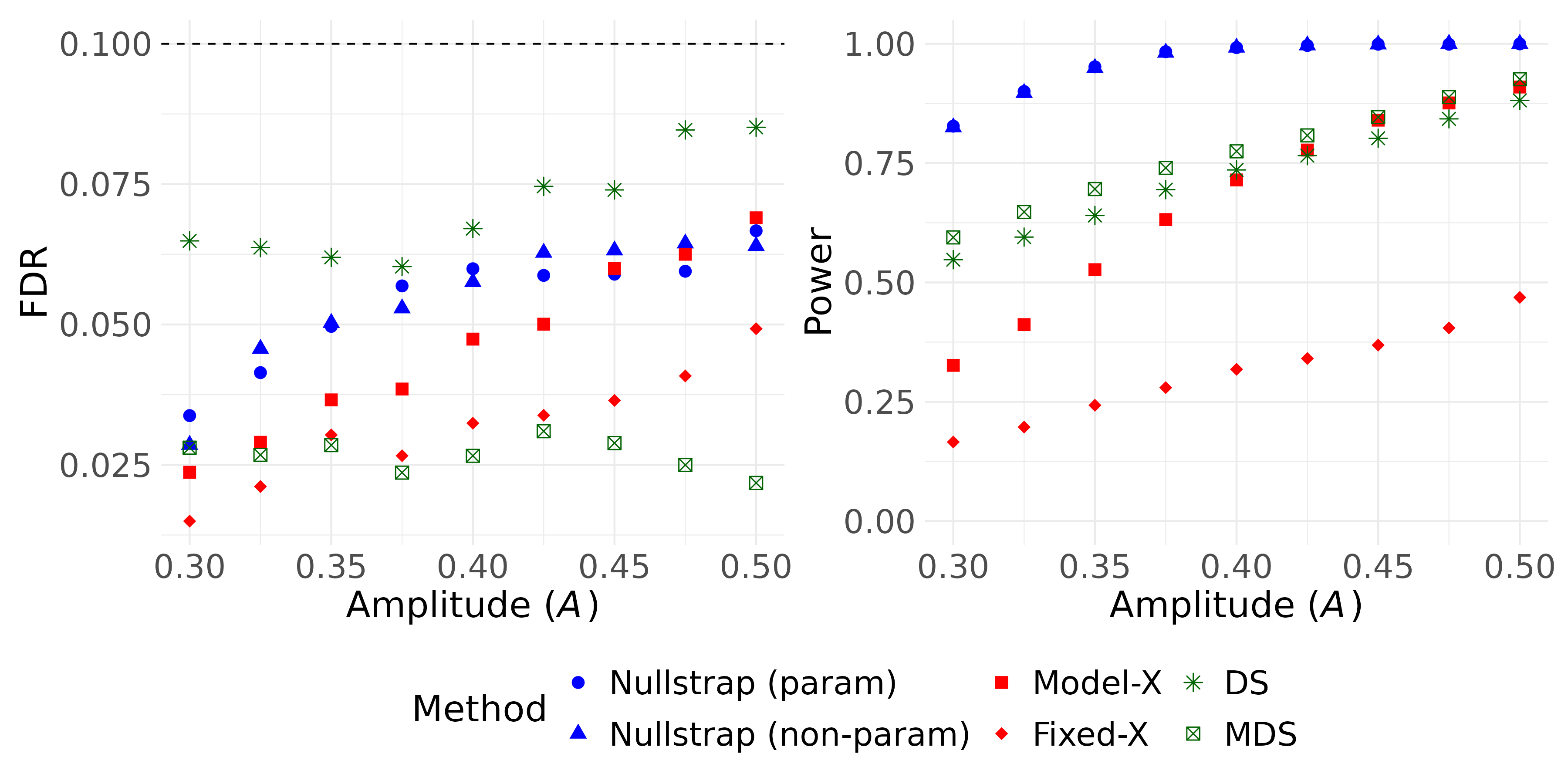}
\caption{Empirical FDR and power vs. signal amplitude ($A$) under Simulation Setting~\ref{sim:setting_diff} (I).}
\centering
\label{fig:amp_power_lap}
\end{figure}

\begin{figure}[H]
\centering
\includegraphics[scale=0.5]{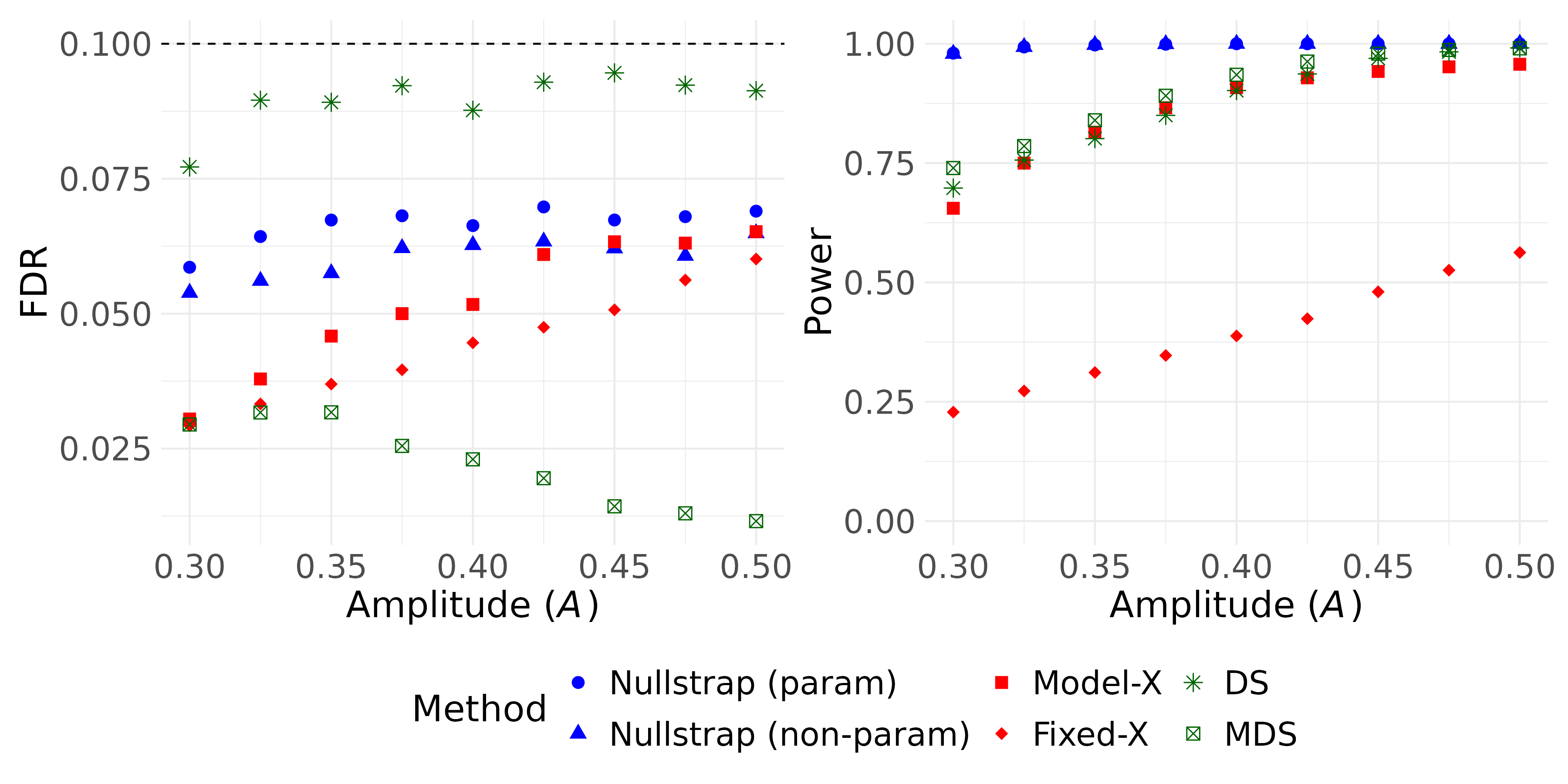}
\caption{Empirical FDR and power vs. signal amplitude ($A$) under Simulation Setting~\ref{sim:setting_diff} (II).}
\centering
\label{fig:amp_power_t10}
\end{figure}

\begin{figure}[H]
\centering
\includegraphics[scale=0.5]{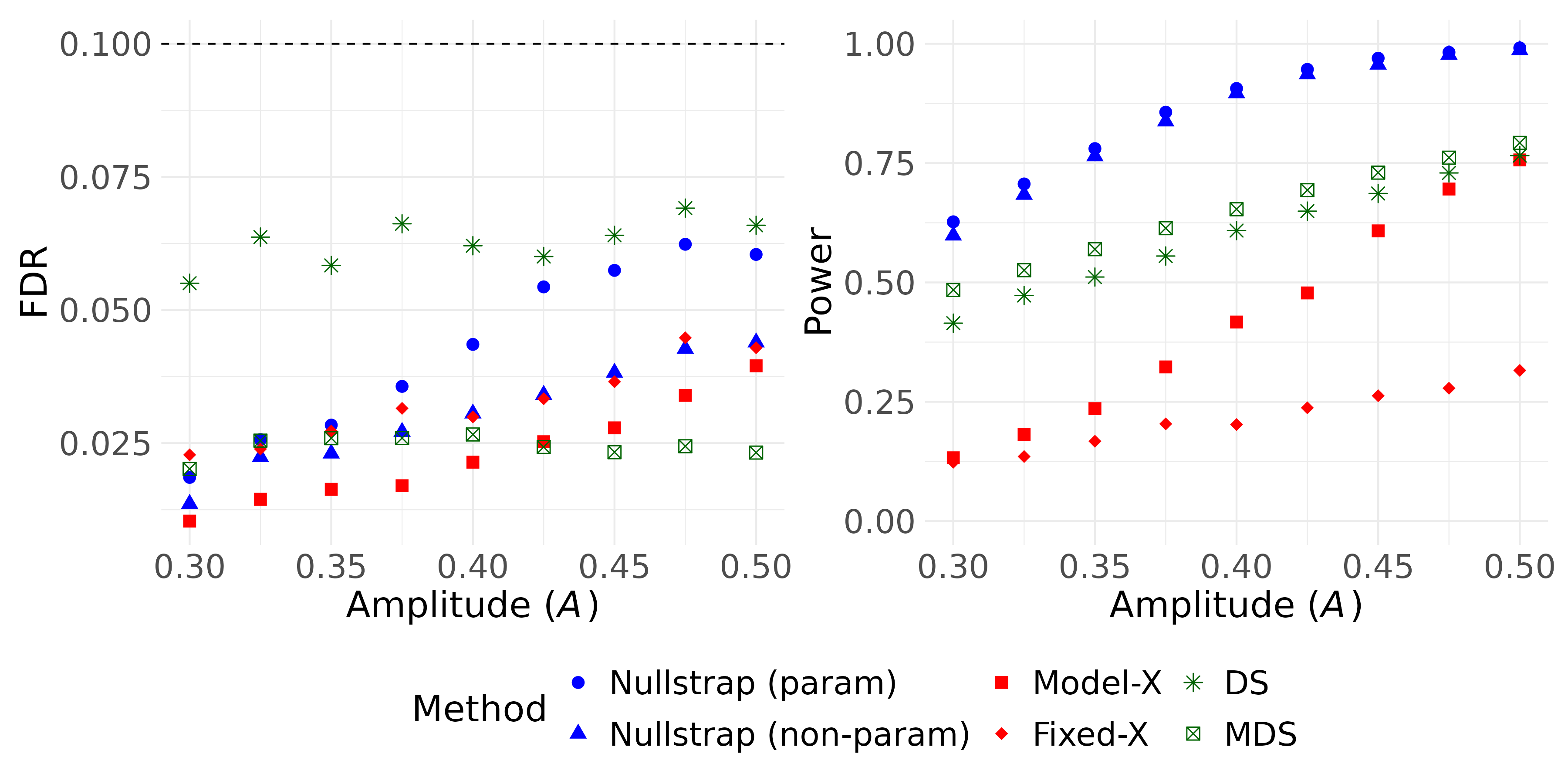}
\caption{Empirical FDR and power vs. signal amplitude ($A$) under Simulation Setting~\ref{sim:setting_diff} (III).}
\centering
\label{fig:amp_power_t3}
\end{figure}

% \begin{simsetting}\label{sim:setting_df}

% We set \( n = 2000 \), \( p = 1000 \), \(q = 0.1\), \(\rho = 0.8\) and \(A=0.3\).  The design matrix \( \bX \) is generated as described in Simulation Setting~1 from the main text. We compare different $t$ distributions with degrees of freedom ($Df$) ranging from $3$ to $10$. To ensure that all distributions have unit variance, we normalize each distribution by dividing it by \(\sqrt{\frac{Df}{Df-2}}\).
% The first \( 30 \) elements of the coefficient vector \( \bbeta \) are randomly assigned values with amplitude \( A \) and random signs, while the remaining \( p - 30 \) elements are set to zero. 
% \end{simsetting}

% \begin{figure}[h]
% \centering
% \includegraphics[width=1.0\textwidth]
% {simulation/result_df_power_lm.png}
% \caption{Empirical FDR and power vs. degree of freedom ($Df$) under Simulation Setting~\ref{sim:setting_df}.}
% \centering
% \label{fig:df_power}
% \end{figure}

% \begin{figure}[h]
% \centering
% \includegraphics[width=0.5\textwidth]{simulation/result_df_aupr_lm_.png}
% \caption{Empirical AUPR vs. degree of freedom ($Df$) under Simulation Setting~\ref{sim:setting_df}.}
% \centering
% \label{fig:df_aupr}
% \end{figure}

\begin{figure}[H]
\centering
\includegraphics[width=1.0\textwidth]{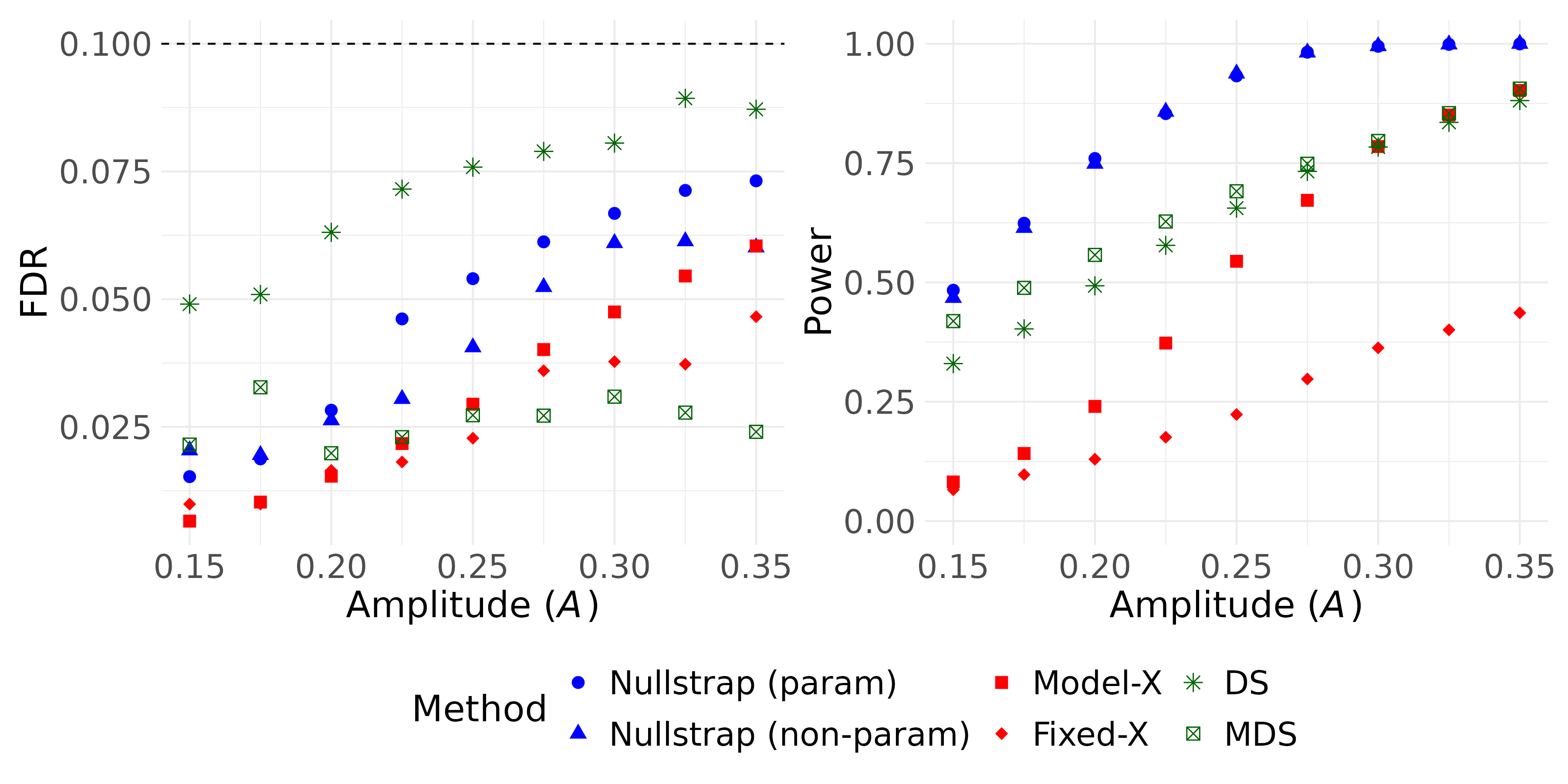}
\caption{Empirical FDR and power vs. signal amplitude ($A$) under Simulation Setting~\ref{sim:setting_gamma}.}
\centering
\label{fig:gamma_power}
\end{figure}

\begin{figure}[H]
\centering
\includegraphics[width=0.5\textwidth]{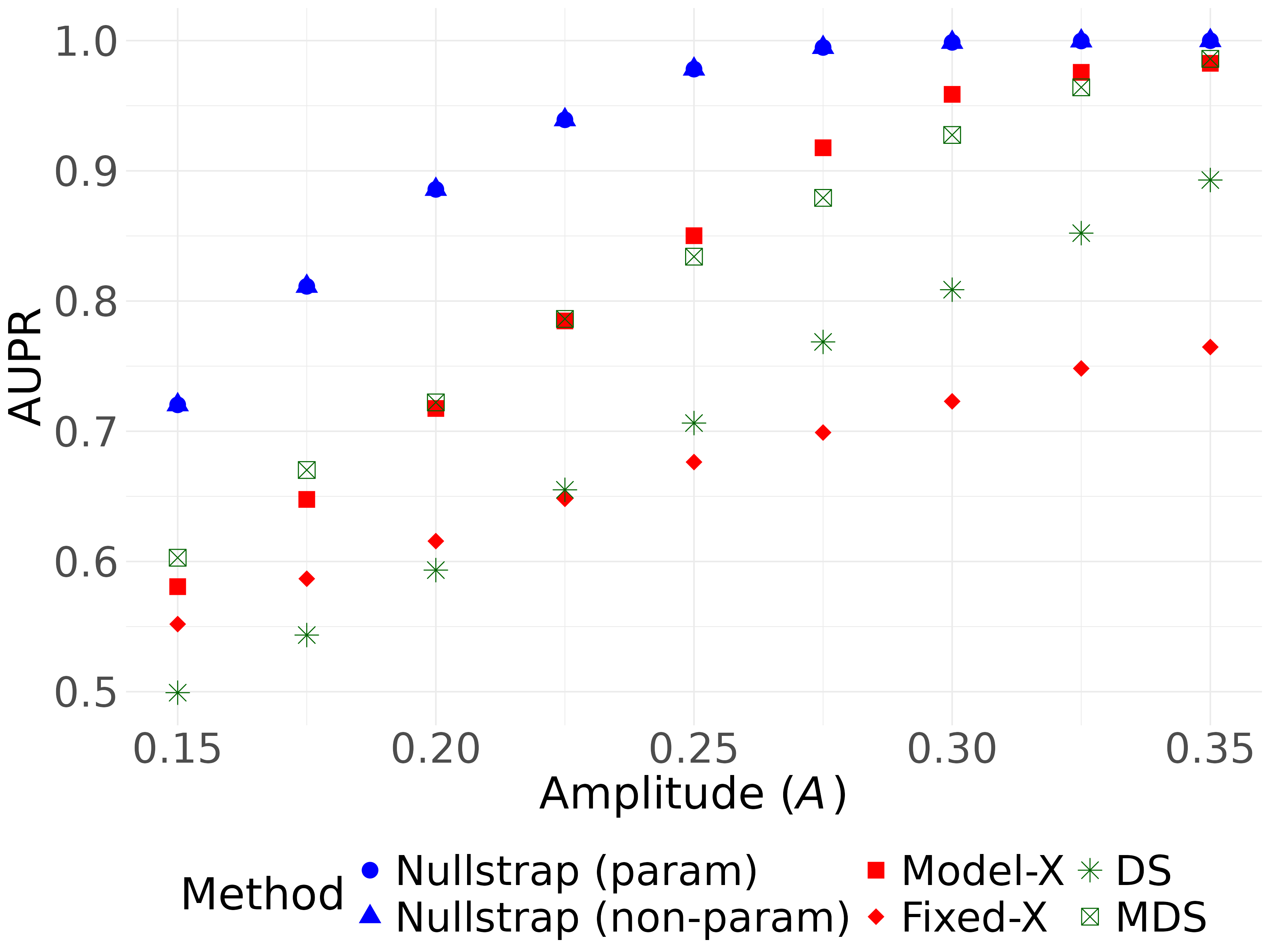}
\caption{Empirical AUPR vs. signal amplitude ($A$) under Simulation Setting~\ref{sim:setting_gamma}.}
\centering
\label{fig:gamma_aupr}
\end{figure}

We next assume the errors follow a centered, non-symmetric Gamma distribution.
\begin{simsetting}\label{sim:setting_gamma}

We set \( n = 2000 \), \( p = 1000 \), and \(\rho = 0.8\).  The design matrix \( \bX \) is generated as described in Simulation Setting~1 from the main text. We consider one simulation parameter for adjustment:

\begin{itemize}
    \item the signal amplitude \( A \in [0.15, 0.35] \).
\end{itemize}

%(1) the signal amplitude \( A \in [0.15, 0.35] \).
We set the distribution of noise as:
\[
\varepsilon_i \sim \text{Gamma}(1,1) - 1
\]
The first \( 30 \) elements of the coefficient vector \( \bbeta \) are randomly assigned values with amplitude \( A \) and random signs, while the remaining \( p - 30 \) elements are set to zero. 
\end{simsetting}
Figures~\ref{fig:gamma_power}--\ref{fig:gamma_aupr} report FDR, power, and AUPR.  
Even with a non-symmetric error distribution, both Nullstrap (param) and Nullstrap (non-param) maintain FDR control and achieve higher power and AUPR, demonstrating robustness to error distribution misspecification.

\clearpage
\section{Nullstrap for generalized linear models}\label{sec:GLM}

In this section, we outline the specific steps for applying Nullstrap to perform variable selection in a high-dimensional generalized linear model (GLM).  Let \(\mathbf{X}=(\mathbf{x}_1,\dots,\mathbf{x}_n)^{\top}\), where each row \(\mathbf{x}_i\in\mathbb{R}^p\).  
Denote by \(f(\,\cdot\mid\mathbf{x};\boldsymbol{\beta},\phi)\) the GLM density, with coefficient vector \(\boldsymbol{\beta}\in\mathbb{R}^p\) and dispersion (nuisance) parameter \(\phi\).

\begin{definition}[Synthetic null data for a GLM]
For a generalized linear model (GLM), Nullstrap defines the synthetic null response \( \tilde{\mathbf{y}} = (\tilde{y}_1, \dots, \tilde{y}_n)\tran \in \mathbb{R}^n \) by
\[
\tilde{y}_i \sim f(\cdot \mid \bx_i; \bbeta_0, \hat{\phi}), \quad i = 1, \dots, n,
\]
where \( \bbeta_0 = (0, \dots, 0)\tran \in \mathbb{R}^p \) is the coefficient vector under the global null hypothesis, and \( \hat{\phi} \) is an estimate of the nuisance parameter \( \phi \) from the original data \( \{\mathbf{y}, \mathbf{X}\} \).
\end{definition}

The LASSO estimator for logistic regression on the original data \( \{\mathbf{y}, \mathbf{X}\} \) is defined as the minimizer of the \( \ell_1 \)-penalized negative log-likelihood:
\[
\hat{\boldsymbol{\beta}} = \operatorname*{argmin}_{\boldsymbol{\beta}} \left\{ \frac{1}{n} \sum_{i=1}^{n} -\log\left(f(y_i \mid \mathbf{x}_i, \boldsymbol{\beta}, \hat{\phi})\right) + \lambda_n \|\boldsymbol{\beta}\|_1 \right\},
\]
where \( \lambda_n \) is a regularization parameter selected via 10-fold cross-validation.

In parallel, we apply the LASSO to the synthetic null data \( \{\tilde{\mathbf{y}}, \mathbf{X}\} \) using the same objective and regularization parameter:
\[
\tilde{\boldsymbol{\beta}} = \operatorname*{argmin}_{\boldsymbol{\beta}} \left\{ \frac{1}{n} \sum_{i=1}^{n} -\log\left(f(\tilde{y}_i \mid \mathbf{x}_i, \boldsymbol{\beta}, \hat{\phi})\right) + \lambda_n \|\boldsymbol{\beta}\|_1 \right\}.
\]

\begin{lemma}\label{lem:glm}
Under the conditions specified in Theorem 2.1 of \cite{Geer2008HIGHDIMENSIONALGL}, Assumption~\ref{assum:gamma_estimator} holds for the LASSO estimator with 
\[
\gamma_{n,p} = \kappa \left( \lambda_n + s\sqrt{\frac{\log p}{n}} \right),
\]
where \( \kappa \) is a constant and \( s = \max(\#\mathcal{S}(F), 1) \), with \( \#\mathcal{S}(F) \) denoting the number of nonzero coefficients in \( \boldsymbol{\beta} \).
\end{lemma}

Lemma~\ref{lem:glm}, based on the result in \cite{Geer2008HIGHDIMENSIONALGL}, establishes the existence of a correction factor \( \gamma_{n,p} \). In practice, we select \( \gamma_{n,p} \) in a data-driven manner using Algorithm~\ref{alg:gamma}.

% We observe that an additional term \( s \) appears in \( \gamma_{n,p} \). 
% This arises because, to our knowledge, current theoretical results only provide a bound for \( \|\hat{\bbeta} - \bbeta\|_1 \), which leads to a larger bound. However, in our simulations, 
%We use Algorithm~\ref{alg:gamma} to estimate $ \kappa$. 
% For a more rigorous theoretical guarantee, it would be necessary to derive results for the bound \( \|\hat{\bbeta} - \bbeta^*\|_{\infty} \), but this remains outside the scope of this paper.
% In this application, the hyperparameter \( \kappa \) in Lemma \ref{lem:glm} is set to \( 0.3\hat{\sigma} \).
\subsection{Simulation results}
As an example of a GLM, consider logistic regression, where the response variable \( Y \) is binary, i.e., \( Y \in \{0, 1\} \). In logistic regression, the conditional distribution of \( Y \) given the predictor variables \( \bx \) follows a Bernoulli distribution:
\[
Y \mid \bx \sim \text{Bernoulli}(p),
\]
where \( p = \mathbb{P}(Y = 1 \mid \bx) \) and the mean of \( Y \) is \( \mu = \mathbb{E}[Y] = p \). The model uses the canonical logit link function:
\[
g(\mu) = \log\left(\frac{\mu}{1 - \mu}\right) = \mu_0 + \bx\tran \boldsymbol{\beta},
\]
where \( \mu_0 \) is the intercept and \( \boldsymbol{\beta} \) is the vector of regression coefficients.

Prior to applying the LASSO, we standardize the columns of \( \mathbf{X} \) so that each variable has unit standard deviation. The regularization parameter \( \lambda_n \) is selected via 10-fold cross-validation.

\begin{simsetting}\label{sim:setting4}
We consider a logistic regression model with a sample size of \( n = 3000 \). The design matrix \( \bX \) is generated as described in Simulation Setting~1 from the main text. Subsequently, \( \bX \) is centered and scaled by dividing each element by \( \sqrt{n} \). The coefficient vector $\bbeta$ is defined in the same manner as in Simulation Setting~\textcolor{red}{1}. We consider three simulation parameters for adjustment: 

\begin{itemize}
    \item (a) the autocorrelation parameter \( \rho \in [0, 0.9] \), 
    \item (b) the signal amplitude \( A \in [6, 12] \), 
    \item (c) the target FDR level \( q \in [0.05, 0.4] \).
\end{itemize}

% (a) the autocorrelation parameter \( \rho \in [0, 0.9] \),  
% (b) the signal amplitude \( A \in [6, 12] \),  
% (c) the target FDR level \( q \in [0.05, 0.4] \). 
 
For each scenario where one parameter varies, the remaining parameters are held constant as:  
\vspace{-10pt}
\begin{equation}\label{eq:para_glm}
\rho = 0.6, \, A = 9, \, q=0.1, \, \text{and } \, p = 500.
\vspace{-10pt}
\end{equation}
The first \( 30 \) elements of the coefficient vector \( \bbeta \) are randomly assigned values with amplitude \( A \) and random signs, while the remaining \( p - 30 \) elements are set to zero. The response vector \(\mathbf{y}\) is generated from a logistic regression model.
\end{simsetting}
% In this application, we set the sample size $n = 3000$ and keep the number of non-zero regression coefficients $s=30$. The design matrix $\bX$ is constructed similarly as in the first case, where it has i.i.d. rows and auto-regressive AR(1) columns. The elements of $\bX$ are still drawn from a multivariate normal distribution $\mathcal{N}(0, \bSigma)$, where $\bSigma$ is constructed using the Toeplitz structure based on a correlation value $\rho \in (0,1)$. To match the scaling adjustments in this scenario, we scale $X$ such that each element has zero mean and a variance of ${n}^{-1}$. The coefficient vector $\bbeta$ is defined in the same manner as in the previous case. 
% The regularization parameter \( \lambda_n \) is selected using cross-validation and the target FDR level is set to $q = 0.1$.

\begin{figure}[H]
\centering
\includegraphics[scale=0.5]{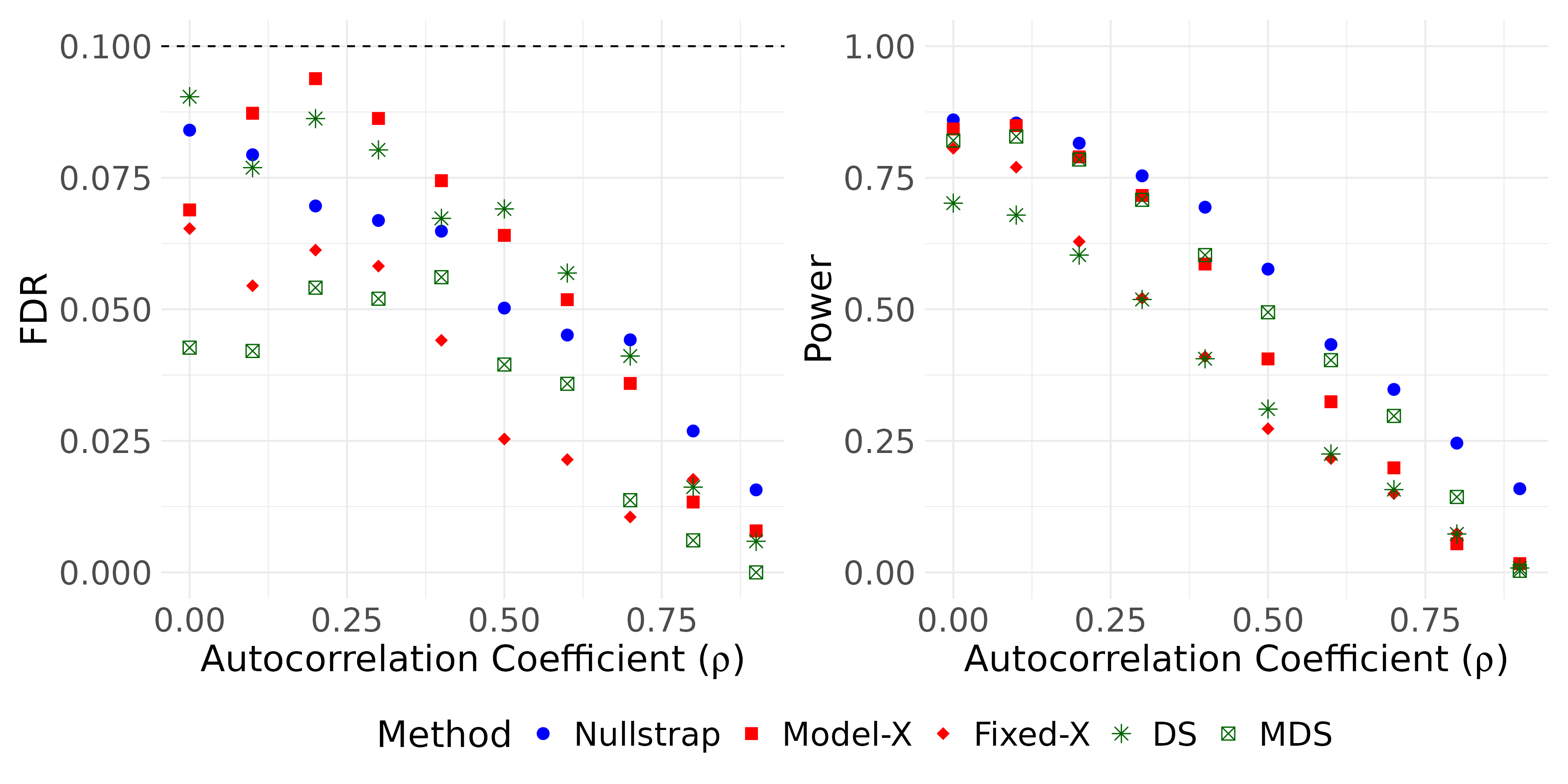}
\caption{Empirical FDR and power vs. autocorrelation ($\rho$) under Simulation Setting~\ref{sim:setting4}.}
\centering
\label{fig:GLM_rho}
\end{figure}

\begin{figure}[H]
\centering
\includegraphics[scale=0.5]{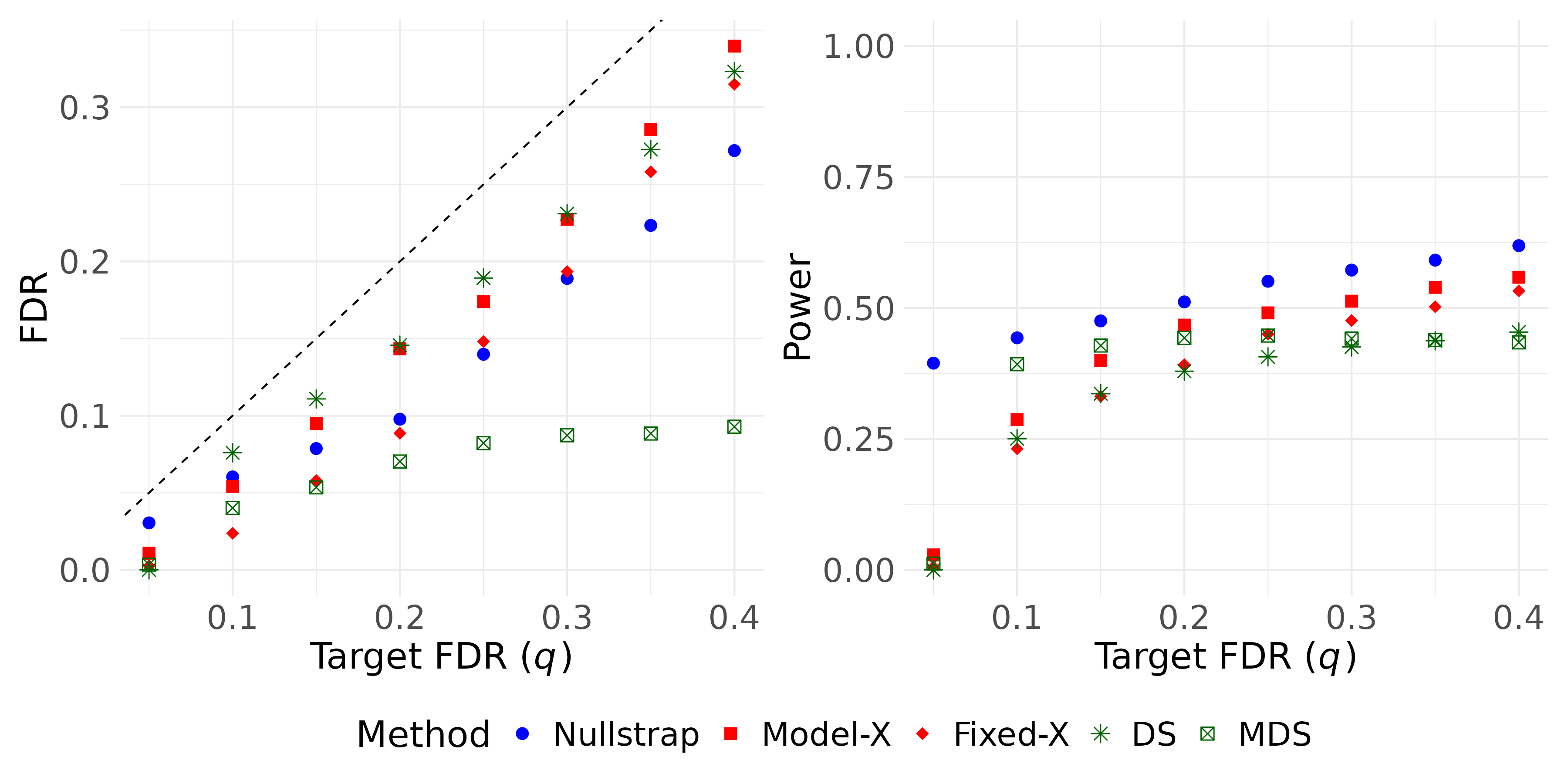}
\caption{Empirical FDR and power vs. target FDR level ($q$) under Simulation Setting~\ref{sim:setting4}.}
\centering
\label{fig:GLM_fdr}
\end{figure}

\begin{figure}
        \centering
        \begin{subfigure}{0.45\textwidth}
            \centering
            \includegraphics[width=\linewidth]{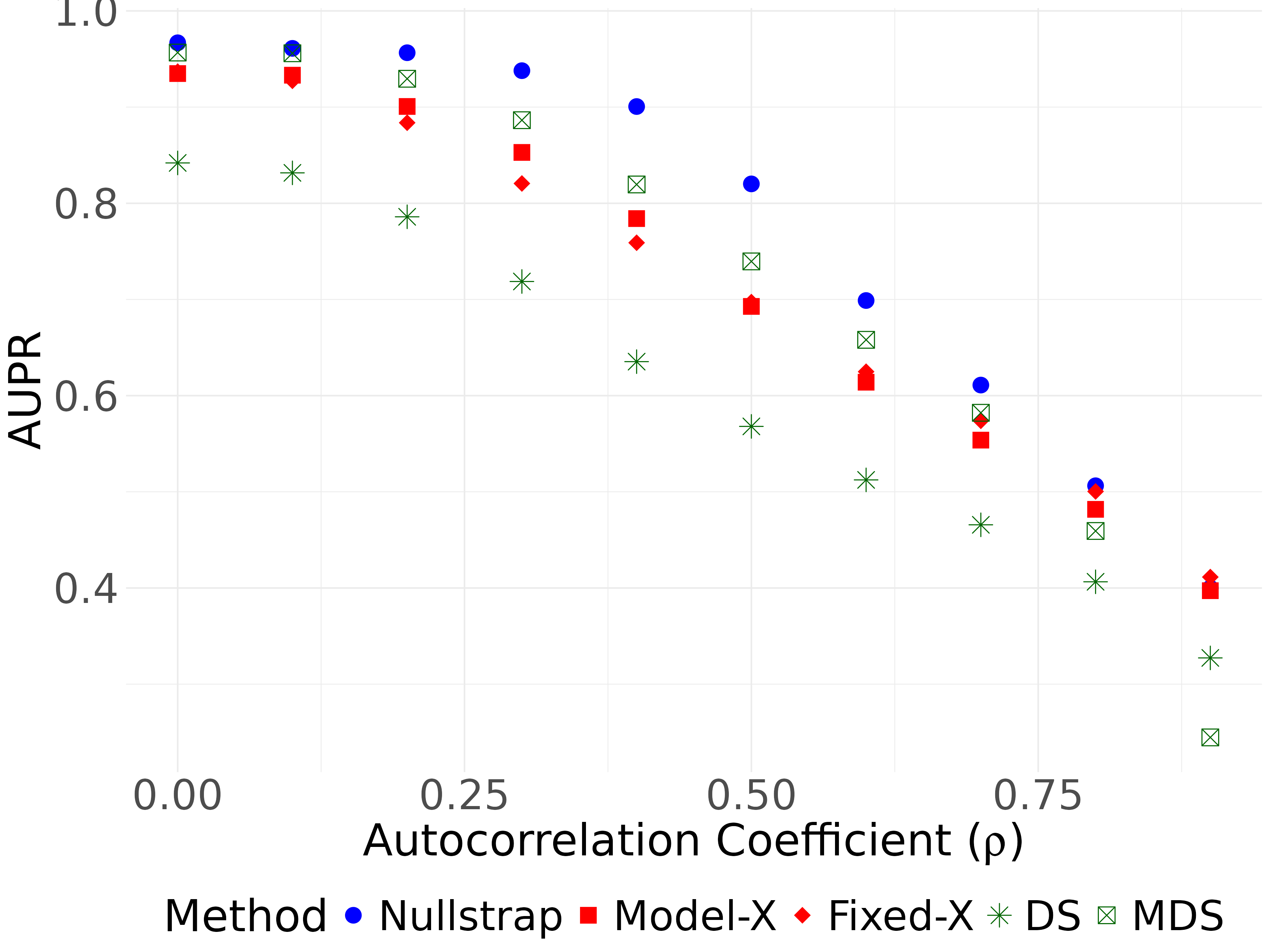} 
            \caption{Empirical AUPR vs. autocorrelation ($\rho$) under Simulation Setting~\ref{sim:setting4}.}
            \label{fig:GLM_rho(aupr)}
        \end{subfigure}
        \hfill
        \begin{subfigure}{0.45\textwidth}
            \centering
            \includegraphics[width=\linewidth]{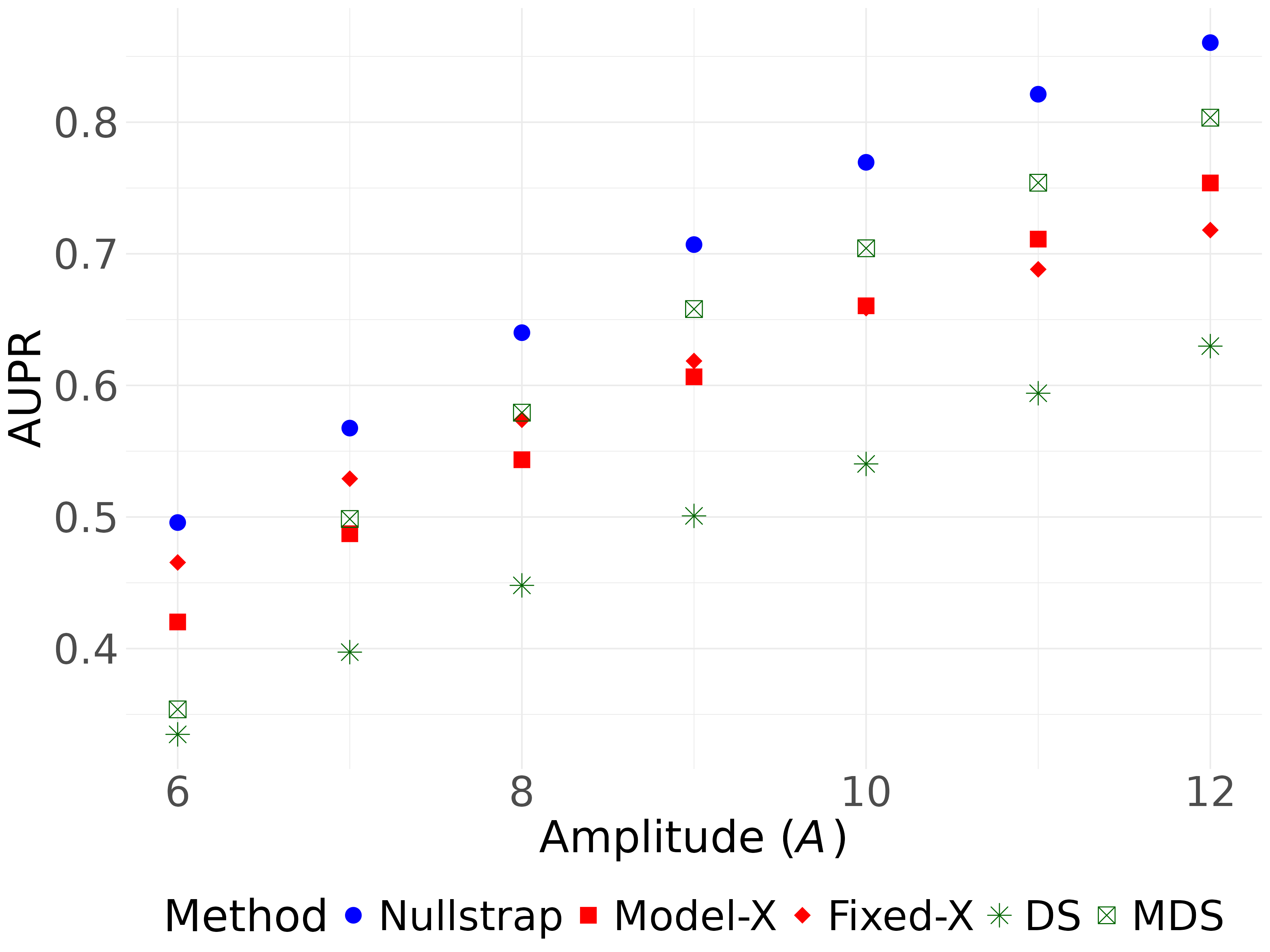} 
            \caption{Empirical AUPR vs. signal amplitude ($A$) under Simulation Setting~\ref{sim:setting4}.}
            \label{fig:GLM_amp(aupr)}
        \end{subfigure}
        \caption{Empirical AUPR for the logistic regression model.}
        \label{fig:AUPR_GLM}
    \end{figure}

% In this study, we focus on varying two parameters across different scenarios: 
% (a)the correlation value $\rho \in [0,0.9]$, and
% (b) the signal amplitude $A \in [6, 12]$.
% For each scenario where one parameter varies, the remaining parameters are held constant at 
% \vspace{-20pt}
% \begin{equation}\label{eq:para_glm}
% \rho = 0.6, A = 9, q=0.1, \text{ and } p = 500.
% \vspace{-20pt}
% \end{equation}
We replicate each setting 100 times. In this application, we continue to compare the same five methods: Fixed-X, Model-X, DS, MDS, and our proposed method, Nullstrap. The empirical FDR and power results are shown in Figures~\ref{fig:GLM_rho}--\ref{fig:GLM_fdr}, and the AUPR results are presented in Figure~\ref{fig:AUPR_GLM}.
% \textcolor{red}{(change AUPR plots layout and caption)}. \textcolor{red}{(delete) In certain specified settings (e.g., \( \rho = 0.4 \)), }
% Nullstrap exhibits an exceptionally minor violation of FDR control, with the observed FDR only marginally exceeding the target value. This deviation is reasonable, considering the asymptotic nature of the theoretical guarantee.
In these results, Nullstrap achieves the highest power and AUPR values across all simulation parameters. 
% From Figure \ref{fig:GLM_rho}, when varying the correlation, the decrease in power slows as the correlation increases, with Nullstrap exhibiting a more gradual decline compared to other methods.
% Figure \ref{fig:GLM_amp} demonstrates that Nullstrap maintains a consistent power of 1 for signal amplitudes above 9, while the power of other methods decreases as the amplitude increases.

\begin{figure}[H]
\centering
\includegraphics[scale=0.5]{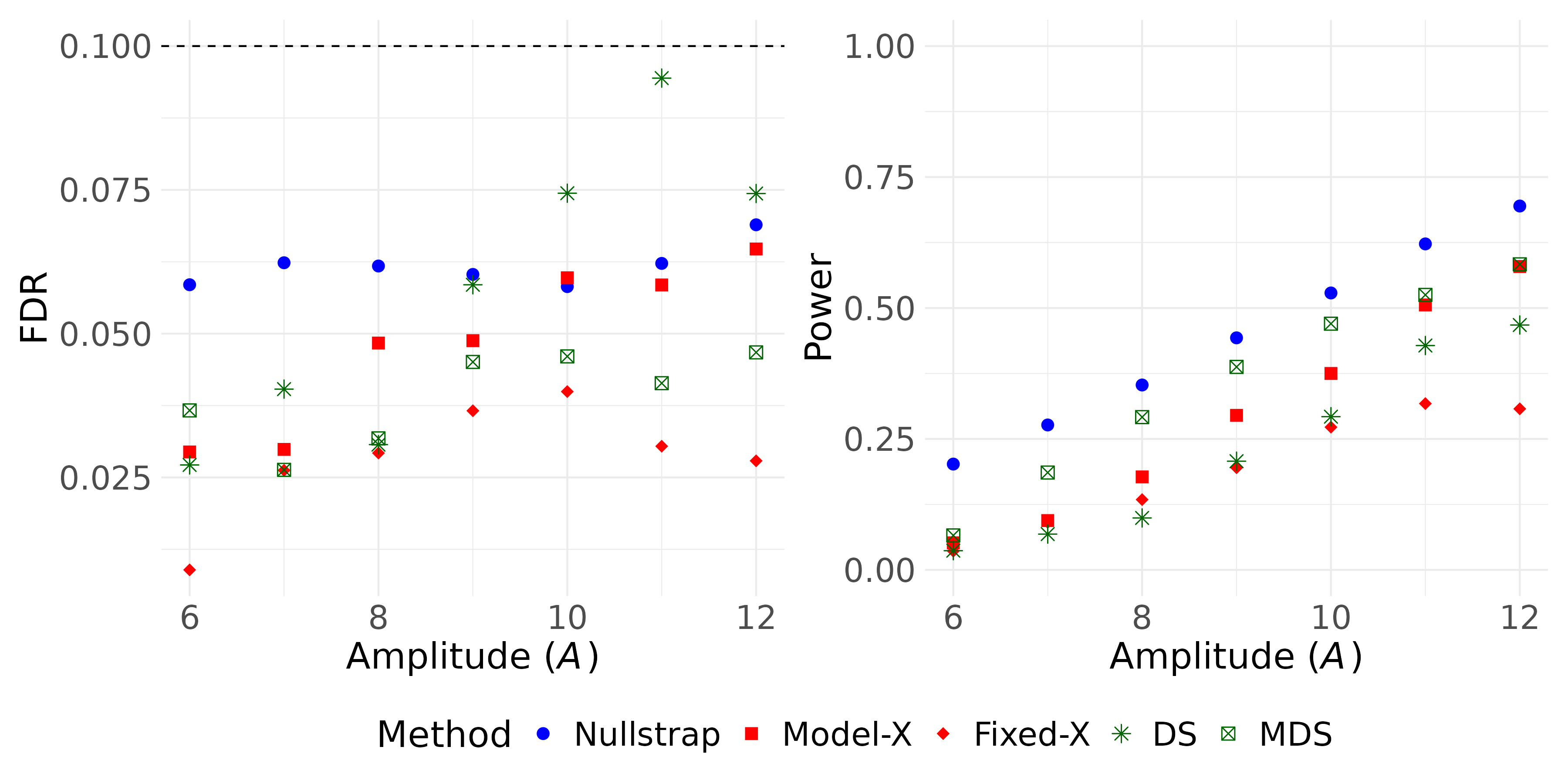}
\caption{Empirical FDR and power vs. signal amplitude ($A$) under Simulation Setting~\ref{sim:setting4}.}
\centering
\label{fig:GLM_amp}
\end{figure}

% \begin{figure}[h]
% 	\centering
% 	\includegraphics[scale=0.4]{simulation/result_rho_aupr_glm.png}
% 	\caption{Empirical AUPR for the generalized linear regression model (Correlation).}
% 	\centering
% 	\label{fig:GLM_rho(aupr)}
% 	\end{figure}
	
% 	\begin{figure}[h]
% 	\centering
% 	\includegraphics[scale=0.4]{simulation/result_amp_aupr_glm.png}
% 	\caption{Empirical AUPR for the generalized linear regression model (Amplitude).}
% 	\centering
% 	\label{fig:GLM_amp(aupr)}
% 	\end{figure}

\begin{figure}[h]
\centering
\includegraphics[scale=0.5]{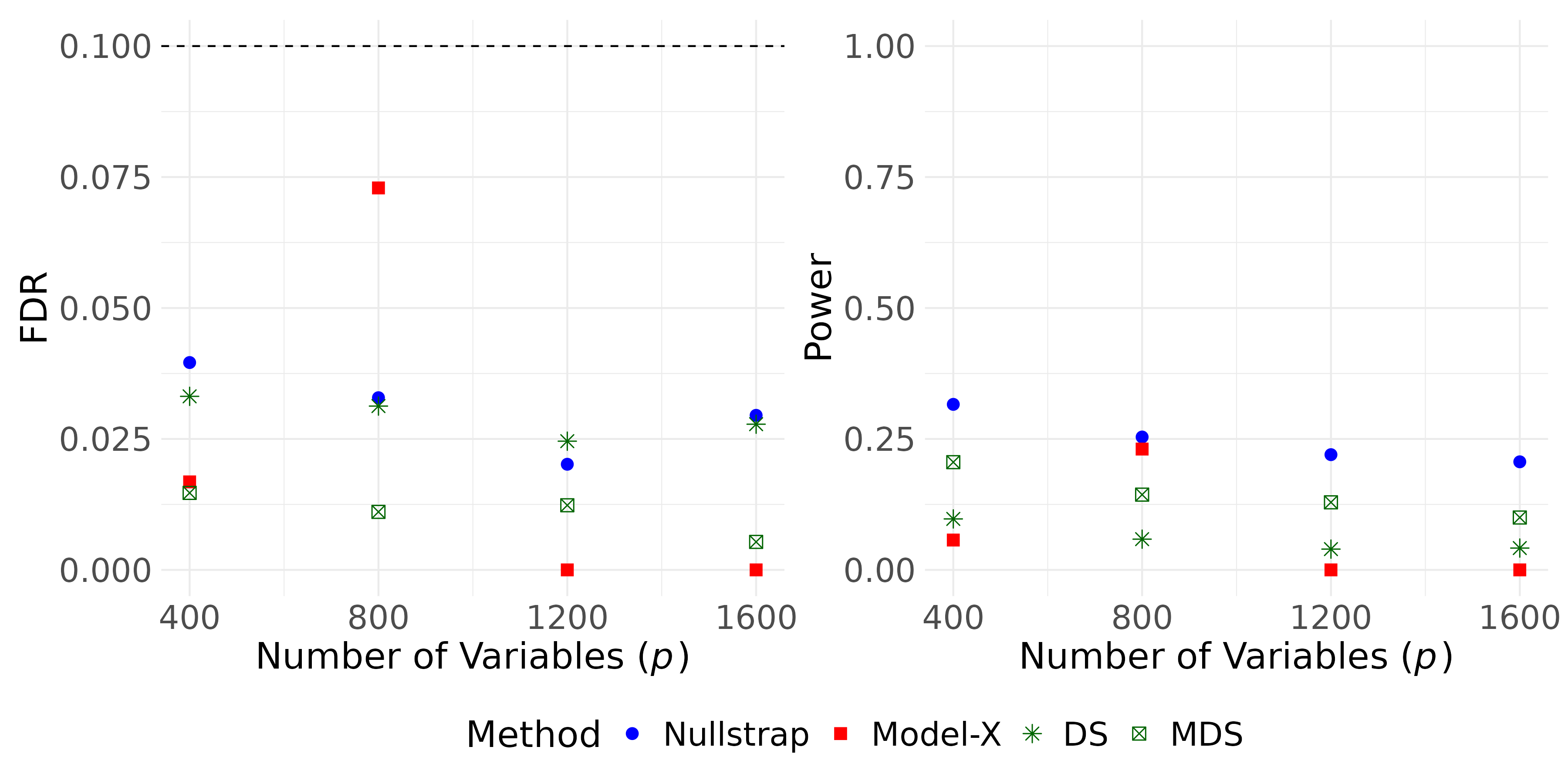}
\caption{Empirical FDR and power vs. number of variables ($p$) under Simulation Setting~\ref{sim:setting5}.}
\centering
\label{fig:GLM_num}
\end{figure}

\begin{figure}[H]
	\centering
	\includegraphics[scale=0.4]{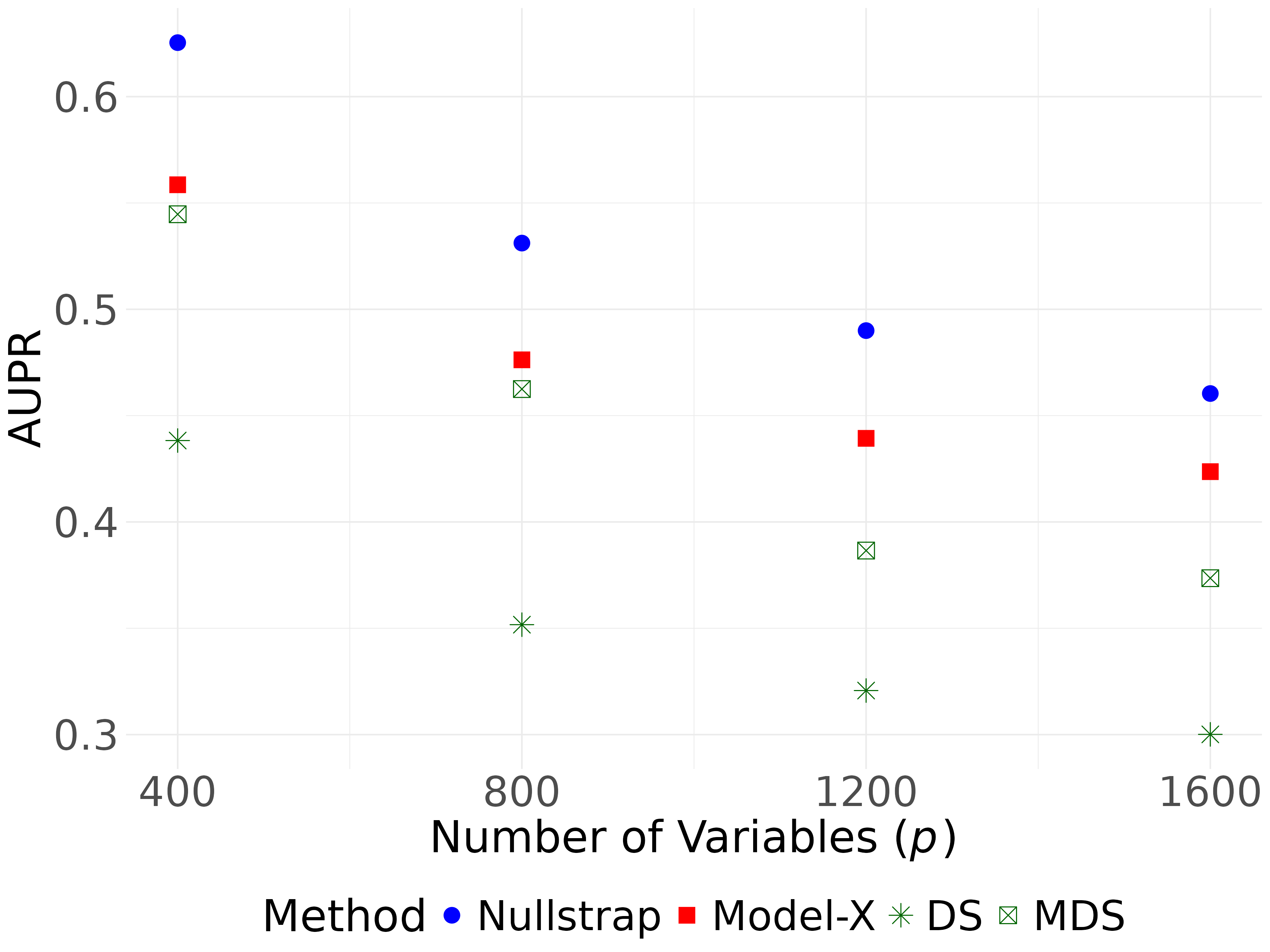}
	\caption{Empirical AUPR vs. number of variables ($p$) under Simulation Setting~\ref{sim:setting5}.}
	\centering
	\label{fig:GLM_num(aupr)}
	\end{figure}

Next, we compare the performance of different methods as the number of variables varies under Simulation Setting~\ref{sim:setting5}. The results are summarized in Figure~\ref{fig:GLM_num} and Figure~\ref{fig:GLM_num(aupr)}. Across all variable counts, Nullstrap consistently outperforms the other methods, achieving the highest power and AUPR.

\begin{simsetting}\label{sim:setting5}
	We set the sample size to \( n = 800 \), with the number of variables \( p \) varying from \( 400 \) to \( 1600 \) in increments of \( 400 \). The remaining parameters are fixed as \( \rho = 0.6 \), \( A = 9 \), and \( q = 0.1 \). The design matrix \( \bX \), the response vector $\by$ and the coefficient vector \( \bbeta \) are generated following the procedure described in Simulation Setting~\ref{sim:setting4}.
\end{simsetting}

\begin{table}[htbp]
\centering
\caption{Comparison of runtimes (in seconds) for the logistic regression model under Simulation Setting~\ref{sim:setting4}, using the default parameter configuration in~\eqref{eq:para_glm}.}
\begin{tabular}{ccccc}
    \toprule
    Nullstrap & Model-X & Fixed-X  & DS & MDS  \\
    \midrule
    6.37 & 23.96 &  14.3  & 1.97 &  81.22\\
    \bottomrule
\end{tabular}
\label{glm_time}
\end{table}

% \begin{table}[htbp][h!]
% \centering
% \begin{tabular}{cccc}
%     \toprule
%     Nullstrap &  Model-X  & DS & MDS  \\
%     \midrule
%     700.54 &   45642.84 &  446.34 &  4683.98\\
%     \bottomrule
% \end{tabular}
% \caption{Comparison of runtimes (s) in Simulation Setting~\ref{sim:setting5}.}
% \label{glm_time}
% \end{table}

Table~\ref{glm_time} summarizes the runtimes of the five methods under Simulation Setting~\ref{sim:setting4}, using the default parameter configuration in~\eqref{eq:para_glm}. As shown, Nullstrap achieves a fast runtime of \( 6.37 \, \text{s} \), outperforming Model-X knockoff (23.96 s), Fixed-X knockoff (14.3 s), and MDS (81.22 s), while also delivering superior statistical performance.

% The fatest method in this comparison is still DS, with a runtime of \( 1.97 \, \text{s} \), and GM remains the slowest, with a runtime of \( 292.25 \, \text{s} \), approximately 70 times slower than our method. This reflects GM is significantly more time-consuming than the other methods. 
% Nullstrap is also nearly 20 times faster than the second-best performing method, MDS, requiring \( 81.22 \, \text{s} \). This balance between runtime and statistical performance makes Nullstrap a highly competitive option for practical use in generalized linear regression models.

\begin{table}[htbp]
\centering
\caption{Comparison of Jaccard index under the default parameter setting~\eqref{eq:para_glm} in Simulation Setting~\ref{sim:setting4}.}
\begin{tabular}{cccc}
    \toprule
    Nullstrap & Model-X &  DS & MDS  \\
    \midrule
      0.732 & 0.000 & 0.085 & 0.699\\
    \bottomrule
\end{tabular}
\label{glm_jac}
\end{table}

Table~\ref{glm_jac} reports the Jaccard index, averaged over 100 replications under Simulation Setting~\ref{sim:setting4}, using the default parameter configuration in~\eqref{eq:para_glm}, as a measure of each method's stability across random seeds. Nullstrap achieves the highest stability with a Jaccard index of \( 0.732 \), followed by MDS at \( 0.699 \), while DS and Model-X exhibit much lower stability, with values of \( 0.085 \) and \( 0.000 \), respectively.

\subsection{Interactions between signal variables}
For the logistic regression model, we also consider a simulation setting in which interactions between signal variables are incorporated into the design matrix, resulting in explicit correlations among its columns.

\begin{simsetting}\label{sim:setting_glm_inter}
	We set \( n = 1000 \), \( p_{\rm{base}} = 20 \), and \( p = p_{\rm{base}} + \frac{p_{\rm{base}}(p_{\rm{base}}-1)}{2} \). The base design matrix \(\bX_{\text{base}}\) is drawn from \(\mathcal{N}(\mathbf{0},\boldsymbol{\Sigma}_{\text{base}})\), where \(\boldsymbol{\Sigma}_{\text{base}}\) is a Toeplitz correlation matrix with autocorrelation parameter \(\rho = 0.6\).
	We then construct interaction terms by computing pairwise products of the first \( p_{\rm{base}} \) variables, forming an interaction matrix \( \bX_{\rm{interact}} \). The first \( 5 \) elements of the coefficient vector \( \bbeta \) are randomly assigned values with amplitude \( A \) and random signs. Additionally, if both variables involved in an interaction term are among the first \( 5 \) variables, their corresponding coefficient is also randomly assigned values with amplitude \( A \) and random signs.  
	Finally, the full design matrix \( \bX \) is formed by concatenating \( \bX_{\rm{base}} \) and \( \bX_{\rm{interact}} \). We consider one simulation parameter for adjustment:  
	\begin{itemize} 
		\item the signal amplitude \( A \in [9, 15] \). 
	\end{itemize}  
     The response vector $\by$ is generated following the procedure described in Simulation Setting~\ref{sim:setting4}.
	\end{simsetting}

% 	We still set the correction factor \( \gamma_{n,p} \) to
% $
% 0.2 \Hat{\sigma} \left(\lambda_n + \sqrt{\frac{\log p}{n}} \right)
% $
% in this simulation setting.

\begin{figure}[H]
\centering
\includegraphics[scale=0.5]{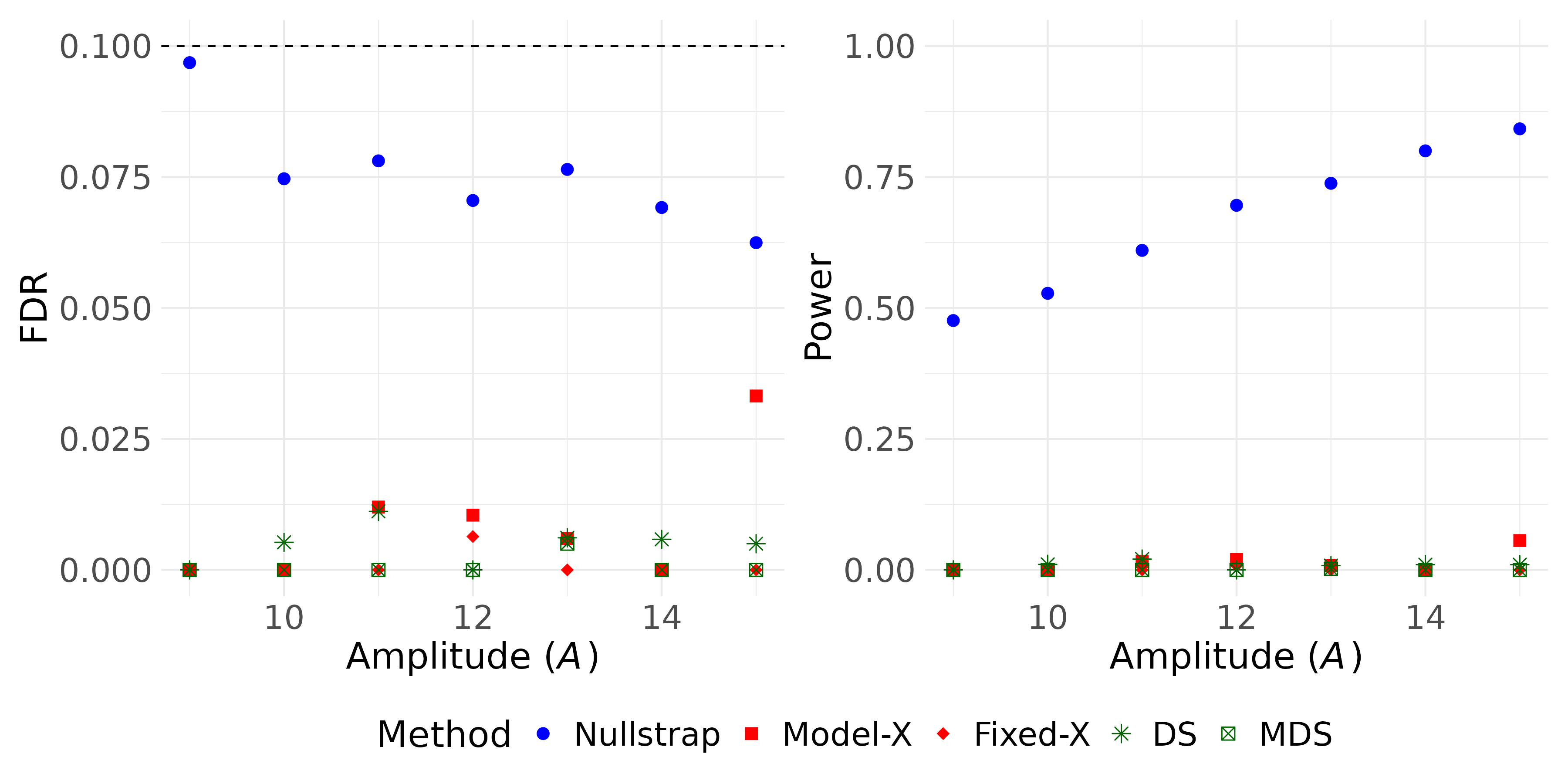}
\caption{Empirical FDR and power vs. signal amplitude ($A$) under Simulation Setting~\ref{sim:setting_glm_inter}.}
\centering
\label{fig:GLM_amp_inter}
\end{figure}

\begin{figure}[H]
\centering
\includegraphics[scale=0.5]{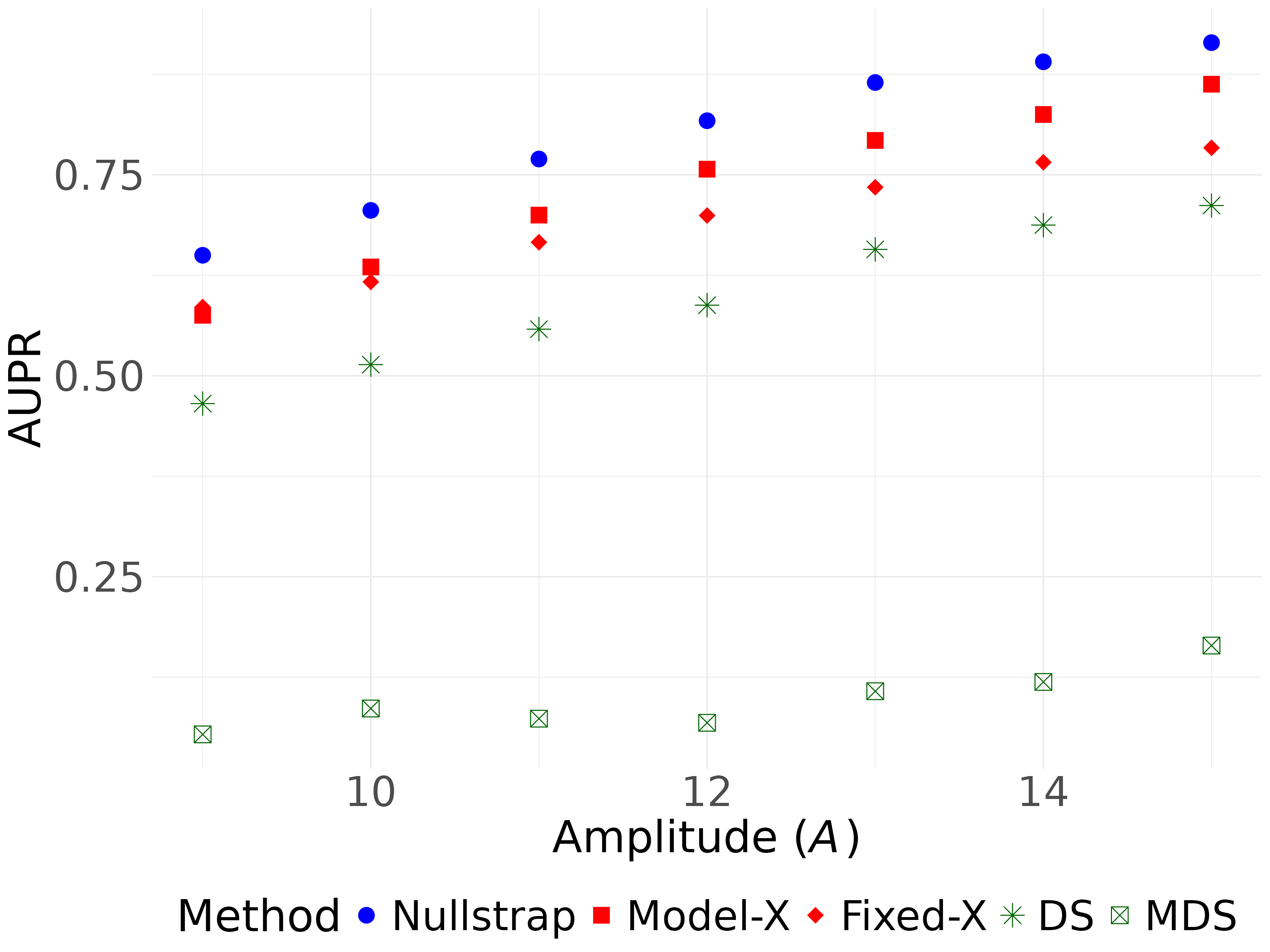}
\caption{Empirical AUPR vs. signal amplitude ($A$) under Simulation Setting~\ref{sim:setting_glm_inter}.}
\centering
\label{fig:GLM_aupr_inter}
\end{figure}

For each scenario under Simulation Setting~\ref{sim:setting_glm_inter}, we compare the FDR, power, and AUPR of the five methods using 100 replications. The empirical FDR and power results are shown in Figure~\ref{fig:GLM_amp_inter}, and the AUPR results are presented in Figure~\ref{fig:GLM_aupr_inter}.
Overall, most methods maintain FDR control across all scenarios. Notably, Nullstrap consistently demonstrates reliable FDR control and, more importantly, achieves higher power and AUPR than the other methods in every case.

% Compared to Nullstrap, Nullstrap-Diff exhibits lower power than DS and MDS, particularly in more challenging scenarios such as those with high autocorrelations and high signal amplitudes. 

\clearpage
\section{Nullstrap for Cox proportional hazards models}\label{sec:Cox}
%\subsection{Problem statement}
Let \( \mathbf{y}=(y_1, \ldots, y_n)\tran \) represent the vector of survival times, and let \( \mathbf{X} = (\mathbf{x}_1, \dots, \mathbf{x}_n)\tran \) denote the \( n \times p \) design matrix. For simplicity, we assume that there is no censoring. However, when censoring is present, Nullstrap can still be constructed if the censoring distribution can be reliably estimated.

In this subsection, we consider the Cox proportional hazards model:
\[
h(t \mid \mathbf{x}) = h_0(t) \exp(\boldsymbol{\beta}\tran \mathbf{x}),
\]
where \( h(t \mid \mathbf{x}) \) is the hazard function at time \( t \) given the $p$ variables in \( \mathbf{x} \), \( h_0(t) \) is the baseline hazard function, and \( \boldsymbol{\beta} = (\beta_1, \dots, \beta_p)\tran \) is a vector of unknown coefficients that quantify the importance of variables in the model. We assume there are no ties in the observed survival times \( y_i \); if ties are present, the method of \citet{breslow1974covariance} can be applied.

The partial log-likelihood for the observed data $\{\by, \bX\}$ is given by:
\[
\ell(\boldsymbol{\beta}; \mathbf{y}, \mathbf{X}) = \frac{1}{n} \sum_{i=1}^n \left\{ \boldsymbol{\beta}^\top \mathbf{x}_i - \log \left[ \sum_{j=1}^n I(y_j \geq y_i) \exp(\boldsymbol{\beta}^\top \mathbf{x}_j) \right] \right\}.
\]
%Let \( \boldsymbol{\beta} = (\beta_1, \dots, \beta_p)\tran \in \mathbb{R}^p \) denote the true coefficient vector.

% Define \( \cS_0 = \{1 \leq j \leq p : \beta_{0j} \neq 0\} \) as the index set of relevant covariates for the hazard function, and \( \cS_0^c = \{1, \dots, p\} \setminus \cS_0 \) as the index set of noise covariates.

\begin{definition}[Synthetic null data for a Cox proportional hazards model]
Nullstrap defines the synthetic null response \( \tilde{\mathbf{y}} = (\tilde{y}_1, \dots, \tilde{y}_n)\tran \) by sampling each \( \tilde{y}_i \) from a Cox proportional hazards model with hazard function \( \hat{h}_0(t) \exp(\boldsymbol{\beta}_0\tran \mathbf{x}_i) \),  
where \( \boldsymbol{\beta}_0 = (0, \dots, 0)\tran \in \mathbb{R}^p \) is the coefficient vector under the global null hypothesis.  
The baseline hazard function \( \hat{h}_0(t) \) is estimated from the original data \( \{\mathbf{y}, \mathbf{X}\} \).
\end{definition}

We consider the following LASSO-type penalized estimators, obtained by maximizing the partial log-likelihood for the original data and the synthetic null data in parallel:
\vspace{-20pt}
$$
\hat{\boldsymbol{\beta}} = \operatorname*{argmin}_{\boldsymbol{\beta} \in \mathbb{R}^p} \left\{ -\ell(\boldsymbol{\beta}; \by, \bX) + \lambda_n \|\boldsymbol{\beta}\|_1 \right\}, 
\quad \text{and} \quad 
\tilde{\boldsymbol{\beta}} = \operatorname*{argmin}_{\boldsymbol{\beta} \in \mathbb{R}^p} \left\{ -\ell(\boldsymbol{\beta}; \tilde{\by}, \bX) + \lambda_n \|\boldsymbol{\beta}\|_1 \right\},
\vspace{-20pt}
$$
where \( \lambda_n \) is the regularization parameter selected via 10-fold cross-validation on the original data and applied consistently to both estimators.

% We also consider the penalized estimator from the partial log-likelihood with t:
% $$

% $$
\begin{lemma}\label{lem:cox}
	Under the conditions specified in Theorem 3.1 of \cite{Huang2013ORACLEIF}, Assumption \ref{assum:gamma_estimator} holds for the LASSO estimator with $\gamma_{n,p} = \kappa ( \lambda_n + \sqrt{\frac{\log p}{n}} )$ and $\kappa$ is a constant.
\end{lemma}
Lemma~\ref{lem:cox}, based on the result in \cite{Huang2013ORACLEIF}, establishes the existence of a correction factor \( \gamma_{n,p} \). In practice, we select \( \gamma_{n,p} \) in a data-driven manner using Algorithm~\ref{alg:gamma}. The baseline hazard function \( h_0(t) \) is estimated using the \texttt{survival} package in R.

% We define \( |\tilde{\beta}^\prime_j| = |\tilde{\beta}_j| + \gamma_{n,p} \) and a threshold \( \tau_q > 0 \) as described in \eqref{eq:tauq},
% where \( q \) represents the target FDR level. Next, we select the variables as
% $
% \widehat{\cS} = \{ j : |\hat{\beta}_j| \geq \tau_q \}.
% $
% By applying Lemma \ref{lem:cox} and Theorem \ref{thm:main}, we conclude that for a given target FDR control \( q \), we obtain \( \text{FDR} \leq q + o(1) \) and \( \text{Power} = 1 + o(1) \), as \( n \to \infty \).

\subsection{Simulation results}

In this simulation, we compare the performance of our method, Nullstrap, with two knockoff filters: Model-X and Fixed-X. DS and MDS are excluded from the comparison due to the lack of available code implementations for the Cox proportional hazards model.  
Before applying the LASSO, we standardize the columns of \( \mathbf{X} \) so that each has unit standard deviation.

\begin{simsetting}\label{setting:cox}
We set the sample size $n = 400$. The design matrix \( \bX \) is generated as described in Simulation Setting~2, with autocorrelation \( \rho \in [0, 0.9] \). Subsequently, \( \bX \) is centered and scaled by dividing each element by \( \sqrt{n} \). The baseline hazard function \( h_0(t) \) is taken to correspond to the Weibull distribution with shape parameter \( 1 \) and scale parameter \( 1 \). The coefficient vector \( \bbeta \) is defined in the same manner as in Simulation Setting~2. We consider four simulation parameters for adjustment:
\begin{itemize}
    \item (a) the autocorrelation parameter \( \rho \in [0, 0.9] \),
    \item (b) the signal amplitude \( A \in [2, 9] \),
    \item (c) the target FDR level \( q \in [0.05, 0.4] \),
    \item (d) the number of variables \( p \in [200, 800] \).
\end{itemize}
% (a) the autocorrelation parameter \( \rho \in [0, 0.9] \),
% (b) the signal amplitude \( A \in [2, 9] \),
% (c) the target FDR level \( q \in [0.05, 0.4] \), and
% (d) the number of variables \( p \in [200, 800] \).
For each scenario where one parameter varies, the remaining parameters are held constant as follows:  
\vspace{-10pt}
\begin{equation}\label{eq:para_cox}
	\rho = 0.4, A = 5, q=0.1, \text{and } p = 200.\vspace{-10pt}
\end{equation}
The first \( 30 \) elements of the coefficient vector \( \boldsymbol{\beta} \) are randomly assigned values with magnitude \( A \) and random signs, while the remaining \( p - 30 \) elements are set to zero. The survival times \( \mathbf{y} \) are then generated from the Cox proportional hazards model.
\end{simsetting}

% (d) the number of variables \( p \in \{500, 1000, \cdots, 3500\} \).  
% For each scenario where one parameter varies, the remaining parameters are held constant as:  
% \vspace{-10pt}
% \begin{equation}\label{eq:para_glm}
% 	\rho = 0.4, A = 5, q=0.1, \text{ and } p = 200.
% \vspace{-10pt}
% \end{equation}
% The first \( 30 \) elements of the coefficient vector   \( \bbeta^* \) are randomly assigned values with amplitude \( A \) and random signs, while the remaining \( p - 30 \) elements are set to zero.
%   and ensure that the number of non-zero regression coefficients remains consistent with the previous two cases. The design matrix $\bX$ and coefficient vector $\bbeta$  are kept consistent with the settings used in the GLM application. Here, $\by$ is considered as data without right-censoring, so all values in the survival status will be 1, meaning the event was observed for all individuals. We set FDR level at 0.1 and each setting is evaluated using 100 replications. We vary three parameters:
% (a) the correlation value $\rho \in [0, 0.9]$,
% (b) the signal amplitude $A \in [2, 9]$,
% (c) the number of variables $p \in [200,800]$.
% In scenarios where we do not vary a specific parameter, we fix the remaining parameters at
% \vspace{-20pt}
% \begin{equation}\label{eq:para_cox}
% \rho = 0.4, A = 5,\text{ and } p = 200.
% \vspace{-20pt}
% \end{equation}

\begin{figure}[H]
\centering
\includegraphics[scale=0.5]{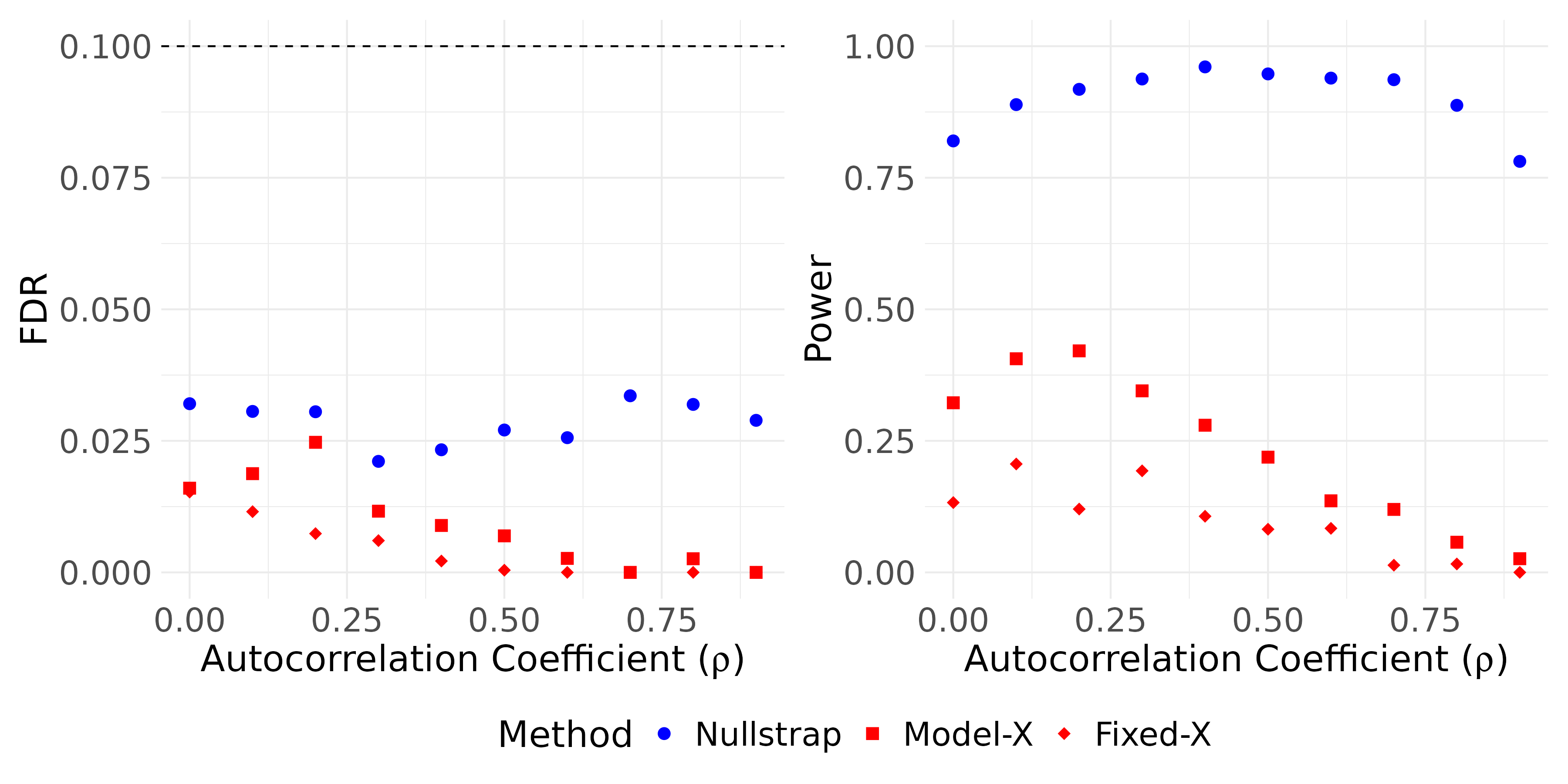}
\caption{Empirical FDR and power vs. autocorrelation ($\rho$) under Simulation Setting~\ref{setting:cox}.}
\centering
\label{fig:Cox_rho}
\end{figure}

\begin{figure}[H]
\centering
\includegraphics[scale=0.5]{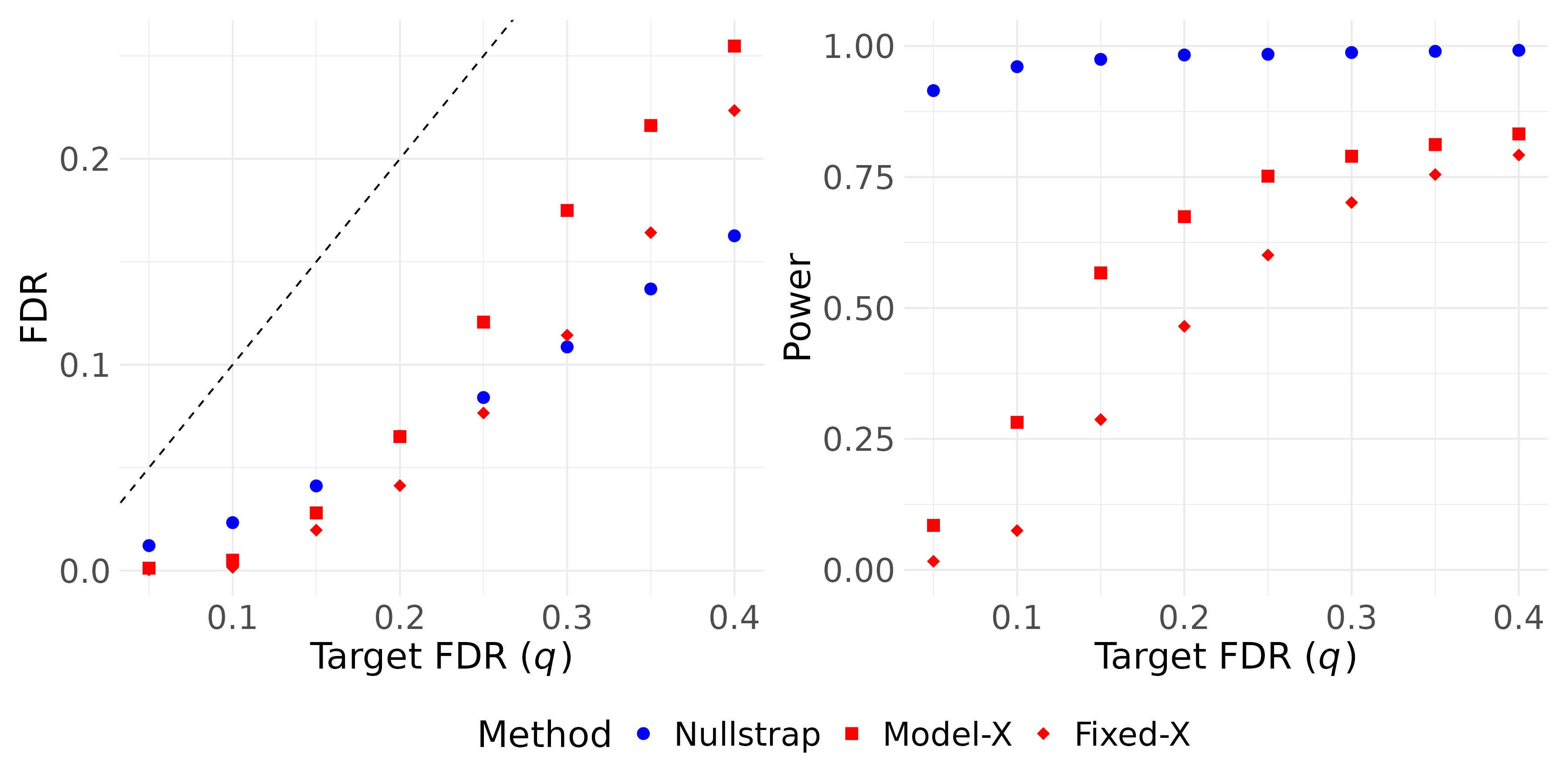}
\caption{Empirical FDR and power vs. target FDR level ($q$) under Simulation Setting~\ref{setting:cox}.}
\centering
\label{fig:Cox_fdr}
\end{figure}

\begin{figure}[H]
\centering
\includegraphics[scale=0.5]{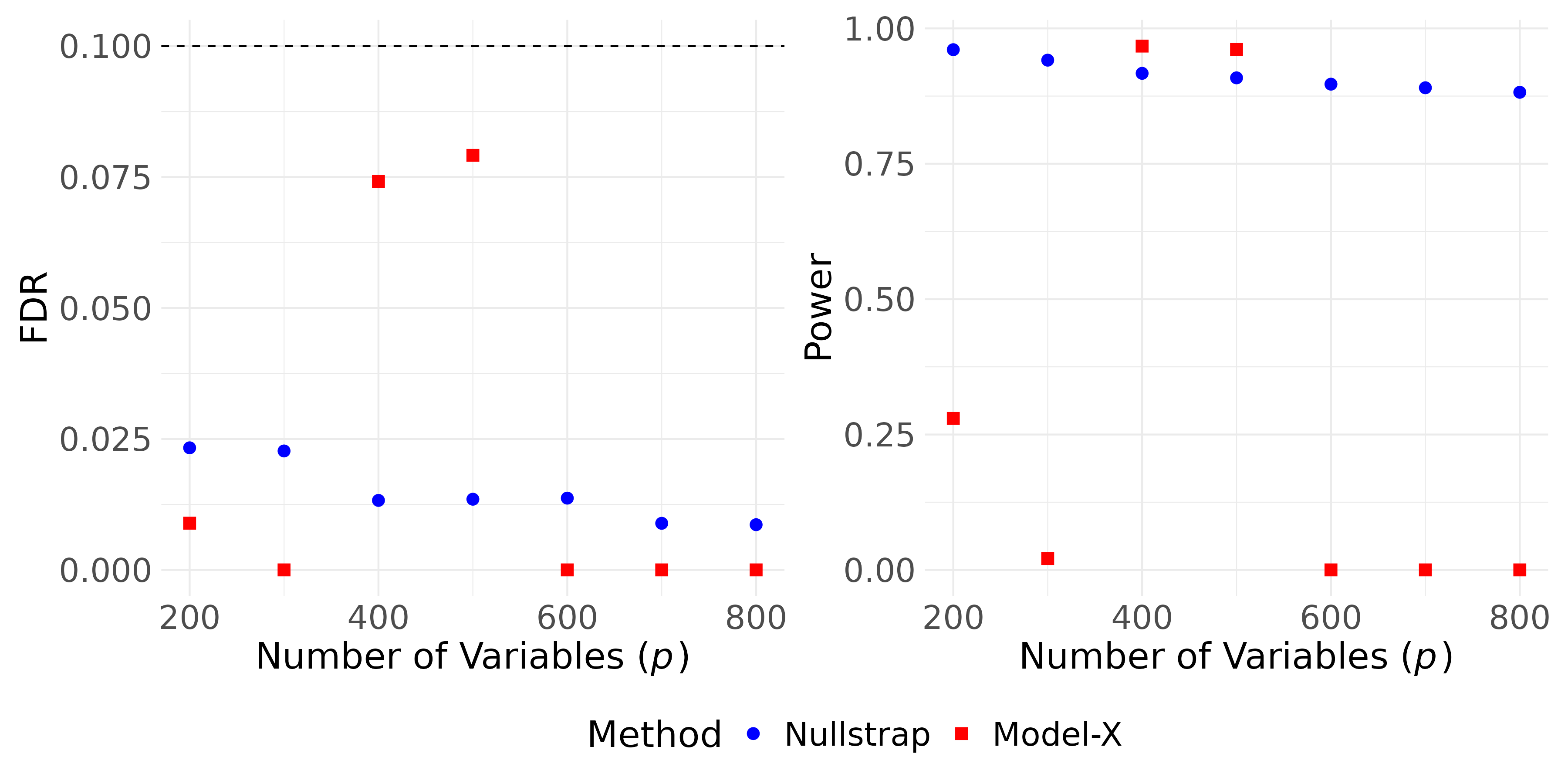}
\caption{Empirical FDR and power vs. number of variables ($p$) under Simulation Setting~\ref{setting:cox}.}
\centering
\label{fig:Cox_num}
\end{figure}

\begin{figure}
        \centering
        \begin{subfigure}{0.45\textwidth}
            \centering
            \includegraphics[width=\linewidth]{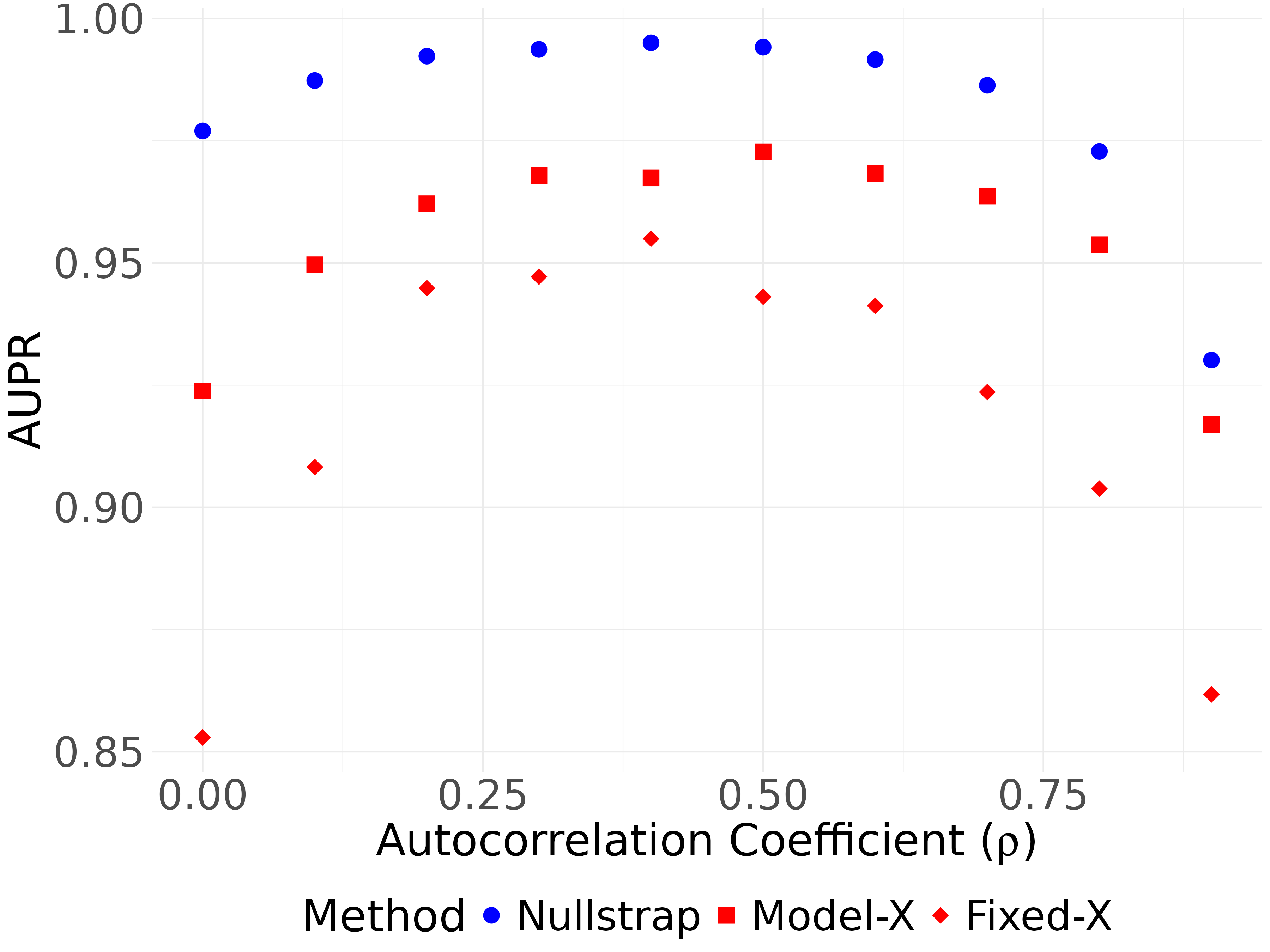} 
            \caption{Empirical AUPR vs. autocorrelation ($\rho$) under Simulation Setting~\ref{setting:cox}.}
            \label{fig:Cox_rho(aupr)}
        \end{subfigure}
        \hfill
        \begin{subfigure}{0.45\textwidth}
            \centering
            \includegraphics[width=\linewidth]{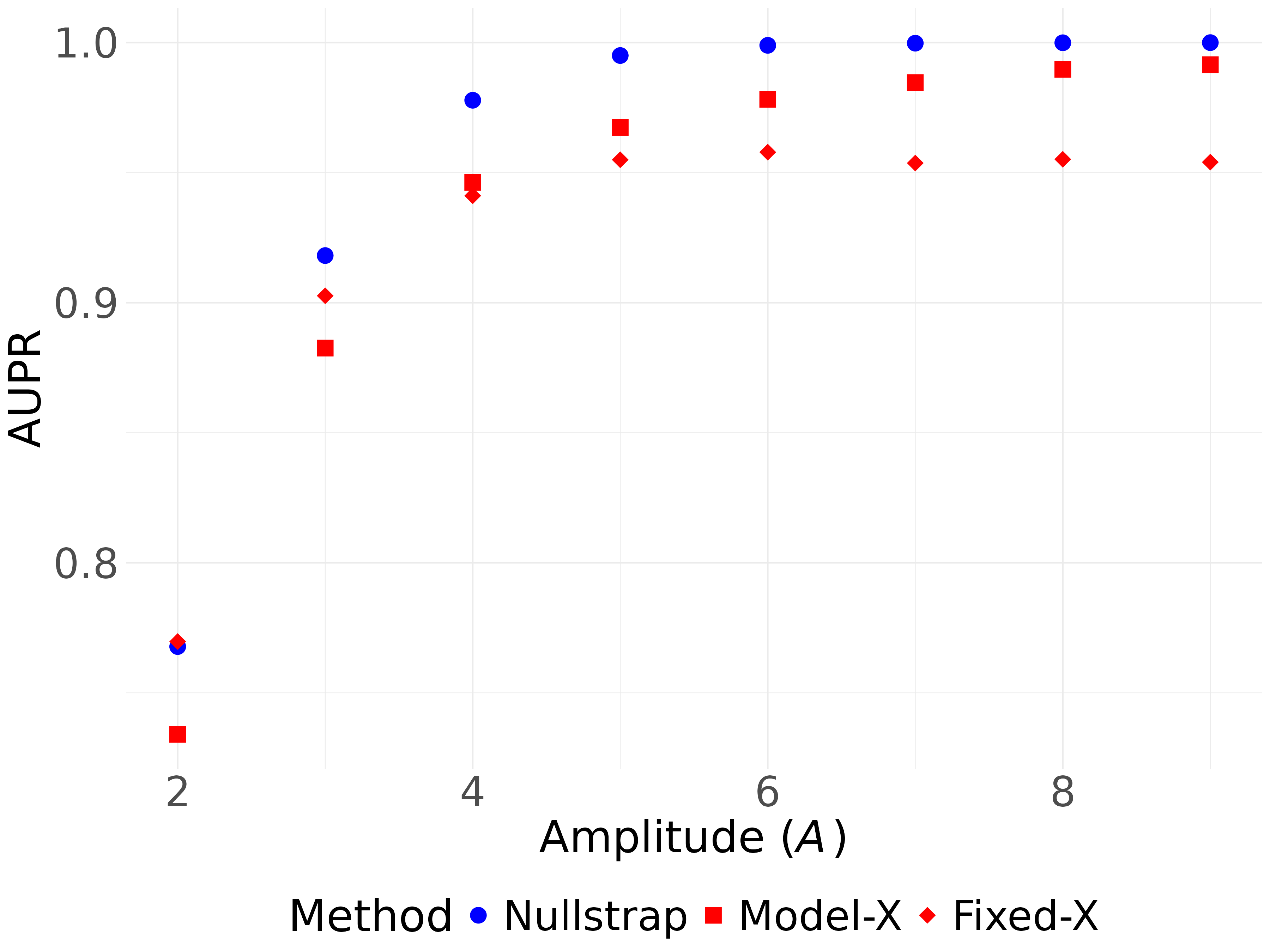} 
            \caption{Empirical AUPR vs. signal amplitude ($A$) under Simulation Setting~\ref{setting:cox}.}
            \label{fig:Cox_amp(aupr)}
        \end{subfigure}
        \vspace{0.5cm}
        \begin{subfigure}{0.45\textwidth}
            \centering
            \includegraphics[width=\linewidth]{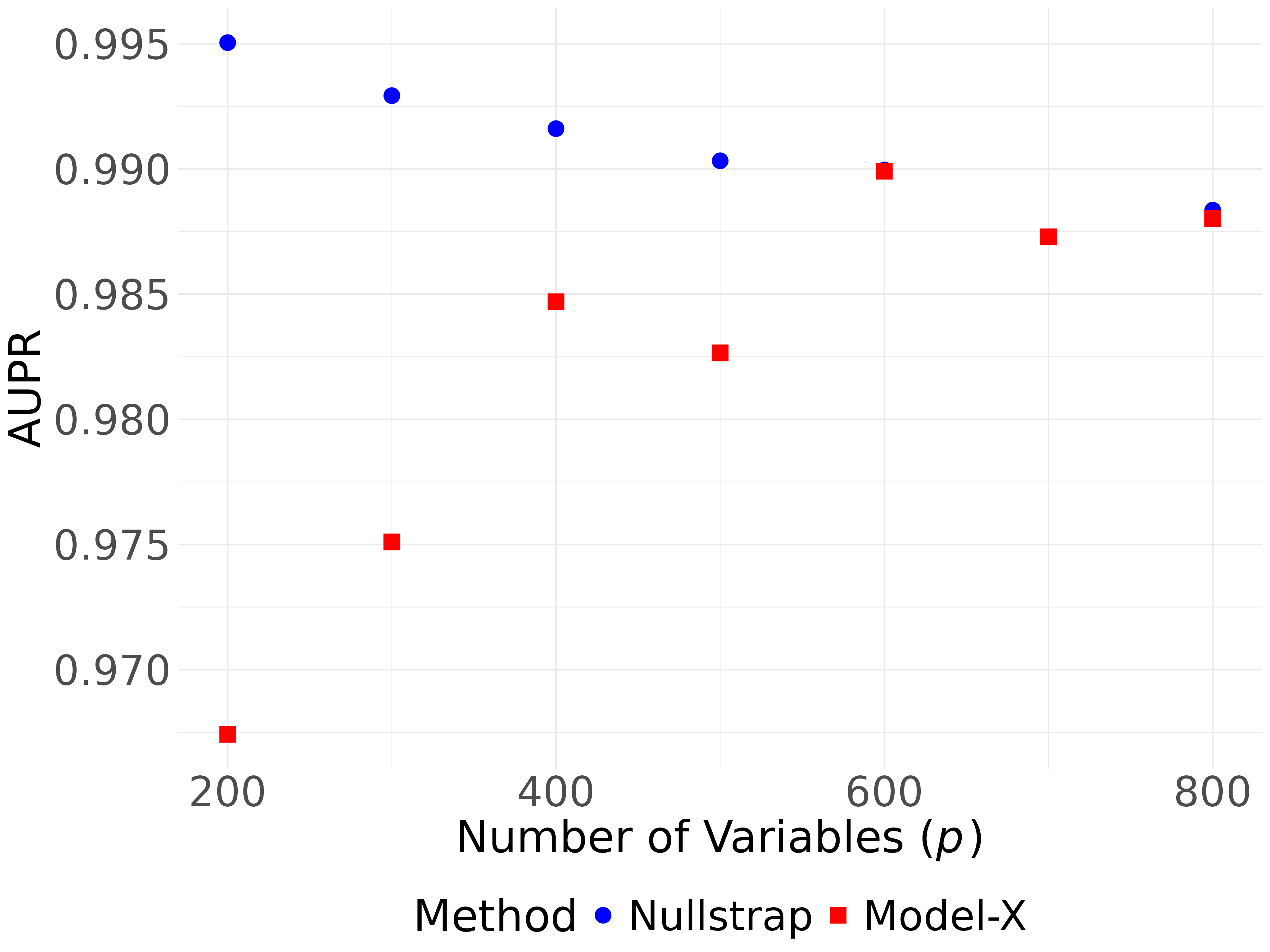}
            \caption{Empirical AUPR vs. number of variables ($p$) under Simulation Setting~\ref{setting:cox}.}
            \label{fig:Cox_num(aupr)}
        \end{subfigure}
        \hfill
        \begin{subfigure}{0.45\textwidth}
        \centering

        \end{subfigure}
        \caption{Empirical AUPR for the Cox proportional hazards model.}
        \label{fig:AUPR_cox}
    \end{figure}

The empirical FDR and power results are presented in Figures~\ref{fig:Cox_rho}--\ref{fig:Cox_num}, while the AUPR results are shown in Figure~\ref{fig:AUPR_cox}.  
Overall, both Model-X and Fixed-X exhibit conservative behavior, leading to low power across scenarios.
% This can be partially attributed to the data-driven threshold used by the knockoff methods, which is excessively large. 
% These results suggest that the knockoff methods lack robustness when applied to the Cox model.

Specifically, in Figure~\ref{fig:Cox_rho}, the power of the two knockoff methods approaches zero as the correlation increases. In contrast, Nullstrap remains significantly more robust to high correlations among variables. In Figure~\ref{fig:Cox_amp}, where the amplitude \( A \) is varied, we observe that once \( A = 7 \), the power of Nullstrap reaches \( 1 \) and remains constant. Moreover, for \( A < 7 \), Nullstrap’s power increases more rapidly than that of the knockoff methods. Figure~\ref{fig:Cox_num} shows that the power and FDR of the Model-X knockoff method collapse to zero when the number of variables \( p > 500 \). By contrast, Nullstrap remains stable, highlighting its scalability and practical utility in high-dimensional settings.

Table~\ref{cox_time} summarizes the runtimes of all methods under Simulation Setting~\ref{setting:cox}, using the default parameter configuration in~\eqref{eq:para_cox}.

\begin{figure}[H]
\centering
\includegraphics[scale=0.5]{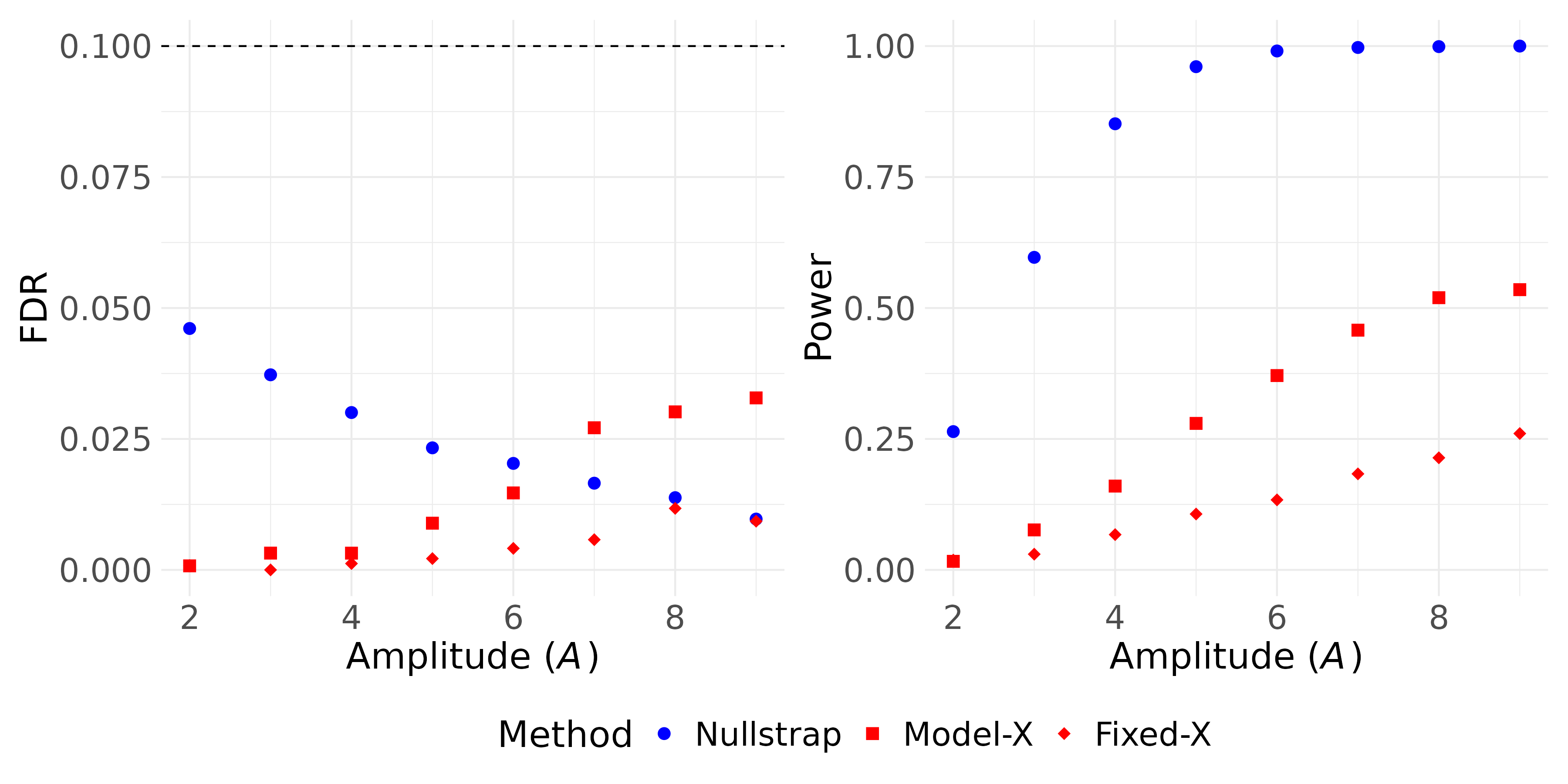}
\caption{Empirical FDR and power vs. signal amplitude ($A$) under Simulation Setting~\ref{setting:cox}.}
\centering
\label{fig:Cox_amp}
\end{figure}

% \begin{figure}[h]
% \centering
% \includegraphics[scale=0.4]{simulation/result_rho_aupr_cox.png}
% \caption{Empirical AUPR for the Cox model (Correlation).}
% \centering
% \label{fig:Cox_rho(aupr)}
% \end{figure}

% \begin{figure}[h]
% \centering
% \includegraphics[scale=0.4]{simulation/result_amp_aupr_cox.png}
% \caption{Empirical AUPR for the Cox model (Amplitude).}
% \centering
% \label{fig:Cox_amp(aupr)}
% \end{figure}

% \begin{figure}[h]
% \centering
% \includegraphics[scale=0.4]{simulation/result_num_aupr_cox.png}
% \caption{Empirical AUPR for the Cox model (Number of Variables).}
% \centering
% \label{fig:Cox_num(aupr)}
% \end{figure}

\begin{table}[htbp]
\centering
\caption{Comparison of runtimes (in seconds) in the Cox model under Simulation Setting~\ref{setting:cox}, using the default parameter configuration in~\eqref{eq:para_cox}.}
\begin{tabular}{ccc}
    \toprule
    Nullstrap & Model-X & Fixed-X \\
    \midrule
    13.71 & 21.41 & 12.61 \\
    \bottomrule
\end{tabular}
\label{cox_time}
\end{table}

% \begin{table}[htbp][h!]
% \centering
% \begin{tabular}{cc}
%     \toprule
%     Nullstrap &  Model-X \\
%     \midrule
%     6659.00 & 22387.68 \\
%     \bottomrule
% \end{tabular}
% \caption{Comparison of runtimes (s) in the Cox model}
% \label{cox_time}
% \end{table}

% The results are consistent with those observed for the generalized linear regression model. Nullstrap remains the fastest method, with a runtime of \( 13.71 \, \text{s} \), which is approximately two times faster than that of Model-X, which requires \( 21.41 \, \text{s} \). These results further demonstrate the stability and efficiency of our method in terms of computational time.
\begin{table}[htbp]
\centering
\caption{Comparison of Jaccard index under Simulation Setting~\ref{setting:cox}, using the default parameter configuration in~\eqref{eq:para_cox}.}
\begin{tabular}{cc}
    \toprule
    Nullstrap & Model-X\\
    \midrule
    0.938 & 0.000 \\
    \bottomrule
\end{tabular}
\label{cox_jac}
\end{table}

% \begin{table}[htbp][h!]
% \centering
% \begin{tabular}{ccc}
%     \toprule
%     Nullstrap &  Model-X\\
%     \midrule
%     0.938 & 0.348 \\
%     \bottomrule
% \end{tabular}
% \caption{Comparison of Jaccard index in the Cox model under Equation~\eqref{eq:para_cox}.}
% \label{cox_jac}
% \end{table}

Table~\ref{cox_jac} reports the Jaccard index, averaged over 100 replications under Simulation Setting~\ref{setting:cox}, using the default parameter configuration in~\eqref{eq:para_cox}. The Jaccard index quantifies each method’s stability across random seeds (noting that Fixed-X knockoff is deterministic). Nullstrap achieves the highest stability with a Jaccard index of \( 0.938 \), while Model-X exhibits no stability, with a value of \( 0.000 \). This stark contrast underscores the robustness of Nullstrap in consistently identifying relevant variables in the Cox proportional hazards model.

\subsection{Interactions between signal variables}
For the Cox model, we also consider a simulation setting in which interactions between signal variables are incorporated into the design matrix, resulting in explicit correlations among its columns.

\begin{simsetting}\label{sim:setting_cox_inter}
	We set \( n = 1000 \), \( p_{\rm{base}} = 30 \), and \( p = p_{\rm{base}} + \frac{p_{\rm{base}}(p_{\rm{base}}-1)}{2} \). The base design matrix \(\bX_{\text{base}}\) is drawn from \(\mathcal{N}(\mathbf{0},\boldsymbol{\Sigma}_{\text{base}})\), where \(\boldsymbol{\Sigma}_{\text{base}}\) is a Toeplitz covariance matrix with autocorrelation parameter \(\rho = 0.4\). 
	We then construct interaction terms by computing pairwise products of the first \( p_{\rm{base}} \) variables, forming an interaction matrix \( \bX_{\rm{interact}} \). The first {\( 5 \)} elements of the coefficient vector \( \bbeta \) are randomly assigned values with amplitude \( A \) and random signs. Additionally, if both variables involved in an interaction term are among the first {\( 5 \)} variables, their corresponding coefficient is also randomly assigned values with amplitude \( A \) and random signs.  
	Finally, the full design matrix \( \bX \) is formed by concatenating \( \bX_{\rm{base}} \) and \( \bX_{\rm{interact}} \) column-wise.  We consider one simulation parameter for adjustment:  
	\begin{itemize} 
		\item the signal amplitude \( A \in [3, 9] \).  
	\end{itemize}  
	\end{simsetting}

% 	We still set the correction factor \( \gamma_{n,p} \) to
% $
% 0.2 \Hat{\sigma} \left(\lambda_n + \sqrt{\frac{\log p}{n}} \right)
% $
% in this simulation setting.

\begin{figure}[H]
\centering
\includegraphics[scale=0.5]{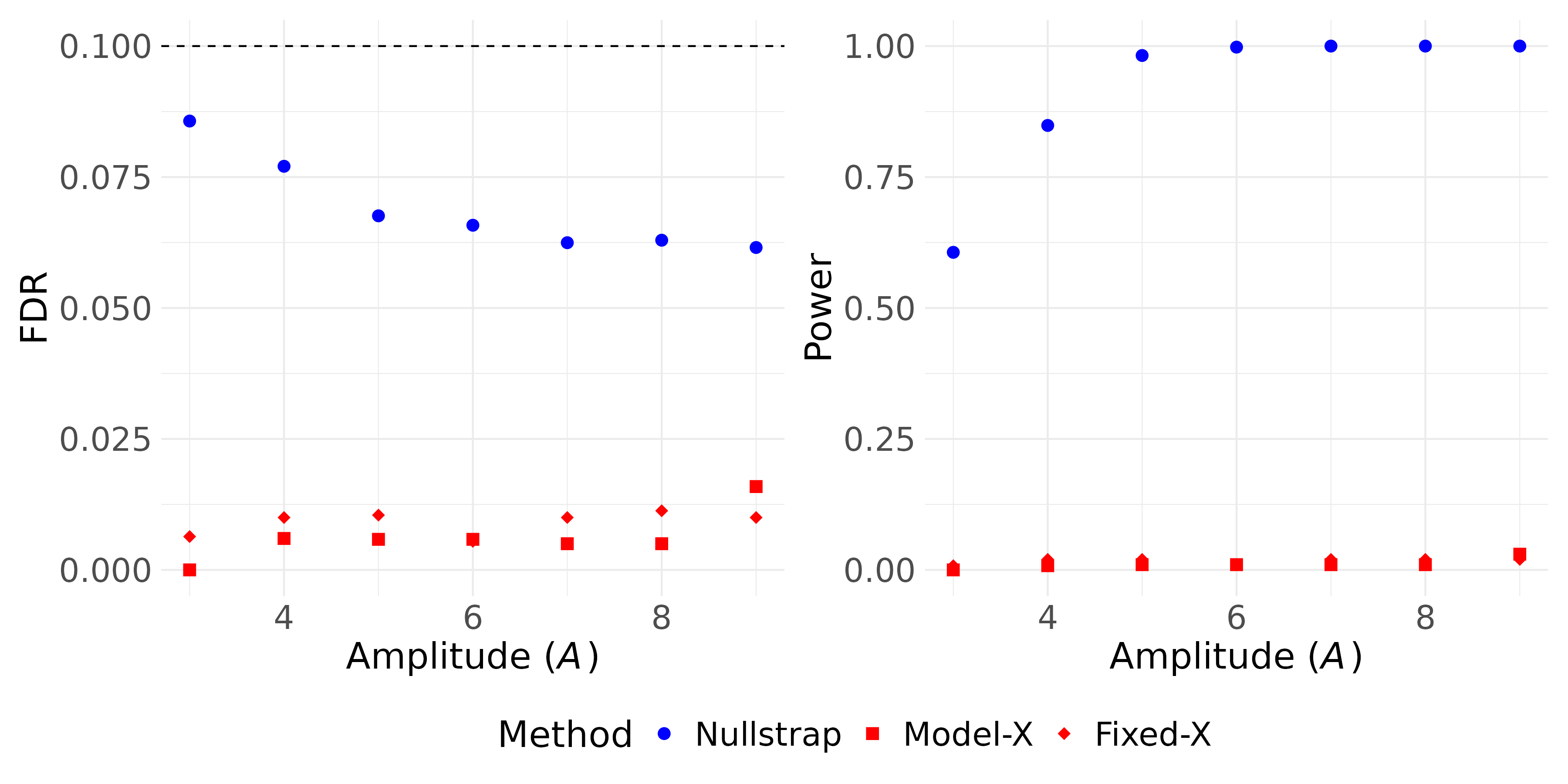}
\caption{Empirical FDR and power vs. signal amplitude ($A$) under Simulation Setting~\ref{sim:setting_cox_inter}.}
\centering
\label{fig:cox_amp_inter}
\end{figure}

\begin{figure}[H]
\centering
\includegraphics[scale=0.5]{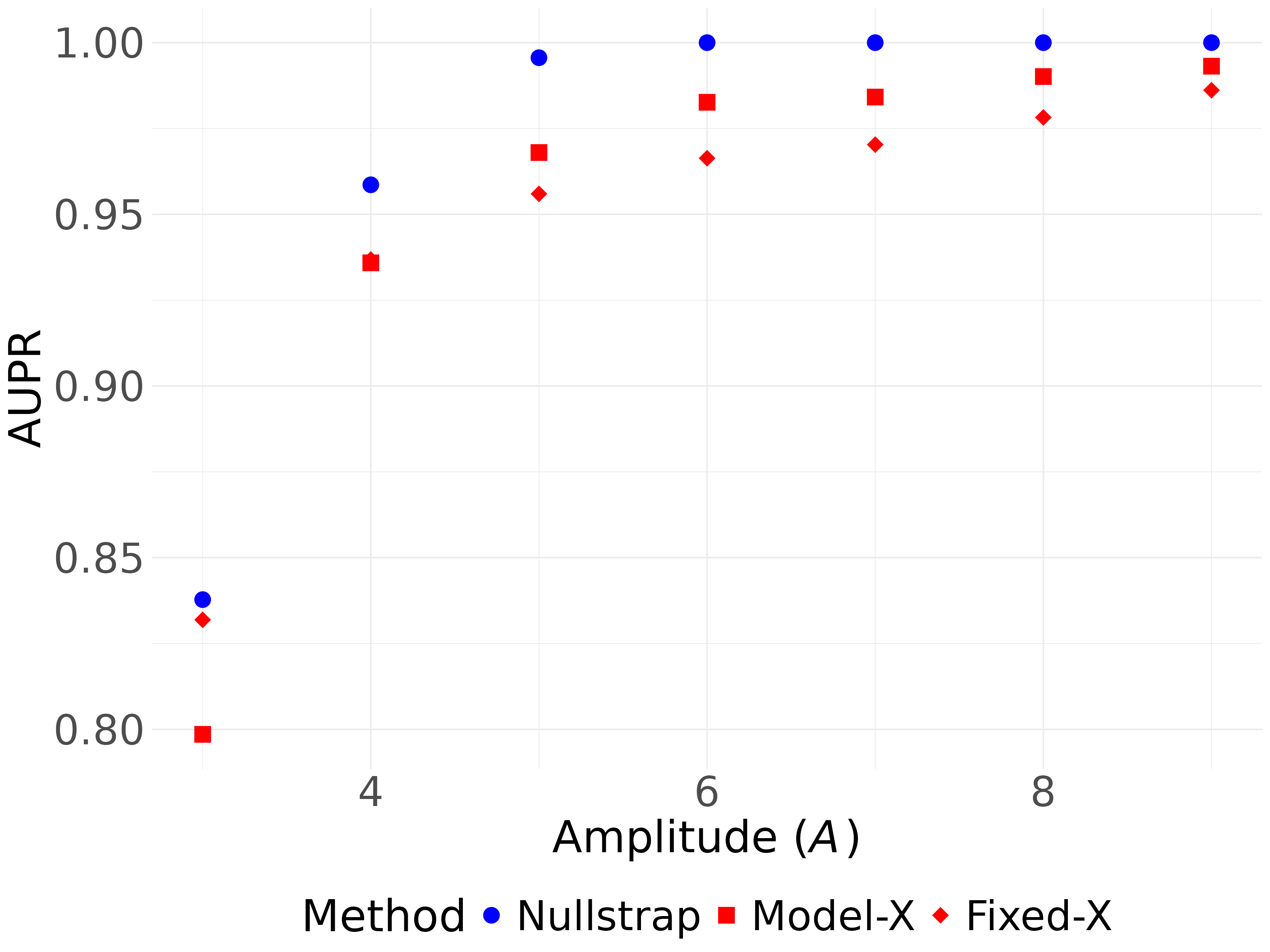}
\caption{Empirical AUPR vs. signal amplitude ($A$) under Simulation Setting~\ref{sim:setting_cox_inter}.}
\centering
\label{fig:cox_aupr_inter}
\end{figure}

For each scenario under Simulation Setting~\ref{sim:setting_cox_inter}, we compare the FDR, power, and AUPR of the three methods using 100 replications. The empirical FDR and power results are shown in Figure~\ref{fig:cox_amp_inter}, while the AUPR results are provided in Figure~\ref{fig:cox_aupr_inter}. Across all scenarios, all methods achieve FDR control; however, Nullstrap not only maintains reliable control but also consistently attains the highest power and AUPR.

% Compared to Nullstrap, Nullstrap-Diff exhibits lower power than DS and MDS, particularly in more challenging scenarios such as those with high autocorrelations and high signal amplitudes. 

\clearpage

\section{Nullstrap for Gaussian graphical models}\label{sec:GGM}
In this section, we outline the specific steps for applying Nullstrap to perform variable selection in a Gaussian graphical model (GMM), $\by \sim \mathcal{N}(0, \bSigma)$, where $\bSigma = \bTheta^{-1}$. 
% to control the FDR using Nullstraps in the case of Gaussian graphical models $	\bY \sim \mathcal{N}(0, \bTheta^{*-1})$. 
In this subsection, we adopt the notation \( \boldsymbol{\Theta} \), consistent with the literature on Gaussian graphical models (GGMs), in place of \( \boldsymbol{\beta} \). Our goal is to estimate the set  
\[
\mathcal{S} := \{ (i, j) \mid i > j,\ \Theta_{ij} \neq 0 \},
\]  
which corresponds to the variable selection problem in a GGM.

Given \( n \) independent and identically distributed observations \( \{\by_{k}\}_{k=1}^{n} \), we define the sample covariance matrix as
$
\hat{\bSigma} = n^{-1} \sum_{k=1}^{n} \by_{k} \by_{k}\tran.
$
We also define the off-diagonal \( \ell_1 \) regularizer
$
\|\bTheta\|_{1, \text{off}} := \sum_{i \neq j} |\Theta_{ij}|,
$
where the sum ranges over all \( i, j = 1, \dots, p \) with \( i \neq j \). We consider estimating \( \bTheta \) by solving the following \( \ell_1 \)-regularized log-determinant program \citep{friedman2008sparse}:
$
\hat{\bTheta} = \arg \min_{\bTheta \succ 0} \left\{ \langle \hat{\bSigma}, \bTheta \rangle - \log \det(\bTheta) + \lambda_n \|\bTheta\|_{1, \text{off}} \right\},
$
where \( \bTheta \succ 0 \) denotes that \( \bTheta \) is positive definite and $\lambda_n$ is the regularization parameter selected by cross-validation.

%To lighten notation, we occasionally drop the superscript \( n \), and simply write \( \hat{\Sigma} \) for the sample covariance. 

\begin{definition}[Synthetic null data for a GGM]
	For a GGM $\by \sim \mathcal{N}(0, \bTheta^{-1})$, Nullstrap defines 
$
	\tilde{{\by}}_{k} \sim  \mathcal{N}(0, \hat{\bTheta}_0^{-1}),
$	
where $\hat{\bTheta}_0^{-1} = \diag(\hat{\bSigma})$ and $\hat{\bSigma}$ is the sample covariance matrix of the original data \( \{\by_{k}\}_{k=1}^{n} \).
\end{definition}
Given synthetic null data \( \{\tilde{\by}_k\}_{k=1}^{n} \), we define the synthetic null covariance matrix as
$
\tilde{\bSigma} = n^{-1} \sum_{k=1}^{n} \tilde{\by}_{k} \tilde{\by}_{k}\tran.
$
Given the same regularization parameter \( \lambda_n > 0 \), we let
$
\tilde{\bTheta} = \arg \min_{\bTheta \succ 0} \left\{ \langle \tilde{\bSigma}, \bTheta \rangle - \log \det(\bTheta) + \lambda_n \|\bTheta\|_{1, \text{off}} \right\}.
$
%In this application, we choose $\epsilon_n = \nu \left( \lambda_n + \sqrt{\frac{\log p}{n}} \right)$, where $\nu$ is a hyper-parameter.
% Let \( \bbeta^* \) be the vector obtained by vectorizing the elements of the matrix \( \bTheta^* \), excluding the diagonal elements. Similarly, let \( \hat{\bbeta} \) and \( \tilde{\bbeta} \) be the vectors formed by vectorizing the elements of the matrices \( \hat{\bTheta} \) and \( \tilde{\bTheta} \), respectively, also excluding the diagonal elements. 
% Recall that \( \bbeta_0 = (0, \dots, 0)\tran \) is the null parameter vector.
\begin{lemma}\label{lem:ggm}
	Under the conditions specified in Corollary 1 of \cite{Ravikumar2008HighdimensionalCE}, Assumption \ref{assum:gamma_estimator} holds for the graphical LASSO estimator with $\gamma_{n,p} = \kappa \left( \lambda_n + \sqrt{\frac{\log p}{n}} \right)$ and $\kappa$ is a constant.
\end{lemma}

Lemma \ref{lem:ggm}, based on the result in \cite{Ravikumar2008HighdimensionalCE}, establishes the existence of a correction factor \( \gamma_{n,p} \). In practice, we select \( \gamma_{n,p} \) in a data-driven manner using Algorithm~\ref{alg:gamma}. 
Next, we define \( |\tilde{\Theta}^\prime_{ij}| \) as \( |\tilde{\Theta}^\prime_{ij}| = |\tilde{\Theta}_{ij}| + \gamma_{n,p} \), and set the threshold \( \tau_q > 0 \) as:
\[
\tau_q = \min \left\{ t > 0 : \frac{\#\{ (i,j) : i > j \text{ and } |\tilde{\Theta}^\prime_{ij}| \geq t \}}{\max\left( \#\{ (i,j) : i > j \text{ and } |\hat{\Theta}_{ij}| \geq t \}, 1 \right)} \leq q \right\},
\]
where \( q \) denotes the target FDR level. Finally, we select the variables as:  
\[
\widehat{\cS}(\tau_q) = \{ (i,j) : i > j \text{ and } |\hat{\Theta}_{ij}| \geq \tau_q \}.
\]
% By applying Lemma \ref{lem:ggm} and Theorem \ref{thm:main}, we conclude that for a given target FDR control \( q \), we obtain \( \text{FDR} \leq q + o(1) \) and \( \text{Power} = 1 + o(1) \), as \( n \to \infty \).

Parameter estimation for the GGM can be performed using different approaches: Nullstrap relies on the graphical LASSO, whereas knockoff-based and data-splitting methods use nodewise regression \citep{meinshausen2006high}.
% For each vertex \( Y_j \), we can express it as a linear combination of the remaining variables, denoted \( \bY_{-j} \), plus an error term:
% $
% Y_j = \bY_{-j}\tran \bbeta^j + \bvarepsilon_j,
% $
% where \( \bbeta^j = -\bTheta_{jj}^{-1} \bTheta_{-j,j} \), and \( \bvarepsilon_j \) is the residual, independent of \( \bY_{-j} \). 
% The conditional independence between two variables \( Y_i \) and \( Y_j \) can be inferred from the precision matrix \( \bTheta \), where \( \Theta_{ij} = 0 \). 
% Therefore, it is natural to first recover the support of each \( \bbeta^j \) using a variable selection method such as the {LASSO} regression, described in Section~\ref{sec:LR}. After estimating the support, the nodewise selection results can be combined to reconstruct the entire graph.  
While Nullstrap can also use nodewise regression, it is slower than the graphical LASSO. In contrast, knockoff methods are not readily applicable to graphical LASSO, highlighting Nullstrap’s broader applicability. Moreover, Nullstrap is compatible with the D-trace LASSO \citep{zhang2014sparse}, which similarly challenges knockoff-based approaches, further demonstrating Nullstrap’s flexibility across model classes.

%Nodewise regression v.s.  GLASSO
%GLASSO provide a positive-definite matrix, 

\subsection{Simulation results}
%To ensure control of the False Discovery Rate (FDR), we additionally add a perturbation term $\epsilon_n$ to $\tilde{\Theta}$ as follows:
%\[
%|\tilde{\Theta}^\prime_{ij}| = |\tilde{\Theta}|_{ij} + \epsilon_n.
%\]
%\begin{lemma}
%	With the inclusion of \(\epsilon_n\), for any fixed threshold \(t > 0\) and null  \((i, j) \notin E(\Theta^*) \), under  Assumption X,  we have
%	\[
%	\mathbb{P}\left( |{\Theta}|_{ij}\geq t \right) \leq \mathbb{P}\left(|\tilde{\Theta}^\prime_{ij}| \geq t \right) + o(p^{-1}).
%	\]
%\end{lemma}
We generate data from a GMM to evaluate the performance of Nullstrap in controlling the FDR. Prior to applying the graphical LASSO, we scale the columns of the data matrix \( \bY = (\by_1, \dots, \by_n)\tran\).
Following the work of \citet{li2021ggm}, we consider the following simulation setting.
\begin{simsetting}\label{setting:ggm}
	We set the dimension of the precision matrix $\bTheta$ as \( p = 200 \). 
We draw $n$ independent samples from a multivariate normal distribution $\mathcal{N}(0, \bTheta^{-1})$, where $\bTheta$ is the precision matrix associated with one of four commonly used graph structures in Gaussian graphical models: band graphs, block graphs, Erd\H{o}s-R\'enyi graphs, and cluster graphs. In specific, we let $\bTheta:= \bTheta^0 + (|\lambda_{\min}(\bTheta^0)| + 0.5)\mathbf{I}$, where $\lambda_{\min}(\bTheta^0)$ is the minimum eigenvalue of $\bTheta^0$, to ensure the precision matrix is positive definite.  The $\bTheta^0$ corresponding to four graph structures are constructed as follows:

\begin{enumerate}
    \item  Band graph: $\Theta^0_{ii} = 1$ for $i = 1, \dots, p$, and the off-diagonal elements $\Theta^0_{ij} = \mathrm{sign}(b) \cdot |b|^{\frac{|i - j|}{10}} \cdot \mathbf{1}\{|i - j| \leq 10\}$ for $i \ne j$, where $b=-0.8$ is edge strength.
    
    \item  Block graph: $\bTheta^0$ is constructed by dividing the matrix into 10 blocks, each containing 20 consecutive nodes. Within each block, all diagonal elements are set to $1$, and all off-diagonal elements are set to $b=-0.8$.
    
    \item  Erd\H{o}s-R\'enyi: $\Theta^0_{ii} = 1$ for $i = 1, \dots, p$, and the off-diagonal elements $\Theta^0_{ij} = \Theta_{ij} \cdot \phi_{ij}$ for $i > j$, where $\Theta_{ij} \sim \mathrm{Bernoulli}\left(\frac{1}{10}\right)$ and $\phi_{ij} \sim \mathrm{Uniform}([-0.6, -0.2] \cup [0.2, 0.6])$, with $\Theta^0_{ij}= \Theta^0_{ji}$ to maintain symmetry.
    
    \item  Cluster graph: $\bTheta^0$ is constructed by dividing the matrix into 5 blocks, each containing 40 consecutive nodes. Each block is constructed as the Erd\H{o}s-R\'enyi graph but $\Theta_{ij} \sim \mathrm{Bernoulli}\left(\frac{1}{2}\right)$.
\end{enumerate}
We consider two parameters for adjustment:
\begin{itemize}
    \item (a) the sample size \( n \in \{1500, 2000,\dots, 4000\} \),
    \item (b) the FDR level \( q \in [0.1, 0.4] \).
\end{itemize}
% (a) the sample size \( n \in [1500, 4000] \) and (b) the FDR level \( q \in [0.1, 0.4] \). 
For each scenario where one parameter varies, the remaining parameters are held constant at:
\begin{equation}\label{eq:para_ggm}
n = 3500,\text{ and } q = 0.2. 
\end{equation}
\end{simsetting}

We replicate each scenario in Simulation Setting~\ref{setting:ggm} 100 times and compare our proposed method, Nullstrap, with four competing approaches: GFC-L, GFC-SL, KO2, and DS. GFC-L and GFC-SL are two methods for high-dimensional Gaussian graphical models introduced by \cite{698516a0-42f0-3d43-8108-2614216a2921}, implemented via the \texttt{SILGGM} R package \citep{zhang2018silggm} with default tuning parameters. KO2, a knockoff-based method proposed by \cite{yu2021false}, is implemented using the R code provided at \url{https://github.com/LedererLab/GGM-FDR}. We exclude MDS and the GGM knockoff filter with sample-splitting-recycling (GKF-Re+) \citep{li2021ggm} from the comparison due to their high computational cost.

\begin{figure}[H]
\centering
\includegraphics[scale=0.5]{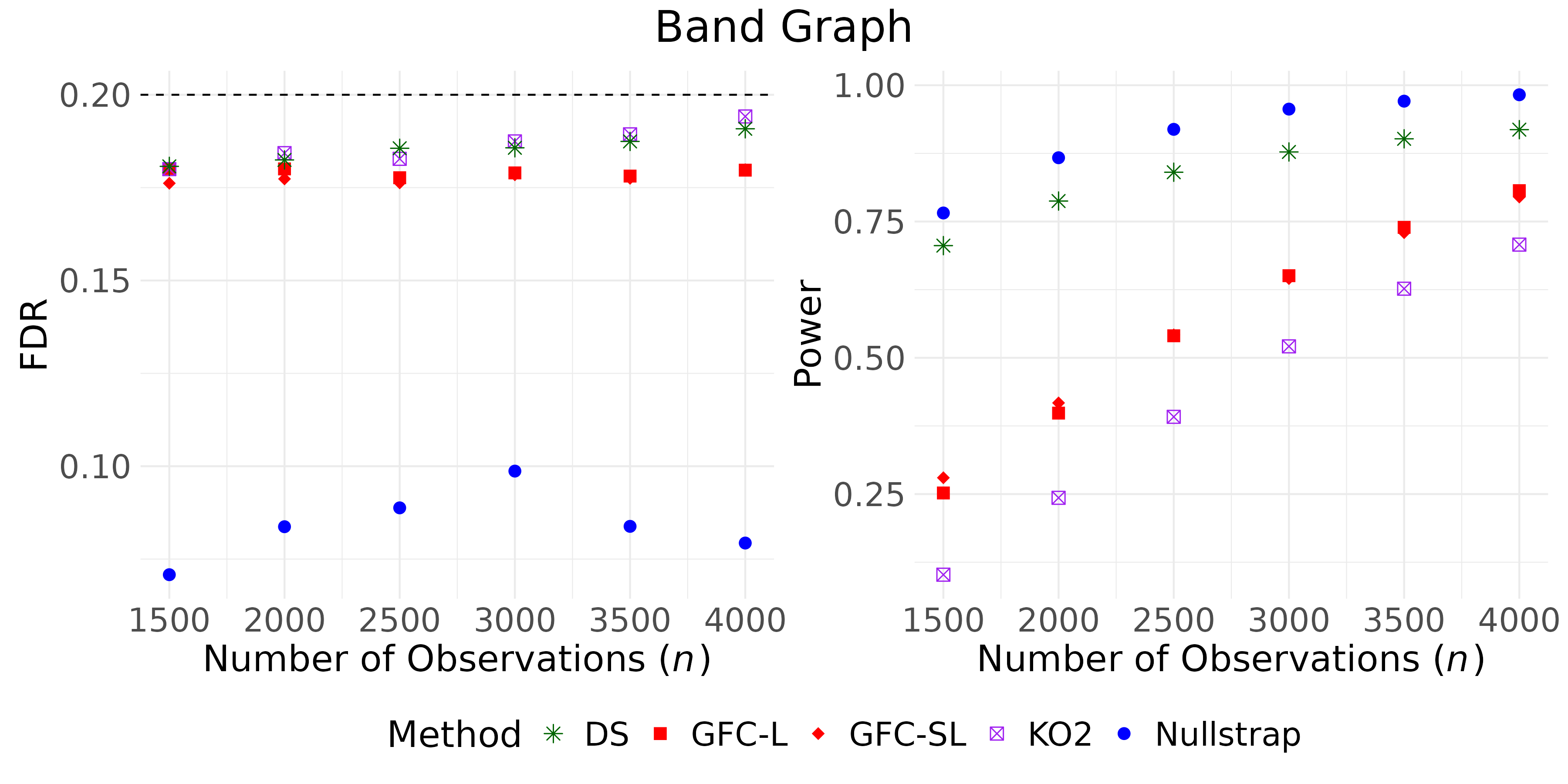}
\caption{Empirical FDR and power vs. number of observations ($n$) with a band graph under Simulation Setting~\ref{setting:ggm}.}
\centering
\label{fig:GGM_n_band}
\end{figure}

\begin{figure}[H]
\centering
\includegraphics[scale=0.5]{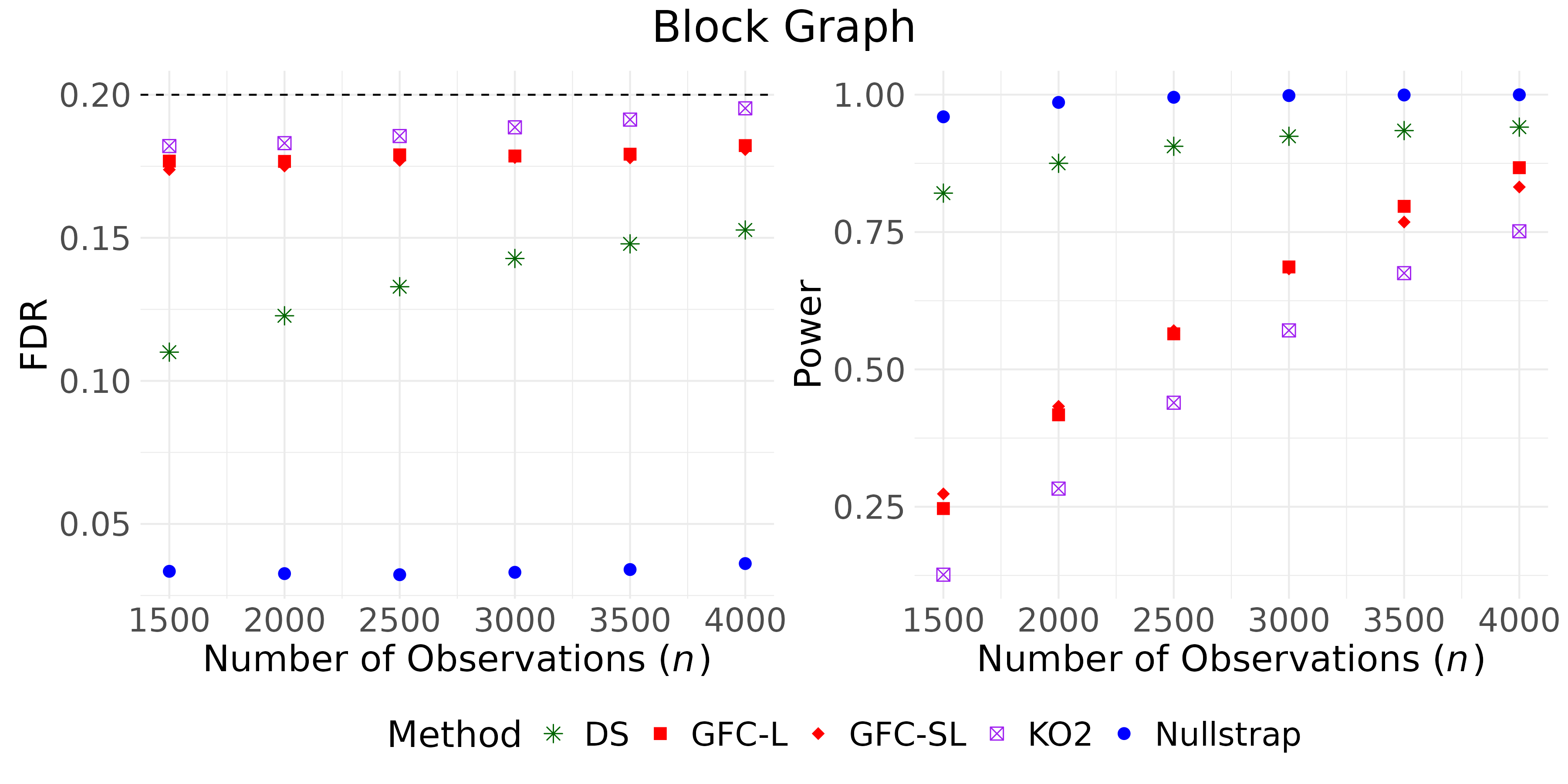}
\caption{Empirical FDR and power vs. number of observations ($n$) with a block graph under Simulation Setting~\ref{setting:ggm}.}
\centering
\label{fig:GGM_n_block}
\end{figure}

\begin{figure}[H]
\centering
\includegraphics[scale=0.5]{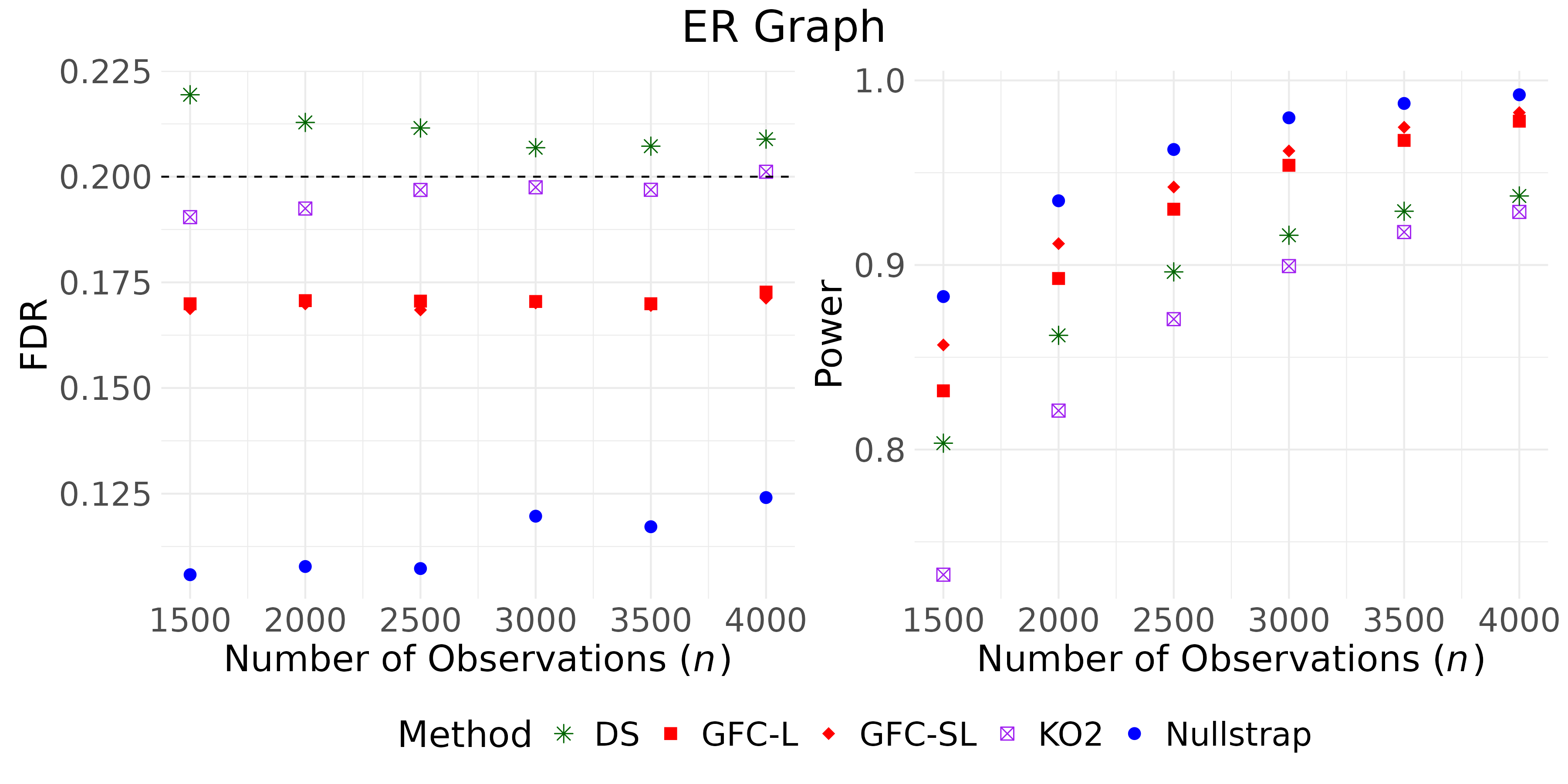}
\caption{Empirical FDR and power vs. number of observations ($n$) with an Erd\H{o}s-R\'enyi graph under Simulation Setting~\ref{setting:ggm}.}
\centering
\label{fig:GGM_n_er}
\end{figure}

\begin{figure}[H]
\centering
\includegraphics[scale=0.5]{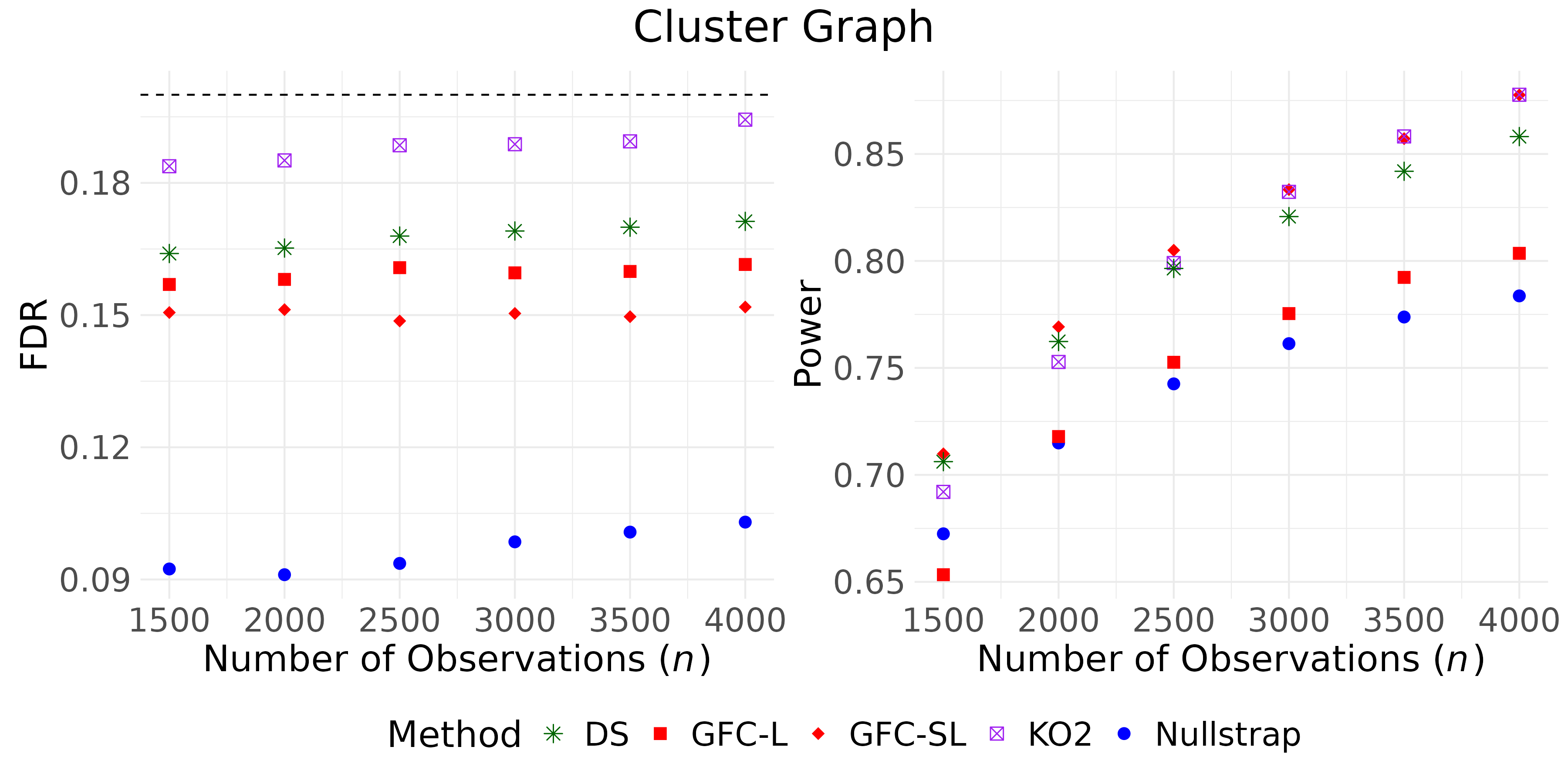}
\caption{Empirical FDR and power vs. number of observations ($n$) with a cluster graph  under Simulation Setting~\ref{setting:ggm}.}
\centering
\label{fig:GGM_n_cluster}
\end{figure}

\begin{figure}[H]
\centering
\includegraphics[scale=0.5]{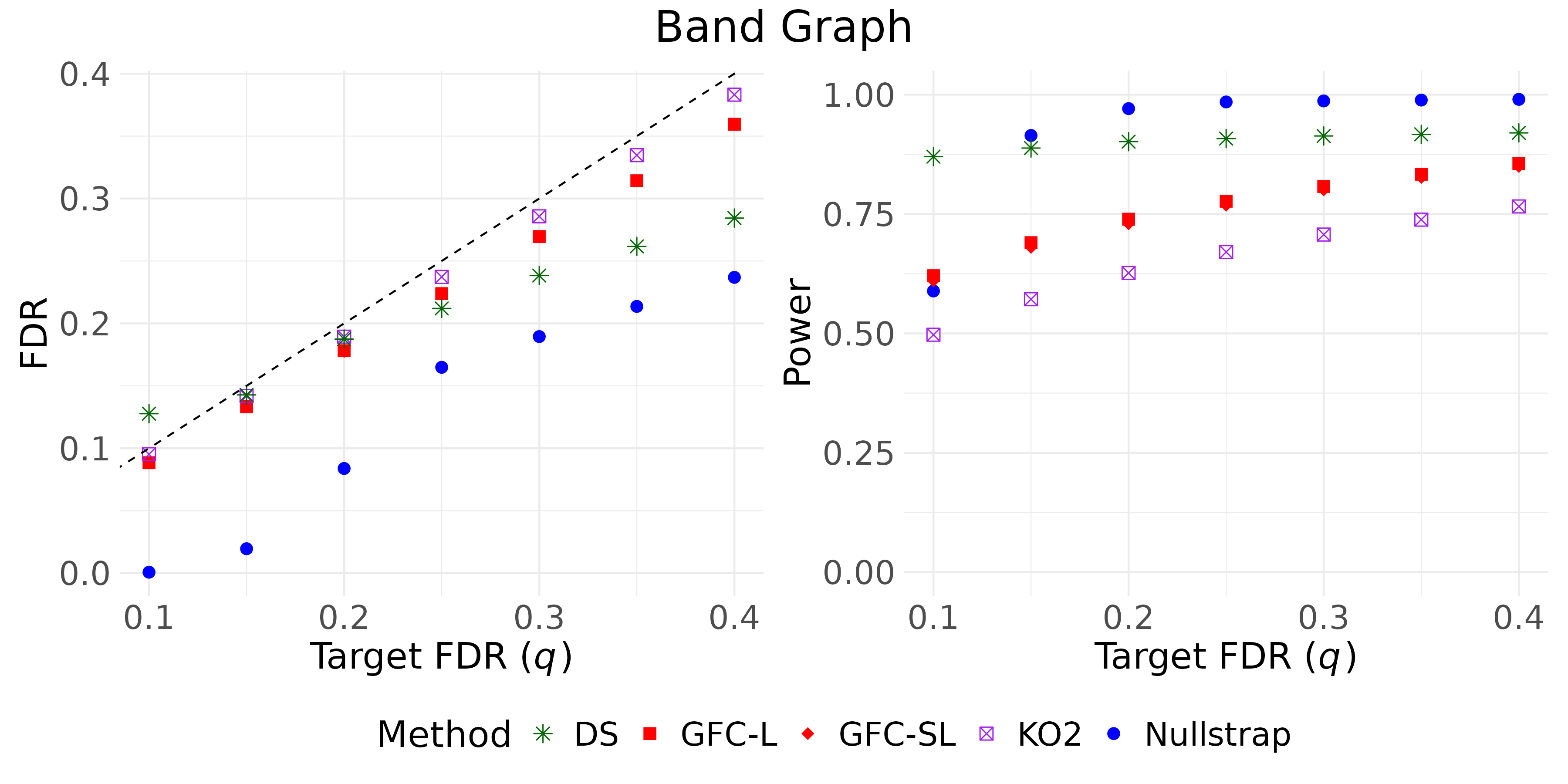}
\caption{Empirical FDR and power vs. target FDR level ($q$) with a band graph under Simulation Setting~\ref{setting:ggm}.}
\centering
\label{fig:GGM_fdr_band}
\end{figure}

\begin{figure}[H]
\centering
\includegraphics[scale=0.5]{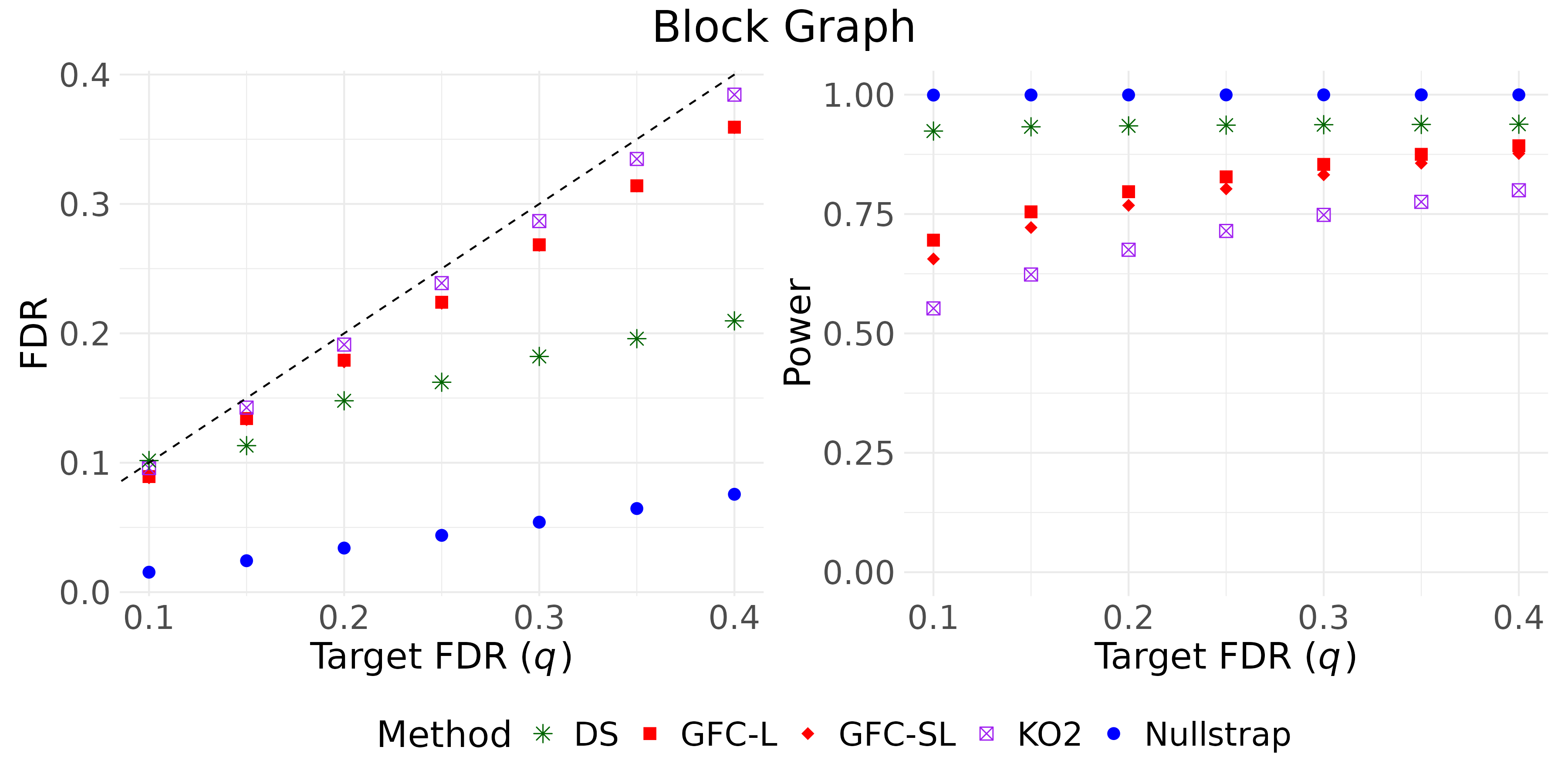}
\caption{Empirical FDR and power vs. target FDR level ($q$) with a  block graph under Simulation Setting~\ref{setting:ggm}.}
\centering
\label{fig:GGM_fdr_block}
\end{figure}

\begin{figure}[H]
\centering
\includegraphics[scale=0.5]{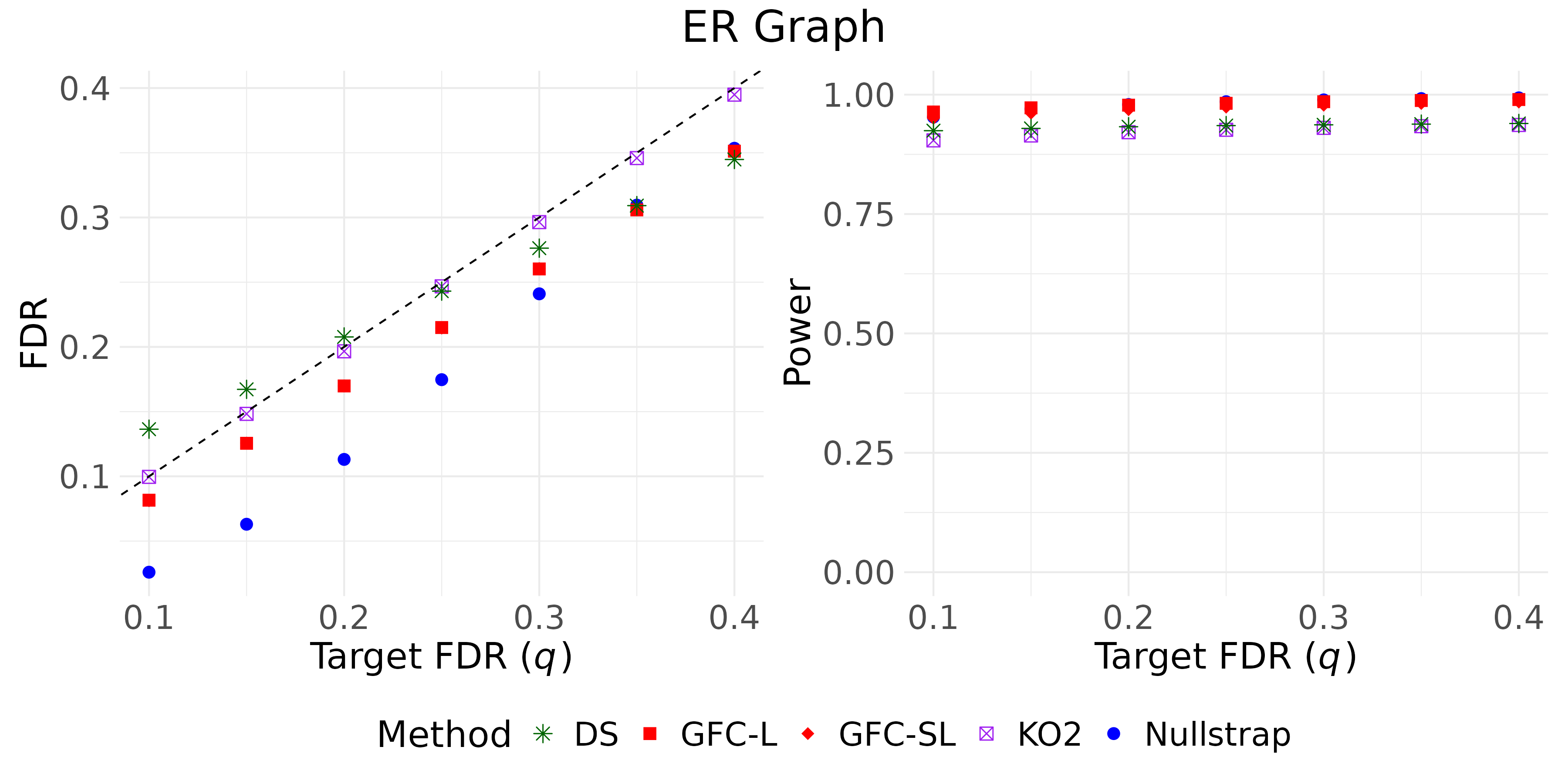}
\caption{Empirical FDR and power  vs. target FDR level ($q$) with a Erd\H{o}s-R\'enyi graph under Simulation Setting~\ref{setting:ggm}.}
\centering
\label{fig:GGM_fdr_er}
\end{figure}

\begin{figure}[H]
\centering
\includegraphics[scale=0.5]{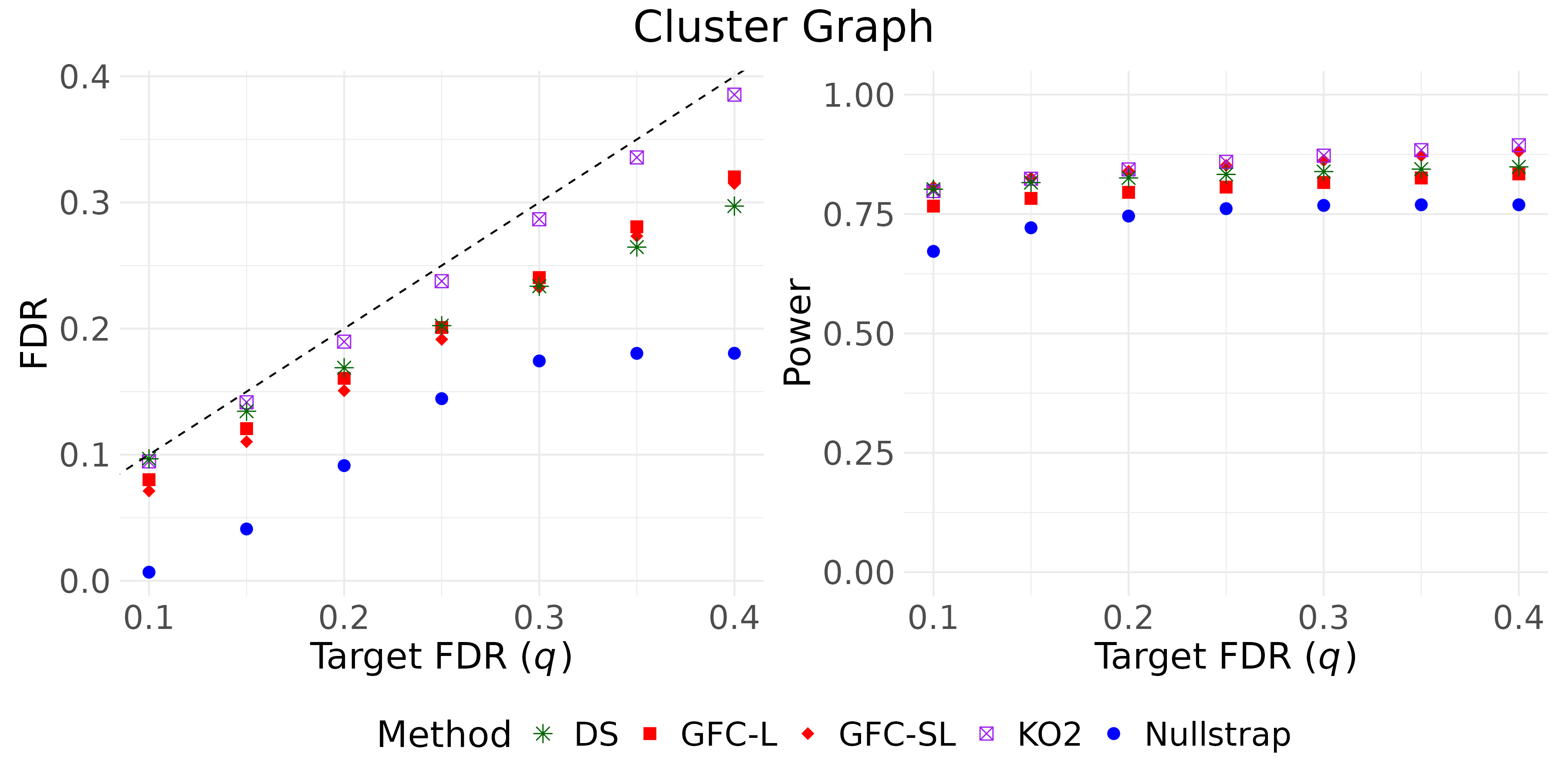}
\caption{Empirical FDR and power vs. target FDR level ($q$) with a  cluster graph under Simulation Setting~\ref{setting:ggm}.}
\centering
\label{fig:GGM_fdr_cluster}
\end{figure}

\begin{figure}[H]
        \centering
        \begin{subfigure}{0.45\textwidth}
            \centering
            \includegraphics[width=\linewidth]{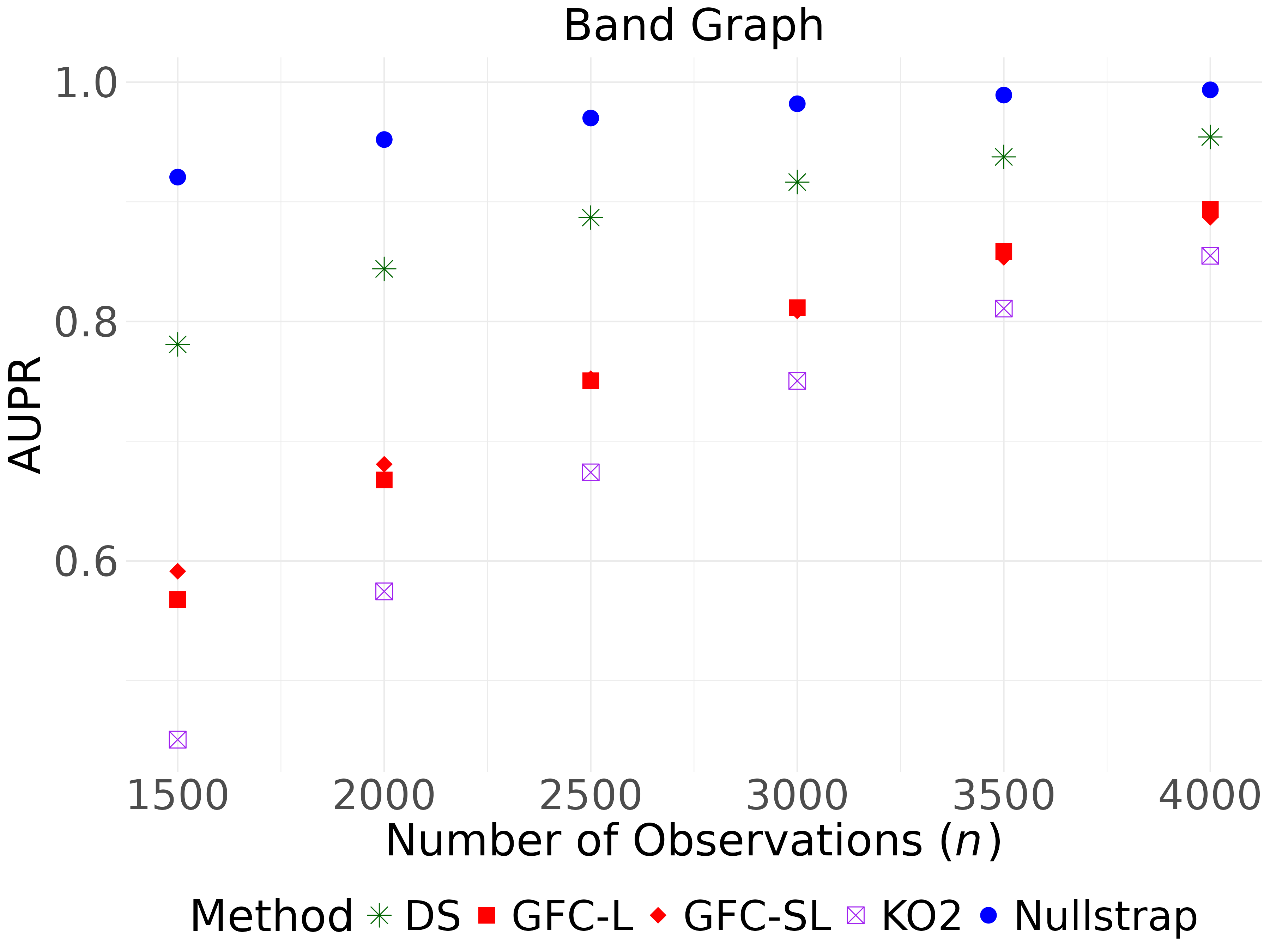}
            \caption{Empirical AUPR vs. number of observations ($n$) with a band graph under Simulation Setting~\ref{setting:ggm}.}
            \label{fig:GGM_n_band(aupr)}
        \end{subfigure}
        \hfill
        \begin{subfigure}{0.45\textwidth}
            \centering
            \includegraphics[width=\linewidth]{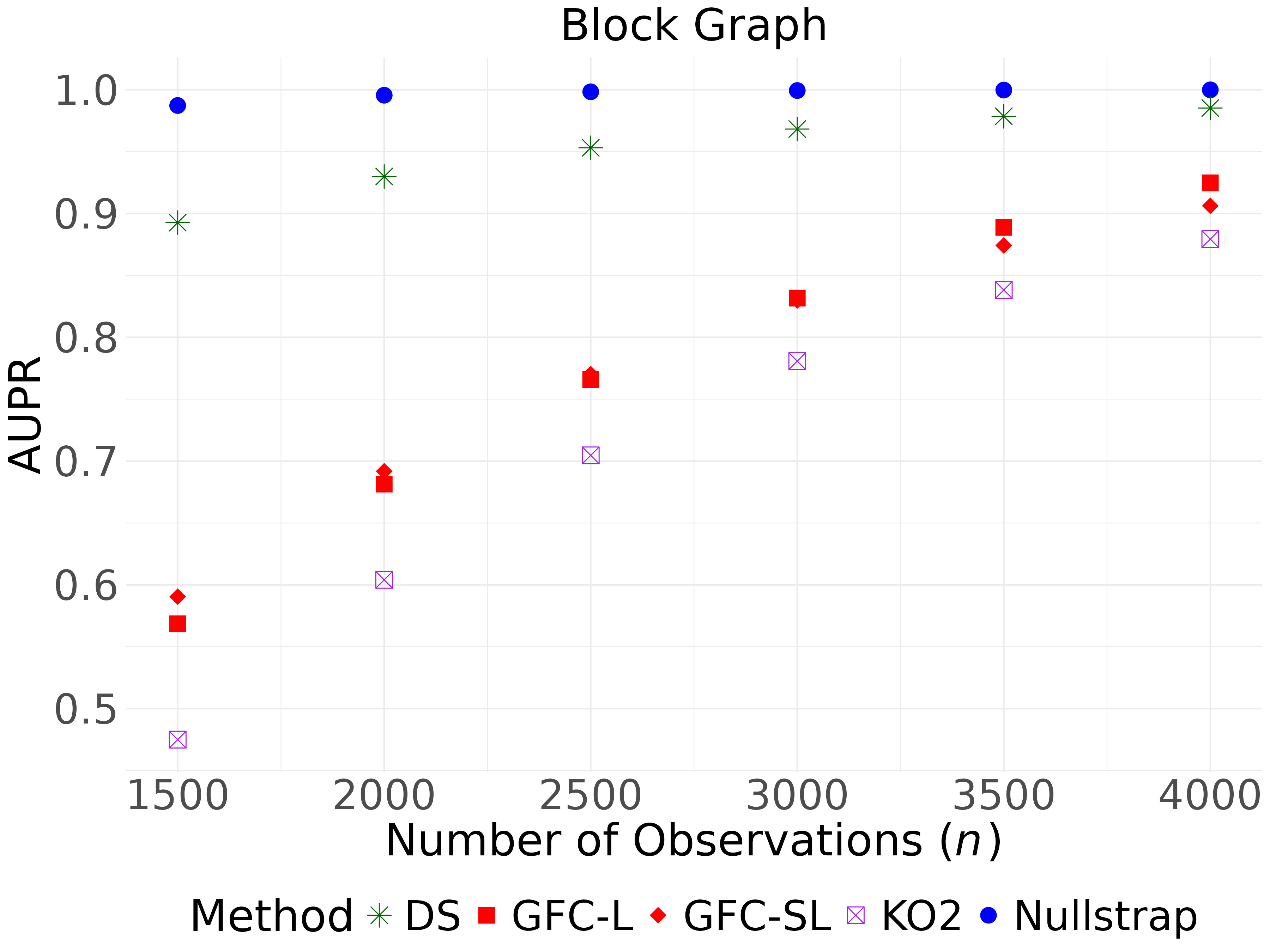} 
            \caption{Empirical AUPR vs. number of observations ($n$) with a block graph under Simulation Setting~\ref{setting:ggm}.}
            \label{fig:GGM_n_block(aupr)}
        \end{subfigure}
        \vspace{0.5cm}
        \begin{subfigure}{0.45\textwidth}
            \centering
            \includegraphics[width=\linewidth]{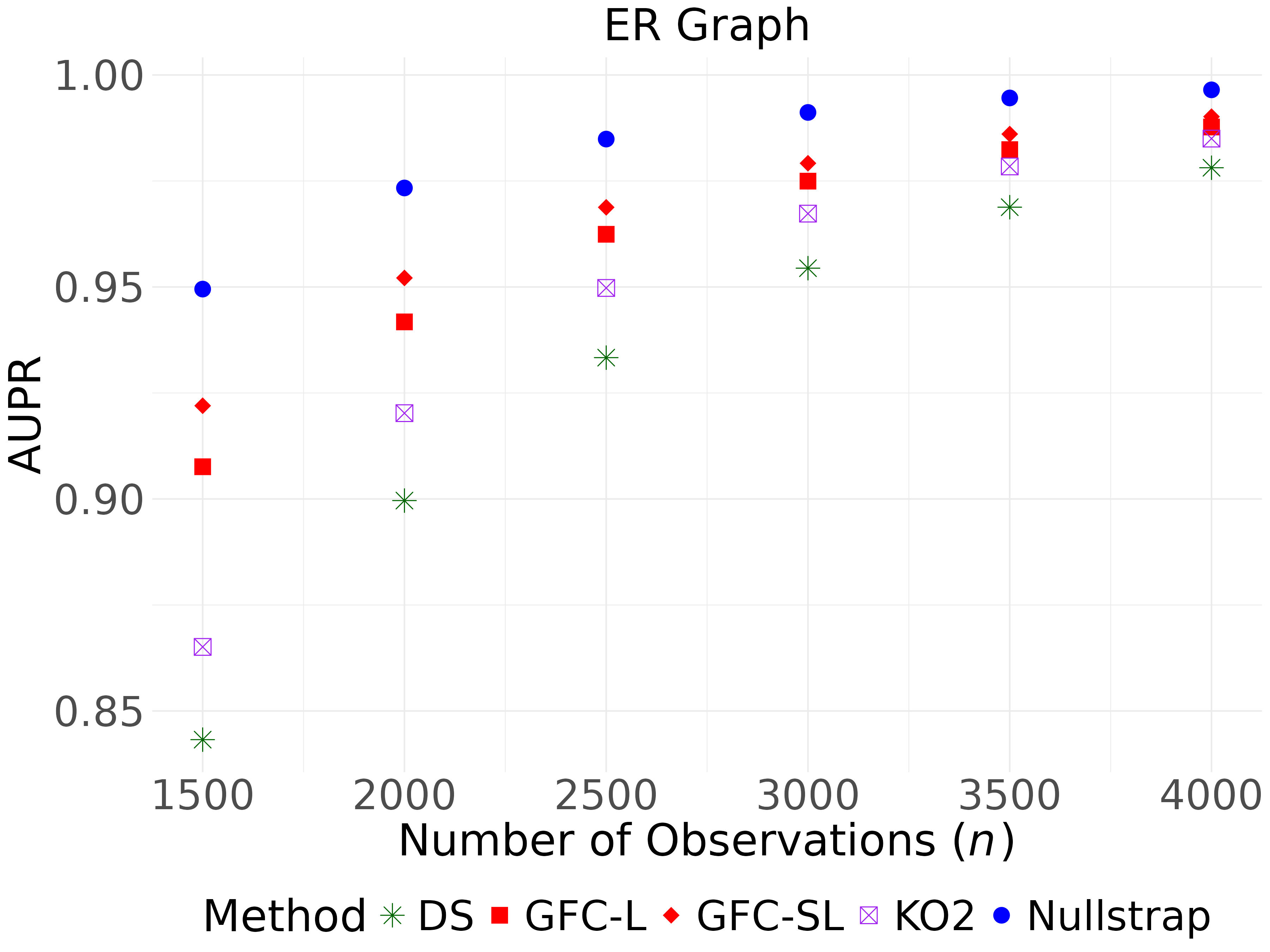}
            \caption{Empirical AUPR vs. number of observations ($n$) with a Erd\H{o}s-R\'enyi graph under Simulation Setting~\ref{setting:ggm}.}
            \label{fig:GGM_n_er(aupr)}
        \end{subfigure}
        \hfill
        \begin{subfigure}{0.45\textwidth}
            \centering
            \includegraphics[width=\linewidth]{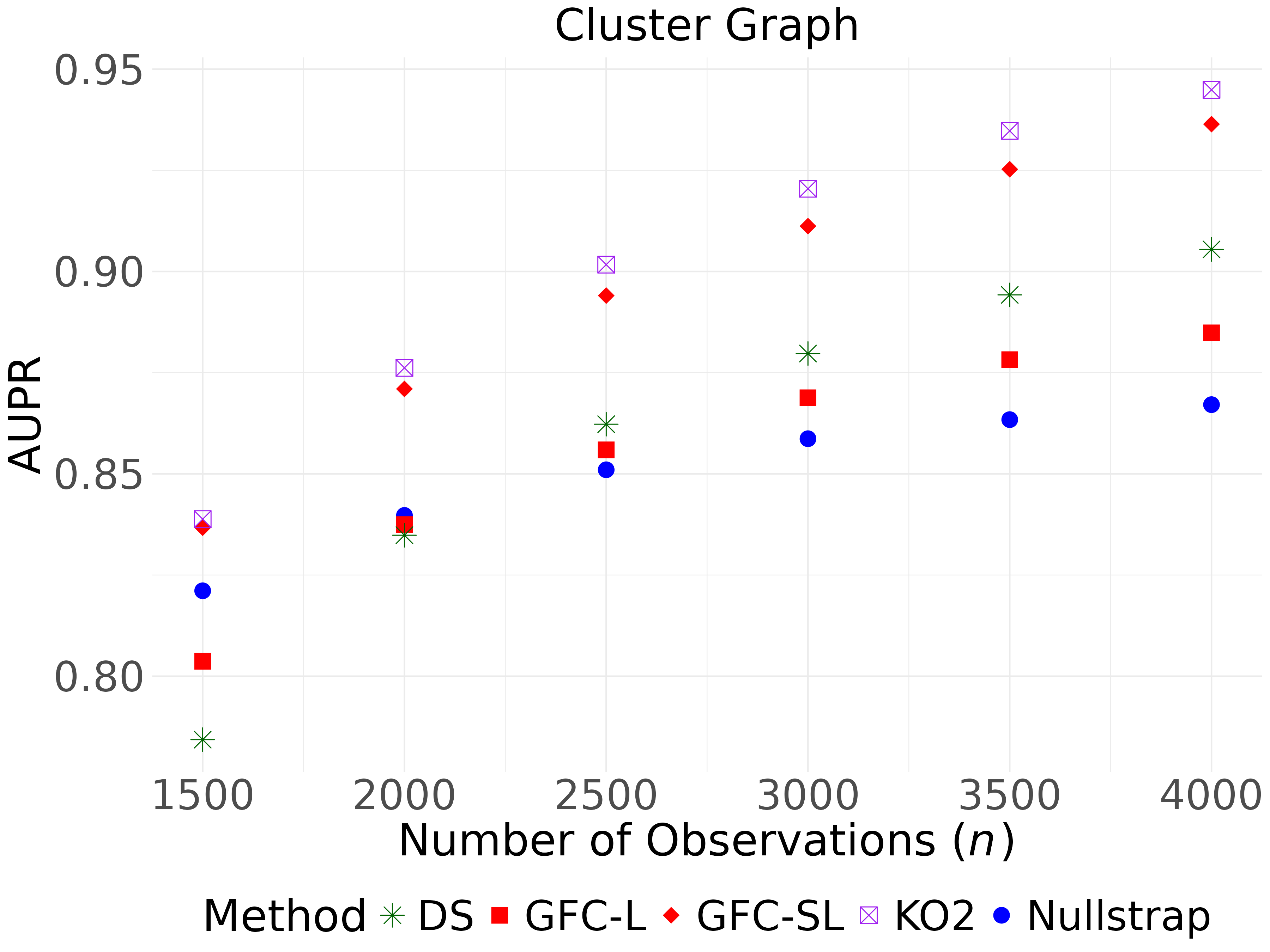}
            \caption{Empirical AUPR vs. number of observations ($n$) with a cluster graph under Simulation Setting~\ref{setting:ggm}.}
            \label{fig:GGM_n_cluster(aupr)}
        \end{subfigure}
        \caption{Empirical AUPR for the GGM.}
        \label{fig:AUPR_ggm}
    \end{figure}

% \begin{figure}[h]
% \centering
% \includegraphics[scale=0.4]{simulation/result_n_band_aupr_ggm.png}
% \caption{Empirical AUPR in band graph for the Gaussian graphical model (Number of Observations).}
% \centering
% \label{fig:GGM_n_band(aupr)}
% \end{figure}

% \begin{figure}[h]
% \centering
% \includegraphics[scale=0.4]{simulation/result_n_block_aupr_ggm.png}
% \caption{Empirical AUPR in block graph for the Gaussian graphical model (Number of Observations).}
% \centering
% \label{fig:GGM_n_block(aupr)}
% \end{figure}

% \begin{figure}[h]
% \centering
% \includegraphics[scale=0.4]{simulation/result_n_er_aupr_ggm.png}
% \caption{Empirical AUPR in Erd\H{o}s-R\'enyi graph for the Gaussian graphical model (Number of Observations).}
% \centering
% \label{fig:GGM_n_er(aupr)}
% \end{figure}

% \begin{figure}[h]
% \centering
% \includegraphics[scale=0.4]{simulation/result_n_cluster_aupr_ggm.png}
% \caption{Empirical AUPR in cluster graph for the Gaussian graphical model (Number of Observations).}
% \centering
% \label{fig:GGM_n_cluster(aupr)}
% \end{figure}

The empirical FDR and power results are shown in Figures~\ref{fig:GGM_n_band}--\ref{fig:GGM_fdr_cluster}, and the AUPR results are presented in Figure~\ref{fig:AUPR_ggm}. All methods except DS maintain FDR control across settings. Notably, Nullstrap demonstrates the most reliable FDR control across all graph types and scenarios, with particularly strong performance in the block and cluster graph structures. In contrast, DS struggles to control FDR, especially at lower target FDR levels.

In terms of AUPR, all five methods perform well overall. Nullstrap achieves the highest AUPR in all graph structures except the cluster graph, with especially strong results in the band and block graphs. A similar pattern is observed for power: Nullstrap consistently outperforms other methods in all settings except the cluster graph. The slightly reduced performance in the cluster graph is likely due to the advantage of nodewise regression over the graphical LASSO for that structure.
%appropriate than graphical LASSO for this structure.
% Power results increase with the parameter values, and our proposed method is particularly excelled at large sample size $n$. In the case of $n = 3500$, the block and Erd\H{o}s-R\'enyi: graphs consistently achieve a power value of $1$. 

\begin{table}[htbp]
\centering
\caption{Comparison of runtimes (in seconds) for the GGM across four graph structures under Simulation Setting~\ref{setting:ggm}, using the default parameter configuration in~\eqref{eq:para_ggm}.}
\begin{tabular}{cccccc}
    \toprule
    Nullstrap & GKF-Re+ & GFC-L & GFC-SL & KO2 & DS \\
    \midrule
     16.32 & 5811.43 & 70.99 &  5.96 & 6.63 & 214.66 \\
    \bottomrule
\end{tabular}
\label{ggm_time}
\end{table}

% We included the method that is competitive in terms of FDR results, the GGM knockoff filter with sample-splitting-recycling (GKF-Re+), which shows competitive performance in terms of FDR control \citep{li2021ggm}. 
Table~\ref{ggm_time} summarizes the total runtimes of each method across four graph structures. While Nullstrap is not the fastest under the specific setting \( n = 3500 \) and \( q = 0.2 \)—with GFC-SL achieving the shortest runtime of \( 5.96 \, \text{s} \)—it still runs efficiently at \( 16.32 \, \text{s} \) and delivers the best statistical performance in most scenarios.

In comparison, DS performs substantially slower for GGMs than for linear models, requiring \( 214.66 \, \text{s} \), which is approximately 13 times slower than Nullstrap. MDS is even slower due to its repeated application of DS. The GKF-Re+ method is the most computationally intensive, with a runtime of \( 5811.43 \, \text{s} \) under the default setting~\eqref{eq:para_ggm}, making it impractical for real-world use.

Overall, Nullstrap demonstrates consistently fast and stable performance across graph structures, underscoring its versatility and suitability for high-dimensional graphical modeling.

\clearpage

% for negative control $\tilde{\bbeta}$ construction, this paper primarily relies on synthetic null data. However, null data and true data are inherently different, and the deeper theoretical relationship between the zero-coefficient estimates derived from them warrants further investigation. To address this discrepancy, we introduce the estimation error correction factor as a potential solution.  
% Alternatively, synthetic data can be generated under the alternative hypothesis, i.e., the true model \( \by \sim F(\cdot \mid \bX; \bbeta, \bnu) \). This approach could leverage a sparse estimator of \( \bbeta \) to generate synthetic alternative data, offering a complementary perspective on bias estimation. %A notable advantage of Nullstrap is that it does not require a sparse estimator of \( \bbeta^* \), making it more broadly applicable.  
% Lastly, this paper estimates the estimation error correction factor using \( \left\|\hat{\bbeta} - \bbeta \right\|_{\infty} \), which requires theoretical derivation and a hyperparameter \( \kappa \). Different model structures, however, may yield varying estimates of \( \left\|\hat{\bbeta} - \bbeta \right\|_{\infty} \). A promising future direction would be the development of data-driven, simulation-based methods for estimation error correction factor estimation. Such methods could eliminate the need for theoretical derivation and manual hyperparameter selection, enhancing the flexibility and robustness of Nullstrap.
\bibliographystyle{chicago}
\setlength{\bibsep}{0pt}

\bibliography{Parallel.bib}

\begin{thebibliography}{}

\bibitem[\protect\citeauthoryear{Barber and Cand{\`e}s}{Barber and
  Cand{\`e}s}{2015}]{barber2015controlling}
Barber, R.~F. and E.~J. Cand{\`e}s (2015).
\newblock Controlling the false discovery rate via knockoffs.
\newblock {\em The Annals of Statistics\/}, 2055--2085.

\bibitem[\protect\citeauthoryear{Barber, Cand{\`e}s, and Samworth}{Barber
  et~al.}{2020}]{barber2020robust}
Barber, R.~F., E.~J. Cand{\`e}s, and R.~J. Samworth (2020).
\newblock Robust inference with knockoffs.
\newblock {\em The Annals of Statistics\/}~{\em 48\/}(3), 1409--1431.

\bibitem[\protect\citeauthoryear{Bates, Cand{\`e}s, Janson, and Wang}{Bates
  et~al.}{2021}]{bates2021metropolized}
Bates, S., E.~Cand{\`e}s, L.~Janson, and W.~Wang (2021).
\newblock Metropolized knockoff sampling.
\newblock {\em Journal of the American Statistical Association\/}~{\em
  116\/}(535), 1413--1427.

\bibitem[\protect\citeauthoryear{Benjamini and Hochberg}{Benjamini and
  Hochberg}{1995}]{benjamini1995controlling}
Benjamini, Y. and Y.~Hochberg (1995).
\newblock Controlling the false discovery rate: a practical and powerful
  approach to multiple testing.
\newblock {\em Journal of the Royal Statistical Society Series B: Statistical
  Methodology\/}~{\em 57\/}(1), 289--300.

\bibitem[\protect\citeauthoryear{Benjamini, Krieger, and Yekutieli}{Benjamini
  et~al.}{2006}]{benjamini2006adaptive}
Benjamini, Y., A.~M. Krieger, and D.~Yekutieli (2006).
\newblock Adaptive linear step-up procedures that control the false discovery
  rate.
\newblock {\em Biometrika\/}~{\em 93\/}(3), 491--507.

\bibitem[\protect\citeauthoryear{Benjamini and Yekutieli}{Benjamini and
  Yekutieli}{2001}]{benjamini2001control}
Benjamini, Y. and D.~Yekutieli (2001).
\newblock The control of the false discovery rate in multiple testing under
  dependency.
\newblock {\em Annals of Statistics\/}, 1165--1188.

\bibitem[\protect\citeauthoryear{Bogdan, Van Den~Berg, Sabatti, Su, and
  Cand{\`e}s}{Bogdan et~al.}{2015}]{bogdan2015slope}
Bogdan, M., E.~Van Den~Berg, C.~Sabatti, W.~Su, and E.~J. Cand{\`e}s (2015).
\newblock Slope—adaptive variable selection via convex optimization.
\newblock {\em The Annals of Applied Statistics\/}~{\em 9\/}(3), 1103.

\bibitem[\protect\citeauthoryear{Breslow}{Breslow}{1974}]{breslow1974covariance}
Breslow, N. (1974).
\newblock Covariance analysis of censored survival data.
\newblock {\em Biometrics\/}, 89--99.

\bibitem[\protect\citeauthoryear{Candes, Fan, Janson, and Lv}{Candes
  et~al.}{2018}]{candes2018panning}
Candes, E., Y.~Fan, L.~Janson, and J.~Lv (2018).
\newblock Panning for gold:‘model-{X}’knockoffs for high dimensional
  controlled variable selection.
\newblock {\em Journal of the Royal Statistical Society Series B: Statistical
  Methodology\/}~{\em 80\/}(3), 551--577.

\bibitem[\protect\citeauthoryear{Dai, Lin, Xing, and Liu}{Dai
  et~al.}{2023a}]{dai2023false}
Dai, C., B.~Lin, X.~Xing, and J.~S. Liu (2023a).
\newblock False discovery rate control via data splitting.
\newblock {\em Journal of the American Statistical Association\/}~{\em
  118\/}(544), 2503--2520.

\bibitem[\protect\citeauthoryear{Dai, Lin, Xing, and Liu}{Dai
  et~al.}{2023b}]{dai2023scale}
Dai, C., B.~Lin, X.~Xing, and J.~S. Liu (2023b).
\newblock A scale-free approach for false discovery rate control in generalized
  linear models.
\newblock {\em Journal of the American Statistical Association\/}~{\em
  118\/}(543), 1551--1565.

\bibitem[\protect\citeauthoryear{Edelstam, Karlsson, Westgren, L{\"o}wbeer, and
  Swahn}{Edelstam et~al.}{2007}]{edelstam2007human}
Edelstam, G., C.~Karlsson, M.~Westgren, C.~L{\"o}wbeer, and M.-L. Swahn (2007).
\newblock Human chorionic gonadatropin ({hCG}) during third trimester
  pregnancy.
\newblock {\em Scandinavian Journal of Clinical and Laboratory
  Investigation\/}~{\em 67\/}(5), 519--525.

\bibitem[\protect\citeauthoryear{Fan and Li}{Fan and
  Li}{2001}]{fan2001variable}
Fan, J. and R.~Li (2001).
\newblock Variable selection via nonconcave penalized likelihood and its oracle
  properties.
\newblock {\em Journal of the American statistical Association\/}~{\em
  96\/}(456), 1348--1360.

\bibitem[\protect\citeauthoryear{Fan, Gao, and Lv}{Fan
  et~al.}{2023}]{fan2023ark}
Fan, Y., L.~Gao, and J.~Lv (2023).
\newblock Ark: Robust knockoffs inference with coupling.
\newblock {\em arXiv preprint arXiv:2307.04400\/}.

\bibitem[\protect\citeauthoryear{Friedman, Hastie, and Tibshirani}{Friedman
  et~al.}{2008}]{friedman2008sparse}
Friedman, J., T.~Hastie, and R.~Tibshirani (2008).
\newblock Sparse inverse covariance estimation with the graphical lasso.
\newblock {\em Biostatistics\/}~{\em 9\/}(3), 432--441.

\bibitem[\protect\citeauthoryear{Ge, Chen, Song, McDermott, Woyshner,
  Manousopoulou, Wang, Li, Wang, and Li}{Ge et~al.}{2021}]{ge2021clipper}
Ge, X., Y.~E. Chen, D.~Song, M.~McDermott, K.~Woyshner, A.~Manousopoulou,
  N.~Wang, W.~Li, L.~D. Wang, and J.~J. Li (2021).
\newblock Clipper: p-value-free {FDR} control on high-throughput data from two
  conditions.
\newblock {\em Genome Biology\/}~{\em 22}, 1--29.

\bibitem[\protect\citeauthoryear{Ge, Zhang, and Zhang}{Ge
  et~al.}{2024}]{ge2024false}
Ge, Y., S.~Zhang, and X.~Zhang (2024).
\newblock False discovery rate control for high-dimensional cox model with
  uneven data splitting.
\newblock {\em Journal of Statistical Computation and Simulation\/}~{\em
  94\/}(7), 1462--1493.

\bibitem[\protect\citeauthoryear{H{\'e}dou, Mari{\'c}, Bellan, Einhaus,
  Gaudilli{\`e}re, Ladant, Verdonk, Stelzer, Feyaerts, Tsai, et~al.}{H{\'e}dou
  et~al.}{2024}]{hedou2024discovery}
H{\'e}dou, J., I.~Mari{\'c}, G.~Bellan, J.~Einhaus, D.~K. Gaudilli{\`e}re,
  F.-X. Ladant, F.~Verdonk, I.~A. Stelzer, D.~Feyaerts, A.~S. Tsai, et~al.
  (2024).
\newblock Discovery of sparse, reliable omic biomarkers with stabl.
\newblock {\em Nature Biotechnology\/}, 1--13.

\bibitem[\protect\citeauthoryear{Huang, Sun, Ying, Yu, and Zhang}{Huang
  et~al.}{2013}]{Huang2013ORACLEIF}
Huang, J., T.~Sun, Z.~Ying, Y.~Yu, and C.-H. Zhang (2013).
\newblock Oracle inequalities for the lasso in the cox model.
\newblock {\em Annals of Statistics\/}~{\em 41 3}, 1142--1165.

\bibitem[\protect\citeauthoryear{Javanmard and Javadi}{Javanmard and
  Javadi}{2019}]{javanmard2019false}
Javanmard, A. and H.~Javadi (2019).
\newblock False discovery rate control via debiased lasso.
\newblock {\em Electronic Journal of Statistics\/}~{\em 13}, 1212--1253.

\bibitem[\protect\citeauthoryear{Jordon, Yoon, and van~der Schaar}{Jordon
  et~al.}{2018}]{jordon2018knockoffgan}
Jordon, J., J.~Yoon, and M.~van~der Schaar (2018).
\newblock {KnockoffGAN}: Generating knockoffs for feature selection using
  generative adversarial networks.
\newblock In {\em International Conference on Learning Representations}.

\bibitem[\protect\citeauthoryear{Li, Yu, and Zhao}{Li
  et~al.}{2023}]{li2023coxknockoff}
Li, D., J.~Yu, and H.~Zhao (2023).
\newblock Coxknockoff: Controlled feature selection for the {Cox} model using
  knockoffs.
\newblock {\em Stat\/}~{\em 12\/}(1), e607.

\bibitem[\protect\citeauthoryear{Li and Maathuis}{Li and
  Maathuis}{2021}]{li2021ggm}
Li, J. and M.~H. Maathuis (2021).
\newblock {GGM knockoff filter: False discovery rate control for Gaussian
  graphical models}.
\newblock {\em Journal of the Royal Statistical Society Series B: Statistical
  Methodology\/}~{\em 83\/}(3), 534--558.

\bibitem[\protect\citeauthoryear{Liu}{Liu}{2013}]{698516a0-42f0-3d43-8108-2614216a2921}
Liu, W. (2013).
\newblock Gaussian graphical model estimation with false discovery rate
  control.
\newblock {\em The Annals of Statistics\/}~{\em 41\/}(6), 2948--2978.

\bibitem[\protect\citeauthoryear{Lounici}{Lounici}{2008}]{lounici2008sup}
Lounici, K. (2008).
\newblock Sup-norm convergence rate and sign concentration property of lasso
  and dantzig estimators.
\newblock {\em Electronic Journal of Statistics\/}~{\em 2}, 90--102.

\bibitem[\protect\citeauthoryear{Ma, Tony~Cai, and Li}{Ma
  et~al.}{2021}]{ma2021global}
Ma, R., T.~Tony~Cai, and H.~Li (2021).
\newblock Global and simultaneous hypothesis testing for high-dimensional
  logistic regression models.
\newblock {\em Journal of the American Statistical Association\/}~{\em
  116\/}(534), 984--998.

\bibitem[\protect\citeauthoryear{Meinshausen and B{\"u}hlmann}{Meinshausen and
  B{\"u}hlmann}{2006}]{meinshausen2006high}
Meinshausen, N. and P.~B{\"u}hlmann (2006).
\newblock High-dimensional graphs and variable selection with the lasso.
\newblock {\em The Annals of Statistics\/}~{\em 34\/}(3), 1436--1462.

\bibitem[\protect\citeauthoryear{Meinshausen and B{\"u}hlmann}{Meinshausen and
  B{\"u}hlmann}{2010}]{meinshausen2010stability}
Meinshausen, N. and P.~B{\"u}hlmann (2010).
\newblock Stability selection.
\newblock {\em Journal of the Royal Statistical Society Series B: Statistical
  Methodology\/}~{\em 72\/}(4), 417--473.

\bibitem[\protect\citeauthoryear{Petraglia, De~Vita, Gallinelli, Aguzzoli,
  Genazzani, Romero, and Woodruff}{Petraglia
  et~al.}{1995}]{petraglia1995abnormal}
Petraglia, F., D.~De~Vita, A.~Gallinelli, L.~Aguzzoli, A.~R. Genazzani,
  R.~Romero, and T.~K. Woodruff (1995).
\newblock {Abnormal concentration of maternal serum activin-A in gestational
  diseases}.
\newblock {\em The Journal of Clinical Endocrinology {\&} Metabolism\/}~{\em
  80\/}(2), 558--561.

\bibitem[\protect\citeauthoryear{Ravikumar, Wainwright, Raskutti, and
  Yu}{Ravikumar et~al.}{2008}]{Ravikumar2008HighdimensionalCE}
Ravikumar, P., M.~J. Wainwright, G.~Raskutti, and B.~Yu (2008).
\newblock High-dimensional covariance estimation by minimizing
  {$\ell_1$}-penalized log-determinant divergence.
\newblock {\em Electronic Journal of Statistics\/}~{\em 5}, 935--980.

\bibitem[\protect\citeauthoryear{Reid, Tibshirani, and Friedman}{Reid
  et~al.}{2016}]{reid2016study}
Reid, S., R.~Tibshirani, and J.~Friedman (2016).
\newblock A study of error variance estimation in lasso regression.
\newblock {\em Statistica Sinica\/}, 35--67.

\bibitem[\protect\citeauthoryear{Ren and Barber}{Ren and
  Barber}{2024}]{ren2024derandomised}
Ren, Z. and R.~F. Barber (2024).
\newblock Derandomised knockoffs: leveraging e-values for false discovery rate
  control.
\newblock {\em Journal of the Royal Statistical Society Series B: Statistical
  Methodology\/}~{\em 86\/}(1), 122--154.

\bibitem[\protect\citeauthoryear{Ren, Wei, and Cand{\`e}s}{Ren
  et~al.}{2023}]{ren2023derandomizing}
Ren, Z., Y.~Wei, and E.~Cand{\`e}s (2023).
\newblock Derandomizing knockoffs.
\newblock {\em Journal of the American Statistical Association\/}~{\em
  118\/}(542), 948--958.

\bibitem[\protect\citeauthoryear{Romano, Sesia, and Cand{\`e}s}{Romano
  et~al.}{2020}]{romano2020deep}
Romano, Y., M.~Sesia, and E.~Cand{\`e}s (2020).
\newblock Deep knockoffs.
\newblock {\em Journal of the American Statistical Association\/}~{\em
  115\/}(532), 1861--1872.

\bibitem[\protect\citeauthoryear{Spector and Janson}{Spector and
  Janson}{2022}]{spector2022powerful}
Spector, A. and L.~Janson (2022).
\newblock Powerful knockoffs via minimizing reconstructability.
\newblock {\em The Annals of Statistics\/}~{\em 50\/}(1), 252--276.

\bibitem[\protect\citeauthoryear{Stelzer, Ghaemi, Han, Ando, H{\'e}dou,
  Feyaerts, Peterson, Rumer, Tsai, Ganio, et~al.}{Stelzer
  et~al.}{2021}]{stelzer2021integrated}
Stelzer, I.~A., M.~S. Ghaemi, X.~Han, K.~Ando, J.~J. H{\'e}dou, D.~Feyaerts,
  L.~S. Peterson, K.~K. Rumer, E.~S. Tsai, E.~A. Ganio, et~al. (2021).
\newblock Integrated trajectories of the maternal metabolome, proteome, and
  immunome predict labor onset.
\newblock {\em Science Translational Medicine\/}~{\em 13\/}(592), eabd9898.

\bibitem[\protect\citeauthoryear{Sur and Cand{\`e}s}{Sur and
  Cand{\`e}s}{2019}]{sur2019modern}
Sur, P. and E.~J. Cand{\`e}s (2019).
\newblock A modern maximum-likelihood theory for high-dimensional logistic
  regression.
\newblock {\em Proceedings of the National Academy of Sciences\/}~{\em
  116\/}(29), 14516--14525.

\bibitem[\protect\citeauthoryear{Tibshirani}{Tibshirani}{1996}]{tibshirani1996regression}
Tibshirani, R. (1996).
\newblock Regression shrinkage and selection via the lasso.
\newblock {\em Journal of the Royal Statistical Society Series B: Statistical
  Methodology\/}~{\em 58\/}(1), 267--288.

\bibitem[\protect\citeauthoryear{van~de Geer}{van~de
  Geer}{2008}]{Geer2008HIGHDIMENSIONALGL}
van~de Geer, S.~A. (2008).
\newblock High-dimensional generalized linear models and the lasso.
\newblock {\em Annals of Statistics\/}~{\em 36}, 614--645.

\bibitem[\protect\citeauthoryear{Xing, Zhao, and Liu}{Xing
  et~al.}{2023}]{xing2023controlling}
Xing, X., Z.~Zhao, and J.~S. Liu (2023).
\newblock Controlling false discovery rate using {Gaussian} mirrors.
\newblock {\em Journal of the American Statistical Association\/}~{\em
  118\/}(541), 222--241.

\bibitem[\protect\citeauthoryear{Yu, Kaufmann, and Lederer}{Yu
  et~al.}{2021}]{yu2021false}
Yu, L., T.~Kaufmann, and J.~Lederer (2021).
\newblock False discovery rates in biological networks.
\newblock In {\em International Conference on Artificial Intelligence and
  Statistics}, pp.\  163--171. PMLR.

\bibitem[\protect\citeauthoryear{Zhang, Ren, and Chen}{Zhang
  et~al.}{2018}]{zhang2018silggm}
Zhang, R., Z.~Ren, and W.~Chen (2018).
\newblock Silggm: An extensive r package for efficient statistical inference in
  large-scale gene networks.
\newblock {\em PLoS computational biology\/}~{\em 14\/}(8), e1006369.

\bibitem[\protect\citeauthoryear{Zhang and Zou}{Zhang and
  Zou}{2014}]{zhang2014sparse}
Zhang, T. and H.~Zou (2014).
\newblock Sparse precision matrix estimation via lasso penalized d-trace loss.
\newblock {\em Biometrika\/}~{\em 101\/}(1), 103--120.

\bibitem[\protect\citeauthoryear{Zou and Hastie}{Zou and
  Hastie}{2005}]{zou2005regularization}
Zou, H. and T.~Hastie (2005).
\newblock Regularization and variable selection via the elastic net.
\newblock {\em Journal of the Royal Statistical Society Series B: Statistical
  Methodology\/}~{\em 67\/}(2), 301--320.

\end{thebibliography}
\end{document}